\pdfoutput=1 
\documentclass[
  aps,%
  prx,
 twocolumn,%
 groupedaddress,% 
 superscriptaddress,%
  showpacs,%
 letterpaper,%
 amsfonts,%
 footinbib,%
  10pt,%
 floatfix,%
 nofootinbib
]{revtex4-2}
 
\usepackage{dcolumn} 
\usepackage{tikz}
\usetikzlibrary{positioning,arrows.meta}

\usepackage{hyperref}
\hypersetup{
    colorlinks=true,
    citecolor=blue,
    linkcolor=blue , 
    filecolor=cyan,      
    urlcolor=magenta,
    pdfcreator = {\LaTeX\ and \flqq hyperref\frqq},
}

\RequirePackage{graphicx,amsmath,amssymb,bm}
\usepackage{bbm}
\usepackage{float}
\usepackage{framed} 
\usepackage{amsthm}
\usepackage{caption}
\usepackage{subcaption}
\usepackage[normalem]{ulem}
\usepackage{comment}
\usepackage{slashed} 
\usepackage{upgreek}
\usepackage{xcolor}
\usepackage{enumitem}
\usepackage{tikz-3dplot}  
\usepackage{yhmath}
\usepackage{mathrsfs}

\graphicspath{ {./Figs/} }

\newcommand{\1}{\mathbbm{1}}

\def\be{\begin{equation}}
\def\ee{\end{equation}} 
\def\bsh{\begin{shaded}}
\def\esh{\end{shaded}} 
\def\bpm{\begin{pmatrix}}
\def\epm{\end{pmatrix}}

\begin{document} 

\title{Locality from Nonlocality: Symmetry Relocalization and Gauging of Lattice Dirac Fermions}

\author{Lei Su} 
\affiliation{Division of Condensed Matter Physics and Materials Science, Brookhaven National Laboratory, Upton, New York 11973, USA} 
 
\begin{abstract}
Spatially nonlocal symmetries need not imply nonlocal gauge dynamics. Nonlocal canonical transformations make certain exact nonlocal compact symmetries of free lattice Dirac fermions onsite while retaining sufficient Hamiltonian locality for gauging. Symmetry relocalization yields exact or emergent gauge locality in 1D, 2D, and 3D, subject to lattice structure, topology, and anomalies.

In 1D, a Hamiltonian-preserving transformation exchanges onsite vector and nonlocal spectral axial charges of a half-filled chain, enabling local gauging of each and exposing their mixed anomaly. Bosonization identifies momentum--winding T-duality, with a parity-projected operator flowing to a noninvertible topological defect. A nonperturbative norm-resolvent bound proves emergent dynamical gauge locality at finite chemical potential, while a doubled Chern-insulator parent reveals the duality's bulk origin and superconducting extensions reveal hidden compact symmetries.

In 2D, the square $\pi$-flux Majorana model shares the Kitaev honeycomb model's single Dirac fermion and odd Pfaffian parity but supports sharp and exponentially local compact charges, the latter with nontrivial Chern topology. Relocalization retains Schwartz locality for asymptotically local gauging, while AIII/AII parents clarify bulk charge origin and parity-anomaly inflow. Two flavors yield compact $U(2)$, a quaternionic invariant, and asymptotically local non-Abelian gauging.

In 3D, an exact mixed vector--axial flux response is derived, while a distinct model supports an exponentially local vector $U(1)$ charge with strong quaternionic topology. A nonlocal map makes the charge and Kramers symmetry onsite while preserving exponential Hamiltonian locality. For a compact $SU(2)$ Weyl doublet, a no-go theorem forbids translation-preserving Gaussian onsite relocalization retaining Weyl content and Hamiltonian locality.
\end{abstract}

\maketitle

\tableofcontents

\section{Introduction}
\label{sec:introduction}

\subsection{Exact nonlocal symmetries behind emergent $U(1)$}

A simple puzzle arises in the Kitaev--Majorana honeycomb lattice. The isotropic Hamiltonian $H_K$ describes one
real Majorana fermion per honeycomb vertex with nearest-neighbor hopping \cite{kitaev2006anyons}. Since the
microscopic degrees of freedom are Majorana rather than complex fermions, there is no conventional onsite
particle-number $U(1)$ symmetry; only fermion parity is manifest. Yet its gapless spectrum contains two Majorana
nodes related by the reality condition, which together form a single complex Dirac fermion in the infrared (IR).
The resulting Dirac theory possesses a continuous number symmetry, $\Psi\mapsto e^{i\theta}\Psi$. \textit{What,
then, is the microscopic origin of this continuum $U(1)_{\rm Dirac}$ symmetry?}

\begin{figure}[tb]
\centering
\includegraphics[width=\columnwidth]{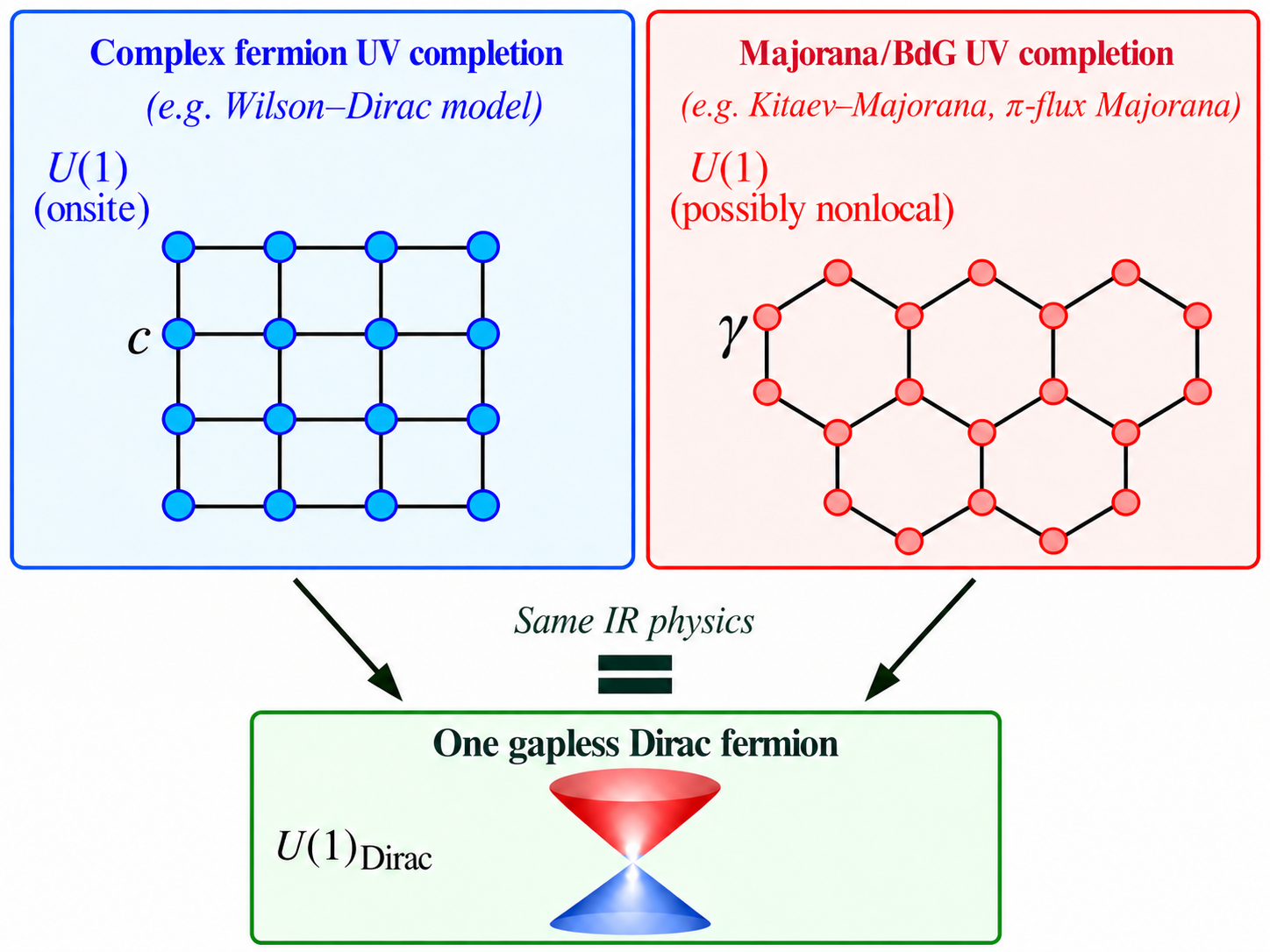}
\caption{Distinct UV completions of a single gapless Dirac fermion in 2D. A local complex-fermion lattice Hamiltonian can realize $U(1)_{\rm Dirac}$ as onsite particle number, while a local Majorana lattice Hamiltonian can realize the same IR symmetry through an exact non-onsite compact charge. Common IR matter and symmetry need not imply identical UV topology or gauge locality.}
\label{fig:uv_dirac_completions}
\end{figure}

One possibility is that the symmetry is merely accidental, emerging only at low energies without an exact
ultraviolet (UV) ancestor. Here, we explore another possibility: the lattice Hamiltonian may already possess an
exact compact global symmetry with the desired IR action, even though its generator is spatially nonlocal in the
original microscopic coordinate. \footnote{Throughout this work, the locality of a global symmetry refers to the
spatial range of its generator in a given lattice description. We call a symmetry \emph{local} if its generator
is a sum of finite-range operators, \emph{quasi-local} if it has rapidly decaying couplings, and \emph{onsite}
if its action factorizes over individual sites. These notions should not be confused with local gauge symmetry,
which involves position-dependent transformations and gauge fields.} This contrasts with the Wilson--Dirac
model, which realizes the same single-Dirac IR theory with an onsite particle-number $U(1)$ symmetry, as
illustrated in Fig.~\ref{fig:uv_dirac_completions}.

A related phenomenon appears in 1D. The half-filled nearest-neighbor complex-fermion chain flows to a massless
Dirac fermion with vector and axial $U(1)$ symmetries. While vector charge is onsite, axial charge is usually
identified only after linearizing around the two Fermi points. The mixed 't Hooft anomaly obstructs realizing
both $U(1)_V$ and $U(1)_A$ as exact compact onsite symmetries of a local lattice Hamiltonian describing a single
massless Dirac fermion in the IR. Nevertheless, an exact compact axial symmetry can exist with a spatially
nonlocal generator. Assigning quantized axial charges to the microscopic momentum modes gives
\begin{equation}
Q_{A,s}=\sum_k s(k)n_k,
\label{eq:intro_qa}
\end{equation}
where $s(k)=\pm1$ and $s(k_F+q)=+1$, $s(-k_F+q)=-1$ near the Fermi points $\pm k_F$. A simple choice is
$s_0(k)=\operatorname{sgn}(\sin k)$, with antiperiodic boundary conditions ensuring that the allowed momenta
avoid its discontinuities at $k=0,\pi$. Since $Q_{A,s}$ is diagonal in momentum occupation numbers, it commutes
exactly with the free Hamiltonian, and its integer spectrum guarantees $e^{2\pi iQ_{A,s}}=1$. Its IR action
reproduces the axial $U(1)_A$ of the Dirac theory. The price is spatial nonlocality: discontinuities in $s(k)$
produce algebraically decaying real-space kernels, with $1/r$ tails for $s_0(k)$.

A complementary 1D example, closer in spirit to the Kitaev--Majorana honeycomb model, is a gapless
superconducting Bogoliubov--de Gennes (BdG) chain. Pairing breaks microscopic particle-number conservation, and
the Hamiltonian can be expressed entirely in terms of Majorana fermions. Despite the absence of a manifest
onsite number $U(1)$ symmetry, suitable gapless BdG chains realize a single massless complex Dirac fermion in
the IR. This contrasts with the critical Kitaev--Majorana chain, whose low-energy theory contains only a single
real Majorana field and lacks a continuous number symmetry. As we show below, these superconducting Dirac chains
admit not only exact spatially nonlocal axial $U(1)$ symmetries but also hidden exact compact vector $U(1)$
symmetries with finite-range or exponentially local generators, whose IR action reproduces the continuum vector
$U(1)$ symmetry.

Taken together, these examples motivate a general strategy. Starting from an internal symmetry of a low-energy
fermionic theory, we seek an exact compact lattice symmetry whose action agrees with the continuum symmetry near
the gapless nodes but extends to the entire microscopic one-particle Hilbert space. For free fermions, which are
our primary focus, one way to achieve this is by assigning quantized charges to conserved spectral sectors, for
example, through suitable partitions of momentum space, subject to additional reality constraints in Majorana
and BdG systems. The resulting generators may be spatially nonlocal, revealing microscopic symmetry structures
that are not apparent from the onsite symmetries alone.

Why, then, are spatially nonlocal symmetries often regarded as undesirable? The difficulty arises when coupling
them to gauge fields. Although charge quantization ensures the $2\pi$ periodicity required for a compact $U(1)$,
it does not guarantee local matter--gauge interactions. In particular, sharp momentum-space partitions typically
produce spectral charge generators with algebraically decaying real-space kernels. Gauging such symmetries can
expose nonlocal UV effects that persist even when the ungauged Hamiltonian has a local continuum limit, as
encountered in SLAC-type lattice formulations
\cite{drellStrongcoupling1976,drellStrongcoupling1976a}.

A particular gauging of our 1D spectral axial charges in Eq.~\eqref{eq:intro_qa} provides a concrete example of
this SLAC-type difficulty. The function $s(k)$ distinguishes right- and left-moving modes near the two Fermi
points while extending their charge assignments throughout the Brillouin zone. If these two sets of modes are
treated as independently periodic momentum bands, their dispersions can develop artificial UV discontinuities. A
gauge field transferring momentum across such a discontinuity can then produce, even for arbitrarily small
momentum transfer $q$, a singular fermion--photon vertex $\Gamma_{\rm seam}^{1}(q)\sim\Delta\epsilon_{\rm
seam}/q$, where $\Delta\epsilon_{\rm seam}$ is the finite jump in the branch dispersion across the ultraviolet
seam. Although this singularity is specific to the chosen regulator, it highlights why an exact compact global
symmetry and a local IR Dirac theory do not, by themselves, guarantee a local gauge theory.

This motivates a central question of our work: \emph{Can a spatially nonlocal global symmetry be made local or
quasi-local through a generally nonlocal canonical transformation}? We call this procedure \emph{symmetry
relocalization}.\footnote{We use ``symmetry relocalization'' rather than ``symmetry localization,'' which has
established meanings in studies of symmetry-protected and symmetry-enriched topological phases
\cite{Else2014,Barkeshli2019}. There, symmetry localization concerns representing global symmetry actions by
operators supported near spatial boundaries or localized excitations, while keeping the underlying spatial
structure fixed. Here, we reorganize the matter degrees of freedom themselves through a generally nonlocal
canonical transformation.} Here, canonical means that the transformation preserves the fermionic anticommutation
relations.  Of particular interest is \emph{onsite symmetry relocalization}, in which the symmetry acts
independently at each site of a new, symmetry-adapted lattice. We focus on this case throughout and often omit
the qualifier ``onsite" for brevity.  The new fermions are generally nonlocal combinations of the original ones,
but no additional degrees of freedom are introduced. To retain the usual continuum interpretation of space, the
momentum reorganization must be smooth and reversible near each gapless node, preserving the long-wavelength
spatial structure. In our constructions, the same canonical transformation is nonlocal on microscopic lattice
fermions but acts locally on the continuum fields:
\begin{equation}
\mathcal W:\quad
\text{UV nonlocality}
\;\rightarrow\;
\text{IR locality}.
\end{equation}
Crucially, making the symmetry local need not preserve the locality of the Hamiltonian.
Establishing whether the resulting description supports well-behaved local gauge couplings is therefore an
additional challenge that we address in this work.

It is instructive to view symmetry relocalization in reverse. Consider a local or quasilocal Hamiltonian $H_0$
with an onsite symmetry generated by $Q_0$. A generally nonlocal canonical transformation $\mathcal W$ produces
\begin{equation}
H'=\mathcal W H_0\mathcal W^{-1},\quad
Q'=\mathcal W Q_0\mathcal W^{-1}.
\label{eq:intro_reverse_relocalization}
\end{equation}
Although $Q'$ may become non-onsite or genuinely spatially nonlocal, $H'$ can remain local or quasilocal, with
$Q'$ still an exact conserved charge. The reverse transformation then reveals the hidden onsite origin of this
symmetry. Particularly interesting is the case where $H'$ has a simple finite-range form while $H_0$ is
quasilocal. Starting from $H'$ and its nonlocal charge $Q'$, symmetry relocalization recovers a description with
an onsite symmetry and sufficiently local dynamics to support compact gauge couplings. Unlike
locality-preserving fermionic quantum cellular automata (QCAs), which map local operators to local operators
with uniformly bounded spreading \cite{gross2012index, farrelly2020review}, our canonical transformations need not preserve
locality in general. The examples developed in this work demonstrate how this greater freedom can reveal hidden
symmetry structures and alternative realizations of locality in lattice fermion systems.

A special case arises when a canonical transformation changes the spatial realization of a symmetry while
leaving the Hamiltonian invariant,
\begin{equation}
\mathcal W H_0\mathcal W^{-1}=H_0.
\end{equation}
We call this \emph{Hamiltonian-preserving relocalization}. Such transformations belong to the
\emph{canonical centralizer} of the Hamiltonian, the group of canonical transformations that leave it
invariant, and thus constitute possibly nonlocal symmetries of the Hamiltonian. At first sight, this
construction may seem unnecessary: if a nonlocal charge can be transformed into an onsite one without changing
the Hamiltonian, why not simply use the onsite representation? The significance becomes apparent when different
exact symmetries compete for locality. A Hamiltonian-preserving relocalization may render one charge onsite
while simultaneously making another, originally onsite charge nonlocal. As we show below, the 1D complex-fermion
chain introduced above exhibits precisely this competition: an exact nonlocal canonical transformation exchanges
the vector and axial charges while preserving the nearest-neighbor Hamiltonian.

The canonical centralizer also helps clarify the relation between microscopic and IR symmetry algebras. Distinct
exact conserved generators may act differently on UV modes while becoming indistinguishable at low energies. Let
$\mathfrak g_{\rm UV}$ denote the Lie algebra of exact conserved quadratic generators with well-defined IR
internal actions, and let $\rho_{\rm IR}$ be the corresponding Lie-algebra homomorphism obtained by restricting
to the gapless modes. Then
\begin{equation}
\mathfrak g_{\rm UV}/\ker\rho_{\rm IR}
\simeq \operatorname{im}\rho_{\rm IR}.
\label{eq:intro_IR_quotient}
\end{equation}  For any generator $B\in\mathfrak g_{\rm UV}$, a
Hamiltonian-preserving canonical transformation $\mathcal W$ that preserves the low-energy subspace and induces
an IR transformation $w_{\rm IR}$ satisfies
\begin{equation}
\rho_{\rm IR}(\mathcal W B\mathcal W^{-1})
=w_{\rm IR}\rho_{\rm IR}(B)w_{\rm IR}^{-1}.
\label{eq:intro_centralizer_conjugation}
\end{equation}
Thus, canonical transformations can reorganize microscopic symmetries while preserving, rotating, or exchanging
their IR actions. Conversely, two generators with identical IR actions need not be related by a
Hamiltonian-preserving canonical transformation. The canonical centralizer also includes lattice translations and point-group operations that preserve the Hamiltonian. Near a gapless node, a microscopic translation acts as a long-wavelength spatial translation accompanied by a node-dependent internal phase, known as an emanant symmetry \cite{cheng2023lieb}. In the half-filled 1D chain, the opposite phases associated with the two Fermi points induce an axial rotation of the continuum Dirac fermion. Here the compact spectral axial $U(1)$ provides an exact microscopic lift of this IR symmetry, although its action away from the Fermi points generally differs from that of lattice translation. The existence of such a compact lift is additional microscopic information not guaranteed by the quotient relation alone.

Onsite symmetry relocalization provides a natural starting point for compact gauging. In the symmetry-adapted coordinate, one can introduce independent sitewise gauge transformations and couple matter to compact lattice gauge fields. The resulting Hamiltonian can be made exactly gauge covariant, even with long-range hopping, but gauge covariance alone does not guarantee gauge locality. A particularly favorable case is \emph{local-to-local relocalization}, in which both Hamiltonians in Eq.~\eqref{eq:intro_reverse_relocalization} are local. Strictly finite-range Hamiltonians admit conventional local gauging, while sufficiently rapidly decaying hopping permits gauge-invariant finite-range approximations with controlled dynamical errors. Transforming the gauged Hamiltonian and Gauss laws back to the original fermions preserves gauge covariance, although the resulting couplings are generally nonlocal in the original lattice geometry. Crucially, our canonical transformations are regular near the physical gapless nodes and therefore act locally on the continuum fermion fields. Consequently, gauge couplings that are nonlocal in the original lattice variables can have local long-wavelength descriptions, with the dynamical gauge-locality estimates providing the necessary control beyond this IR field identification.

Beyond exact finite-spacing locality, we distinguish several notions of emergent gauge locality.
\emph{Emergent background gauge locality} requires that coupling to smooth, nondynamical gauge fields
approaches the expected local continuum coupling. The stronger notion of \emph{emergent dynamical gauge
locality} concerns fluctuating gauge fields, for which agreement restricted to low-energy fermionic modes is
insufficient. We use \emph{asymptotic gauge locality} to describe controlled comparisons, as the lattice spacing
$a\to0$, between a potentially nonlocal microscopic gauging and a family of local gauge regulators. The strength
of such comparisons depends on the required notion of convergence.

A strong criterion is \emph{operator-norm local approximability}. On a fixed physical volume, this requires a
family of finite-range gauge-invariant Hamiltonians $H_{a,L(a)}$ whose physical interaction ranges vanish as
$a\to0$ and which satisfy
\begin{equation}
\|H_a-H_{a,L(a)}\|\rightarrow0.
\label{eq:intro_operator_norm_locality}
\end{equation}
This controls the difference uniformly over all normalized states, including UV states. A weaker criterion is
\emph{norm-resolvent gauge locality}. Let $H_a^{\rm loc}$ be a local gauge regulator and
$c_a=\inf\operatorname{Spec}(H_a^{\rm loc})$ its ground-state energy. After subtracting this common reference
energy, we require
\begin{equation}
\bigl\|(H_a-c_a-z)^{-1}
-(H_a^{\rm loc}-c_a-z)^{-1}\bigr\|
\rightarrow0,
\label{eq:intro_norm_resolvent_locality}
\end{equation}
for fixed $z\in\mathbb C\setminus\mathbb R$. Unlike operator-norm convergence of the Hamiltonians,
norm-resolvent convergence allows substantial UV differences, as high-energy contributions are suppressed in the
resolvents. Both criteria provide controlled routes to emergent dynamical gauge locality without requiring
strict microscopic locality.  

Another important aspect of our work is the interplay between symmetry relocalization and 't Hooft anomalies
\cite{hooft1980naturalness}. A 't Hooft anomaly arises when coupling a global symmetry to background gauge
fields fails to preserve gauge invariance, signaling an obstruction to gauging the symmetry within an autonomous
local theory. Such anomalies constrain symmetry relocalization: an anomalous symmetry structure cannot be made
simultaneously onsite and locally gaugeable while preserving all the relevant symmetries. However, a mixed
anomaly between two symmetries does not necessarily obstruct local-to-local relocalization of either symmetry
individually; rather, it obstructs their simultaneous realization in a locally gaugeable form. The mixed
vector--axial $U(1)$ anomaly in 1D and the parity anomaly in 2D provide examples of these competing
requirements. As we show below, exact lattice symmetries and their responses to background fluxes allow us to
identify certain anomaly signatures and connect them to their IR counterparts.

A complementary perspective is to realize an anomalous autonomous theory as a boundary sector of a local
higher-dimensional parent. In this setting, anomaly inflow \cite{callanAnomalies1985} from a symmetry-protected topological (SPT) bulk
\cite{chen2013} compensates the boundary anomaly, allowing the combined bulk--boundary theory to maintain gauge
invariance. We construct examples in which an autonomous lattice theory admits an \emph{exact} correspondence with a boundary sector of a higher-dimensional topological parent, and a nonlocal symmetry of the autonomous theory corresponds to the boundary action of an onsite symmetry of the parent. This provides two complementary approaches to gauging: one through
symmetry relocalization within the autonomous theory, and the other through gauge coupling inherited from the
higher-dimensional parent. Importantly, integrating out the bulk can yield a gauge-invariant effective boundary
description, but this need not be equivalent to an autonomous local boundary gauge theory. Comparing these
constructions clarifies the distinction between microscopic gauge responses, their dependence on the chosen
realization, and the universal IR anomaly-inflow structure.

A central insight of this work is that a compact symmetry can be spatially nonlocal in the UV yet admit local gauge dynamics in the IR. A generally nonlocal canonical transformation can make such an exact symmetry onsite while retaining sufficient Hamiltonian locality for compact gauging. In our constructions, these transformations are regular near the physical gapless nodes and therefore act locally on the continuum fermion fields, even when the resulting gauge couplings are nonlocal in the original lattice variables. Gauge locality must nevertheless be established dynamically and remains subject to anomaly constraints. It can be realized exactly at finite lattice spacing or through controlled asymptotically local gauge regulators, while higher-dimensional bulk completions clarify the origin of boundary symmetries and anomaly inflow. This emergence of local symmetry and gauge dynamics from microscopic nonlocality motivates the title \emph{Locality from Nonlocality}.

\subsection{Lattice fermions and recent developments}

The questions raised above concerning the lattice realization of compact symmetries are closely related to the
longstanding constraints on lattice chiral fermions and their symmetries. The Nielsen--Ninomiya theorem
\cite{nielsenAbsence1981,nielsenAbsence1981a,nielsennogo1981} establishes that, under the standard
assumptions of locality, translation invariance, and conventional chiral structure, lattice Weyl fermions must
occur with compensating doublers. Friedan's charge-resolved refinement \cite{friedanproof1982a} further shows
that, for sufficiently regular compact internal symmetries, the net chirality must vanish separately within each
symmetry sector, even when the associated charge bundle is topologically nontrivial.   Appendix~\ref{app:friedan_charge_doubling} reviews the geometric argument and its extension to Majorana fermions, emphasizing the role of nontrivial charge-bundle topology in our constructions.

Different lattice formulations address the constraints of fermion doubling by modifying or relaxing different
assumptions. Wilson fermions explicitly break chiral symmetry at finite lattice spacing
\cite{wilson1975quarks}; staggered fermions reduce the number of fermion species while preserving a remnant
chiral symmetry \cite{kogutHamiltonian1975a,susskindLattice1977}; domain-wall fermions introduce an additional
dimension, with chiral boundary modes closely related to anomaly inflow
\cite{kaplanmethod1992}; and SLAC-type formulations relax spatial
locality through nonlocal kinetic operators \cite{drellStrongcoupling1976,drellStrongcoupling1976a}.

The Ginsparg--Wilson relation, realized explicitly by overlap fermions, provides another approach to lattice
chiral symmetry. Rather than imposing the conventional continuum chiral transformation at finite lattice
spacing, this framework admits an exact modified chiral symmetry adapted to the lattice Dirac operator
\cite{ginspargremnant1982,neubergerExactly1998,neubergerVectorlike1998,luscherExact1998}. The associated
index and fermion measure reproduce the continuum chiral anomaly. Lüscher established a nonperturbative
construction of anomaly-free Abelian chiral gauge theories with exact lattice gauge invariance
\cite{luscherAbelian1999a}, and subsequently extended the construction to arbitrary compact gauge groups,
including non-Abelian theories, to all orders in perturbation theory \cite{luscherLattice2000}. More recently,
Craig proposed a nonperturbative gauge-invariant fermion measure for a class of non-Abelian chiral gauge
theories, including the Standard Model, using overlap fermions \cite{craiggaugeinvariant2026}. Related proposals
by Hsu and Witten retain continuum Weyl fermions coupled to gauge fields interpolated from lattice variables,
offering alternative approaches to gauge-invariant formulations of anomaly-free chiral theories
\cite{hsuStandard2026,wittenNote2026}. For a review of lattice chiral fermions, see
Ref.~\cite{kaplanChiral2012}.

Recent Hamiltonian approaches have emphasized exact non-onsite symmetries of lattice fermions. Quantized vector,
axial, and valley charges can generate UV symmetry algebras substantially larger than the internal symmetry
algebras of their low-energy fields, including generalized Onsager structures
\cite{chatterjee2025quantized,pace2025parity,misumiTastesplitting2026, kourkoulouOnsager2026}. Ref.~\cite{gioia2025exact}
constructed models with a single Weyl fermion protected by an exact non-onsite, noncompact chiral symmetry, as
well as a Weyl doublet whose exact $U(1)$ symmetries generate an IR $SU(2)$ and an enlarged UV Onsager algebra.
These constructions are consistent with the locality-preserving no-go constraints of Fidkowski and Xu
\cite{fidkowskinogo2023}. They also reveal connections to continuum anomalies and Lieb--Schultz--Mattis (LSM)
constraints, although an enlarged non-Abelian UV symmetry algebra does not, by itself, imply an anomaly
\cite{lew-smithInfiniteOrder2026}.

Another direction employs enlarged microscopic local Hilbert spaces. Modified Villain formulations, rotor
variables, and bosonization provide exact lattice descriptions of compact bosons, chiral symmetries, dualities,
and anomalies
\cite{sulejmanpasicAbelian2019,fazzaLattice2023,gorantlaModified2021,fidkowskiNoninvertible2025a,luLattice2026}.
In 1D, recent constructions realize exact vector and axial symmetries together with their mixed anomaly,
including applications to anomaly-free chiral gauge theories such as the 3450 model
\cite{seifnashriExactly2026, thorngrenChiral2026a, luFermionic2026,dharanikota2026}. Related developments include emanant
symmetries connecting microscopic lattice transformations to IR internal symmetries and LSM constraints
\cite{cheng2023lieb}.

Our approach complements these developments by retaining finite-dimensional fermionic matter Hilbert spaces and
local or quasilocal free Hamiltonians while allowing exact compact symmetry generators to be spatially nonlocal.
Its distinguishing ingredient is a generally nonlocal canonical transformation that reorganizes the existing
fermionic degrees of freedom without introducing new ones. Unlike locality-preserving symmetry disentanglers,
such transformations need not preserve the locality of arbitrary operators. This broader framework allows us to
investigate hidden compact symmetries, their spectral and topological structures, and the conditions under which
their gauging admits an exact or emergent local realization.

\subsection{Main results and organization}

Our results are organized by dimension around three questions: which IR symmetries admit exact compact
microscopic charges; when nonlocal symmetries can be made onsite while retaining sufficient locality for
gauging; and how these constructions illuminate the roles of charge-bundle topology and anomalies.

\paragraph{1D: Vector--axial relocalization and the mixed anomaly.}
For the half-filled nearest-neighbor complex-fermion chain, we construct the exact compact spectral axial charge $Q_{A,0}$ of Eq.~\eqref{eq:intro_qa}. A Hamiltonian-preserving Gaussian canonical transformation $\mathcal W$ exchanges the centered vector and axial charges,
\begin{equation}
\mathcal W Q_V\mathcal W^{-1}=Q_{A,0}.
\label{eq:intro_1d_H_preserving_exchange}
\end{equation}
The vector charge is onsite in the original lattice coordinate, while the axial charge becomes onsite in the symmetry-adapted coordinate. The transformation preserves the Hamiltonian and its nearest-neighbor structure, allowing conventional local gauging of either compact $U(1)$ in its respective frame. Although nonlocal on the lattice, $\mathcal W$ acts pointwise on the continuum Dirac fields.

The construction extends to inversion-odd spectral signs $s(-k)=-s(k)$ and a family of finite-range
deformations. Each sign defines an exactly conserved compact charge $Q_{A,s}$ admitting a Hamiltonian-preserving
Gaussian onsite relocalization. Signs agreeing with $s_0(k)=\operatorname{sgn}(\sin k)$ near the Fermi points
have the same IR axial action but different UV embeddings. Their compact gauge theories are unitarily equivalent
when matter operators, gauge links, Gauss laws, and observables are transported together.

At fixed physical chemical potential, however, the axial frame has algebraically decaying hopping, and
shortest-path compact gauging remains nonlocal at every finite cutoff. We prove a nonperturbative norm-resolvent
comparison with a strictly local nearest-neighbor gauge regulator, with error $O(a[1+\log(\ell/a)])$ as $a\to0$
at fixed physical chemical potential, gauge coupling, Fermi velocity, and circumference $\ell$. The estimate
incorporates dynamical holonomy and Gauss-law constraints on the reduced physical Hilbert space without a
weak-coupling expansion. It establishes emergent dynamical gauge locality without requiring operator-norm
convergence of the Hamiltonians on UV states.

The compact charges also expose the mixed $U(1)_V$--$U(1)_A$ anomaly on the lattice. Their Fourier densities
reproduce the IR Schwinger term, while insertion of $2\pi$ vector or axial flux pumps two units of the
complementary charge. A smooth current representative recovers the continuum Ward identities, and the
translation-selected axial subgroup reproduces the LSM momentum pump. The mixed anomaly precludes simultaneously
anomaly-free onsite gauging of both compact symmetries in an autonomous local regulator for a single Dirac
fermion.

Under bosonization, $\mathcal W$ acts as a chiral reflection and realizes momentum--winding T-duality of the
$c=1$ compact boson at the XX radius $R=\sqrt2$. In the APBC fermion sectors, it exchanges the compact
magnetization and spectral winding charges, $Q_M=Q_V$ and $Q_W=Q_{A,0}/2$. On the periodic XX spin chain,
parity projection produces a noninvertible spectral operator $\mathcal D$ satisfying $\mathcal D^2=1+U$,
with $U=e^{i\pi Q_M}$. Its IR action exchanges momentum and winding as $(m,w)\mapsto(2w,m/2)$ on even-$m$
sectors and annihilates odd-$m$ states, realizing a noninvertible T-duality topological defect. Unlike the
lattice duality of Ref.~\cite{pace2024lattice}, our construction uses commuting compact charges and an
algebraically nonlocal spectral transformation. The same $\mathcal W$ exchanges Su--Schrieffer--Heeger (SSH)
dimerization and superconducting pairing, producing a self-dual Ising critical line. The spectral
transformation also acts independently on a single Majorana chain, whose parity projection realizes the
Kramers--Wannier topological defect in the IR.

An exact 2D bulk--edge construction provides a higher-dimensional realization of the vector--axial
duality. A selected edge sector of a doubled Chern-insulator SPT parent reproduces the entire nearest-neighbor chain and its compact spectral
axial charge at finite circumference. An onsite particle--hole transformation in one Chern sector induces the
nonlocal exchange on the unfolded edge, while $C_\pm=\pm1$ account for the mixed anomaly and flux index two.
Although the ungauged correspondence and spectral-flow indices agree exactly, bulk-induced and autonomous
edge gauging need not have identical gauge responses, since the parent permits charge transfer between edge
and bulk. The correspondence extends to uniform vector holonomy away from edge--bulk thresholds, but not to
arbitrary gauge backgrounds.

We also exhibit gapless superconducting BdG chains with exact compact axial $U(1)$ symmetries generated by spatially nonlocal operators and hidden compact vector $U(1)$ symmetries with exponentially local generators, and contrast them with a locality-preserving Majorana-shift construction. These realizations share the continuum Dirac symmetries and mixed anomaly but differ in their gapless class-D Pfaffian parities \cite{kitaev2001unpaired}.

\paragraph{2D: Compact charges, charge topology, and gauge locality.}
We first contrast lattice realizations of a single massless complex Dirac fermion. The Kitaev--Majorana
honeycomb Hamiltonian $H_K$ admits an exact compact but spatially nonlocal spectral Dirac-number charge, whose
natural onsite relocalization encounters UV seams and does not supply a local microscopic gauging. The
Wilson--Dirac Hamiltonian $H_W$, by contrast, has onsite particle number and conventional local gauging;
unfolding its Majorana components gives a finite-range honeycomb representative $H_{\rm hc}^{W}$ with the same
property. Although their IR Dirac theories agree, $H_K$ and $H_{\rm hc}^{W}$ have opposite class-D Pfaffian
parities \cite{kitaevPeriodic2009,schnyder2008classification,ryu2010tenfold,chiu2016classification}.

Our central 2D construction is the square-lattice $\pi$-flux Majorana Hamiltonian $H_{\rm sq}$. It shares the odd class-D Pfaffian parity of $H_K$ while supporting both a sharp spectral charge $Q_{s_P}$ and an exponentially local compact charge $Q_{\rm loc}$ without modifying the finite-range Hamiltonian. These charges have the same IR Dirac-number action but differ in their microscopic locality and crystalline properties. The smooth charge has positive-charge Chern number $C_Q=1$, and at the massless Dirac point we obtain
$\mathcal P_D(H_{\rm sq})=(-1)^{C_Q}=-1$. 
More generally, we prove that a single Majorana-related pair of Dirac nodes away from particle--hole-invariant momenta (PHIM) necessarily has odd class-D Pfaffian parity in the presence of a globally regular, orientation-reversing parity that fixes the nodes and preserves their IR Dirac-number assignments. This result requires no smooth microscopic compact charge and applies to both the square and Kitaev models, whereas the even-Pfaffian Wilson--Dirac completion does not satisfy the requisite parity conditions. For gapped charge-preserving descendants, the physical class-D Chern number $\nu_{\rm Maj}$ and the occupied positive-charge Chern number $C_{+,\mathrm{occ}}$ satisfy $\nu_{\rm Maj}=2C_{+,\mathrm{occ}}-C_Q$.

Both compact charges admit exact, generally nonlocal canonical relocalizations that make them onsite while
carrying $H_{\rm sq}$ to a globally smooth massless Bloch Hamiltonian with Schwartz-decaying hopping. These
transformations are regular near the physical Dirac nodes and therefore act locally on the continuum fermion
fields. Compact Wilson-line gauging admits finite-range gauge-invariant approximants with vanishing physical
interaction range and controlled operator-norm and dynamical errors, establishing asymptotic dynamical gauge locality. The nonzero
Chern number $C_Q$ rules out a strictly finite-range representative of this compact charge and
obstructs globally smooth periodic Gaussian onsite trivializations within the specified smooth same-torus
class. It also governs spectral flow of the boundary-compressed charge. Projective reflections, competing parity- and $U(1)$-preserving masses, and
boundary spectra distinguish this charge topology from Pfaffian parity and the parity anomaly.

Exact 3D bulk completions provide a bulk realization of the square-model symmetry. A finite-range
class-AIII parent $\widetilde{\mathcal H}_\chi$ has winding number $\nu_3=1$ and axion angle $\theta=\pi$, with a selected
surface sector reproducing $H_{\rm sq}$ at zero gauge background. Its sharp surface charge descends from onsite
bulk particle number, and the Hamiltonian-preserving rotation $W_P$ gives $Q_{\rm loc}$. Onsite bulk gauging
produces an exactly gauge-covariant bulk--boundary theory, but its magnetic-flux zero-mode response differs at
finite cutoff from that of an autonomously gauged surface regulator. A complementary parent $\mathcal H_*$,
possessing both AIII chiral and Kramers symmetries, connects AIII and AII descendants through a uniformly
bulk-gapped path that preserves the selected surface states and compact charge. Thus exact agreement of ungauged
surface theories can coexist with distinct bulk symmetry protections and microscopic gauge responses.

With two identical square-Majorana flavors, we construct exact compact $U(2)$ symmetries with sharp spectral and exponentially local generators. The smooth $SU(2)$ representation defines a quaternionic multiplicity bundle with invariant $\nu_{SU(2)}=C_Q\pmod2=1$, although the undeformed doubled Hamiltonian has even class-D Pfaffian parity in the primitive-cell convention. This invariant remains defined at the gapless point and survives an explicit $SU(2)$-preserving perturbation that breaks the displayed central $U(1)$. Its nontriviality obstructs globally smooth, periodic, Majorana-compatible onsite trivialization in the original doubled-cell Bloch geometry. Nevertheless, a patched nonlocal canonical relocalization makes the full $U(2)$ onsite while retaining Schwartz locality of the Hamiltonian. Compact non-Abelian gauging then admits controlled finite-range gauge-invariant approximants and asymptotic dynamical gauge locality. For fully gapped quadratic descendants preserving this specified $SU(2)$ bundle structure, the full Chern number satisfies $\nu_{\rm full}\in4\mathbb Z+2$. Explicit massive descendants also satisfy the magnetic-translation LSM constraint, while their mod-two index agrees with the IR parity anomaly of a single fundamental $SU(2)$ Dirac doublet.

\paragraph{3D: Compact symmetries, charge topology, and anomalies.}
For a two-band Weyl semimetal whose opposite-chirality nodes form one massless 3D Dirac fermion, we identify an
exact compact spectral axial charge with algebraic spatial tails alongside onsite vector charge. Unlike in 1D,
the axial $U(1)_A$ has a pure perturbative gauge anomaly as well as the mixed $AVV$ anomaly. Its onsite
relocalization cannot, within the constructions studied, preserve both Hamiltonian locality and the intended IR
axial action. Independently, finite-volume vector-flux insertion and momentum-slice Chern numbers yield the
exact response $\Delta Q_A=2n_\Phi$ for $n_\Phi$ magnetic-flux units. We therefore focus on the anomaly-free
vector $U(1)_V$ for compact local gauging.

From a class-AII Wilson--Dirac projector, we construct a distinct massless 3D Dirac model with an exponentially
local compact vector charge and strong quaternionic positive-charge index $\nu_0=1$, despite vanishing first
Chern numbers. The topology gives a boundary charge-spectrum Dirac cone, distinct from physical surface-energy
modes and mass-interface states of the associated class-DIII descendants. For the uniform massless Hamiltonian,
an exact nonlocal Gaussian relocalization makes both vector $U(1)_V$ and Kramers time reversal onsite while
preserving exponential Hamiltonian locality, allowing compact vector gauging with quasilocal couplings. The
transformation is smooth at the physical Dirac point, so its action on the continuum fermion field is local.
A corresponding locality result for arbitrary spatially varying mass interfaces remains open.

Finally, we construct an exponentially local exact compact $SU(2)$ symmetry whose IR content is one independent
fundamental Weyl doublet, with strong quaternionic multiplicity index $\nu_0=1$. The continuum doublet carries
the mod-two Witten anomaly. Consistently, a general no-go theorem excludes translation-preserving
Majorana-canonical Gaussian relocalizations that make the full $SU(2)$ onsite while retaining the original Weyl
content without additional gapless crossings and keeping the transformed Bloch Hamiltonian periodic and
continuously differentiable, even if the transformation itself is spatially nonlocal. This establishes a
Hamiltonian-locality obstruction within the stated class; it does not independently compute the gauge-Pfaffian
holonomy associated with the Witten anomaly.

\paragraph{Organization.}
The main text follows the same sequence. Sec.~\ref{sec:1d_relocalization} develops the 1D duality, with exact
and emergent compact gauging in Sec.~\ref{sec:1d_exact_axial_gauging}, anomaly diagnostics and LSM pumping in
Sec.~\ref{sec:1d_anomaly}, its bulk--edge origin in Sec.~\ref{sec:1d_bulk_edge_spt}, and superconducting
variants in Sec.~\ref{sec:1d_variants}.

Sec.~\ref{sec:2d_u1} develops the 2D constructions, beginning with the honeycomb and Wilson--Dirac comparison
and the square-Majorana model in Sec.~\ref{sec:2d_sq}. Charge-bundle topology and Pfaffian parity are discussed
in Sec.~\ref{sec:2d_sq_CQ}, followed by symmetry relocalization and gauge locality in
Sec.~\ref{sec:2d_sq_gauge_locality}, and crystalline symmetries, mass terms, and the parity anomaly in
Sec.~\ref{sec:2d_sq_gapping_channels}. Exact AIII/AII bulk completions appear in Sec.~\ref{sec:2d_sq_3d_parent},
and the compact $U(2)$ extension in Sec.~\ref{sec:2d_su2}.

Sec.~\ref{sec:3d_vectorlike} turns to 3D: axial spectral symmetry and flux response in Sec.~\ref{sec:3d_dirac},
quaternionic vector-charge topology and its onsite relocalization in Sec.~\ref{sec:3d_strong_charge}, and the
anomalous $SU(2)$ Weyl doublet and locality obstruction in Sec.~\ref{sec:3d_weyl_doublet}.
Sec.~\ref{sec:discussion} discusses the implications and open questions. The appendices supply a charge-resolved
doubling and symmetry-bundle review (Appendix~\ref{app:friedan_charge_doubling}), the 1D bosonization, gauging,
and anomaly details (Appendices~\ref{app:bosonization_tduality} and \ref{app:1d_gauge_locality}), the 2D
topology and gauge-locality proofs (Appendix~\ref{app:2d_gauge_locality}), the bulk--surface constructions
(Appendix~\ref{app:2d_surface_inflow}), and the 3D flux, charge-topology, and no-go derivations
(Appendix~\ref{app:3d_anomaly}).

\section{1D: Vector--axial relocalization and the mixed anomaly}
\label{sec:1d_relocalization}

\subsection{Exact compact axial symmetry and relocalizations}
\paragraph{Free complex-fermion Hamiltonian.}
Consider a nearest-neighbor hopping chain of $N=2M=4m\in4\mathbb Z$ sites, each hosting a spinless complex fermion,
\begin{equation}
\begin{aligned}
H_0&=-t\sum_j(c_j^\dagger c_{j+1}+\mathrm{H.c.})\\
&=\sum_k\epsilon(k)n_k, \quad
\epsilon(k)=-2t\cos k.
\end{aligned}
\label{eq:1d_H0}
\end{equation}
Here $t>0$, $c_k=N^{-1/2}\sum_j e^{-ikj}c_j$, and $n_k=c_k^\dagger c_k$. At half filling, the Fermi points are
$k_R=k_F\simeq \pi/2$ and $k_L=-k_F$. For a small momentum displacement $\kappa$,
\begin{equation}
\begin{aligned}
\epsilon(k_F+\kappa)&=v_F\kappa+O(\kappa^3),\\
\epsilon(-k_F+\kappa)&=-v_F\kappa+O(\kappa^3)
\end{aligned}
\quad
v_F=2t.
\end{equation}
With the site coordinate $x_j=j$ in lattice units, the continuum expansion is
\begin{equation}
c_j\simeq e^{ik_Fx_j}\psi_R(x_j)
+e^{-ik_Fx_j}\psi_L(x_j),
\end{equation}
with low-energy Hamiltonian
\begin{equation}
H_{\rm IR}=\int dx\,
\left[-iv_F\psi_R^\dagger\partial_x\psi_R
+iv_F\psi_L^\dagger\partial_x\psi_L\right].
\end{equation}
Continuum bilinears are normal ordered relative to the half-filled sea. We define centered charge
\begin{equation}
Q_V=\sum_j\left(n_j-\tfrac12\right) = \sum_kn_k -M.
\end{equation}
The constant shift does not affect the $U(1)_V$ action on fermion operators and places the half-filled sector at
$Q_V=0$. Restricting to excitations
near the two Fermi points then gives the continuum vector charge
\begin{equation}
Q_V^{\rm IR}=\int dx\,
\left(\psi_R^\dagger\psi_R+\psi_L^\dagger\psi_L\right).
\end{equation}
The corresponding continuum axial charge is
\begin{equation}
Q_A^{\rm IR}=\int dx\,
\left(\psi_R^\dagger\psi_R-\psi_L^\dagger\psi_L\right).
\end{equation}
An exact compact microscopic lift of the axial charge is
\begin{equation}
Q_{A,0}=\sum_k s_0(k)n_k,
\quad
s_0(k)=\operatorname{sgn}(\sin k).
\label{eq:1d_QA0}
\end{equation}
Under periodic boundary conditions (PBC), the two exceptional momenta $k=0,\pi$, where $\sin k=0$, require
an additional prescription; assigning opposite signs gives a balanced spectral charge. To avoid this ambiguity,
we work with antiperiodic boundary conditions (APBC), $c_{j+N}=-c_j$. In this case, $\sum_k s_0(k)=0$, so the charge can
equivalently be written as $Q_{A,0}=\sum_k s_0(k)(n_k-1/2)$. It obeys
\begin{equation}
[H_0,Q_{A,0}]=0,\ \
\operatorname{Spec}(Q_{A,0})\subset\mathbb Z,
\ \
e^{2\pi iQ_{A,0}}=1.
\end{equation}
Here $\operatorname{Spec}$ denotes the spectrum of an operator. Since $s_0(k)=+1$ near $k_F$ and $-1$ near
$-k_F$, the IR restriction of $Q_{A,0}$ reproduces $Q_A^{\rm IR}$ in the continuum limit. The faithfully acting
symmetry group is
\begin{equation}
\frac{U(1)_V\times U(1)_{A,0}}{\mathbb Z_2},
\quad
e^{i\pi Q_V}=e^{i\pi Q_{A,0}}=(-1)^F,
\end{equation}
where $(-1)^F$ is fermion parity.

The exact compact axial symmetry comes at the cost of spatial nonlocality. In real space,
\begin{equation}
Q_{A,0}=\sum_{j,\ell}c_j^\dagger S_{j\ell}c_\ell,
\label{eq:1d_axial_charge}
\end{equation}
where the discrete Fourier transform of the square-wave symbol $s_0(k)=\operatorname{sgn}(\sin k)$ is
\begin{equation}
S_{j\ell}=
\begin{cases}
\displaystyle\frac{2i}{N\sin[\pi(j-\ell)/N]},
&j-\ell\ \mathrm{odd},\\[1mm]
0,&j-\ell\ \mathrm{even}.
\end{cases}
\label{eq:1d_axial_finite_kernel}
\end{equation}
For fixed odd $r=j-\ell$, the thermodynamic limit is $S_{j\ell}\to 2i/(\pi r)$, exhibiting the algebraic
nonlocality of the axial charge. Our construction differs from the quantized axial charge of
Ref.~\cite{chatterjee2025quantized}, which does not commute with the microscopic vector charge. By contrast,
$Q_{A,0}$ is both compact and commuting with $Q_V$, at the cost of spatial nonlocality. As we show below, this
distinction enables a symmetric exact microscopic gauging construction.
\paragraph{Branch relocalization.}
Since $Q_V$ is onsite, coupling $U(1)_V$ to a compact $U(1)$ gauge field is straightforward. Our goal is to
construct an analogous coupling for the axial symmetry. As discussed in the Introduction, we first introduce a
purely algebraic \emph{branch relocalization}, which separates the two low-energy branches and renders the axial
symmetry onsite. Introduce a dimensionless reduced momentum
$
-\pi\leq p<\pi
$
and define
\begin{equation}
r_p
=
c_{k_F+p/2},
\quad
\ell_p
=
c_{-k_F+p/2}.
\label{eq:1d_rp_lp}
\end{equation}
The reduced Brillouin zone contains $M = N/2$ momenta. We define $r_X=M^{-1/2}\sum_p e^{ipX}r_p$, and similarly
for $\ell_X$, where $X$ labels cells with physical spacing $2$. Each cell contains two modes, $r_X$ and
$\ell_X$, obtained by reorganizing momentum sectors rather than grouping neighboring microscopic sites. Thus,
these branch cells should not be confused with ordinary two-site cells in the original lattice. Writing
$n_{R,X}=r_X^\dagger r_X$ and $n_{L,X}=\ell_X^\dagger\ell_X$, the charges become
\begin{equation}
Q_V=\sum_X(n_{R,X}+n_{L,X})-M,
\end{equation}
and
\begin{equation}
Q_{A,0}=\sum_X(n_{R,X}-n_{L,X}).
\label{eq:1d_branch_charges}
\end{equation}
Both global symmetries are therefore onsite in the branch representation. However, the Hamiltonian becomes
\begin{equation}
H_0
=
\sum_p
2t\sin\frac p2
\left(
r_p^\dagger r_p
-
\ell_p^\dagger\ell_p
\right).
\label{eq:1d_branch_H}
\end{equation}
The dispersion is discontinuous across the seam of the reduced Brillouin circle, since $\sin( \pm \pi/2)=\pm 1$.
Consequently, its Fourier transform has algebraically decaying $1/r$ hopping amplitudes. Although these
long-range hoppings can be coupled to a compact gauge field through a Wilson-line prescription, gauge covariance
alone does not ensure IR locality. In particular, gauge interactions can couple low-energy excitations to modes
near the Brillouin-zone seam, whose virtual contributions need not decouple and may induce nonlocal IR
interactions. This is the familiar SLAC-fermion problem
\cite{drellStrongcoupling1976,drellStrongcoupling1976a}. Fortunately, in our case the seam can be removed by
a suitable momentum-dependent canonical rotation. We first perform a particle--hole transformation on the
left-moving branch and introduce
\begin{equation}
d_p\equiv\ell_{-p}^\dagger
=c^\dagger_{-k_F-p/2},
\quad
\Psi_A(p)=
\begin{pmatrix}r_p\\d_p\end{pmatrix}.
\label{eq:1d_axial_particle_hole_main}
\end{equation}
Up to a normal-ordering constant, the Hamiltonian is
\begin{equation}
H_0=\sum_p\Psi_A^\dagger(p)
h_{\rm br}(p)\Psi_A(p),
\ \
h_{\rm br}(p)=2t\sin\frac p2\,\sigma_z.
\label{eq:1d_Hdiag_A}
\end{equation}
The exact charges become
\begin{equation}
Q_{A,0}=\sum_p\Psi_A^\dagger(p)\Psi_A(p)-M,
\label{eq:1d_QA_identity}
\end{equation}
and
\begin{equation}
Q_V=\sum_p\Psi_A^\dagger(p)\sigma_z\Psi_A(p).
\label{eq:1d_QV_sigma_z}
\end{equation}
Thus the axial symmetry is already ordinary fermion number in the particle--hole basis. The remaining task is to
restore Hamiltonian locality. At the reduced Brillouin-zone boundary, the two branches are exchanged: $h_{\rm
br}(\pi)=\sigma_xh_{\rm br}(-\pi)\sigma_x$. A Hadamard transformation diagonalizes this exchange, leaving a
relative boundary sign that can be removed by a half-momentum phase. Combining these operations, we define
\begin{equation}
\begin{aligned}
\Phi(p)&=\mathcal R(p)\Psi_A(p),\\
\mathcal R(p)
&=\begin{pmatrix}
1&0\\0&ie^{ip/2}
\end{pmatrix}
\frac1{\sqrt2}
\begin{pmatrix}
1&1\\1&-1
\end{pmatrix}\\
&=\frac1{\sqrt2}
\begin{pmatrix}
1&1\\
ie^{ip/2}&-ie^{ip/2}
\end{pmatrix}.
\end{aligned}
\label{eq:1d_Rp}
\end{equation}
Although $\mathcal R(p)$ is not itself periodic, it satisfies $\mathcal R(p+2\pi)\sigma_x=\mathcal R(p)$. The
transformed Hamiltonian is
\begin{equation}
\begin{aligned}
h_{\rm loc}(p)
&=\mathcal R(p)h_{\rm br}(p)
\mathcal R^\dagger(p)\\
&=-t
\begin{pmatrix}
0&1-e^{-ip}\\
1-e^{ip}&0
\end{pmatrix}.
\end{aligned}
\label{eq:1d_hloc_exact}
\end{equation}
Thus the seam is removed and the Hamiltonian becomes a periodic, first-degree Laurent polynomial. The
transformation is spatially nonlocal in the original branch coordinates, but the resulting Hamiltonian is
strictly finite range. To unfold the two components into a single chain, set $q=p+\pi$ and define
\begin{equation}
\eta_{2m+a}
=\frac1{\sqrt M}\sum_p
e^{imq}\Phi_{a+1}(p),
\quad a=0,1.
\label{eq:1d_eta_unfold}
\end{equation}
The corresponding two-site Bloch Hamiltonian gives exactly
\begin{equation}
H_0=-t\sum_n
\left(\eta_n^\dagger\eta_{n+1}
+\mathrm{H.c.}\right).
\label{eq:1d_eta_H}
\end{equation}
The axial charge is onsite:
\begin{equation}
Q_{A,0}=\sum_n\eta_n^\dagger\eta_n-M.
\label{eq:1d_eta_QA}
\end{equation}
Here, $n$ can be viewed as the new relocalized coordinate. The unfolding also gives the simple momentum-space
identification
\begin{equation}
\begin{aligned}
\eta_{k_F+p/2}&=r_p =c_{k_F+p/2}, \\
\eta_{-k_F+p/2}& =d_p = c^\dagger_{-k_F-p/2}.
\end{aligned}
\label{eq:1d_eta_momentum_identification}
\end{equation}
Consequently, the spectator vector charge becomes
\begin{equation}
Q_V=\sum_k s_0(k)
\left(\eta_k^\dagger\eta_k-\frac12\right),
\quad
s_0(k)=\operatorname{sgn}(\sin k).
\label{eq:1d_QV_after_R}
\end{equation}
Thus the relocalization makes $U(1)_{A,0}$ onsite while transferring the nonlocality to $U(1)_V$, without
sacrificing the strict locality of the Hamiltonian. If $\mathcal W$ denotes the nonlocal axial relocalization
constructed above, then after identifying the $\eta$ chain with the original lattice convention one has
\begin{equation}
\begin{aligned}
\mathcal WH_0\mathcal W^{-1}&=H_0,\\
\mathcal WQ_{A,0}\mathcal W^{-1}&=Q_V,\\
\mathcal WQ_V\mathcal W^{-1}&=Q_{A,0}.
\end{aligned}
\label{eq:1d_exact_H_preserving_VA_exchange}
\end{equation}
Thus the nearest-neighbor free chain admits an exact nonlocal canonical automorphism which preserves the
Hamiltonian and exchanges the two compact $U(1)$ symmetries, as alluded to in the introduction. The axial
symmetry can therefore be gauged in this coordinate frame.

\begin{figure}[tb]
\centering
\includegraphics[width=\columnwidth]{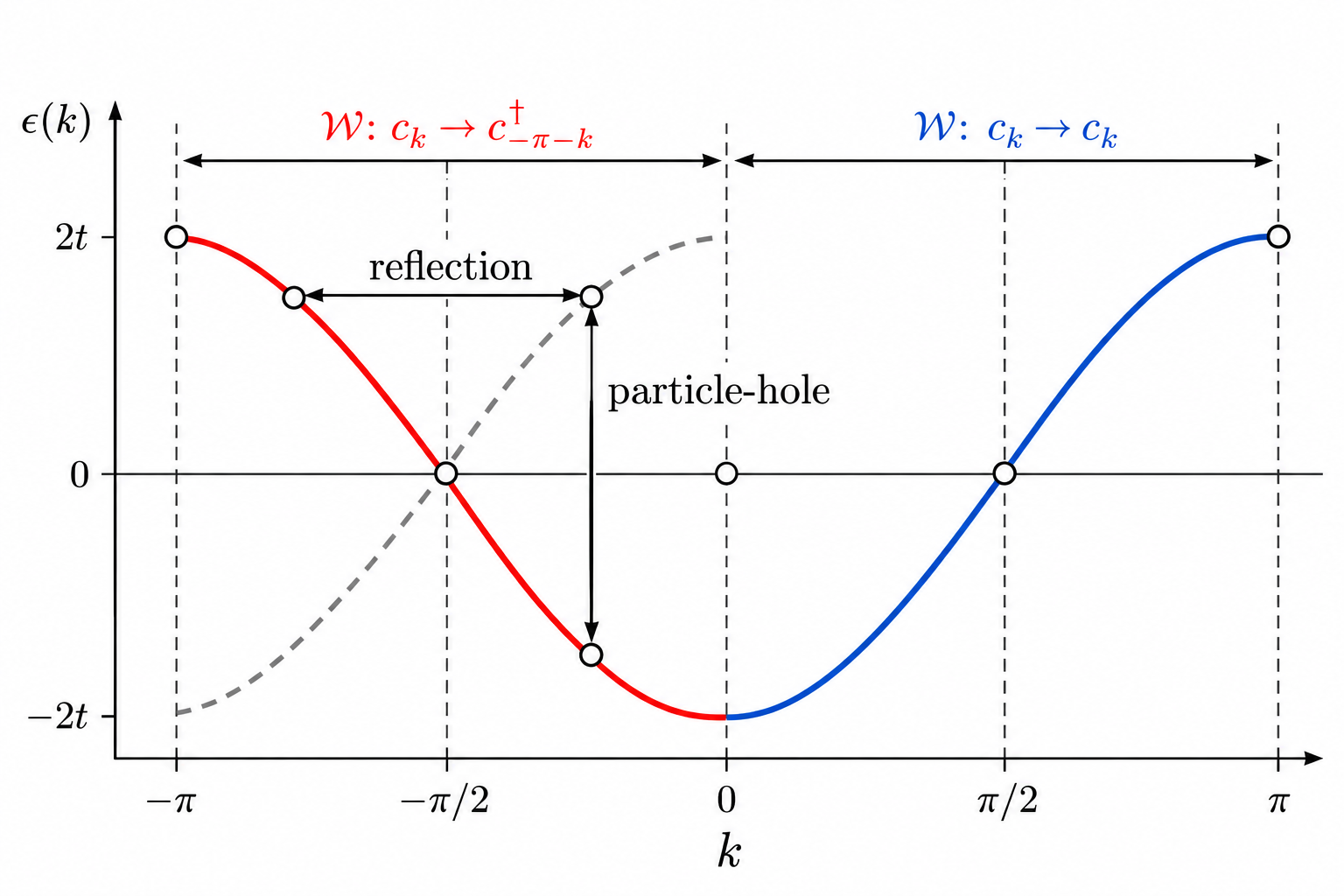}
\caption{Hamiltonian-preserving vector--axial exchange.
For the dispersion $\epsilon(k)=-2t\cos k$, the canonical
transformation $\mathcal W$ in Eq.~\eqref{eq:1d_W_explicit}
leaves the positive-momentum modes unchanged (blue) and maps
$c_k\mapsto c_{-\pi-k}^\dagger$ for $-\pi<k<0$ (red).
Momentum reflection reverses the single-particle energy
(gray dashed curve), while particle--hole conjugation
compensates for this reversal. Consequently,
$\mathcal W H_0\mathcal W^{-1}=H_0$ and
$\mathcal W$ exchanges $Q_V\leftrightarrow Q_{A,0}$.
Thus the same nearest-neighbor Hamiltonian realizes
onsite $U(1)_V$ in the original $c$ representation
and onsite $U(1)_A$ in the relocalized $\eta$ representation.}
\label{fig:relocalization}
\end{figure}

\paragraph{Direct particle--hole transformation.}
The preceding seam-sewing construction provides an intuitive picture that extends naturally to higher dimensions, as discussed in the next section. For $H_0$, however, a more direct algebraic derivation is available. With $N=4m$ as assumed above, $M$ is even. The APBC momentum grid is invariant under $k\mapsto-k$ and $k\mapsto\pi-k$ modulo $2\pi$, and excludes $0,\pi,\pm\pi/2$. The key observation is that negative-momentum modes can be paired with modes of opposite single-particle energy. For $-\pi<k<0$, define $\bar k=-\pi-k$, so that $\epsilon(\bar k)=-\epsilon(k)$. Equation~\eqref{eq:1d_eta_momentum_identification} motivates the canonical transformation
\begin{equation}
\eta_k\equiv\mathcal Wc_k\mathcal W^{-1}
=\begin{cases}
c_k,&0<k<\pi,\\
c_{\bar k}^\dagger,&-\pi<k<0.
\end{cases}
\label{eq:1d_W_direct_action}
\end{equation}
An explicit implementation is
\begin{equation}
\mathcal W=\prod_{-\pi<k<-\pi/2}
\left[
e^{i\pi n_{\bar k}}
\exp\!\left\{\frac\pi2
\left(c_k^\dagger c_{\bar k}^\dagger-c_{\bar k}c_k\right)
\right\}\right].
\label{eq:1d_W_explicit}
\end{equation}
Each factor is an even Gaussian unitary acting on a disjoint momentum pair.\footnote{For APBC with $N=4m+2$, the mode $k_*=-\pi/2$ is fixed by $k\mapsto\bar k$. Writing $\mathcal W_{\rm pair}$ for the product over the remaining pairs, the extension is
\begin{equation}
\mathcal W=
(c_{k_*}+c_{k_*}^\dagger)
(-1)^{\sum_{q\ne k_*}n_q}
\mathcal W_{\rm pair}.
\label{eq:1d_W_fixed_mode}
\end{equation}
The parity string leaves the transformation of all other modes unchanged. This implementer is parity odd, but remains an exact canonical transformation.}
The pairing exponential maps $c_k\mapsto-c_{\bar k}^\dagger$ and $c_{\bar k}\mapsto c_k^\dagger$, while the parity factor removes the relative minus sign. Thus, Eq.~\eqref{eq:1d_W_explicit} implements Eq.~\eqref{eq:1d_W_direct_action} exactly, preserving the fermionic anticommutation relations and fermion parity. The momentum-space exchange is illustrated in Fig.~\ref{fig:relocalization}. Writing $m_k=n_k-1/2$, we have $m_k\mapsto m_k$ for positive momenta and $m_k\mapsto-m_{\bar k}$ for negative momenta. Together with $\epsilon(\bar k)=-\epsilon(k)$, this establishes the exact Hamiltonian-preserving vector--axial exchange in Eq.~\eqref{eq:1d_exact_H_preserving_VA_exchange}, without an additive energy shift.

The relocalized real-space fermions follow directly:
\begin{equation}
\eta_j=\frac1{\sqrt N}\sum_ke^{ikj}\eta_k
=\mathcal Wc_j\mathcal W^{-1}.
\label{eq:1d_eta_fourier_W}
\end{equation}
Their relation to the microscopic fermions is nonlocal,
\begin{equation}
\eta_j=\sum_{\ell}F_{j\ell}
\left[c_{\ell}+(-1)^j c_{\ell}^\dagger\right].
\label{eq:1d_eta_microscopic_raw}
\end{equation}
Here $F_{j\ell}=\frac1N\sum_{0<k<\pi}e^{ik(j-\ell)}$ has algebraic tails. Nevertheless, the momentum reflection in the left-moving branch preserves the continuum spatial coordinate. Defining the slowly varying fields $\eta_{R,L}$ by the analogous Fermi-point expansion of $\eta_j$, with $x=j$ in lattice units, we obtain
\begin{equation}
\eta_R(x)=\psi_R(x),
\quad
\eta_L(x)=\psi_L^\dagger(x).
\end{equation}
Thus $\mathcal W$ acts pointwise on the continuum Dirac fields despite its microscopic nonlocality.  

\paragraph{Vector--axial duality and bosonization.}
The Hamiltonian-preserving transformation $\mathcal W$ exchanges the commuting compact vector and axial charges and their microscopic locality. Its pointwise chiral particle--hole action bosonizes to $\phi\leftrightarrow\theta$, realizing momentum--winding T-duality of the $c=1$ compact-boson conformal field theory (CFT) at $K=1$, as detailed in Appendix~\ref{app:bosonization_tduality}.

In the spin-$1/2$ XX chain, the Jordan--Wigner (JW) transformation identifies the onsite magnetization and nonlocal spectral winding charges as
\begin{equation}
Q_M=Q_V,\quad Q_W=\frac12Q_{A,0}.
\label{eq:1d_XX_spectral_winding}
\end{equation}
Both are exactly conserved and compact, with $[Q_M,Q_W]=0$. For $N\in4\mathbb Z$ with fermionic APBC, their exchange is exact:
\begin{equation} 
\mathcal WQ_M\mathcal W^{-1}
=2Q_W,\ \
\mathcal WQ_W\mathcal W^{-1}
=\frac12Q_M. 
\label{eq:1d_XX_exact_momentum_winding}
\end{equation}
On the periodic XX spin chain, parity projection produces a noninvertible spectral operator $\mathcal D$ satisfying $\mathcal D^2=1+U$, where $U$ is the spin-flip symmetry. Labeling compact-boson states by their momentum and winding quantum numbers $(m,w)$, its nonzero IR action exchanges these numbers as $(m,w)\mapsto(2w,m/2)$ in the even-parity sectors, while annihilating odd-parity states. It thus realizes a noninvertible T-duality topological defect at the XX radius $R=\sqrt2$, despite its algebraically nonlocal microscopic action. This spectral construction differs from the lattice duality of Ref.~\cite{pace2024lattice}, which has a different microscopic charge algebra and locality structure.

The emergence of noninvertible defects from nonlocal transformations is not limited to the $c=1$ compact boson. An analogous construction applies to the $c=1/2$ Ising theory. To isolate a single Majorana sector, we use the fact that the same $\mathcal W$ exchanges local SSH dimerization and superconducting pairing, yielding a self-dual Hamiltonian with one critical Majorana species $\chi$ and a gapped spectator $\chi'$, despite breaking both continuous $U(1)$ symmetries. Since $\mathcal W$ acts nontrivially only on $\chi$, the spectral construction applies independently to a single Majorana chain. It exchanges the Ising field and bond couplings exactly, including the parity-dependent boundary term. On the periodic transverse-field Ising chain, parity projection produces a noninvertible operator $\mathcal D_{\rm I}$ satisfying $\mathcal D_{\rm I}^2=1+U_{\rm I}$. At criticality, this operator flows to the conventional Kramers--Wannier topological defect~\cite{seiberg2024majorana}. Although the microscopic spectral transformation is algebraically nonlocal and differs from the usual lattice order--disorder duality, its IR limit reproduces the same topological defect. The spin-sector structure and comparisons with related dualities~\cite{li2023non,chatterjee2025quantized,su2025z2} are also discussed in Appendix~\ref{app:bosonization_tduality}.

\subsection{Exact and emergent axial gauge locality}
\label{sec:1d_exact_axial_gauging}

\subsubsection{Exact finite-spacing axial gauge locality}
\label{sec:1d_exact_finite_spacing_gauge_locality}

The onsite vector charge $Q_V$ can be gauged directly in the original microscopic coordinate. With $\nu_j=[1-(-1)^j]/2$, choose the integer-valued local vector charge density
\begin{equation}
q_{V,j}=c_j^\dagger c_j-\nu_j,
\quad Q_V=\sum_jq_{V,j}.
\label{eq:1d_vector_density}
\end{equation}
The Peierls substitution then gives the conventional staggered Schwinger regulator. The vector--axial duality established above provides an equally local gauging of the originally nonlocal axial symmetry, now in the relocalized $\eta$ representation.

The corresponding onsite axial charge density is
\begin{equation}
q_{A,j}=\eta_j^\dagger\eta_j-\nu_j,
\quad Q_{A,0}=\sum_jq_{A,j}.
\label{eq:1d_eta_density}
\end{equation}
Introduce compact axial links and their conjugate integer-valued electric fields,
\begin{equation}
U_j^A=e^{i\vartheta_j^A},
\quad [E_m^A,U_n^A]=\delta_{mn}U_n^A.
\end{equation}
For either $I=V,A$, $\vartheta_j^I$ denotes a spatial link angle, $\varphi_{I,j}$ a temporal scalar potential, and $\Phi_I=\sum_j\vartheta_j^I$ the spatial holonomy.  

The axial symmetry admits the strictly local dynamical gauge Hamiltonian
\begin{equation} 
H_A[U^A,E^A]
=-t\sum_j\left(\eta_j^\dagger U_j^A\eta_{j+1}
+\mathrm{H.c.}\right)+\frac{g_A^2}{2}\sum_j(E_j^A)^2,
\label{eq:1d_local_axial_gauge}
\end{equation}
with Gauss-law generators
\begin{equation}
\mathcal G_{A,j}=E_j^A-E_{j-1}^A-q_{A,j}.
\label{eq:1d_local_Gauss}
\end{equation}
Each nearest-neighbor hopping term requires only one gauge link, eliminating any Wilson-line path ambiguity. Moreover, $\mathcal W$ establishes an exact unitary equivalence between the separately gauged vector and axial theories, including their Gauss laws and physical Hilbert spaces, upon identifying the gauge-link Hilbert spaces and couplings; see Appendix~\ref{app:1d_tduality_gauged_intertwiner}.

As established above, $\mathcal W$ acts pointwise on the continuum Dirac fields, with $\eta_R(x)=\psi_R(x)$ and $\eta_L(x)=\psi_L^\dagger(x)$. Thus the onsite transformation $\eta_j\mapsto e^{i\alpha}\eta_j$ induces
\begin{equation}
\psi_R(x)\mapsto e^{i\alpha}\psi_R(x),
\quad
\psi_L(x)\mapsto e^{-i\alpha}\psi_L(x),
\end{equation}
precisely the continuum axial $U(1)$ action. Gauge locality, however, is realized in the relocalized $\eta$ representation: the gauge fields in Eq.~\eqref{eq:1d_local_axial_gauge} reside on the $\eta$-chain links, and their couplings and gauge transformations are generally nonlocal when expressed in the original $c$ variables. Nevertheless, the pointwise IR action of $\mathcal W$ ensures that this local axial gauging retains its intended continuum spatial interpretation.

The gauged $\eta$ chain is exactly the conventional Hamiltonian staggered-fermion regulator of the single-flavor Schwinger model. Restoring physical units, the continuum scaling is
$t=v_F/(2a)$, with electric energy
$(g_A^2a/2)\sum_n E_n^2$.  Its continuum limit is well established \cite{banks1976strong,carroll1976lattice,hamer1982massive}. More precisely, the conventional symmetry-improved mass parametrization is $m_{\rm cont}=m_{\rm lat}+g_A^2a/8$, up to cutoff corrections \cite{dempsey2022discrete}, where $m_{\rm lat}$ multiplies $\sum_n(-1)^n\eta_n^\dagger\eta_n$. For our Hamiltonian, $m_{\rm lat}=0$, so the finite-spacing mass offset vanishes as $a\to0$ at fixed $g_A$.

Thus, in terms of the original Dirac fermion, local $\eta$-number gauging realizes axial minimal coupling, with continuum Lagrangian
\begin{equation}
{\cal L}_{\rm IR}
=\bar\Psi i\gamma^\mu
\left(\partial_\mu+iA_\mu^A\gamma^5\right)\Psi
-\frac{1}{4g_A^2}
F_{A,\mu\nu}F_A^{\mu\nu},
\label{eq:1d_IR_axial_Schwinger}
\end{equation}
where $\Psi=(\psi_R,\psi_L)^T$ and the IR velocity is normalized to unity. Particle--hole conjugation of one chirality identifies this massless axial-gauge theory with the ordinary vector-coupled single-flavor Schwinger model.

\subsubsection{Spectral axial charges and finite-range deformations}
\label{sec:1d_general_s_gauge_locality}

The exact compact axial symmetry and its local gauging admit two independent extensions: varying the inversion-odd spectral sign and deforming the kinetic dispersion. Both preserve Hamiltonian locality in suitable symmetry-adapted coordinates, and the two freedoms can be combined. Further details of their finite-volume constructions and locality constraints are given in Appendix~\ref{app:1d_general_s_details}.

For the nearest-neighbor Hamiltonian $H_0$, every sign $s(k)\in\{\pm1\}$ satisfying $s(-k)=-s(k)$ on the APBC momentum grid defines a conserved compact charge $Q_{A,s}=\sum_k s(k)(n_k-1/2)$. To realize the continuum axial symmetry, we require $s(k)=s_0(k)$ near both Fermi points. For cutoff-dependent profiles, these matching regions must contain the momentum scales retained in the IR limit. Away from the Fermi points, the spectral sign can be modified by exchanging energy-degenerate modes at $k$ and $-k$, since $\epsilon(k)=\epsilon(-k)$. Define $\mathcal A_s=\{k\in(0,\pi):s(k)=-1\}$ and introduce the number-conserving Gaussian unitary
\begin{equation}
\mathcal U_s=\prod_{k\in\mathcal A_s}
\exp\!\left[\frac\pi2
\left(c_{-k}^\dagger c_k-c_k^\dagger c_{-k}\right)\right].
\label{eq:1d_s_U_swap}
\end{equation}
The factors act on disjoint momentum pairs and commute. For $k\in\mathcal A_s$,
\begin{equation}
\mathcal U_sc_k\mathcal U_s^{-1}=c_{-k},
\quad
\mathcal U_sc_{-k}\mathcal U_s^{-1}=-c_k.
\label{eq:1d_s_U_action}
\end{equation}
Consequently, $\mathcal U_s$ preserves $H_0$ while mapping $Q_{A,0}$ to $Q_{A,s}$. Conjugating the reference vector--axial exchange gives
\begin{equation}
\mathcal W_s=\mathcal U_s\mathcal W\mathcal U_s^{-1},
\end{equation}
which preserves $H_0$ and exchanges $Q_V\leftrightarrow Q_{A,s}$. To obtain a local representation, define the adapted fermions $\xi_j^{(s)}=\mathcal U_s\eta_j\mathcal U_s^{-1}$, where $\eta_j=\mathcal Wc_j\mathcal W^{-1}$. In these variables, $Q_{A,s}$ becomes onsite and $H_0$ remains nearest-neighbor, admitting strictly local compact gauging.

Independently, the reference vector--axial transformation preserves any real, inversion-even, $\pi$-antiperiodic finite-range dispersion. A family retaining exactly two simple Fermi points and the original Fermi velocity is
\begin{equation}
\epsilon_\lambda(k)
=-2t\cos k\,P_\lambda(\cos^2 k),
\label{eq:1d_general_dispersion_main}
\end{equation}
where $P_\lambda$ is a real polynomial satisfying $P_\lambda(0)=1$ and $P_\lambda(x)>0$ for $0\le x\le1$. These dispersions contain only odd cosine harmonics, giving finite-range odd-distance hopping after relocalization while keeping the axial charge onsite. Since $\epsilon_\lambda(k)=\epsilon_\lambda(-k)$, the same momentum-pair swaps $\mathcal U_s$ preserve the deformed Hamiltonian. The spectral-sign and dispersion extensions can therefore be combined, with exact compact gauge invariance obtained by dressing each finite-range hopping term with the corresponding oriented Wilson product.

The Gauss-law derivation of the discrete-chiral mass reflection extends to these odd-distance hopping Hamiltonians, yielding the same symmetry-centered bare mass $m_{\rm lat}^*=-g_A^2a/8$ as in the conventional staggered regulator~\cite{dempsey2022discrete}. At $m_{\rm lat}=0$, the displacement from this point vanishes as $a\to0$ at fixed $g_A$. Since the additional hopping terms contribute only higher-derivative corrections near the Dirac points, these models are expected to share the massless Schwinger continuum limit, although this does not constitute a separate continuum-limit theorem.

Finally, $\epsilon(k+\pi)=-\epsilon(k)$ is necessary and sufficient for the specified vector--axial transformation to preserve the Hamiltonian. For real-analytic, inversion-even dispersions with exactly two simple Fermi points, a spectral-invariant argument further establishes this condition as necessary for smooth, periodic Hamiltonian locality under Gaussian rotations at fixed reduced momentum. This constraint does not exclude momentum-rearranging or more general canonical relocalizations.

\subsubsection{Emergent dynamical gauge locality at finite chemical potential}
\label{sec:1d_mu_gauge_locality}

The preceding exact compact-gauging constructions concern the half-filled kinetic Hamiltonian $H_0$. Introducing a chemical potential $\mu$ conjugate to the centered vector charge $Q_V$ gives $H_\mu=H_0-\mu Q_V$. Although every compact axial charge $Q_{A,s}$ remains conserved, a nonzero chemical potential generally destroys Hamiltonian locality in the symmetry-adapted representation. For the reference sign $s=s_0$, the vector--axial exchange maps $Q_V$ to the spectral axial charge, giving the relocalized dispersion
\begin{equation}
\epsilon_\mu(k)
=-2t\cos k-\mu s_0(k),
\quad
s_0(k)=\operatorname{sgn}(\sin k).
\label{eq:1d_mu_relocalized_dispersion}
\end{equation}
The discontinuities of $s_0(k)$ at $k=0,\pi$ produce algebraically decaying hopping amplitudes. Thus, for $\mu\ne0$, the relocalized Hamiltonian contains long-range hopping, and dressing these terms with shortest-path Wilson products does not yield a strictly local finite-spacing gauge theory. The same limitation applies to the general spectral-sign constructions, for which the chemical-potential term becomes $-\mu Q_{A,0}[\xi^{(s)}]$.

Agreement with the continuum Dirac dispersion near the Fermi points does not by itself resolve this UV
nonlocality. As alluded to in the Introduction, an artificial discontinuity in the UV dispersion can
generate a singular fermion--photon vertex, $\Gamma_{\rm seam}^{1}(q)\sim\Delta\epsilon_{\rm seam}/q$, even for
arbitrarily small momentum transfer $q$. Establishing dynamical gauge locality therefore requires controlling
the complete coupled matter--gauge system, rather than its free low-energy spectrum alone.

We accomplish this by comparing the nonlocal gauging with a strictly local regulator obtained by replacing
$s_0(k)$ with $\sin k$:
\begin{equation}
\epsilon_\mu^{\rm loc}(k)
=-2t_a\cos k-\mu\sin k.
\label{eq:1d_mu_local_dispersion}
\end{equation}
The two dispersions have the same linearized Dirac theory near $k=\pm\pi/2$. Unlike the spectral sign, $\sin k$
generates only nearest-neighbor hopping and admits conventional compact Peierls gauging. A Gauss-law-preserving
unitary transformation maps the local regulator exactly to the compact axial gauging of the nearest-neighbor
chain $H_0$, with a renormalized hopping amplitude
\begin{equation}
t_a'=\sqrt{t_a^2+\frac{\mu^2}{4}}.
\end{equation}
This transformation shifts the dynamical holonomy while leaving the electric energy and Gauss-law constraints
unchanged, as shown in Appendix~\ref{app:1d_mu_stability}.

To formulate the continuum comparison, restore the lattice spacing $a$ and physical circumference $\ell=Na$. We
take $a\to0$ with $v_F=2at_a$, $\mu$, $g_A$, and $\ell$ fixed. Let $H_a^{\rm nl}$ and $H_a^{\rm loc}$ denote the
compact gaugings of Eqs.~\eqref{eq:1d_mu_relocalized_dispersion} and
\eqref{eq:1d_mu_local_dispersion}, respectively, with identical electric energies and Gauss-law constraints.
Writing $c_a=\inf\operatorname{Spec}(H_a^{\rm loc})$ for the ground-state energy, we prove that, for every fixed
$E_*>0$ and sufficiently small $a$,
\begin{equation}
\begin{aligned}
&\bigl\|
(H_a^{\rm nl}-c_a+E_*)^{-1} -(H_a^{\rm loc}-c_a+E_*)^{-1}
\bigr\|\\
&\quad\le
Ca\left[1+\log\frac{\ell}{a}\right].
\end{aligned}
\label{eq:1d_mu_stability_main}
\end{equation}
The estimate holds in operator norm on the full physical matter--gauge Hilbert space, including dynamical
electric fields and compact holonomy, and is nonperturbative in $g_A$. Using the same shortest-path compact
gauging in every adapted $\xi^{(s)}$ frame transports both comparison Hamiltonians and their Gauss laws
unitarily. The estimate is therefore uniform over the inversion-odd spectral-sign family, independently of the
additional node-matching condition required for a common IR axial interpretation.

The norm-resolvent estimate is stronger than matching low-energy dispersions, but weaker than operator-norm
convergence of the Hamiltonians themselves. Indeed, defining
\begin{equation}
B_a:=H_a^{\rm nl}-H_a^{\rm loc},
\label{eq:1d_mu_B_main}
\end{equation}
we find, for fixed $\mu\neq0$, $\|B_a\|\gtrsim|\mu|\ell/a$. Thus the regulators remain distinguishable on
suitable UV states. Their resolvents nevertheless become close because high-energy contributions are suppressed
by an inverse-energy factor.

The underlying mechanism is quantified by a relative-form estimate. With $L_a=H_a^{\rm loc}-c_a\ge0$,
Appendix~\ref{app:1d_mu_stability} establishes
\begin{equation}
\begin{aligned}
|\langle\Psi,B_a\Psi\rangle|
&\le
\epsilon_a\langle\Psi,L_a\Psi\rangle
+\beta_a\|\Psi\|^2,\\
\epsilon_a=O(a), & \quad \beta_a=O\!\left(
a\left[1+\log\frac{\ell}{a}\right]
\right),
\end{aligned}
\label{eq:1d_mu_relative_form_main}
\end{equation}
for every physical state $|\Psi\rangle$ with finite local energy, uniformly in the lattice spacing $a$. The
Hamiltonian difference is therefore small relative to the excitation energy, even though its unrestricted
operator norm diverges. In particular, for any fixed excitation-energy cutoff $E<\infty$, let
$P_{a,E}=\mathbf1_{[0,E]}(L_a)$ project onto the physical states of the local regulator with energies at most
$E$ above its ground state. Then
\begin{equation}
\|P_{a,E}B_aP_{a,E}\|
\le\epsilon_a E+\beta_a \rightarrow0.
\label{eq:1d_mu_low_energy_norm_main}
\end{equation}
Thus the Hamiltonian difference vanishes in operator norm within any fixed low-energy sector. More generally,
norm-resolvent closeness also implies convergence of continuously energy-filtered real-time dynamics, as
established in Appendix~\ref{app:1d_mu_stability}.

We thus establish emergent dynamical gauge locality: despite its algebraically long-ranged hopping at finite
lattice spacing, the exactly relocalized and compactly gauged Hamiltonian becomes asymptotically norm-resolvent
equivalent to a strictly local compact gauge regulator. This result controls the fully dynamical matter--gauge
system, rather than merely the free spectrum or prescribed external gauge backgrounds.

The fixed-physical-chemical-potential scaling is essential. Since $|\mu|/t_a \to 0$, the theorem does not apply
at fixed nonzero $\mu/t_a$. It transfers any independently established continuum limit of the local comparison
family, under compatible continuum identifications, to the specified nonlocal gauging, but does not itself
construct that limit. Nor does it extend the finite-$\mu$ estimate to the deformed dispersions above, arbitrary
long-range hopping, or alternative gauging prescriptions without separate analysis.

\subsection{Microscopic mixed anomaly, Ward identities, and LSM pumping}
\label{sec:1d_anomaly}

We derive the mixed $U(1)_V$--$U(1)_A$ anomaly in three complementary ways: an exact finite-lattice Schwinger
term, quantized large-gauge spectral flow with its matched local Ward identities, and the translation-induced
LSM momentum pump. The fixed microscopic spectral charge, its gauge-covariant extension, and its local IR
current representative must be distinguished.

\subsubsection{Exact lattice Schwinger term}

We return to the half-filled nearest-neighbor Hamiltonian of Eq.~\eqref{eq:1d_H0}, with the vector charge $Q_V$
and an axial charge $Q_{A,s}$. We impose APBC and take $N=2M\in4\mathbb Z$. The Fermi sea $|\mathrm{FS}\rangle$
occupies precisely the momenta $|k|<\pi/2$.

Introduce the centered vector density and a symmetrically defined axial density,
\begin{equation}
\begin{aligned}
\rho_V(q)
&=\sum_k c_{k+q}^\dagger c_k
-\frac N2\delta_{q,0},\\
\rho_{A,s}(q)
&=\frac12\sum_k
[s(k+q)+s(k)]c_{k+q}^\dagger c_k.
\end{aligned}
\label{eq:1d_anom_symmetric_densities}
\end{equation}
The latter corresponds to the real-space density
\begin{equation}
\rho_{A,s;j}=\frac12\sum_l
\left[
c_j^\dagger(S_s)_{jl}c_l
+c_l^\dagger(S_s)_{lj}c_j
\right],
\label{eq:1d_anom_real_density}
\end{equation}
where $(S_s)_{jl}$ is the Fourier transform of $s(k)$. This symmetric prescription fixes a definite lattice
density algebra; other short-distance prescriptions may differ by improvement terms. The zero-momentum
components satisfy $\rho_{A,s}(0)=Q_{A,s}$ and $\rho_V(0)=Q_V$. The uniform centering in $\rho_V(q)$ differs
from the staggered background in Eq.~\eqref{eq:1d_vector_density} only by a fixed c-number Fourier component,
which leaves the current algebra unchanged.

Using the fermionic anticommutation relations, we obtain the exact finite-lattice operator identity
\begin{equation}
[\rho_V(q),\rho_{A,s}(-q)]
=\frac12\sum_k
[s(k-q)-s(k+q)]n_k.
\label{eq:1d_anom_exact_commutator}
\end{equation}
The right-hand side is an operator rather than a central element. In particular, the full Fock-space trace of
this commutator vanishes, as required in finite dimensions.

For $s(k)=s_0(k)=\operatorname{sgn}(\sin k)$ and $0<q<\pi/2$, only momenta within angular distance $q$ of the
sign discontinuities at $k=0,\pi$ contribute:
\begin{equation}
[\rho_V(q),\rho_{A,0}(-q)]
=-\sum_{|k|<q}n_k
+\sum_{\operatorname{dist}(k,\pi)<q}n_k,
\label{eq:1d_anom_seam_commutator}
\end{equation}
where $\operatorname{dist}(k,\pi)$ denotes angular distance on the momentum circle. For $q=2\pi m/N$, each sum
contains exactly $2m$ momenta. The first set is fully occupied and the second empty in $|\mathrm{FS}\rangle$.
Hence
\begin{equation}
\langle\mathrm{FS}|
[\rho_V(q),\rho_{A,0}(-q)]
|\mathrm{FS}\rangle
=-2m=-\frac{Nq}{\pi}.
\label{eq:1d_anom_schwinger_exact}
\end{equation}
The coefficient has no finite-size corrections.

More strongly, the commutator acts centrally on a restricted low-energy sector. Let $P_{\mathrm{IR}}$ project
onto Fock states obtained from $|\mathrm{FS}\rangle$ by particle--hole excitations confined to small
neighborhoods of $k=\pm\pi/2$. Occupations near the UV sign discontinuities at $k=0,\pi$ are then frozen.
Provided the momentum transfer $q$ is sufficiently small that the contributing intervals in
Eq.~\eqref{eq:1d_anom_seam_commutator} do not overlap the excitation windows, we have
\begin{equation}
P_{\mathrm{IR}}
[\rho_V(q),\rho_{A,0}(-q)]P_{\mathrm{IR}}
=-\frac{Nq}{\pi}P_{\mathrm{IR}}.
\label{eq:1d_anom_projected_central}
\end{equation}
Thus, an exact microscopic operator commutator becomes a central term upon restriction to the low-energy sector.

The same IR central term holds for any inversion-odd sign $s(k)$ that agrees with $s_0(k)$ in neighborhoods of
both Fermi points. Indeed, shifting momenta in the Fermi-sea expectation of
Eq.~\eqref{eq:1d_anom_exact_commutator} gives
\begin{equation}
\langle[\rho_V(q),\rho_{A,s}(-q)]\rangle_{\mathrm{FS}}
=\frac12\sum_k s(k)
\left(n_{k+q}^{\mathrm{FS}}
-n_{k-q}^{\mathrm{FS}}\right).
\end{equation}
Only the occupation discontinuities at the Fermi points contribute. For sufficiently small $q$, their signs
coincide with those of $s_0$, yielding the same coefficient $-Nq/\pi$. The projected identity likewise holds
when the excitation windows lie within these constant-sign neighborhoods. The IR Schwinger term is therefore
independent of the UV choice of spectral sign.

Restoring the physical length $\ell=Na$ and momentum $p=q/a$, we obtain the continuum Fourier algebra
\begin{equation}
\langle[\rho_V(p),\rho_A(-p)]\rangle
=-\frac{\ell p}{\pi}.
\end{equation}
With our Fourier convention, this gives the distributional Schwinger term
\begin{equation}
[j_V^0(x),j_A^0(y)]_{\mathrm{IR}}
=-\frac{i}{\pi}\partial_x\delta(x-y).
\label{eq:1d_anom_continuum_schwinger}
\end{equation}
Using $j_V^1=v_Fj_A^0$ for the massless Dirac theory recovers the familiar density--spatial-current commutator.

The corresponding anomaly matrix is
\begin{equation}
\mathcal A_{IJ}
=\sum_{\lambda=R,L}
\chi_\lambda q_{I,\lambda}q_{J,\lambda}
=\begin{pmatrix}
0&2\\
2&0
\end{pmatrix}_{IJ},
\label{eq:1d_anomaly_matrix}
\end{equation}
where $\chi_R=+1$, $\chi_L=-1$ are the branch chiralities, and $q_{V,\lambda}=1$, $q_{A,\lambda}=\chi_\lambda$
are their vector and axial charges. Equation~\eqref{eq:1d_anom_schwinger_exact} therefore realizes $-(\mathcal
A_{VA}/2\pi)\ell p$ exactly on the finite lattice. The Schwinger term diagnoses the perturbative mixed
vector--axial anomaly. It does not, by itself, obstruct gauging either anomaly-free $U(1)$ individually; rather,
the mixed anomaly constrains the simultaneous preservation of both symmetries after gauging.

\subsubsection{Complementary gauge responses and local Ward identities}

\paragraph{Vector gauging and exact continuity.}
In the original $c$ coordinate, the vector symmetry is onsite. Using the local density $q_{V,j}$ of
Eq.~\eqref{eq:1d_vector_density}, its Peierls coupling to $U_j^V=e^{i\vartheta_j^V}$ and the temporal potential
$\varphi_{V,j}$ is
\begin{equation}
H_V[U^V,\varphi_V]
=-t\sum_j(c_j^\dagger U_j^Vc_{j+1}+\mathrm{H.c.})
+\sum_j\varphi_{V,j}q_{V,j},
\label{eq:1d_vector_x}
\end{equation}  
Under $c_j\mapsto e^{i\alpha_j}c_j$, the background fields transform as $U_j^V\mapsto
e^{i(\alpha_j-\alpha_{j+1})}U_j^V$ and $\varphi_{V,j}\mapsto\varphi_{V,j}-\dot\alpha_j$. The gauge-invariant
bond current
\begin{equation}
J^V_{j+1/2}=-it\left(
c_j^\dagger U_j^Vc_{j+1}-\mathrm{H.c.}
\right)
\end{equation}
obeys the exact continuity equation
\begin{equation}
\frac{d}{dt}q_{V,j}
+J^V_{j+1/2}-J^V_{j-1/2}=0.
\label{eq:1d_anom_exact_vector_continuity}
\end{equation}
Thus $U(1)_V$ has an exact local lattice Ward identity, and its dynamical gauging admits the usual local vector
Gauss law.

\paragraph{Covariant axial charge and large vector flux.}
Define the covariant shift $(T_V\psi)_j=U_j^V\psi_{j+1}$, including the APBC boundary sign, and its Hermitian
difference $D_V=(T_V-T_V^\dagger)/(2i)$. The covariant extension of the sharp axial charge is
\begin{equation}
\begin{aligned}
& Q_{A,0}^{\mathrm{cov}}[U^V]=c^\dagger S_Vc,\\
S_V&=\operatorname{sgn}(D_V)
=\frac{D_V}{\sqrt{D_V^2}}\quad(0\notin\operatorname{Spec}(D_V)).
\end{aligned}
\label{eq:1d_anom_covariant_sharp}
\end{equation}
Here $c^\dagger Bc=\sum_{jl}c_j^\dagger B_{jl}c_l$. We take $\operatorname{sgn}(0)=0$ on the kernel, or use
one-sided spectral limits when studying a crossing. Away from zero modes $S_V^2=1$, so the charge is compact and
integer-valued; at zero field it reduces to $Q_{A,0}$. For every static background with vanishing scalar
potential,
\begin{equation}
[H_V[U^V,0],Q_{A,0}^{\mathrm{cov}}[U^V]]=0,
\label{eq:1d_anom_covariant_conservation}
\end{equation}
because $T_V+T_V^\dagger$ commutes with every spectral function of $D_V$. Although $D_V$ is local, its spectral
sign $S_V$ is generally spatially nonlocal. In a time-dependent vector background, the exact Heisenberg
derivative of the covariant charge is determined by the gauge-covariant variation of this spectral projector,
rather than a local continuity equation.\footnote{For $0\notin\operatorname{Spec}(D_V)$, define
$\varphi_V=\operatorname{diag}(\varphi_{V,j})$, $\mathcal E^V=\operatorname{diag}(\mathcal E_j^V)$, and
$\mathscr D_t^VB=\dot B+i[\varphi_V,B]$. Then $\dot Q_{A,0}^{\mathrm{cov}} =c^\dagger(\mathscr D_t^VS_V)c$, with
$\mathscr D_t^VD_V =(\mathcal E^VT_V+T_V^\dagger\mathcal E^V)/2$. Since $S_V=2P_+(D_V)-1$,
\begin{equation}
\mathscr D_t^VS_V
=\frac{1}{\pi i}\oint_{\Gamma_+}dz\,
(z-D_V)^{-1}(\mathscr D_t^VD_V)(z-D_V)^{-1},
\end{equation}
where $\Gamma_+$ encloses the positive spectrum. This holds between zero crossings, where $S_V$ jumps.}

For a uniform vector holonomy $\Phi_V=\sum_j\vartheta_j^V$, choose $U_j^V=e^{i\Phi_V/N}$. Then
\begin{equation}
Q_{A,0}^{\mathrm{cov}}(\Phi_V)
=\sum_k\operatorname{sgn}\!\left[\sin\!\left(k+\frac{\Phi_V}{N}\right)\right]n_k.
\label{eq:1d_anom_flat_sharp}
\end{equation}
The fixed microscopic $Q_{A,0}$ commutes with $H_V(\Phi_V)$ at every flux and stays conserved during
uniform-flux evolution. The covariant charge instead jumps when $D_V$ acquires zero eigenvalues. With APBC, two
eigenvalues of $D_V$ cross zero simultaneously at the sole exceptional flux $\Phi_V=\pi$ in $0<\Phi_V<2\pi$.
The crossing modes are $k_-=-\pi/N$ and $k_+=\pi-\pi/N$. Hence
\begin{equation}
Q_{A,0}^{\mathrm{cov}}(\pi^+)-Q_{A,0}^{\mathrm{cov}}(\pi^-)
=2(n_{k_-}-n_{k_+}).
\label{eq:1d_anom_exact_sharp_jump}
\end{equation}

The initial half-filled sea has $n_{k_-}=1$ and $n_{k_+}=0$, and its momentum occupations remain fixed
throughout uniform-flux evolution. For a general spectral sign $s$, its covariant extension along a uniform flux
path requires an off-grid continuation. The full-cycle response, however, depends only on the original
momentum-grid values and is independent of this choice. For any $s(k)$ agreeing with $s_0(k)$ near the Fermi
points, it is
\begin{equation}
\begin{aligned}
\Delta\langle Q_{A,s}^{\mathrm{cov}}\rangle_{\mathrm{FS}}
&=\sum_{k\in\mathrm{FS}}
\left[s\left(k+\frac{2\pi}{N}\right)-s(k)\right]\\
&=s\left(\frac{\pi}{2}+\frac{\pi}{N}\right)
-s\left(-\frac{\pi}{2}+\frac{\pi}{N}\right)\\
&=2,\quad(\Phi_V:0\to2\pi).
\end{aligned}
\label{eq:1d_vector_IR_flow}
\end{equation}
For $s=s_0$, this response arises entirely from the jump at $\Phi_V=\pi$. Other spectral signs may exhibit
different intermediate jumps depending on their off-grid continuation, but the integrated response is unchanged.

The same index follows from the large vector gauge unitary
\begin{equation}
\mathscr U_{V}=\exp\!\left[\frac{2\pi i}{N}\sum_{j=0}^{N-1}j\,q_{V,j}\right].
\end{equation}
With $\delta k=2\pi/N$, it acts as
$\mathscr U_{V}c_k\mathscr U_{V}^{-1}=c_{k+\delta k}$ and maps the
$2\pi$-holonomy Hamiltonian to the zero-holonomy Hamiltonian. Its inverse adjoint action gives
\begin{equation}
\mathscr U_{V }^{-1}Q_{A,s}
\mathscr U_{V}-Q_{A,s}
=\sum_k[s(k+\delta k)-s(k)]n_k.
\label{eq:1d_large_gauge_QA}
\end{equation}
Its filled-sea expectation reproduces Eq.~\eqref{eq:1d_vector_IR_flow}. The resulting relative phase
$e^{2i\alpha}$ between axial rotation and large vector gauge transport is a monodromy of the filled-sea sector,
not a central operator identity on the full Fock space. This is the large-gauge manifestation of the mixed
anomaly already diagnosed by the Schwinger term, rather than an independent global anomaly.

This reveals complementary IR and UV manifestations of the mixed anomaly at finite lattice size. Under
vector-flux insertion, zero-energy spectral flow occurs at the Fermi points, $k+\Phi_V/N=\pm\pi/2$, whereas the
covariant axial charge for $s_0$ registers the same index through jumps at the UV sign discontinuities
$k+\Phi_V/N=0,\pi$. Other spectral signs may redistribute these UV jumps, but their integrated response is fixed
by the common IR behavior.

\paragraph{Axial gauging and complementary vector spectral flow.}
The Hamiltonian-preserving relocalization exchanges the onsite and spectral charges, as established in
Eqs.~\eqref{eq:1d_eta_H}, \eqref{eq:1d_eta_density}, and \eqref{eq:1d_QV_after_R}. In the $\eta$ coordinate,
$Q_{A,0}=\sum_jq_{A,j}$ is onsite while $Q_V$ is spectral. Using the axial links $U_j^A$ already introduced in
Eq.~\eqref{eq:1d_local_axial_gauge}, the background Peierls Hamiltonian (without dynamical electric energy) is
\begin{equation}
H_A[U^A,\varphi_A]=-t\sum_j
(\eta_j^\dagger U_j^A\eta_{j+1}+\mathrm{H.c.})+\sum_j\varphi_{A,j}q_{A,j},
\label{eq:1d_anom_dual_gauging}
\end{equation}
The gauge-invariant bond current
\begin{equation}
J^A_{j+1/2}=-it\left(
\eta_j^\dagger U_j^A\eta_{j+1}-\mathrm{H.c.}
\right)
\end{equation}
obeys the exact continuity equation
\begin{equation}
\frac{d}{dt}q_{A,j}
+J^A_{j+1/2}-J^A_{j-1/2}=0.
\label{eq:1d_anom_dual_exact_continuity}
\end{equation}
For dynamical axial links, the corresponding local Gauss-law generator was defined in
Eq.~\eqref{eq:1d_local_Gauss}. The spectator $Q_V$ is sharp and nonlocal in this frame. Defining
$D_A=(T_A-T_A^\dagger)/(2i)$, $(T_A\psi)_j=U_j^A\psi_{j+1}$, and
$Q_V^{\mathrm{cov}}[U^A]=\eta^\dagger\operatorname{sgn}(D_A)\eta$, the preceding vector-flux argument applies
verbatim. In particular,
\begin{equation}
\Delta\langle Q_V^{\mathrm{cov}}\rangle_{\mathrm{FS}}=2
\quad(\Phi_A:0\to2\pi).
\label{eq:1d_anom_dual_flux}
\end{equation}
The fixed $Q_V$ remains conserved under uniform axial-flux evolution; its covariant extension supplies the
spectral-flow index.

\paragraph{From spectral charges to local Ward identities.}
At finite $N$, the sharp covariant charge jumps at $\Phi_V=\pi$, so its flux derivative is distributional rather
than the smooth continuum response $\dot\Phi_V/\pi$. A local IR Ward identity requires a matched composite
current, not an identification of the sharp spectral sign with a differentiable observable.

One matching prescription smooths the UV jumps without changing the charges at the Fermi points. Let $F_\delta$
be smooth and odd on $[-1,1]$, equal to $\operatorname{sgn}z$ for $|z|\ge\delta$, $0<\delta<1$, and define
\begin{equation}
\begin{aligned}
f_\delta(k)&=F_\delta(\sin k),\\
Q_{A,\delta}^{\mathrm{cov}}[U^V]&=c^\dagger F_\delta(D_V)c.
\end{aligned}
\label{eq:1d_anom_smooth_matching}
\end{equation}
The $C^\infty$ symbol has rapidly decreasing (Schwartz) hopping tails, agrees exactly with $s_0$ near both Fermi
points, but is not an exactly compact charge. For uniform flux,
\begin{equation}
\frac{dQ_{A,\delta}^{\mathrm{cov}}}{dt}
=\frac{\dot\Phi_V}{N}\sum_k f_\delta'\!\left(k+\frac{\Phi_V}{N}\right)n_k.
\label{eq:1d_anom_smooth_flux_derivative}
\end{equation}
At fixed smooth $f_\delta$, taking the filled-sea expectation and then $N=\ell/a\to\infty$ yields
\begin{equation} 
\frac{d}{dt}\langle Q_{A,\delta}^{\mathrm{cov}}\rangle_{\mathrm{FS}}
\approx \frac{\dot\Phi_V}{2\pi}\int_{-\pi/2}^{\pi/2}f_\delta'(k)\,dk =\frac{\dot\Phi_V}{\pi}. 
\label{eq:1d_anom_Wilsonian_bridge}
\end{equation}
The coefficient depends only on $f_\delta(\pi/2)-f_\delta(-\pi/2)=2$ and holds for any smooth periodic symbol
with these endpoint values. The sharp and smoothed operators differ at finite cutoff, even though their
integrated one-cycle responses agree. Cutoff-dependent smoothing additionally requires control of the grid
spacing relative to $\delta$.

A strictly local axial-density representative is obtained by taking $f(k)=\sin k$. In the vector-gauged $c$ representation, define the gauge-invariant bond operators
\begin{equation}
\begin{aligned}
&\rho^A_{j+1/2}
=\frac{c_j^\dagger U_j^Vc_{j+1}-\mathrm{H.c.}}{2i},\\
&K_{j+1/2}
=\frac{c_j^\dagger U_j^Vc_{j+1}+\mathrm{H.c.}}2,\\
&C_j^V=c_j^\dagger U_j^VU_{j+1}^Vc_{j+2}.
\end{aligned}
\label{eq:1d_anom_local_bond_operators}
\end{equation}
Here $\sum_j\rho^A_{j+1/2}=c^\dagger D_Vc$, which reduces to $\sum_k\sin k\,n_k$ at zero field. This local axial-density representative is distinct from the exact compact spectral charge $Q_{A,0}$. Its associated spatial current,
\begin{equation}
J_j^A=t\left[q_{V,j}+\nu_j
-\frac{C_{j-1}^V+C_{j-1}^{V\dagger}}2\right],
\label{eq:1d_anom_local_axial_current}
\end{equation}
obeys the exact lattice Ward identity
\begin{equation}
\begin{aligned}
&\frac{d\rho^A_{j+1/2}}{dt}
+J_{j+1}^A-J_j^A
=\mathcal E_j^V K_{j+1/2},\\
&\mathcal E_j^V
=\dot\vartheta_j^V+\varphi_{V,j}-\varphi_{V,j+1}.
\end{aligned}
\label{eq:1d_anom_local_Ward}
\end{equation}
Here $J_j^A$ denotes the local axial current in the vector-gauged $c$ representation, rather than the axial gauge current in the $\eta$ representation. The electric field $\mathcal E_j^V$ receives contributions from the explicit time dependence of the link and the scalar-potential commutator. Summing over links yields
\begin{equation}
\frac{d}{dt}(c^\dagger D_Vc)
=\sum_j\mathcal E_j^V K_{j+1/2}.
\label{eq:1d_anom_local_integrated_Ward}
\end{equation}
In the half-filled free sea,
\begin{equation}
\langle K_{j+1/2}\rangle_{\mathrm{FS}}
=\frac{1}{N}\sum_{|k|<\pi/2}\cos k
=\frac{1}{N\sin(\pi/N)}
\rightarrow\frac1\pi.
\label{eq:1d_anom_kinetic_coefficient}
\end{equation}
For smooth backgrounds, $\vartheta_j^V=aA_1^V$, $\varphi_{V,j}=A_0^V$, and $\mathcal E_j^V=aF_{01}^V(x_{j+1/2})+O(a^2)$, where $F_{01}^V=\partial_0A_1^V-\partial_1A_0^V$. Matching the local composite current and taking free low-energy matrix elements gives the continuum Ward identities
\begin{equation} 
\partial_\mu j_V^\mu=0,\ \
\partial_\mu j_{A,\mathrm{ren}}^\mu
=\frac1\pi\epsilon^{\mu\nu}\partial_\mu A_\nu^V, 
\label{eq:1d_anom_vector_Ward_continuum}
\end{equation}
with $\epsilon^{01}=+1$. This identifies the free IR/adiabatic anomaly coefficient; general interacting gauge backgrounds require the usual derivative expansion and composite-operator matching. In particular, the result does not imply that the sharp spectral operator $\operatorname{sgn}(D_V)$ becomes smooth.

The axial-adapted frame has the exact mirror construction. Replacing $(c,U^V)$ by $(\eta,U^A)$ in the local bond
operators defines $\widetilde\rho^V_{j+1/2}$, $\widetilde K_{j+1/2}$, and the corresponding spatial current
$\widetilde J_j^V$. They obey
\begin{equation}
\begin{aligned}
& \frac{d\widetilde\rho^V_{j+1/2}}{dt}
+\widetilde J_{j+1}^V-\widetilde J_j^V
 =\mathcal E_j^A\widetilde K_{j+1/2},\\
& \mathcal E_j^A =\dot\vartheta_j^A+\varphi_{A,j}-\varphi_{A,j+1}.
\end{aligned}
\label{eq:1d_anom_dual_local_Ward}
\end{equation}
Since $\langle\widetilde K\rangle_{\mathrm{FS}}\to1/\pi$, the complementary continuum Ward identities are
\begin{equation} 
\partial_\mu j_A^\mu =0,\ \
\partial_\mu j_{V,\mathrm{ren}}^\mu
=\frac1\pi\epsilon^{\mu\nu}\partial_\mu A_\nu^A.
\label{eq:1d_anom_axial_Ward_continuum}
\end{equation}
Both $U(1)_V$ and $U(1)_A$ are individually gaugeable in their respective localizations, but their mixed anomaly
prevents simultaneous anomaly-free gauging in a standalone one-Dirac regulator.

\subsubsection{Translation-selected axial symmetry and LSM pumping}

The relation between LSM constraints and IR chiral anomalies was established in, e.g., Ref.~\cite{cho2017anomaly};
Cheng and Seiberg~\cite{cheng2023lieb} named the internal action induced by translation an emanant
symmetry. Here that action lifts to an exact compact spectral axial $U(1)$, and its flux response reproduces the
LSM momentum pump.

Consider a translation-invariant chain at filling $\nu$, with Fermi momenta $\pm k_F$, where $k_F=\pi\nu$. Let
$P_{\rm chain}=\sum_k k n_k$ denote the microscopic momentum modulo $2\pi$. For a spectral sign $s(k)=\pm1$
agreeing with the axial charge assignments near the Fermi points, define
\begin{equation}
q_s(k)=k-k_Fs(k),
\quad
P_s=\sum_kq_s(k)n_k.
\end{equation}
The microscopic momentum then admits the exact decomposition
\begin{equation}
P_{\rm chain}=P_s+k_FQ_{A,s}
\pmod{2\pi}.
\label{eq:1d_Pfactor}
\end{equation}
Indeed, near either Fermi point, $k=\pm k_F+p$ and $s(k)=\pm1$, so $q_s(k)=p$. Accordingly, the microscopic
translation phase separates as $e^{ik}=e^{ip}e^{\pm ik_F}$, where the first factor represents long-wavelength
spatial translation and the second acts internally on the Dirac branches. With the convention that translation
acts on annihilation operators as $Tc_kT^{-1}=e^{ik}c_k$, we have $T=e^{-iP_{\rm chain}}$, and hence
\begin{equation}
T\xrightarrow{\rm IR}
e^{-iP_{\rm IR}}e^{-ik_FQ_A^{\rm IR}}.
\label{eq:1d_Tfactor}
\end{equation}
The internal factor is the translation-selected emanant symmetry, whose compact microscopic lift is
$C_{T,s}=e^{-ik_FQ_{A,s}}$. Equation~\eqref{eq:1d_Pfactor} is exact, whereas the identification of $P_s$ with
ordinary spatial momentum applies to the IR Dirac branches.

The axial relocalization provides a distinct microscopic realization of this structure. At half filling, the
exact compact axial charge becomes onsite while the Hamiltonian remains nearest-neighbor in the symmetry-adapted
lattice, whose coordinate recovers physical space in the IR. The translation-selected axial transformation is
therefore onsite in this representation, although the original microscopic one-site translation becomes
nonlocal. Thus relocalization changes the microscopic locality of the symmetries without altering their IR
action or the associated LSM anomaly.

The LSM momentum pump follows from the large vector gauge unitary $\mathscr U_{V}$ defined above.
With $Tn_jT^{-1}=n_{j+1}$, one obtains the exact operator relation
\begin{equation}
T\mathscr U_{V}T^{-1}
=e^{-i2\pi (Q_V+M)/N}\mathscr U_{V}.
\end{equation}
In a sector with $N_V=Q_V+M=N\nu$, the state transformed by $\mathscr U_{V}$ consequently acquires
an additional translation phase $e^{-i2\pi\nu}$, corresponding to the momentum shift
\begin{equation}
\Delta P_{\rm chain}=2\pi\nu
\pmod{2\pi}.
\label{eq:1d_LSM_momentum_shift}
\end{equation}
This is the standard LSM momentum pump \cite{yamanakaNonperturbative1997, oshikawa2000}.

Our exact spectral calculation provides a complementary derivation. The same large-gauge identification shifts
the occupied momenta by $2\pi/N$. For the filled two-Fermi-point sea, Eq.~\eqref{eq:1d_vector_IR_flow}
gives an axial-charge shift $\Delta Q_{A,s}=2$ under large-gauge identification. Using the exact
decomposition in Eq.~\eqref{eq:1d_Pfactor}, we find
\begin{equation}
\begin{aligned}
\Delta P_s
&=\Delta P_{\rm chain}-k_F\Delta Q_{A,s}\\
&=2\pi\nu-2k_F=0
\pmod{2\pi}.
\end{aligned}
\end{equation}
Thus the momentum pump of this state is entirely accounted for by the spectral axial contribution,
\begin{equation}
\Delta P_{\rm chain}
=k_F\Delta Q_{A,s}
=2k_F=2\pi\nu
\pmod{2\pi}.
\end{equation}
Here the axial charge change refers to the large-gauge identification, not to nonconservation of the fixed
spectral charge during uniform-flux evolution. In the IR, this identifies the LSM momentum pump with the
restriction of the mixed $U(1)_V$--$U(1)_A$ anomaly to the translation-selected axial transformation.

\subsection{Bulk--edge origin of vector--axial duality}
\label{sec:1d_bulk_edge_spt}

The vector--axial duality of the 1D chain has an exact 2D bulk--edge realization. We construct a
nearest-neighbor fermionic SPT parent whose normalizable edge states reproduce $H_0$ and its compact spectral
axial charge. An onsite particle--hole transformation of the parent induces the nonlocal vector--axial exchange
upon unfolding. The correspondence extends to uniform gauge holonomies and reproduces the mixed-anomaly spectral
flow. Crucially, the selected edge subspace need not remain separately gauge covariant, even though the complete
parent is.

\subsubsection{Exact bulk--edge realization and onsite origin}
\label{sec:1d_bulk_edge_exact}

\paragraph{Doubled Chern-insulator parent.}
Consider two decoupled two-orbital Chern insulators~\cite{qi2006}, labeled by $\chi=\pm1$, with Bloch
Hamiltonians
\begin{equation}
\begin{aligned}
h_\chi(k_x,k_y)=2t\big[&\chi\sin k_x\,\sigma_x
-\sin k_y\,\sigma_y\\
&+(\cos k_x+\cos k_y-1)\sigma_z\big].
\end{aligned}
\label{eq:1d_bulk_edge_parent}
\end{equation}
Here $\mathbf k=(k_x,k_y)$ is the 2D crystal momentum, $t>0$, and $\sigma_i$ act on the two orbitals. Writing
$h_\chi=2t\,\mathbf d_\chi\cdot\boldsymbol\sigma$ gives
\begin{equation}
|\mathbf d_\chi|^2
=1+2(1-\cos k_x)(1-\cos k_y)\ge1,
\end{equation}
so the bulk is gapped by at least $2t$ about zero energy. Let $P_\chi(k_x,k_y)$ project onto the negative-energy
band. With the convention
\begin{equation}
C_\chi=\frac{1}{2\pi i}
\int_{\mathrm{BZ}}dk_x\,dk_y\,
\mathrm{tr}\left(
P_\chi[\partial_{k_x}P_\chi,\partial_{k_y}P_\chi]
\right),
\end{equation}
the Chern numbers are $C_\chi=\chi$, with the chosen momentum-space orientation. We retain this
projector-curvature normalization throughout.

\begin{figure}[tb]
\centering
\includegraphics[width=\columnwidth]{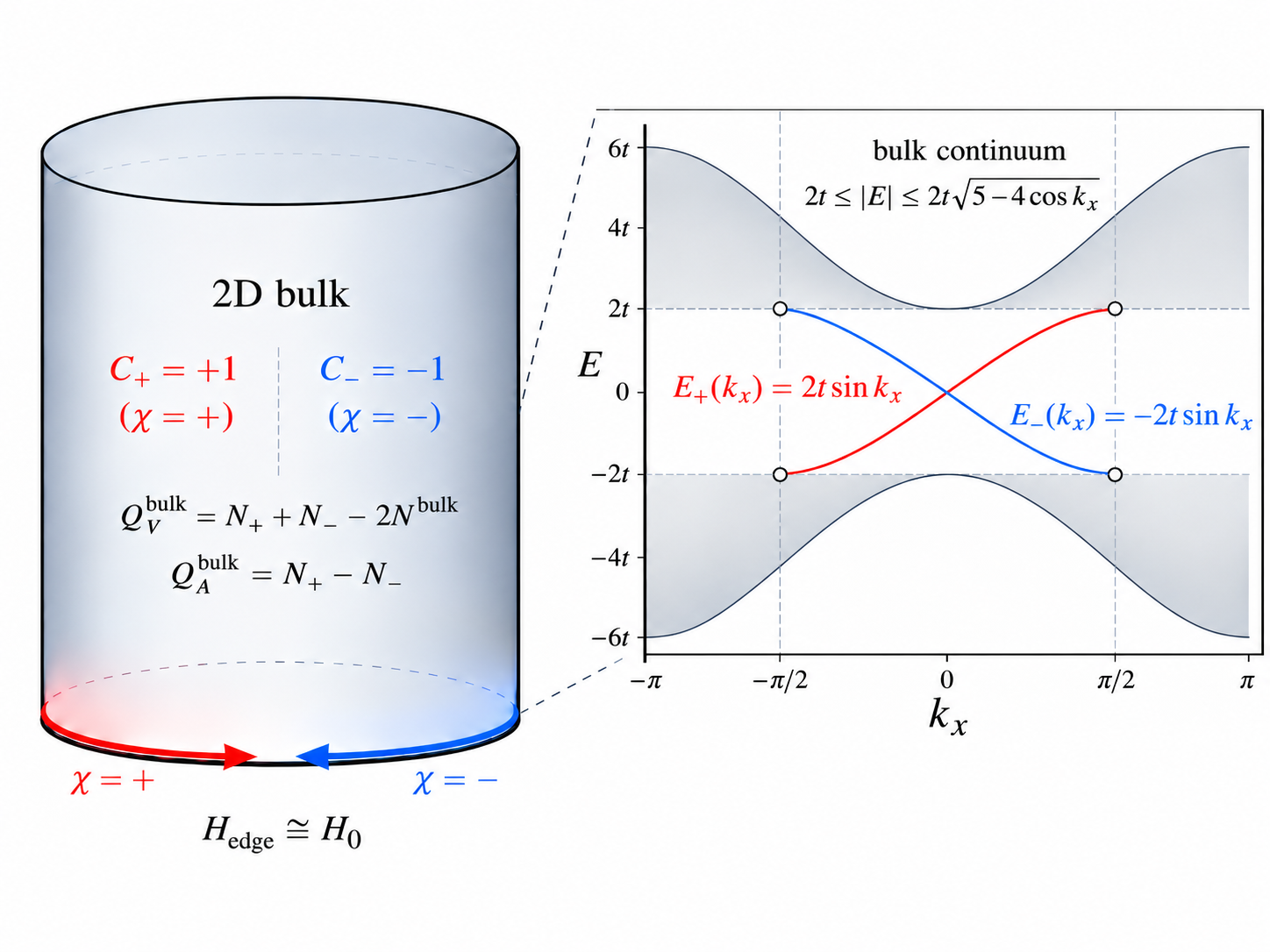}
\caption{Bulk--edge origin of vector--axial duality.
A doubled Chern insulator with $C_+=+1$ (red) and $C_-=-1$ (blue) is placed on a half-cylinder
with one boundary. The sectors support counterpropagating edge modes with exact dispersions
$E_\chi(k_x)=2t\chi\sin k_x$, normalizable for $|k_x|<\pi/2$. The right panel shows the
band structure. Gray regions denote the bulk continuum,
$2t\le |E|\le 2t\sqrt{5-4\cos k_x}$; open circles mark where the edge modes cease to be
normalizable. After unfolding, the boundary Hamiltonian is exactly the 1D chain $H_0$.
The centered bulk vector and axial charges are onsite and restrict to the 1D charges
$Q_V$ and $Q_{A,0}$. Particle--hole conjugation of the negative-Chern sector induces
$\mathcal W$ on the boundary.}
\label{fig:bulk_completion}
\end{figure}

The parent has onsite centered vector and axial charges
\begin{equation}
Q_V^{\rm bulk}=N_++N_--2N^{\rm bulk},
\ \
Q_A^{\rm bulk}=N_+-N_-,
\end{equation}
Here $N^{\rm bulk}$ counts primitive bulk unit cells, each carrying two orbitals per Chern sector, and $N_\chi$
counts fermions in sector $\chi$. Each sector has charges $(q_V,q_A)=(1,\chi)$. The conserved axial pseudospin
$S_z^{\rm bulk}=Q_A^{\rm bulk}/2$ has one-particle generator $\mu_z/2$, where $\mu_i$ act on the two Chern
sectors (flavors); it need not be physical spin.

Introducing the primitive background connections $B_\pm=A_V\pm A_A$, the long-wavelength bulk response is
\begin{equation}
S_{\rm resp}
=\frac{1}{4\pi}\int_{M_3}
\left(B_+\wedge dB_+-B_-\wedge dB_-\right).
\label{eq:1d_bulk_edge_response}
\end{equation}
Here $M_3$ denotes the $(2{+}1)$D bulk spacetime with a $(1{+}1)$D boundary. The Chern numbers yield $\mathcal
A_{VA}=\mathcal A_{AV}=2$ and $\mathcal A_{VV}=\mathcal A_{AA}=0$, matching Sec.~\ref{sec:1d_anomaly}. The
primitive-sector $K$-matrix is $\operatorname{diag}(1,-1)$: the parent is an invertible fermionic SPT, not an
intrinsically topologically ordered phase. Its faithful symmetry group is $[U(1)_V\times U(1)_A]/\mathbb Z_2$,
so compact vector and axial fluxes cannot be chosen independently without respecting this quotient.

\paragraph{Exact edge-bound-state subspace.}
Place the parent on a half-cylinder with $y\ge1$, an open boundary at $y=1$, and APBC along the $x$ direction of
circumference $N\in4\mathbb Z$. At fixed edge momentum $k_x$, write
\begin{equation}
h_{\chi,0}(k_x)
=2t\left[\chi\sin k_x\,\sigma_x
+(\cos k_x-1)\sigma_z\right],
\label{eq:1d_bulk_edge_hopping}
\end{equation}
and $T=t(\sigma_z+i\sigma_y)$, so that
\begin{equation}
h_\chi(k_x,k_y)=h_{\chi,0}(k_x)+Te^{ik_y}+T^\dagger e^{-ik_y}.
\end{equation}
With
\begin{equation}
u=\frac{1}{\sqrt2}\binom11,
\quad
v=\frac{1}{\sqrt2}\binom1{-1},
\end{equation}
the identities $T^\dagger u=0$ and $Tu=2tv$ give the exact normalized edge eigenstate
\begin{equation}
\psi_{\chi,k_x}(y)
 =\sqrt{1-\lambda(k_x)^2}\, \lambda(k_x)^{y-1}u
\label{eq:1d_bulk_edge_exact_mode}
\end{equation}
with $
\lambda(k_x) =1-\cos k_x$ and $E_\chi(k_x)=2t\chi\sin k_x$.
The mode is normalizable precisely for $|k_x|<\pi/2$. This termination has one edge branch per Chern sector,
with no additional in-gap branches; together the normalizable modes span the spectral subspace with energies in
$(-2t,2t)$.

Each branch contains exactly $N/2$ modes on the APBC grid. Let $a_{\chi,k_x}$ annihilate the corresponding edge
state. The restricted Hamiltonian and axial charge are
\begin{equation}
\begin{aligned}
H_{\rm edge}
&=\sum_{\chi,k_x}
2t\chi\sin k_x\,a_{\chi,k_x}^\dagger a_{\chi,k_x},\\
Q_A^{\rm edge}
&=\sum_{k_x}
\left(a_{+,k_x}^\dagger a_{+,k_x}
-a_{-,k_x}^\dagger a_{-,k_x}\right),
\end{aligned}
\label{eq:1d_bulk_edge_restriction}
\end{equation}
where $|k_x|<\pi/2$. Since $\partial_{k_x}E_\chi(k_x)=2t\chi\cos k_x$, the sector pseudospin is locked to the
propagation direction: $\chi=+$ is right-moving and $\chi=-$ is left-moving.

Unfold the two branches into the complete one-dimensional Brillouin zone through
\begin{equation}
c_k=
\begin{cases}
a_{+,k-\pi/2},&0<k<\pi,\\
a_{-,k+\pi/2},&-\pi<k<0.
\end{cases}
\label{eq:1d_bulk_edge_unfolding}
\end{equation}
The shifts preserve the APBC momentum grid because $N\in4\mathbb Z$. Both edge dispersions become $-2t\cos k$,
yielding the exact finite-size identities
\begin{equation}
\begin{aligned}
H_{\rm edge}
&=\sum_k(-2t\cos k)n_k=H_0,\\
Q_A^{\rm edge}
&=\sum_k\operatorname{sgn}(\sin k)n_k
=Q_{A,0},\\
Q_V^{\rm edge}
&=\sum_kn_k-\frac N2=Q_V.
\end{aligned}
\label{eq:1d_bulk_edge_chain_identity}
\end{equation}
Thus the compact spectral axial charge of the 1D chain is exactly the restriction of an onsite bulk symmetry.
The induced edge pseudospin is
\begin{equation}
S_z^{\rm edge}
=\frac12Q_{A,0}
=\frac12\sum_k\operatorname{sgn}(\sin k)n_k.
\end{equation}
The sharp spectral sign follows from unfolding the truncated edge branches; it is not an additional symmetry
imposed on the boundary. Fig.~\ref{fig:bulk_completion} illustrates the exact dispersions.

\paragraph{Onsite bulk origin of the vector--axial duality.}
The parent has an onsite particle--hole transformation acting only on its negative-Chern sector:
\begin{equation}
\mathcal C_-\Psi_{+,\mathbf r}\mathcal C_-^{-1}
=\Psi_{+,\mathbf r},\ \
\mathcal C_-\Psi_{-,\mathbf r}\mathcal C_-^{-1}
=\sigma_x\Psi_{-,\mathbf r}^{\dagger T}.
\label{eq:1d_bulk_edge_ph}
\end{equation}
Here $\Psi^{\dagger T}$ denotes a column of creation operators. The identity
\begin{equation}
h_\chi(k_x,k_y)
=-\sigma_xh_\chi^T(-k_x,-k_y)\sigma_x
\label{eq:1d_bulk_edge_ph_identity}
\end{equation}
implies $\mathcal C_-H_{\rm bulk}\mathcal C_-^{-1}=H_{\rm bulk}$. The transformation is onsite and canonically
implementable at finite volume. Since the negative-Chern sector contains $2N^{\rm bulk}$ complex orbital modes,
$N_-\mapsto2N^{\rm bulk}-N_-$, and hence
\begin{equation}
\begin{aligned}
\mathcal C_-Q_V^{\rm bulk}\mathcal C_-^{-1}
&=Q_A^{\rm bulk},\\
\mathcal C_-Q_A^{\rm bulk}\mathcal C_-^{-1}
&=Q_V^{\rm bulk}.
\end{aligned}
\label{eq:1d_bulk_edge_exchange_bulk}
\end{equation}
On the half-infinite cylinder, this is understood as an onsite canonical automorphism, with extensive charges
defined relative to a reference sea.

Since the edge wavefunction is real, even in $k_x$, and satisfies $\sigma_xu=u$, the transformation preserves
the edge-mode algebra and acts as
\begin{equation}
a_{+,k_x}\longmapsto a_{+,k_x},
\quad
a_{-,k_x}\longmapsto a_{-,-k_x}^\dagger.
\end{equation}
After unfolding,
\begin{equation}
c_k\longmapsto
\begin{cases}
c_k,&0<k<\pi,\\
c_{-\pi-k}^\dagger,&-\pi<k<0.
\end{cases}
\label{eq:1d_bulk_edge_half_zone_duality}
\end{equation}
This preserves $H_0$ and exchanges $Q_{A,0}$ with $Q_V^{\rm edge}$. Writing $P_\pm=(1\pm S_0)/2$, where
$S_0(k)=\operatorname{sgn}(\sin k)$, its real-space action is
\begin{equation}
\begin{aligned}
c_j&\longmapsto
(P_+c)_j+(-1)^j(P_-c)_j^\dagger\\
&=  \sum_{\ell} F_{j\ell}
\left[c_{\ell}+(-1)^j c_{\ell}^\dagger\right]
=\eta_j,
\end{aligned}
\label{eq:1d_bulk_edge_realspace_duality}
\end{equation}
where Eq.~\eqref{eq:1d_eta_microscopic_raw} is used. Thus an onsite bulk symmetry induces precisely the
Hamiltonian-preserving relocalization $\mathcal W$ of the 1D chain, including its algebraically decaying kernel.

\paragraph{Additional symmetries and their edge realizations.}
The parent also has discrete symmetries that may become nonlocal after unfolding. With the Chern-sector Pauli
matrices $\mu_i$, the four-band Bloch Hamiltonian is
\begin{equation}
\begin{aligned}
h(k_x,k_y)=2t\big[&\sin k_x\,\mu_z\sigma_x
-\sin k_y\,\sigma_y\\
&+(\cos k_x+\cos k_y-1)\sigma_z\big].
\end{aligned}
\label{eq:1d_bulk_edge_four_band}
\end{equation}
It has two time-reversal implementations,
\begin{equation}
\mathcal T_+=\mu_xK,\quad
\mathcal T_-=i\mu_yK=\mu_z\mathcal T_+,
\label{eq:1d_bulk_edge_two_TR}
\end{equation}
where $K$ denotes complex conjugation. Their squares on the one-particle space are $+1$ and $-1$, respectively,
with $\mathcal T_-^2=(-1)^F$ on Fock space. Both preserve vector number and reverse axial charge. Since the edge
wavefunctions are real and even in $k_x$, these transformations preserve the exact edge subspace, acting as
\begin{equation}
\mathcal T_+a_{\chi,k_x}\mathcal T_+^{-1}
=a_{-\chi,-k_x},\ \
\mathcal T_-a_{\chi,k_x}\mathcal T_-^{-1}
=\chi\,a_{-\chi,-k_x}.
\label{eq:1d_bulk_edge_TR_actions}
\end{equation}
The edge Hamiltonian is $h_{\rm edge}(k_x)=2t\sin k_x\,\mu_z$. Its crossing at $k_x=0$ is a Kramers pair under
$\mathcal T_-$, although this momentum is excluded by the finite-size APBC grid. A vector-charge-preserving mass
$m_x\mu_x+m_y\mu_y$ is allowed by $\mathcal T_+$ but forbidden by $\mathcal T_-$ and by the conserved axial
$U(1)$. Thus either $\mathcal T_-$ together with $U(1)_V$, or the conserved axial $U(1)$, protects the edge
crossing against quadratic masses. With $C_+=1$ and $C_-=-1$, the parent is a spin-conserving quantum spin Hall
representative with pseudospin Chern number $C_s=(C_+-C_-)/2=1$.

In the periodic bulk, additional unitary symmetries are
\begin{equation}
\mathcal M_x=\mu_x\mathcal R_x,\ \
\mathcal M_y=\mu_x\sigma_z\mathcal R_y,\ \
\mathcal I=\sigma_z\mathcal R_x\mathcal R_y,
\label{eq:1d_bulk_edge_reflections}
\end{equation}
where $\mathcal R_x$ and $\mathcal R_y$ reverse the corresponding coordinates. The edge-preserving reflection
$\mathcal M_x$ induces
\begin{equation}
\mathcal M_xa_{\chi,k_x}\mathcal M_x^{-1}
=a_{-\chi,-k_x}.
\label{eq:1d_bulk_edge_edge_reflection}
\end{equation}
The other two operations reverse the transverse coordinate and need not preserve the chosen boundary. The Bloch
Hamiltonian also anticommutes with the spectral chiral operator $\Gamma=\mu_x\sigma_x$, which reverses
one-particle energies rather than defining a conserved charge. Particle--hole conjugation of the positive-Chern
sector, $\mathcal C_+$, is another onsite Bogoliubov symmetry, acting on the centered charges as $(Q_V^{\rm
bulk},Q_A^{\rm bulk})\mapsto(-Q_A^{\rm bulk},-Q_V^{\rm bulk})$.

After unfolding, the edge-preserving symmetries act on the 1D modes as
\begin{equation}
\begin{aligned}
\mathcal T_+:\quad c_k&\longmapsto c_{-k},\\
\mathcal T_-:\quad c_k&\longmapsto s_0(k)c_{-k},\\
\mathcal M_x:\quad c_k&\longmapsto c_{-k}.
\end{aligned}
\label{eq:1d_bulk_edge_unfolded_symmetries}
\end{equation}
Although $\mathcal T_+$ and $\mathcal M_x$ induce the same momentum permutation, their different antiunitary
character distinguishes their real-space actions: $\mathcal T_+c_j\mathcal T_+^{-1}=c_j$, whereas $\mathcal
M_xc_j\mathcal M_x^{-1}=c_{-j}$. By contrast,
\begin{equation}
\mathcal T_-c_j\mathcal T_-^{-1}=-(S_0c)_j,
\label{eq:1d_bulk_edge_nonlocal_TR}
\end{equation}
which has algebraically decaying spatial tails. Thus an onsite time-reversal symmetry of the parent can become
nonlocal after unfolding.

The momentum offsets in the unfolding also give the exact relation
\begin{equation}
P_{\rm edge}
=P_{\rm chain}-\frac{\pi}{2}Q_{A,0}
=P_s
\pmod{2\pi},
\label{eq:1d_bulk_edge_translation_relation}
\end{equation}
where $P_{\rm edge}=\sum_{\chi,k_x}k_x\,a_{\chi,k_x}^\dagger a_{\chi,k_x}$ and $P_{\rm chain}=\sum_k k n_k$.
Hence physical edge translation differs from translation of the unfolded chain by the same axial factor
appearing in Eq.~\eqref{eq:1d_Pfactor}. This provides an exact higher-dimensional realization of the
translation-selected symmetry discussed in the LSM analysis.

\subsubsection{Gauge backgrounds and induced boundary coupling}
\label{sec:1d_bulk_edge_gauge}

The onsite vector--axial exchange extends to arbitrary compact lattice backgrounds. For compact link angles $A_{V,e},A_{A,e}$, the combinations
$B_{\chi,e}=A_{V,e}+\chi A_{A,e}$ give Peierls factors $U_{\chi,e}=e^{iB_{\chi,e}}$. These
link angles are the positive oriented continuum line integrals, as above. Together with temporal couplings to
centered onsite densities, the onsite particle--hole transformation gives
\begin{equation}
\mathcal C_-H_{\rm bulk}[B_+,B_-]\mathcal C_-^{-1}
=H_{\rm bulk}[B_+,-B_-].
\label{eq:1d_bulk_edge_background_exchange}
\end{equation}
Thus $(A_V,A_A)\mapsto(A_A,A_V)$ is an exact finite-lattice equivalence for arbitrary compact backgrounds,
independently of whether the edge subspace is preserved.

To examine the induced boundary coupling, let $P_0$ project onto the normalizable zero-background edge modes of
Eq.~\eqref{eq:1d_bulk_edge_exact_mode}, and $Q_0=1-P_0$ onto the complementary subspace. The edge penetration
depth
\begin{equation}
\xi_y(k_x)=\frac{1}{-\log(1-\cos k_x)}
\end{equation}
diverges as $|k_x|\to\pi/2$. On the APBC circumference, $\xi_{\max}\sim N/\pi$, while the smallest separation
from the bulk continuum is
\begin{equation}
\Delta_{\rm edge-bulk}
=2t\left[1-\cos\left(\frac{\pi}{N}\right)\right]
\sim\frac{t\pi^2}{N^2}.
\end{equation}
The selected edge band is therefore not uniformly isolated as $N$ grows. Moreover, generic gauge transformations
mix edge and complementary states, $Q_0G(g)P_0\neq0$, so the projected transformation $P_0G(g)P_0$ need not be
unitary. The full parent remains exactly gauge covariant, but its fixed-band projection does not.

The boundary matter coupling induced by the parent can be obtained explicitly. Consider static backgrounds
$A_{I,0}=A_{I,y}=0$ and $A_{I,x}(j,y)=A_{I,x}(j)$, with arbitrary dependence along the boundary, where $I=V,A$.
Write $\vartheta_j^I=aA_{I,x}(j+\tfrac12)$ and $U_{\chi,j}=e^{i(\vartheta_j^V+\chi\vartheta_j^A)}$. The
longitudinal hopping matrix is $T_{x,\chi}=t(\sigma_z-i\chi\sigma_x)$. Since the edge spinor satisfies
$u^\dagger\sigma_z u=0$ and $u^\dagger\sigma_xu=1$, only the $\sigma_x$ hopping channel contributes after
projection.

To evaluate the projected hopping, recall the unfolding $k=k_x+\chi\pi/2$ of
Eq.~\eqref{eq:1d_bulk_edge_unfolding}, where $k_x$ is the parent edge momentum and $k$ belongs to $\mathcal
I_+=(0,\pi)$ or $\mathcal I_-=(-\pi,0)$. The transverse decay factor in Eq.~\eqref{eq:1d_bulk_edge_exact_mode}
becomes $\lambda(k)=1-|\sin k|$. Writing $f_y(k)=\sqrt{1-\lambda(k)^2}\,\lambda(k)^{y-1}$, the edge-filtered
fields and their longitudinal bond operators are
\begin{equation}
\begin{aligned}
d_{\chi,y}(j)
&=\frac1{\sqrt N}\sum_{k\in\mathcal I_\chi}
e^{ikj}f_y(k)c_k,\\
\mathcal B_{\chi,j}
&=\sum_{y\ge1}d_{\chi,y}^\dagger(j)d_{\chi,y}(j+1).
\end{aligned}
\end{equation}
Substituting the Fourier expansion, the transverse sum gives $\mathcal F(k,k')=\sum_{y\ge1}f_y(k)f_y(k')$, while
the longitudinal Fourier transform determines the bond kernels:
\begin{align}
\mathcal F(k,k')
&=\frac{\sqrt{[1-\lambda(k)^2][1-\lambda(k')^2]}}
{1-\lambda(k)\lambda(k')},
\label{eq:app_1d_bulk_edge_overlap}\\
\mathcal K_\chi(r,s)
&=\frac1{N^2}\sum_{k,k'\in\mathcal I_\chi}
 e^{ikr-ik's}\mathcal F(k,k').
\label{eq:app_1d_bulk_edge_G_kernels}
\end{align}
The vector and axial bond operators are $\mathcal B_{V,j}=\mathcal B_{+,j}+\mathcal B_{-,j}$ and $\mathcal
B_{A,j}=\mathcal B_{+,j}-\mathcal B_{-,j}$, with corresponding kernels $\mathcal K_V=\mathcal K_++\mathcal K_-$
and $\mathcal K_A=\mathcal K_+-\mathcal K_-$.

Combining the opposite axial Peierls phases of the two Chern sectors, the projected Hamiltonian takes the form
\begin{equation}
H_{PP}[\vartheta_V,\vartheta_A]
=-t\sum_j\bigl[e^{i\vartheta_j^V}\mathcal C_j
+\mathrm{H.c.}\bigr],
\end{equation}
with
\begin{equation}
\begin{aligned}
&\mathcal C_j
=\cos\vartheta_j^A\,\mathcal B_{V,j}
+i\sin\vartheta_j^A\,\mathcal B_{A,j},\\
&\mathcal B_{I,j}
=\sum_{m,n}\mathcal K_I(r,s)c_m^\dagger c_n,
\end{aligned}
\label{eq:1d_bulk_edge_simultaneous_realspace}
\end{equation}
where $(r,s)=(m{-}j,n{-}j{-}1)$ and $I=V,A$. Since the edge states are normalized, $\mathcal F(k,k)=1$, ensuring
$H_{PP}[0,0]=H_0$ exactly.

For pure vector or axial backgrounds, Eq.~\eqref{eq:1d_bulk_edge_simultaneous_realspace} reduces to
\begin{align}
H_{PP}[\vartheta,0]
&=-t\sum_j\bigl[e^{i\vartheta_j}\mathcal B_{V,j}
+\mathrm{H.c.}\bigr],\\
H_{PP}[0,\vartheta]
&=-t\sum_j\bigl[\cos\vartheta_j\,\mathcal B_{V,j}+i\sin\vartheta_j\,\mathcal B_{A,j} \nonumber
+\mathrm{H.c.}\bigr].
\label{eq:1d_bulk_edge_vector_axial_parallel}
\end{align}
A uniform vector link yields the ordinary Peierls coupling in $c_j$, while a uniform axial link gives
$-2t\sum_k\cos[k+s_0(k)\vartheta]n_k$. For nonuniform backgrounds, the transverse overlaps generate nonlocal
bond kernels, so the fixed-band projection differs from autonomous Peierls gauging in either the $c$ or $\eta$
coordinate.

The nonlocal kernels nevertheless recover the local continuum gauge vertices. Near either Fermi point of the
unfolded chain, $k=\chi\pi/2+a\kappa$ and $k'=\chi\pi/2+a\kappa'$ give
\begin{equation}
\mathcal F(k,k')
=1-\frac{a^4}{8}(\kappa^2-\kappa'^2)^2+O(a^6).
\label{eq:app_1d_bulk_edge_form_factor_IR}
\end{equation}
Thus the transverse overlap approaches unity at fixed physical momenta, and smooth backgrounds recover the local
vector and axial couplings despite the nonlocal finite-cutoff kernels.

The projected Hamiltonians also retain the exact vector--axial exchange induced by the bulk. The relocalization
$\mathcal W$ of Eq.~\eqref{eq:1d_exact_H_preserving_VA_exchange} leaves $\mathcal B_{+,j}$ unchanged and, using
the even, real edge wavefunction with $\sigma_xu=u$, gives $\mathcal W\mathcal B_{-,j}\mathcal W^{-1} =\mathcal
B_{-,j}^\dagger-\gamma_N$, where $\gamma_N=i/[N\sin(\pi/N)]$. Consequently,
\begin{equation}
\mathcal WH_{PP}[\vartheta_V,\vartheta_A]\mathcal W^{-1}
=H_{PP}[\vartheta_A,\vartheta_V]
+C_N[\vartheta_V-\vartheta_A],
\label{eq:1d_bulk_edge_projected_duality}
\end{equation}
with
\begin{equation}
C_N[\vartheta]
=-\frac{2t}{N\sin(\pi/N)}\sum_j\sin\vartheta_j.
\end{equation}
In particular, $\mathcal WH_{PP}[\vartheta,0]\mathcal W^{-1} =H_{PP}[0,\vartheta]+C_N[\vartheta]$. The scalar
term arises from particle--hole normal ordering. This exchange relates the bulk-induced projected background
Hamiltonians, not independently gauged 1D regulators.

The fixed-band Hamiltonian $H_{PP}$ does not include all processes induced by a nonuniform background. An edge
fermion can also scatter into complementary bulk states and return, producing a self-energy in the edge Green's
function. At fixed Fermi velocity $v_F=2at_a$, the bulk excitation gap is $\Delta_{\rm bulk}=2t_a=v_F/a$,
whereas bounded physical backgrounds produce hopping perturbations of $O(1)$. The self-energy is therefore
$O(a)$ at fixed low energy, so $H_{PP}$ captures the continuum edge matter coupling.

This suppression does not make the complementary bulk contribution negligible in the gauge response. Let
$D[A]=\partial_\tau+h_{\rm bulk}[A]$ be the Euclidean fermion operator, with blocks $D_{XY}=XD[A]Y$ for
$X,Y\in\{P_0,Q_0\}$. When the complementary block $D_{QQ}$ is invertible on $\operatorname{Ran}Q_0$, the
Gaussian determinant satisfies
\begin{equation}
\det D_{\rm full}
=\det D_{QQ}\,\det(D_{PP}-D_{PQ}D_{QQ}^{-1}D_{QP}),
\label{eq:app_1d_bulk_edge_gaussian_factorization}
\end{equation}
with consistent regularization. The second factor is the Schur-complement effective edge determinant,
incorporating virtual edge--bulk transitions. The first retains the complementary bulk contribution, which can
carry the topological gauge response responsible for anomaly inflow even when the low-energy edge self-energy
vanishes. Thus the continuum edge coupling emerges from the projected Hamiltonian, while gauge covariance and
anomaly inflow belong to the complete parent.

\subsubsection{Mixed-anomaly inflow and boundary determinants}
\label{sec:1d_bulk_edge_anomaly_inflow}

Uniform vector holonomies admit an exact finite-size bulk--edge correspondence using the instantaneous,
background-dependent edge subspace. For $B_{\chi,x}=\Phi_V/N$, $B_{\chi,y}=0$, and $\delta=\Phi_V/N$, the edge
modes of $h_\chi(k_x+\delta,k_y)$ are
\begin{equation}
\begin{aligned}
\psi_{\chi,k_x;\Phi_V}(y)
&=\sqrt{1-\lambda_\Phi(k_x)^2}\,
\lambda_\Phi(k_x)^{y-1}u,\\
\lambda_\Phi(k_x)&=1-\cos(k_x+\delta),\\
E_{\chi,k_x}(\Phi_V)&=2t\chi\sin(k_x+\delta).
\end{aligned}
\label{eq:1d_bulk_edge_flat_flux_modes}
\end{equation}
Normalizability requires $\cos(k_x+\delta)>0$, selecting $N/2$ momenta in each sector away from exceptional
holonomies. Under the unfolding $k=k_x+\chi\pi/2$, the dispersions become $-2t\cos(k+\delta)$, while
$\chi=\operatorname{sgn}[\sin(k+\delta)]$. Therefore,
\begin{equation}
\begin{aligned}
H_{\rm edge}(\Phi_V)
&=\sum_k[-2t\cos(k+\delta)]\,
c_k^\dagger(\Phi_V)c_k(\Phi_V),\\
Q_{A,\rm edge}(\Phi_V)
&=\sum_k\operatorname{sgn}[\sin(k+\delta)]\,
c_k^\dagger(\Phi_V)c_k(\Phi_V).
\end{aligned}
\label{eq:1d_bulk_edge_flat_flux_match}
\end{equation}
Thus the instantaneous edge Hamiltonian and restricted bulk axial charge reproduce the Peierls-coupled
autonomous chain and its covariant spectral charge exactly, without taking a continuum limit.

This equality extends to the static fermionic determinant. For $|\Phi_V|<\pi$, the instantaneous edge subspace
is invariant under the bulk Hamiltonian, and its spectrum coincides with that of the autonomous chain. With
identical thermal boundary conditions and Gaussian normalization, their holonomy-dependent determinants
therefore satisfy
\begin{equation}
\log\frac{Z_{\rm edge,a}(\Phi_V)}{Z_{\rm edge,a}(0)}
=\log\frac{Z_{\rm aut,a}(\Phi_V)}{Z_{\rm aut,a}(0)}.
\label{eq:1d_bulk_edge_flat_det_est}
\end{equation}
At fixed physical circumference, inverse temperature, and Fermi velocity $v_F=2at_a$, both approach the
continuum Dirac determinant. The exact finite-lattice equality concerns static flat backgrounds, however, and
does not imply equivalence of flux-insertion dynamics.

Indeed, for $N\in4\mathbb Z$ with APBC, $\Phi_V=\pi\pmod{2\pi}$ is exceptional. The endpoint momenta
$k_x+\delta=\pm\pi/2$ simultaneously reach the bulk continuum, where their edge wavefunctions cease to be
normalizable. The edge-subspace rank temporarily drops from $N$ to $N-2$. Across this holonomy, one state per
Chern sector enters the continuum and another emerges. The full parent remains gauge covariant, but its
normalizable edge states cannot be identified through a globally regular fixed-rank subspace over the entire
flux cycle.

The exact lattice dispersions also determine the zero-energy spectral flow. During a $2\pi$ vector-flux cycle,
each sector has one crossing at $k_x=-\pi/N$ and $\Phi_V=\pi$, with $\partial_{\Phi_V}E_\chi=2t\chi/N$. An axial
flux instead shifts $k_x\mapsto k_x+\chi\Phi_A/N$, giving one positive-slope crossing per sector. More
generally, inserting flux of type $J=V,A$ gives a crossing slope with sign $\chi q_{J,\chi}$. Weighting this
oriented crossing by the measured charge $q_{I,\chi}$ yields
\begin{equation}
\mathcal A_{IJ}^{\rm edge}
=\sum_{\chi=\pm1}C_\chi q_{I,\chi}q_{J,\chi}
=\mathcal A_{IJ}.
\label{eq:1d_bulk_edge_flux_index}
\end{equation}
Here $C_\chi=\chi$ and $(q_{V,\chi},q_{A,\chi})=(1,\chi)$, so $\mathcal A_{VV}=\mathcal A_{AA}=0$ and $\mathcal
A_{VA}=\mathcal A_{AV}=2$. These exact lattice Chern numbers and finite-size crossings reproduce the autonomous
chain anomaly matrix
\eqref{eq:1d_anomaly_matrix} and its flux-response indices
\eqref{eq:1d_vector_IR_flow} and \eqref{eq:1d_anom_dual_flux}.
The oriented index reverses on the opposite physical edge.

The matching spectral-flow indices have different UV realizations. In the autonomous chain, large-gauge
monodromy reassigns the covariant spectral charge within a fixed $N$-mode Hilbert space. In the parent, the
edge--bulk transfer at $|E|=2t$ provides the corresponding UV spectral-charge reassignment of
Eq.~\eqref{eq:1d_anom_exact_sharp_jump}. This transfer is distinct from the zero-energy crossings: the parent
can exchange charge between boundary and bulk while conserving its onsite axial charge. Thus the lattice
spectral-flow indices agree exactly, although this does not establish equality of the complete microscopic gauge
responses.

The full parent couples locally to compact gauge backgrounds and satisfies $Z_{\rm full}[A^g]=Z_{\rm full}[A]$ under admissible gauge transformations. Under $A_I\mapsto A_I-d\lambda_I$, the mixed bulk response in Eq.~\eqref{eq:1d_bulk_edge_response} has boundary variation
\begin{equation}
\delta S_{\rm resp}
=-\frac{1}{2\pi}\int_{\partial M_3}
\left(\lambda_VdA_A+\lambda_AdA_V\right),
\end{equation}
which cancels the consistent boundary anomaly. This symmetric anomaly convention differs from the vector-preserving convention of the axial Ward identity derived above by a local boundary counterterm proportional to $\int_{\partial M_3}A_V\wedge A_A$. Although the autonomous chain and the selected edge sector have identical ungauged Hamiltonians, their couplings to background gauge fields need not coincide. The counterterm redistributes the mixed anomaly between the vector and axial Ward identities without changing its anomaly class.

Beyond static flat backgrounds, an exact factorization of the full microscopic determinant into separate boundary and bulk contributions is generally unavailable. For bounded, smooth backgrounds $B_\pm=A_V\pm A_A$ preserving the bulk gap and varying on a scale $R\gg v_F/\Delta_{\rm bulk}$, we nevertheless conjecture that, after appropriate local counterterms and normalization,
\begin{equation}
Z_{\rm full}^{\rm ren}[B_\pm]
\approx
Z_{\partial,\rm Dirac}[B_\pm|_{\partial M_3}]
Z_{\rm bulk,top}[B_\pm].
\label{eq:1d_bulk_edge_IR_det_conjecture}
\end{equation}
Here $Z_{\partial,\rm Dirac}$ is the regulated continuum boundary Dirac determinant coupled to the restricted backgrounds, while $Z_{\rm bulk,top}$ is the invertible bulk topological amplitude determined by the mixed response in Eq.~\eqref{eq:1d_bulk_edge_response} and its global completion. The two factors have compensating gauge variations, so their product is gauge invariant. Globally, the product is understood as the pairing of the boundary anomaly-line section with its dual bulk amplitude. The factorization, including controlled errors and global phase matching, remains conjectural.

\subsubsection{Autonomous versus bulk-induced dynamical gauging}
\label{sec:1d_bulk_edge_dynamical}

The preceding Schur complement eliminates complementary fermionic degrees of freedom at fixed gauge backgrounds, rather than a dynamical bulk gauge field. For smooth external vector and axial connections, the continuum boundary matter action is
\begin{equation}
S_{\partial,\rm Dirac}
=\int_{\partial M_3}\!d^2x\,
\bar\Psi\gamma^\alpha
(\partial_\alpha+iA_{V,\alpha}
+i\gamma^5A_{A,\alpha})\Psi.
\label{eq:1d_bulk_edge_background_dirac}
\end{equation}
Its anomalous boundary determinant is completed by the mixed bulk response. Both connections remain external at this stage.

To compare dynamical gauge theories, gauge $U(1)_V$ throughout the 2D parent while keeping $A_A$ external. Denote the dynamical compact vector connection by $a_V$. Compact links, integer electric fields, and plaquette terms define a local lattice gauge Hamiltonian. The continuum Maxwell action is $S_{\rm Maxwell}=(4g_{2+1}^2)^{-1}\int_{M_3}f_{V,MN}f_{V,MN}$, where $f_V=da_V$. Since $C_++C_-=0$, the vector self--Chern--Simons term cancels, leaving the mixed response
\begin{equation}
S_{\rm mix}=\frac{i}{\pi}\int_{M_3}A_A\wedge da_V.
\label{eq:1d_bulk_edge_dynamical_mixed_response}
\end{equation}

In a homogeneous half-space, integrating out the interior photon at fixed boundary connection $a_{V,\alpha}$ gives, at Gaussian order and neglecting compact-monopole effects,
\begin{equation}
\begin{aligned}
S_{\partial,g}^{(2)}[a_V]
&=\frac{1}{2g_{2+1}^2}\int\!\frac{d^2\mathbf p}{(2\pi)^2}\,
a_{V,\alpha}(-\mathbf p)|\mathbf p|
P_T^{\alpha\beta}(\mathbf p)a_{V,\beta}(\mathbf p),\\
P_T^{\alpha\beta}(\mathbf p)
&=\delta^{\alpha\beta}
-\frac{p^\alpha p^\beta}{\mathbf p^2}.
\end{aligned}
\label{eq:1d_bulk_edge_induced_photon}
\end{equation}
Equivalently, the boundary Maxwell kernel is $(4g_{2+1}^2)^{-1}\int d^2x\,f_{V,\alpha\beta}(-\partial^2)^{-1/2}f_{V,\alpha\beta}$. This mixed-dimensional, or reduced, $\mathrm{QED}_{3,2}$ coupling is related to the nonlocal Schwinger model of Ref.~\cite{frasertaliente2025nonlocal}. The overall coefficient depends on whether the photon propagates in a half-space or a two-sided geometry.

The Gaussian approximation excludes monopole events associated with the compact bulk gauge field. These events change the vector magnetic flux by $2\pi$ and, through Eq.~\eqref{eq:1d_bulk_edge_dynamical_mixed_response}, carry axial charge two. Exact global $U(1)_A$ therefore forbids undressed monopole insertions, although monopoles accompanied by matter excitations of compensating axial charge may be allowed. Their possible effects on the IR dynamics lie beyond Gaussian reduction.

By contrast, autonomous vector gauging in the $c$ chain has a local 1D Maxwell kernel $\mathbf p^2P_T/g_{1+1}^2$. Integrating out the massless Dirac fermion gives the transverse vacuum polarization $\Pi^{\alpha\beta}(\mathbf p)=P_T^{\alpha\beta}(\mathbf p)/\pi$. The resulting inverse transverse photon propagators are
\begin{equation}
\begin{aligned}
D_{\rm aut,T}^{-1}(\mathbf p)
&=\frac{\mathbf p^2}{g_{1+1}^2}+\frac1\pi,\\
D_{\rm ind,T}^{-1}(\mathbf p)
&=\frac{|\mathbf p|}{g_{2+1}^2}+\frac1\pi.
\end{aligned}
\label{eq:1d_bulk_edge_autonomous_reduced_comparison}
\end{equation}
The autonomous theory develops the Schwinger mass $m_{\rm Sch}^2=g_{1+1}^2/\pi$, whereas the bulk-induced theory retains a nonanalytic, gapless continuum response at Gaussian order.

The two constructions also differ in their realization of the spectator axial symmetry. In autonomous vector gauging, the mixed anomaly gives $\partial_\mu j_A^\mu=F_{01}^V/\pi$, with a dynamical vector electric field, so the standalone gauge theory does not preserve continuous global $U(1)_A$. In the bulk-induced theory, the same boundary anomaly is canceled by inflow from Eq.~\eqref{eq:1d_bulk_edge_dynamical_mixed_response}. The complete parent preserves global $U(1)_A$, with axial charge transferred between boundary matter and bulk gauge-field configurations. This compensation must be retained when constructing the reduced boundary theory. Thus identical ungauged boundary Hamiltonians can yield inequivalent dynamical gauge theories, differing both in their IR responses and in the realization of the spectator symmetry. This distinction cannot be attributed merely to a choice of anomaly counterterm.

A complementary distinction arises when $U(1)_A$ is gauged and $U(1)_V$ remains external, particularly at finite physical chemical potential. The deformation $H_0-\mu Q_V$ remains an exact ungauged edge Hamiltonian: the chemical potential couples to onsite bulk vector number and shifts the Fermi points without changing the edge wavefunctions, zero-background projector, or fixed unfolding. Autonomous axial gauging nevertheless encounters the UV seam discussed in Sec.~\ref{sec:1d_mu_gauge_locality}. Bulk-induced gauging instead couples the local parent before projection, avoiding a direct assignment of Wilson lines to the seam hopping. For bounded, smooth axial backgrounds, the edge-state overlap approaches unity near the Fermi points in the continuum limit at fixed physical $\mu$ [Eqs.~\eqref{eq:app_1d_bulk_edge_overlap} and \eqref{eq:app_1d_bulk_edge_form_factor_IR}], yielding a regular IR vertex without the autonomous seam-crossing $1/q$ singularity. Although the fixed-band projection is not separately gauge covariant, the complete bulk-induced effective action inherits gauge invariance from the parent, including contributions from complementary bulk states. The higher-dimensional construction thus provides a regular, gauge-consistent IR coupling without establishing autonomous 1D gauge locality or equivalence of the dynamical gauge theories.

\subsection{Superconducting symmetry relocalization and compact gauging}
\label{sec:1d_variants}
In the preceding discussion, the IR complex Dirac fermion arose from a nearest-neighbor complex-fermion chain
$H_0$ whose vector symmetry is already onsite. We now consider a superconducting realization of the same IR
complex Dirac fermion for which ordinary microscopic particle number is not conserved. The model nevertheless
possesses an exact compact vector symmetry with exponentially decaying spatial couplings. An analytic Bogoliubov
transformation renders this symmetry onsite while preserving exponential locality of the Hamiltonian, allowing a
local compact gauging in the transformed variables and a quasi-local realization in the original superconducting
variables.

\subsubsection{Hidden $U(1)$ charges and analytic Bogoliubov relocalization}
\label{sec:1d_sc_deformation}

Consider the finite-range Hamiltonian
\begin{equation}
H_{\rm SC}=-t\sum_j(c_j^\dagger c_{j+1}+\mathrm{H.c.})
+i\Delta\sum_j(c_jc_{j+2}-\mathrm{H.c.}),
\label{eq:1d_SC_H}
\end{equation}
with $t>0$, real $\Delta\ne0$, and APBC on $N\in4\mathbb Z$ sites. No nontrivial continuous symmetry generated
by an onsite site-diagonal charge $\sum_j\alpha_j(n_j-1/2)$ survives: the hopping imposes
$\alpha_{j+1}=\alpha_j$, whereas the pairing imposes $\alpha_{j+2}=-\alpha_j$, forcing every $\alpha_j$ to
vanish.

With $\Psi_k=(c_k,c_{-k}^\dagger)^T$, write 
\begin{equation}
H_{\rm SC}=\frac12\sum_k\Psi_k^\dagger h_{\rm SC}(k)\Psi_k.
\end{equation}
The
BdG matrix and its dispersion are
\begin{equation}
\begin{aligned}
h_{\rm SC}(k)&=-2\cos k\,[t\tau_z+2\Delta\sin k\,\tau_x],\\
D(k)&=\sqrt{t^2+4\Delta^2\sin^2 k},
\end{aligned}
\label{eq:app_1d_SC_BdG}
\end{equation}
and
\begin{equation}
E_\pm(k)=\pm2|\cos k|D(k).
\label{eq:app_1d_SC_dispersion}
\end{equation}
Since $D(k)>0$ is real analytic and $\pi$-periodic, the only nodes lie at $k=\pm\pi/2$. They constitute one
massless complex Dirac fermion with Fermi velocity $v_F=2\sqrt{t^2+4\Delta^2}$.

Despite the pairing, the Hamiltonian commutes with the Nambu-space generator
\begin{equation}
q_V(k)=\frac{t\tau_z+2\Delta\sin k\,\tau_x}{D(k)},
\quad q_V(k)^2=\mathbf1.
\label{eq:app_1d_SC_qsm}
\end{equation}
It satisfies the BdG compatibility condition $q_V(-k)=-\tau_xq_V(k)^*\tau_x$. Its centered second quantization
therefore defines an exact compact $U(1)_V$ charge $Q_V$, whose real-space bilinears decay exponentially by
analyticity. Here $Q_V$ denotes the hidden conserved vector charge, not the nonconserved ordinary $c$-fermion
number. The sharp axial charge,
\begin{equation}
Q_{A,0}=\sum_k s_0(k)c_k^\dagger c_k,
\quad s_0(k)=\operatorname{sgn}(\sin k),
\label{eq:app_1d_SC_QA}
\end{equation}
has the scalar Nambu generator $q_A(k)=s_0(k)\mathbf1$. Accordingly,
\begin{equation}
[H_{\rm SC},Q_V]=[H_{\rm SC},Q_{A,0}]
=[Q_V,Q_{A,0}]=0.
\label{eq:app_1d_SC_commuting}
\end{equation}
Thus both compact symmetries are exact, although their microscopic locality differs: $Q_V$ is exponentially
local, whereas $Q_{A,0}$ has algebraic tails.

To make $Q_V$ onsite, define the analytic Bogoliubov rotation
\begin{equation}
\varphi(k)=\arctan\frac{2\Delta\sin k}{t},
\quad V(k)=e^{-i\varphi(k)\tau_y/2}.
\label{eq:app_1d_SC_V}
\end{equation}
The angle is odd and periodic, and $V(-k)=\tau_xV(k)^*\tau_x$. Direct conjugation yields
\begin{equation}
\begin{aligned}
&V^\dagger(k)q_V(k)V(k)=\tau_z,\\
&V^\dagger(k)h_{\rm SC}(k)V(k)=\epsilon_{\rm odd}(k)\tau_z,\\
&\epsilon_{\rm odd}(k)=-2\cos k\,D(k).
\end{aligned}
\label{eq:app_1d_SC_diagonalize}
\end{equation}
Let $u(k)=\cos[\varphi(k)/2]$ and $v(k)=\sin[\varphi(k)/2]$, and introduce the canonical fermions $\widetilde
c_k=u(k)c_k+v(k)c_{-k}^\dagger$. Then
\begin{equation}
H_{\rm SC}=\sum_k\epsilon_{\rm odd}(k)
\widetilde c_k^\dagger\widetilde c_k,
\label{eq:app_1d_SC_number_conserving}
\end{equation}
while the charges take the form
\begin{equation}
Q_V=\sum_j(\widetilde c_j^\dagger\widetilde c_j-\tfrac12),\quad
Q_{A,0} =\sum_k s_0(k)\widetilde c_k^\dagger\widetilde c_k.
\label{eq:app_1d_SC_charges_after_V}
\end{equation}
Analyticity of $V(k)$  makes this canonical transformation
exponentially local. Near $k=\pm\pi/2$, the two charges reduce to the standard vector and axial charges of the
IR Dirac fermion. Thus the hidden $U(1)_V$ acts as ordinary Dirac number in the IR, despite its particle--hole
mixing in the original $c$ variables.

The transformed dispersion is even and $\pi$-antiperiodic, $\epsilon_{\rm odd}(k+\pi)=-\epsilon_{\rm odd}(k)$,
because $D(k)$ is even and $\pi$-periodic. Hence the Hamiltonian contains only odd-distance hoppings, with
exponentially decaying amplitudes; for $\Delta\ne0$ it is not generally finite range. These properties also
permit an exact Hamiltonian-preserving vector--axial exchange, as in Sec.~\ref{sec:1d_relocalization}. In the
$\widetilde c$ variables, define
\begin{equation}
\mathcal W_{\rm odd}:\quad
\widetilde c_k\longmapsto
\begin{cases}
\widetilde c_k,&0<k<\pi,\\
\widetilde c_{\pi-k}^\dagger,&-\pi<k<0,
\end{cases}
\label{eq:app_1d_SC_W}
\end{equation}
where momenta are understood modulo $2\pi$. Pulling this transformation back through the analytic Bogoliubov map
$\mathcal B$ gives $\mathcal W_{\rm SC}=\mathcal B^{-1}\mathcal W_{\rm odd}\mathcal B$, with
\begin{equation}
\begin{aligned}
\mathcal W_{\rm SC}H_{\rm SC}\mathcal W_{\rm SC}^{-1}&=H_{\rm SC},\\
\mathcal W_{\rm SC}Q_V\mathcal W_{\rm SC}^{-1}&=Q_{A,0},\\
\mathcal W_{\rm SC}Q_{A,0}\mathcal W_{\rm SC}^{-1}&=Q_V.
\end{aligned}
\label{eq:app_1d_SC_W_original}
\end{equation}
This is a direct analog of Eq.~\eqref{eq:1d_exact_H_preserving_VA_exchange}. Unlike $\mathcal B$, the exchange
$\mathcal W_{\rm SC}$ is spatially nonlocal. It nevertheless provides a separate symmetry-adapted relocalization
for the axial charge. The two compact symmetries can therefore be gauged individually, but not simultaneously as
an anomaly-free standalone Dirac theory. The inversion-odd spectral-sign family of
Sec.~\ref{sec:1d_general_s_gauge_locality} persists under the same deformation. For any $s(-k)=-s(k)$ on the
APBC grid, $Q_{A,s}=\sum_ks(k)c_k^\dagger c_k$ commutes with the pairing, and its scalar Nambu generator is
unchanged by $V(k)$. Composing the corresponding axial relocalization with $\mathcal B$ makes $Q_{A,s}$ onsite
without destroying exponential locality of the odd-distance hopping. Its transported compact gauging remains
exactly gauge covariant and exponentially local. Signs that coincide with $s_0$ near the Fermi points have the
same IR axial action. This statement does not extend the finite-$\mu$ norm-resolvent theorem of
Sec.~\ref{sec:1d_mu_gauge_locality} to chemically deformed superconducting dispersions.

Finally, the superconducting model $H_{\rm SC}$ can be continuously deformed into the original nearest-neighbor
chain $H_0$ by reducing the pairing:
\begin{equation}
\begin{aligned}
h_\lambda(k)&=-2\cos k\left[
 t\tau_z+2(1-\lambda)\Delta\sin k\,\tau_x\right],\\
q_{V,\lambda}(k)&=
\frac{t\tau_z+2(1-\lambda)\Delta\sin k\,\tau_x}
{\sqrt{t^2+4(1-\lambda)^2\Delta^2\sin^2 k}},
\quad 0\le\lambda\le1.
\end{aligned}
\label{eq:app_1d_SC_interpolate}
\end{equation}
Throughout this interpolation the Hamiltonian remains finite range, the vector charge remains exponentially
local, $Q_{A,0}$ is unchanged, and the two Dirac nodes at $k=\pm\pi/2$ persist. The transformed dispersion
remains even and $\pi$-antiperiodic, so the exact Hamiltonian-preserving vector--axial exchange survives. At
$\lambda=1$, $q_{V,1}(k)=\tau_z$, recovering the original chain with onsite vector $U(1)$.

\subsubsection{Compact gauging and quasi-local Gauss laws}
\label{sec:1d_sc_vector_gauging}

We gauge the hidden $U(1)_V$ in the $\widetilde c$ representation, where its charge is onsite, and then
transport the complete matter--gauge Hamiltonian back to the original superconducting variables. This couples a
compact gauge field to an exact symmetry of $H_{\rm SC}$, rather than to its nonconserved microscopic particle
number.

\paragraph{Compact gauging and quasi-local pullback.}
The distinction is explicit in the global symmetry action. In the original Nambu basis, $\Psi_k\mapsto
e^{i\alpha q_V(k)}\Psi_k$ gives
\begin{equation}
c_k\mapsto
\left(\cos\alpha+i\frac{t}{D(k)}\sin\alpha\right)c_k +i\frac{2\Delta\sin k}{D(k)}\sin\alpha\,c_{-k}^\dagger.
\label{eq:app_1d_SC_vector_action}
\end{equation}
For $\Delta\ne0$ the symmetry mixes particles and holes, reducing to ordinary particle-number rotation only at
$\Delta=0$.

Using the coefficients $u(k)$ and $v(k)$ above, let $u_r$ and $v_r$ denote their Fourier coefficients. The
canonical change of variables has the exact real-space form
\begin{equation}
\begin{aligned}
\widetilde c_k&=u(k)c_k+v(k)c_{-k}^\dagger,\\
\widetilde c_j&=\sum_l\left(u_{j-l}c_l+v_{j-l}c_l^\dagger\right),
\end{aligned}
\label{eq:app_1d_SC_real_bogo}
\end{equation}
where both kernels decay exponentially. For $|\Delta|\ll t$,
\begin{equation}
\widetilde c_j=c_j+\frac{i\Delta}{2t}
(c_{j-1}^\dagger-c_{j+1}^\dagger)
+O(\Delta^2/t^2).
\label{eq:app_1d_SC_real_bogo_weak}
\end{equation}

Introduce compact vector links $U_j^V=e^{i\vartheta_j^V}$ and their conjugate integer electric fields $E_j^V$,
with $[E_j^V,U_l^V]=\delta_{jl}U_l^V$. Using the staggered background convention of
Sec.~\ref{sec:1d_exact_finite_spacing_gauge_locality}, define
\begin{equation}
q_{V,j}=\widetilde c_j^\dagger\widetilde c_j-\nu_j,
\quad \nu_j=\frac{1-(-1)^j}{2},
\quad Q_V=\sum_jq_{V,j},
\label{eq:app_1d_SC_vector_density}
\end{equation}
and impose the exact Gauss laws
\begin{equation}
\mathcal G_{V,j}=E_j^V-E_{j-1}^V-q_{V,j},
\quad[\mathcal G_{V,j},\mathcal G_{V,l}]=0.
\label{eq:app_1d_SC_vector_Gauss}
\end{equation}
The subtraction $\nu_j$ makes the local charges integer valued and centers $Q_V$ at half filling.

Because $\epsilon_{\rm odd}(k)$ is analytic, even, and $\pi$-antiperiodic, its Fourier expansion contains real,
exponentially decaying odd-distance hoppings:
\begin{equation}
H_{\rm SC}=\sum_j\sum_{\substack{r>0\\r\ {\rm odd}}}
J_r(\widetilde c_j^\dagger\widetilde c_{j+r}+\mathrm{H.c.}),
\quad |J_r|\le Ce^{-\kappa r},
\label{eq:app_1d_SC_odd_real_H}
\end{equation}
where $J_r$ is the corresponding Fourier coefficient and $J_1=-t+O(\Delta^2/t)$. For each $r>0$, let
$W_{j,r}^V=U_j^V\cdots U_{j+r-1}^V$. The compact vector gauge Hamiltonian is
\begin{equation}
\begin{aligned}
H_V[U^V,E^V]&=
\sum_j\sum_{\substack{r>0\\r\ {\rm odd}}}J_r
(\widetilde c_j^\dagger W_{j,r}^V\widetilde c_{j+r}
+\mathrm{H.c.})\\
& +\frac{g_V^2}{2}\sum_j
(E_j^V)^2.
\end{aligned}
\label{eq:app_1d_SC_vector_gauged_H}
\end{equation}
Physical states obey $\mathcal G_{V,j}=0$. Each hopping term is covariant under
\begin{equation}
\widetilde c_j\longmapsto e^{i\alpha_j}\widetilde c_j,
\quad U_j^V\longmapsto e^{i(\alpha_j-\alpha_{j+1})}U_j^V.
\label{eq:app_1d_SC_vector_local_action}
\end{equation}
Consequently, the full Hamiltonian is exactly gauge invariant. Exponential decay of $J_r$ makes the matter
couplings exponentially local, while truncating to any finite odd hopping range preserves exact compact gauge
invariance. On a finite APBC ring, boundary-crossing Wilson lines include the prescribed fermionic
spin-structure sign.

In the original superconducting variables, the same Gauss-law densities become quasi-local rather than onsite:
\begin{equation}
q_{V,j}=\sum_{l,m}
(u_{j-l}^*c_l^\dagger+v_{j-l}^*c_l)
(u_{j-m}c_m+v_{j-m}c_m^\dagger) -\nu_j.
\label{eq:app_1d_SC_vector_density_original}
\end{equation}
They contain normal and anomalous bilinears with exponentially decaying coefficients. In particular,
Eq.~\eqref{eq:app_1d_SC_real_bogo_weak} gives
\begin{equation}
\begin{aligned}
q_{V,j}={}&n_j-\nu_j+\frac{i\Delta}{2t}\big[
 c_j^\dagger c_{j-1}^\dagger
-c_j^\dagger c_{j+1}^\dagger\\
&\quad-c_{j-1}c_j+c_{j+1}c_j\big]
+O(\Delta^2/t^2).
\end{aligned}
\label{eq:app_1d_SC_vector_density_weak}
\end{equation}
The Gauss generators continue to commute exactly after this canonical change of variables. Their associated
gauge transformations are likewise non-onsite:
\begin{equation}
\begin{aligned}
c_j\longmapsto{}&e^{i\alpha_j}c_j
+\frac{i\Delta}{2t}\Big[
(e^{i\alpha_j}-e^{-i\alpha_{j-1}})c_{j-1}^\dagger\\
&\quad-(e^{i\alpha_j}-e^{-i\alpha_{j+1}})c_{j+1}^\dagger
\Big]+O(\Delta^2/t^2).
\end{aligned}
\label{eq:app_1d_SC_vector_gauge_action_original}
\end{equation}
For constant $\alpha_j=\alpha$, this agrees with the global transformation in
Eq.~\eqref{eq:app_1d_SC_vector_action} to the displayed order.

The gauged hopping may also be pulled back exactly through Eq.~\eqref{eq:app_1d_SC_real_bogo}. Its first-order
expansion illustrates the resulting link-dressed anomalous terms:
\begin{equation}
\begin{aligned}
&H_{V,{\rm hop}}[U]={}-t\sum_j
(c_j^\dagger U_j^Vc_{j+1}+\mathrm{H.c.})\\
&\ \ +\frac{i\Delta}{2}\sum_j\Big[
 c_j^\dagger U_j^Vc_{j+2}^\dagger
+c_{j-1}U_j^Vc_{j+1}
-\mathrm{H.c.}\Big] \\
&\ \
+O(\Delta^2/t).
\end{aligned}
\label{eq:app_1d_SC_vector_hop_weak}
\end{equation}
At $U_j^V=1$, the anomalous terms recover the pairing in Eq.~\eqref{eq:1d_SC_H}. For general links, they are
gauge covariant under the transported, non-onsite transformation, not under independent onsite phase rotations
of the original $c_j$ operators.

\paragraph{Physical electromagnetic interpretation.}
If the hidden $U(1)_V$ is identified with electromagnetic charge, $Q_{\rm em}=eQ_V$, with signed fermion charge $e$ (negative for electrons), gauging in the $\widetilde c$ frame and transforming back gives the same matter--gauge system:
\begin{equation}
H_{\rm SC}[A]=\mathcal B^{-1}H_V[A]\mathcal B.
\label{eq:1d_SC_EM_transport}
\end{equation}
The original $c_j$ fermions are not eigenoperators of the physical charge, so the electromagnetic density, Gauss
laws, and currents contain anomalous as well as normal bilinears. In the IR, the Dirac field has charge $e$ and
the usual vector minimal coupling. The pairing does not require a charge-two Higgs field for this $U(1)_V$; such
a field would instead be needed to gauge ordinary $c$-particle number, which is not conserved by $H_{\rm SC}$.

The same staggered background yields the discrete-chiral transformed-family identity of
Sec.~\ref{sec:1d_exact_finite_spacing_gauge_locality}, with $\eta_j$ replaced by $\widetilde c_j$ and $g_A$ by
$g_V$. Its symmetry-selected bare staggered mass is $m_{\rm lat}^{*}=-g_V^2a/8$. This finite-spacing identity
extends to odd-distance hopping independently of the $J_r$, but by itself does not establish the interacting
continuum limit of the exponentially local regulator. Nor does gauging $U(1)_V$ amount to simultaneous gauging of the mixed-anomalous $U(1)_V\times U(1)_A$ pair.

\paragraph{Translation and the hidden-vector LSM anomaly.}
The analytic Bogoliubov transformation preserves one-site translation, $T\widetilde c_jT^{-1}=\widetilde
c_{j+1}$. On the $N$-site APBC ring, let $\widetilde n_j=\widetilde c_j^\dagger\widetilde c_j$ and $\widetilde
N_V=\sum_j\widetilde n_j=Q_V+N/2$. The hidden-vector large-gauge unitary satisfies the exact translation algebra
\begin{equation}
\begin{aligned}
&\mathscr U_V
=\exp\!\left[\frac{2\pi i}{N}
\sum_{j=0}^{N-1}j\,\widetilde n_j\right],\\
&T\mathscr U_VT^{-1}
=e^{-2\pi i\widetilde N_V/N}\mathscr U_V.
\end{aligned}
\label{eq:1d_SC_LSM_algebra}
\end{equation}
The operator agrees with the main-text notation $\mathscr U_V$: for $N\in4\mathbb Z$, the background shift $\widetilde n_j\mapsto q_{V,j}=\widetilde n_j-\nu_j$ changes its exponent by an integer multiple of $2\pi i$.
In the neutral sector $Q_V=0$, the translation phase is $-1$: $2\pi$ vector-flux insertion pumps $\Delta
P=\pi\pmod{2\pi}$, as in Eq.~\eqref{eq:1d_LSM_momentum_shift}. Hence a translation- and $U(1)_V$-preserving
local deformation cannot produce a unique symmetric gapped ground state under the usual LSM assumptions. In the
original $c_j$ variables, the large-gauge transformation is non-onsite but exponentially quasi-local. If $Q_V$
is identified with electromagnetic charge, this is electromagnetic flux insertion, not flux for the nonconserved
$c$-particle number. The mixed-anomaly interpretation of Sec.~\ref{sec:1d_anomaly} is unchanged.

\subsubsection{Majorana-shift comparison and class-D Pfaffian parity}
\label{sec:1d_majorana_shift}

To prepare for the 2D examples, consider another superconducting representative of the same massless Dirac theory,
\begin{equation}
\begin{aligned}
H_S={}&-t\sum_j\left(c_j^{\dagger}c_j-\tfrac12\right)
-\frac t2\sum_j\left(c_j^\dagger c_{j+2}+\mathrm{H.c.}\right)\\
&-\frac t2\sum_j\left(c_jc_{j+2}+\mathrm{H.c.}\right).
\end{aligned}
\label{eq:1d_shift_HS_complex}
\end{equation}
Writing $c_j=(a_j+ib_j)/2$ in terms of Majorana fermions reveals a simple construction: $H_S$ is obtained from the nearest-neighbor chain by translating only one Majorana component,
\begin{equation}
H_S:=\mathcal S H_0\mathcal S^{-1}
=-\frac{it}{2}\sum_j a_j(b_j+b_{j+2}),
\label{eq:1d_shift_HS}
\end{equation}
where
\begin{equation}
\mathcal S:\quad
a_j\longmapsto a_j,\qquad
b_j\longmapsto b_{j+1}.
\label{eq:1d_shift_S}
\end{equation}
Thus $H_S$ has exactly the same single-particle spectrum as $H_0$ and flows to the same single complex Dirac fermion in the IR.

The vector and axial charges are transported as
\begin{equation}
\begin{aligned}
Q_{V,S}
&:=\mathcal S Q_V\mathcal S^{-1}
=\frac i2\sum_j a_jb_{j+1},\\
Q_{A,S}
&:=\mathcal S Q_{A,0}\mathcal S^{-1}
=Q_{A,0}.
\end{aligned}
\label{eq:1d_shift_QS}
\end{equation}
Unlike the Hamiltonian-preserving vector--axial relocalization above, the relevant transformation here is $\mathcal S^{-1}$, which makes $Q_{V,S}$ onsite while mapping $H_S$ back to $H_0$. The Majorana shift $\mathcal S$ is a strictly locality-preserving Gaussian automorphism, or fermionic QCA. Related Gaussian insulator--superconductor dualities and symmetry transmutations were studied by Cobanera and Ortiz~\cite{cobanera2015equivalence}. Up to the shift orientation and convention choices, the bond charge $Q_{V,S}$ also coincides with the quantized axial charge in the staggered-fermion construction of Ref.~\cite{chatterjee2025quantized}.

Although $\mathcal S$ preserves locality, it changes a microscopic topological diagnostic. For a 1D BdG Hamiltonian with particle--hole symmetry, $A(k)=h(k)\tau_x$ is antisymmetric at the PHIM $k=0,\pi$. We define the class-D Pfaffian parity
\begin{equation}
\mathcal P_D(H)
:=
\operatorname{sgn}\!\left[
\operatorname{Pf}A(0)\operatorname{Pf}A(\pi)
\right].
\label{eq:1d_gapless_PD_definition}
\end{equation}
For a fully gapped class-D superconductor, this is Kitaev's $\mathbb Z_2$ Majorana number~\cite{kitaev2001unpaired,kitaevPeriodic2009}. Our models are gapless, with nodes away from the PHIM, so the sign remains well defined in a fixed microscopic Nambu convention but serves only as a gapless diagnostic, not a gapped-phase invariant. For $H_{\rm SC}$ of Eq.~\eqref{eq:1d_SC_H}, the pairing vanishes at $k=0,\pi$, yielding
\begin{equation}
\mathcal P_D(H_{\rm SC})
=\mathcal P_D(H_0)
=\operatorname{sgn}[(-2t)(2t)]
=-1,
\label{eq:app_1d_SC_PD}
\end{equation}
where $H_0$ is viewed in BdG form.

The Majorana-shifted Hamiltonian has the opposite parity. In the Nambu basis $\Psi_k=(c_k,c_{-k}^\dagger)^T$, its Bloch Hamiltonian can be written as
\begin{equation}
\begin{aligned}
q_S(k)&=\cos k\,\tau_z+\sin k\,\tau_y,\\
h_S(k)&=\epsilon(k)\,q_S(k),
\end{aligned}
\label{eq:1d_shift_BdG}
\end{equation}
with $q_S(k)^2=1$. At the PHIM, $q_S(0)=\tau_z$ and $q_S(\pi)=-\tau_z$. Since $\epsilon(0)=-2t$ and $\epsilon(\pi)=2t$, both $h_S(0)$ and $h_S(\pi)$ equal $-2t\,\tau_z$. Their Pfaffians therefore have the same sign, giving
\begin{equation}
\mathcal P_D(H_S)
=-\mathcal P_D(H_0)
=+1.
\label{eq:1d_shift_main_PS}
\end{equation}
The Majorana shift thus reverses the class-D Pfaffian parity without changing the single-particle spectrum or the IR Dirac theory.

This reversal reflects the nontrivial fermionic QCA topology of $\mathcal S$. Translating one Majorana mode per unit cell realizes the Majorana-shift QCA, with $\operatorname{ind}_{\rm fQCA}(\mathcal S)=\sqrt2$ for the chosen orientation and its reciprocal for the opposite orientation~\cite{fidkowski2019interacting}. Consequently, $\mathcal S$ cannot be implemented by a parity-preserving finite-depth circuit~\cite{gross2012index,farrelly2020review}. On an even periodic ring, the cyclic permutation of one Majorana species has real determinant $-1$ and admits a Pin, but not a Spin, lift. Its finite-system implementation therefore requires care with fermion parity and boundary spin structure.

The comparison shows that a strictly locality-preserving Majorana shift can distinguish UV fermionic realizations with identical Dirac spectra and transported mixed $U(1)_V$--$U(1)_A$ anomalies. The class-D Pfaffian parity detects a microscopic distinction invisible to these IR data. This mechanism differs both from the nonlocal Hamiltonian-preserving vector--axial exchange above and from the 2D obstruction associated with nontrivial charge-bundle topology discussed below.

\section{2D: Compact symmetries, charge topology, and gauge locality}
\label{sec:2d_u1}

Having established \emph{Locality from Nonlocality} in complex-fermion and superconducting chains, we return to the
puzzle posed in the Introduction by the isotropic Kitaev--Majorana honeycomb Hamiltonian $H_K$. Its two nodes,
related by Majorana reality, together describe a single massless complex Dirac fermion, despite the absence of
an onsite Dirac-number symmetry. As in the 1D spectral axial construction, an exact compact Dirac-number $U(1)$
can nevertheless be defined through a sharp partition of momentum space. For $H_K$, however, the natural
relocalization that renders this charge onsite introduces UV seams in the transformed Bloch Hamiltonian,
preventing a globally continuous representation suitable for manifestly local compact gauging. Thus exact
compact symmetry and onsite relocalizability alone do not guarantee gauge locality.

The complex-fermion Wilson--Dirac model $H_W$ on the square lattice provides a contrasting realization. Its
parity-breaking Wilson mass removes the unwanted doublers while retaining a single massless Dirac cone, and its
onsite particle-number symmetry admits strictly local compact gauging at finite lattice spacing. Decomposing the
complex fermions into two Majorana sectors and unfolding them yields a finite-range honeycomb Majorana
Hamiltonian $H_{\rm hc}^{W}$. Although $H_K$ and $H_{\rm hc}^{W}$ realize the same single-Dirac IR theory on the
honeycomb lattice, they have opposite class-D Pfaffian parities. Their comparison demonstrates that neither the
microscopic realization of compact symmetry nor its gauge locality is determined by the IR Dirac spectrum alone.

The UV difficulties of $H_K$ motivate deformations that preserve its odd class-D Pfaffian parity while
introducing the spectral pairing needed for smooth charge-on-site relocalization and controlled gauge locality.
An auxiliary honeycomb representative $H_{\rm hc}$ exhibits these properties, while our main construction is the
related square-lattice $\pi$-flux Majorana Hamiltonian $H_{\rm sq}$. In analogy with the 1D spectral
construction, we introduce both a parity-symmetric sharp spectral charge $Q_{s_P}$, with algebraically decaying
spatial couplings, and an exponentially local compact charge $Q_{\rm loc}$. Both realize Dirac number in the IR,
but differ in their microscopic locality and crystalline-symmetry properties. Crucially, both admit onsite
relocalizations that preserve smoothness of the transformed Bloch Hamiltonian and yield Schwartz-decaying
hopping amplitudes. Their compact gaugings consequently admit asymptotically local finite-range regulators with
controlled operator-norm errors.

\subsection{Kitaev--Majorana and Wilson--Dirac models}

\subsubsection{Kitaev--Majorana and exact spectral Dirac number}

Let $\mathbf a_1$ and $\mathbf a_2$ denote the primitive translations of the triangular Bravais lattice
underlying the honeycomb, with $\mathbf R=m\mathbf a_1+n\mathbf a_2$ labeling its two-site unit cells. Each cell
contains two Majorana operators $\gamma_{\alpha,\mathbf R}=\gamma_{\alpha,\mathbf R}^\dagger$, $\alpha=A,B$,
satisfying $\{\gamma_{\alpha,\mathbf R},\gamma_{\beta,\mathbf R'}\}=2\delta_{\alpha\beta}\delta_{\mathbf
R\mathbf R'}$. The Kitaev--Majorana Hamiltonian describing one real Majorana fermion per honeycomb vertex with
nearest-neighbor hopping is
\begin{equation}
\begin{aligned}
H_K
&=-\frac{it}{2}\sum_{\mathbf R}\gamma_{A,\mathbf R}
\left(
\gamma_{B,\mathbf R}
+\gamma_{B,\mathbf R+\mathbf a_1}
+\gamma_{B,\mathbf R+\mathbf a_2}
\right)
\\
&=\frac i4\gamma^T A_K\gamma .
\end{aligned}
\label{eq:2d_majorana_H}
\end{equation}
Here $A_K$ is the real antisymmetric Majorana hopping matrix in real space. Writing $u=\mathbf k\cdot\mathbf
a_1$ and $v=\mathbf k\cdot\mathbf a_2$, the Bloch Hamiltonian in the $(A,B)$ sublattice basis is
\begin{equation}
\begin{aligned}
h_K(u,v)&=t\left[\operatorname{Im}f_K\,\sigma_x+\operatorname{Re}f_K\,\sigma_y\right],\\
f_K(u,v)&=1+e^{iu}+e^{iv}.
\end{aligned}
\label{eq:2d_hK}
\end{equation}
The only zeros are $\mathbf K_\pm$ at $(u,v)=\pm(2\pi/3,-2\pi/3)$, understood modulo $2\pi$. The Majorana
reality relation
\begin{equation}
\gamma_\alpha(\mathbf k)^\dagger=\gamma_\alpha(-\mathbf k)
\label{eq:2d_reality}
\end{equation}
identifies the two nodes as one complex Dirac fermion.

As in the 1D spectral axial construction, an exact compact Dirac-number charge can be defined by an
inversion-odd partition of momentum space. A simple choice is $s_{K,0}(u,v)=\operatorname{sgn}(\sin u)$, which
distinguishes the two Majorana-reality-related Dirac nodes but singles out a lattice direction. Another natural
choice, preserving the honeycomb reflection and $C_3$ symmetries, is
\begin{equation}
s_{K,P}(u,v)=\operatorname{sgn}\!\left[
\sin\frac u2\sin\frac v2\sin\frac{u-v}{2}
\right].
\label{eq:2d_sKP}
\end{equation}
The argument of the sign function is $2\pi$-periodic in both $u$ and $v$ and odd under momentum inversion.
Consequently, $s_{K,P}(-u,-v)=-s_{K,P}(u,v)$, as required by Majorana reality. Away from its discontinuity
lines, $s_{K,P}$ is invariant under the honeycomb reflection $(u,v)\mapsto(v-u,v)$ and the $2\pi/3$ rotation
$(u,v)\mapsto(-v,u-v)$. The two choices assign opposite charge signs at the Dirac nodes; reversing the overall
sign of $s_{K,0}$ makes their IR charge assignments identical.

For either inversion-odd spectral sign $s(\mathbf k)=\pm1$, the corresponding Majorana complex structure is
\begin{equation} 
J_s(\mathbf k)=-i\,s(\mathbf k)\mathbf1,\ \
J_s^2=-\mathbf1,\ \ [J_s,A_K]=0. 
\label{eq:2d_Js}
\end{equation}
At finite volume, we choose momentum grids avoiding the sign seams and PHIM; exceptional modes require a
separate Majorana-compatible prescription. For example, the reflection- and rotation-symmetric spectral sign
$s_{K,P}$ is well defined on tori with $L_1=4m$, $L_2=4m+2$, and APBC along both primitive directions: neither
coordinate vanishes modulo $2\pi$, and $u=v$ never occurs on the grid. This finite-size choice ensures the sharp
charge is unambiguous; it need not preserve every point-group symmetry of the infinite lattice.

For simplicity, we take $L_1$ and $L_2$ to be even, so the number of honeycomb $A/B$ unit cells is
$N=L_1L_2\in4\mathbb Z$, containing $2N$ Majorana operators. The centered compact charge associated with $J_s$
is
\begin{equation}
Q_s=\frac i4\gamma^T J_s\gamma.
\label{eq:2d_Qs}
\end{equation}
It has integer eigenvalues and satisfies $e^{2\pi iQ_s}=1$. For our preferred choice $s=s_{K,P}$, the
positive-charge domain $D_+=\{0<v<u<2\pi\}$ contains the Dirac node $\mathbf K_+=(4\pi/3,2\pi/3)$, while its
inversion partner contains $\mathbf K_-$. Thus $Q_{s_{K,P}}$ provides an exact compact microscopic lift of the
IR Dirac-number $U(1)$. More generally, any inversion-odd spectral sign agreeing with $s_{K,P}$ near both Dirac
nodes defines an exact compact charge with the same IR action, although its UV locality and crystalline-symmetry
properties may differ.

The sharp spectral charge is spatially nonlocal in the original honeycomb geometry. In fact, $H_K$ admits no
finite-range or exponentially local compact quadratic generator of unit Dirac charge with any finite translation
period. More strongly, this obstruction cannot be overcome merely by allowing a spatially nonlocal canonical
transformation: within the finite-folding Majorana-Gaussian framework, no relocalization can render the compact
Dirac-number symmetry onsite while preserving a periodic, continuously differentiable ($C^1$) Bloch Hamiltonian
(Appendix~\ref{app:2d_HK_finite_no_go}). These results distinguish the obstruction to locality of the charge
itself from the stronger obstruction to simultaneous symmetry and Hamiltonian locality. They do not, however,
rule out more general canonical reorganizations or establish a universal obstruction to compact gauging of
$H_K$.

For comparison, any fully gapped Majorana Hamiltonian with Hermitian single-particle matrix $h=iA$ admits the
spectrally flattened charge
\begin{equation}
q_H=\operatorname{sgn}h=h(h^2)^{-1/2},
\quad J_H=-iq_H.
\label{eq:2d_spectral_flattened_charge}
\end{equation}
For a gapped finite-range Hamiltonian, the associated Majorana complex structure $J_H$ is exponentially local.
\footnote{The spectral-flattening construction was also employed by Ref.~
\cite{kobayashiInamuraShiozaki2026} to realize non-onsite
compact symmetries of a gapped $p+ip$ superconductor and identify their Majorana-layer lattice anomaly.} Unlike
$J_s$, however, $J_H$ distinguishes positive- and negative-energy states rather than assigning a common
Dirac-number charge to both bands near each node. It becomes singular at the massless nodes and therefore does
not provide the desired smooth compact charge for $H_K$. We now contrast $H_K$ with a representative admitting
exact local compact gauging.

\subsubsection{Wilson--Dirac and its unfolded Majorana representative}
\label{sec:2d_hcstar}

The Wilson--Dirac model provides a manifestly local realization of a single massless Dirac fermion with onsite
compact $U(1)$ symmetry. On an ordinary square lattice with site coordinate $\mathbf r$, consider a two-orbital
complex fermion $f_{\mathbf r}=(f_{1,\mathbf r},f_{2,\mathbf r})^T$ with Bloch Hamiltonian
\begin{equation}
h_W(\mathbf p)
=t\big[\sin p_1\,\sigma_x+\sin p_2\,\sigma_z+M_W(\mathbf p)\sigma_y\big],
\label{eq:2d_hW_exact}
\end{equation}
where $M_W(\mathbf p)=2-\cos p_1-\cos p_2$ is the Wilson mass. In real space,
\begin{equation}
H_W= \sum_{\mathbf r}f_{\mathbf r}^{\dagger}(2t\sigma_y)f_{\mathbf r} +\sum_{\mathbf r,j=1,2}
\left(f_{\mathbf r}^{\dagger}T_jf_{\mathbf r+\hat{\mathbf e}_j}
+\mathrm{H.c.}\right),
\label{eq:2d_HW_realspace}
\end{equation}
with nearest-neighbor hopping matrices
\begin{equation}
T_1=-\frac{t}{2}(\sigma_y+i\sigma_x),
\quad
T_2=-\frac{t}{2}(\sigma_y+i\sigma_z).
\end{equation}
Thus $H_W$ is a finite-range complex-fermion Hamiltonian, rather than a continuum symbol introduced through
relocalization. The Wilson mass vanishes at $\mathbf p=0$ but gaps the three would-be Dirac doublers at
$(\pi,0)$, $(0,\pi)$, and $(\pi,\pi)$, leaving a single massless Dirac fermion \cite{wilson1975quarks}. Near
$\mathbf p=0$,
\begin{equation}
h_W(\mathbf p)
=t(p_1\sigma_x+p_2\sigma_z)
+O(|\mathbf p|^2).
\end{equation}
Its compact $U(1)$ symmetry is generated by ordinary onsite particle number,
\begin{equation}
Q_D=\sum_{\mathbf r,\alpha}
\left(f_{\alpha,\mathbf r}^{\dagger}f_{\alpha,\mathbf r}
-\frac12\right),
\label{eq:2d_QD_complex}
\end{equation}
and therefore admits exact local compact gauging at finite lattice spacing.

The Wilson mass explicitly breaks continuum parity. Under $(p_1,p_2)\mapsto(-p_1,p_2)$ with $P=\sigma_z$, the
kinetic terms transform correctly, whereas the Wilson term changes sign under conjugation by $P$:
\begin{equation}
\sigma_z h_W(p_1,p_2)\sigma_z
-h_W(-p_1,p_2)=-2tM_W(\mathbf p)\sigma_y.
\label{eq:2d_W_parity_breaking}
\end{equation}
Parity is therefore broken at finite lattice spacing and recovered only in the massless Dirac limit. Upon
coupling to a dynamical compact $U(1)$ gauge field, gauge invariance does not forbid the Dirac mass
$\bar\Psi\Psi$, and radiative corrections can generate an additive mass shift of order $e^2$ at fixed physical
gauge coupling. Thus the gauged Hamiltonian provides an exact local lattice regulator of one-flavor
Dirac--Maxwell theory, but obtaining massless QED$_3$ requires tuning the renormalized Dirac mass; it does not
follow automatically from the vanishing bare mass.

To compare this manifestly gauge-local model with the Kitaev--Majorana honeycomb Hamiltonian, we first express
$H_W$ in Majorana variables and then unfold the resulting sectors onto the honeycomb lattice. Write each complex
fermion as
\begin{equation}
f_{\alpha,\mathbf r}
=\frac12\left(a_{\alpha,\mathbf r}-ib_{\alpha,\mathbf r}\right).
\end{equation}
The Wilson--Dirac single-particle matrix can be written in real space as $iB$, where $B$ is real, antisymmetric,
and finite range. Consequently, up to a normal-ordering constant,
\begin{equation}
H_W=\frac{i}{4}\left(a^TBa+b^TBb\right),
\label{eq:2d_W_two_real}
\end{equation}
while the compact charge becomes
\begin{equation}
Q_D=-\frac{i}{2}\sum_{\mathbf r,\alpha}
a_{\alpha,\mathbf r}b_{\alpha,\mathbf r}.
\label{eq:2d_QD}
\end{equation}
Thus the complex Wilson--Dirac model consists of two identical Majorana sectors, with the onsite $U(1)$
particle-number symmetry acting as an $SO(2)$ rotation between them.

We now unfold these sectors into a single Majorana system on the microscopic honeycomb lattice. Let $\mathbf
r=(m_1,n_2)$ label the Wilson--Dirac lattice, whose unit cell spans two microscopic lattice spacings along
$\mathbf a_1$. The unfolding places the two Majorana sectors on alternating microscopic sites, with a common staggered
sign in each Wilson cell:
\begin{equation}
\begin{aligned}
a_{\alpha,(m_1,n_2)}
&=(-1)^{m_1}\gamma_{\alpha,(2m_1,n_2)},\\
b_{\alpha,(m_1,n_2)}
&=(-1)^{m_1}\gamma_{\alpha,(2m_1+1,n_2)}.
\end{aligned}
\label{eq:2d_ab_blocking}
\end{equation}
The factor $(-1)^{m_1}$ shifts the horizontal momentum identification to $p_1=2u-\pi$; without it the
unfolding would instead give $p_1=2u$ and move the zero-energy nodes to different momenta. Because it
multiplies both $a$ and $b$, the onsite bilinear $a_{\alpha,\mathbf r}b_{\alpha,\mathbf r}$ and the compact
charge $Q_D$ of Eq.~\eqref{eq:2d_QD} are unchanged. On a periodic finite torus this staggered convention is
single valued for an even number of Wilson cells along $m_1$; an odd number changes the corresponding boundary
twist. Writing $\mathbf R=m\mathbf a_1+n\mathbf a_2$, the resulting finite-range honeycomb Majorana Hamiltonian is
\begin{equation}
\begin{aligned}
H_{\rm hc}^{W}
&={}-
\frac{it}{2}
\sum_{\mathbf R}
\gamma_{A,\mathbf R}
\Bigg[
2\gamma_{B,\mathbf R}
+\gamma_{B,\mathbf R-2\mathbf a_1}
\\
&\hspace{22mm}
-\frac12\gamma_{B,\mathbf R+\mathbf a_2}
-\frac12\gamma_{B,\mathbf R-\mathbf a_2}
\Bigg]
\\
&+
\frac{it}{4}
\sum_{\mathbf R}
\Big[
-\gamma_{A,\mathbf R}\gamma_{A,\mathbf R+\mathbf a_2}
+\gamma_{B,\mathbf R}\gamma_{B,\mathbf R+\mathbf a_2}
\Big].
\end{aligned}
\label{eq:2d_hcstar_micro}
\end{equation}
In momentum space, we use the two-component Majorana Bloch basis
\begin{equation}
\gamma_{\mathbf k}
=
\begin{pmatrix}
\gamma_{A,\mathbf k}\\
\gamma_{B,\mathbf k}
\end{pmatrix},
\quad
\gamma_{\mathbf k}^{\dagger}
=
\gamma_{-\mathbf k}^{T},
\end{equation}
and write
\begin{equation}
H_{\rm hc}^{W}
=
\frac{i}{4}
\sum_{\mathbf k}
\gamma_{-\mathbf k}^{T}
A_{\rm hc}^{W}(\mathbf k)
\gamma_{\mathbf k},
\quad
h_{\rm hc}^{W}(\mathbf k)
=
iA_{\rm hc}^{W}(\mathbf k).
\end{equation}
The Pauli matrices below act in the $A/B$ orbital space, equivalently the honeycomb-sublattice space. With
$
u=\mathbf k\cdot\mathbf a_1$ and $
v=\mathbf k\cdot\mathbf a_2$, 
the Bloch Hamiltonian is
\begin{equation}
\begin{aligned}
h_{\rm hc}^{W}(u,v)
=
t\bigg[
&-\sin(2u)\,\sigma_x
+\sin v\,\sigma_z
\\
&+\left(
2+\cos2u-\cos v
\right)\sigma_y
\bigg].
\end{aligned}
\label{eq:2d_hcstar}
\end{equation}
It reproduces Eq.~\eqref{eq:2d_hW_exact} under the momentum identification $ p_1=2u-\pi$ and $p_2=v$, with
$(u,v)$ restricted to one half of the honeycomb Brillouin zone. In particular,
\begin{equation}
h_{\rm hc}^{W}(u+\pi,v)
=
h_{\rm hc}^{W}(u,v).
\label{eq:2d_hcstar_extra_period}
\end{equation}
The single Wilson--Dirac cone at $\mathbf p=0$ therefore appears in the unfolded Brillouin zone as two Majorana
nodes,
\begin{equation}
(u,v)
=
\left(\frac\pi2,0\right),
\quad
\left(-\frac\pi2,0\right).
\end{equation}
These nodes are related by Majorana reality, Eq.~\eqref{eq:2d_reality}, and together represent only one complex
Dirac fermion.

The unfolding also provides a common representation in which to compare two exact compact realizations of the IR
Dirac-number symmetry: the local charge $Q_D$ inherited from the Wilson--Dirac model and a distinct sharp
spectral charge $Q_s$ of $H_{\rm hc}^{W}$. In the doubled-cell basis, let $\tau_i$ act on the $(a,b)$ sectors
and retain $p_1=2u-\pi$. Their Majorana complex structures are
\begin{equation}
\begin{aligned}
J_D(\mathbf p)
&=-i\tau_y\otimes\mathbf1_2,\\
J_s(\mathbf p)
&=-i\left[
-\sin\frac{p_1}{2}\,\tau_x
+\cos\frac{p_1}{2}\,\tau_y
\right]\otimes\mathbf1_2.
\end{aligned}
\label{eq:2d_Js_JD_blocked}
\end{equation}
Here $\mathbf1_2$ denotes the identity in the orbital space. Both complex structures have the same IR action at
$\mathbf p=0$, but differ in their spatial locality. The question is whether the nonlocal spectral charge can be
mapped to the onsite charge without changing the Hamiltonian.

This is possible because Eq.~\eqref{eq:2d_W_two_real} is proportional to the identity in the $(a,b)$ sector. An
explicit Majorana-canonical transformation $\mathcal W$, whose single-particle matrix in the doubled-cell basis
is
\begin{equation}
\mathcal W(p_1)
=
-i\left[
-\sin\frac{p_1}{4}\,\tau_x
+\cos\frac{p_1}{4}\,\tau_y
\right],
\label{eq:2d_W_exchange}
\end{equation}
commutes with the Hamiltonian and exchanges the two complex structures. The identity on the orbital space is
implicit. Using $\mathcal W$ also for its induced Majorana-canonical transformation on the fermionic
Hilbert space, we have
\begin{equation}
\begin{aligned}
\mathcal W H_{\rm hc}^{W}\mathcal W^{-1}
&=H_{\rm hc}^{W},
\\
\mathcal W J_s\mathcal W^{-1}
&=J_D,
\\
\mathcal W J_D\mathcal W^{-1}
&=J_s.
\end{aligned}
\label{eq:2d_hcstar_exchange}
\end{equation}
The transformation is spatially nonlocal but preserves the Hamiltonian exactly. Thus $J_D$ is the relocalized
form of $J_s$, obtained through a Hamiltonian-preserving canonical transformation. This is distinct from the
preceding unfolding, which merely identifies the Wilson--Dirac and honeycomb representations of the same
quadratic Hamiltonian.

The construction parallels the 1D vector--axial relocalization of Sec.~\ref{sec:1d_relocalization}, but with an
important distinction. In 1D, the transformation exchanges the distinct vector and axial symmetries. Here,
$U(1)_s$ and $U(1)_D$ are two microscopic realizations of the same IR Dirac-number symmetry, and $\mathcal W$
exchanges them without modifying the energy spectrum.

The relocalized onsite charge permits conventional compact gauging. Since $Q_D$ and $H_W$ are strictly local in
the Wilson--Dirac coordinate $\mathbf r$, compact link variables and local Gauss operators define an exact gauge
theory at finite lattice spacing. Transporting the complete matter--gauge system through $\mathcal W^{-1}$ then
provides an exact gauging of the original nonlocal $U(1)_s$, with gauge locality understood in the
symmetry-adapted Wilson--Dirac coordinate rather than necessarily in the original microscopic honeycomb
geometry.

This locality also persists at finite chemical potential. Indeed, the same Hamiltonian-preserving transformation
gives
\begin{equation}
\mathcal W(H_{\rm hc}^{W}-\mu Q_s)\mathcal W^{-1}
=H_{\rm hc}^{W}-\mu Q_D,
\label{eq:2d_hcstar_mu_exchange}
\end{equation}
which becomes the strictly local Hamiltonian $H_W(\mu)=H_W-\mu Q_D$ in the Wilson--Dirac coordinate for any
finite $\mu$. Thus the exact exchange of compact charges is compatible with finite-spacing gauge locality in the
symmetry-adapted representation, without requiring locality of $\mathcal W$ itself.

Finally, the two identical Majorana sectors admit infinitely many finite-range compact generators with the same
IR Dirac-number action. Pairing $a_{\alpha,\mathbf R}$ with a translated $b_{\alpha,\mathbf R+\mathbf r}$
defines an exactly conserved compact charge $Q_{\mathbf r}$ for each lattice displacement $\mathbf r$. These
charges generally do not commute; together with their commutators, they generate a generalized Onsager algebra
that exhausts the finite-range, translation-invariant, orbital-independent, traceless relative-sector quadratic
symmetries (Appendix~\ref{app:2d_hcW_onsager}).

\subsubsection{Distinct class-D regulator sectors}
\label{sec:2d_classD_types}

The Kitaev--Majorana Hamiltonian $H_K$ and the Wilson--Dirac-derived honeycomb Hamiltonian $H_{\rm hc}^{W}$
describe the same single massless Dirac fermion in the IR, but belong to distinct microscopic class-D Pfaffian
sectors. Because both are defined on the same honeycomb Bravais lattice, they can be compared using a common
two-site unit cell and Bloch coordinates $(u,v)$. At the four PHIM,
\begin{equation}
\Lambda\in\{(0,0),(\pi,0),(0,\pi),(\pi,\pi)\},
\end{equation}
the Majorana Bloch matrix $A(\Lambda)$ is real antisymmetric. Fixing the translation sewing and a common
orientation of the periodic Majorana basis, we define the Pfaffian parity
\begin{equation}
\mathcal P_D(H)=
\prod_\Lambda\operatorname{sgn}\operatorname{Pf}A(\Lambda).
\label{eq:2d_classD_pfaffian}
\end{equation}
For $H_K$, the values of $\operatorname{Re}f_K$ at the PHIM are $3,1,1,-1$, giving
\begin{equation}
\mathcal P_D(H_K)=-1.
\label{eq:2d_PK}
\end{equation}
For $H_{\rm hc}^{W}$, the corresponding coefficients of $t\sigma_y$ are $2,2,4,4$, yielding
\begin{equation}
\mathcal P_D(H_{\rm hc}^{W})=+1.
\label{eq:2d_Pstar}
\end{equation}
Thus the two models have opposite Pfaffian parities despite their identical single-Dirac IR content.
\footnote{Within a fixed translation-invariant class-D lattice framework,
the Pfaffian parity is unchanged under continuous deformations that preserve nonzero gaps at all PHIM, even if
the spectrum is gapless elsewhere. Since $H_K$ and $H_{\rm hc}^{W}$ have opposite Pfaffian parities, any such
interpolation between them must encounter a zero-energy crossing at a PHIM. A related explicit nodal transition
is constructed in Appendix~\ref{app:2d_interpolation_paths}.} Their distinction is compatible with the parity
constraint proved in Appendix~\ref{app:2d_dirac_pfaffian}: the even-Pfaffian Wilson--Dirac completion does not
possess the requisite regular, charge-preserving lattice reflection fixing the two non-PHIM nodes.

For fully gapped 2D class-D Hamiltonians, this parity is related to the integer Chern invariant. Let $h(\mathbf
k)=iA(\mathbf k)$ be the Hermitian Bloch matrix and
\begin{equation}
P_{\rm occ}(\mathbf k)=
\mathbf1_{(-\infty,0)}\bigl(h(\mathbf k)\bigr)
\end{equation}
the negative-energy spectral projector. The occupied bundle $\mathcal O=\operatorname{Im}P_{\rm occ}$ has first
Chern number
\begin{equation}
\begin{aligned}
\nu_{\rm Maj}
&=\left\langle c_1(\mathcal O),[T^2]\right\rangle\\
&=\frac{1}{2\pi i}\int_{T^2}
\operatorname{Tr}\left(
P_{\rm occ}\,dP_{\rm occ}\wedge dP_{\rm occ}
\right)
\in\mathbb Z.
\end{aligned}
\label{eq:2d_majorana_chern_integral}
\end{equation}
Here $c_1(\mathcal O)$ is the first Chern class of the occupied bundle, paired with the fundamental class
$[T^2]$ of the Brillouin torus. We call $\nu_{\rm Maj}$ the Majorana Chern number, following the spectral
invariant of the Kitaev honeycomb model \cite{kitaev2006anyons}, to distinguish it from the charge-bundle Chern
number $C_Q$ introduced below. For a fully gapped class-D Hamiltonian, the PHIM Pfaffian product determines the
Chern parity,
\begin{equation}
\mathcal P_D=(-1)^{\nu_{\rm Maj}}.
\label{eq:2d_pfaffian_chern_parity}
\end{equation}
The integer $\nu_{\rm Maj}$ gives the strong 2D class-D classification \cite{kitaevPeriodic2009}, while
additional weak indices may arise when lattice translations are preserved.

Although $H_K$ and $H_{\rm hc}^{W}$ are gapless, their Dirac nodes lie away from the PHIM, so the Pfaffian
products above remain well defined. They consequently constrain the Chern parity of any fully gapped class-D
descendants obtained without closing a gap at a PHIM: $\nu_{\rm Maj}$ must be odd for descendants of $H_K$ and
even for those of $H_{\rm hc}^{W}$. The familiar field-gapped Kitaev phase has $\nu_{\rm Maj}=\pm1$
\cite{kitaev2006anyons}. After gauging fermion parity, odd Majorana Chern number gives non-Abelian vortices
carrying unpaired Majorana zero modes, whereas even Majorana Chern number gives Abelian vortex sectors. This is
the odd/even distinction of Kitaev's sixteen-fold way, related to the weak- and strong-pairing phases of
spinless $p+ip$ superconductors \cite{kitaev2006anyons,readGreen2000}.

Importantly, class-D Pfaffian parity does not determine symmetry relocalizability or gauge locality. The exact
local compact gauging of $H_{\rm hc}^{W}$ follows from its Wilson--Dirac origin and the fixed-shift spectral
pairing in Eq.~\eqref{eq:2d_hcstar_extra_period}, rather than its even Pfaffian parity. By contrast, $H_K$ has
no nontrivial fixed momentum shift preserving its dispersion and obeys the finite-folding locality obstructions
established in Appendix~\ref{app:2d_HK_finite_no_go}. These examples indicate that symmetry relocalizability
depends on microscopic spectral structure, not solely on the class-D invariant.

To show that odd Pfaffian parity does not itself obstruct symmetry relocalization or gauge locality, consider an
auxiliary finite-range four-bond honeycomb Hamiltonian $H_{\rm hc}$, with Bloch symbol $h_{\rm
hc}=t[\operatorname{Im}f_{\rm hc}\,\sigma_x+\operatorname{Re}f_{\rm hc}\,\sigma_y]$ and structure factor
\begin{equation}
f_{\rm hc}(u,v)=1+e^{iv}
+\frac1{\sqrt3}\left(e^{iu}-e^{i(v-u)}\right).
\end{equation}
Its two Majorana-reality-related nodes at $(\pm\pi/2,\pi)$ describe one massless complex Dirac fermion, as in
$H_K$, with the same odd Pfaffian parity, $\mathcal P_D(H_{\rm hc})=-1$. Nevertheless, $H_{\rm hc}$ admits both
sharp and exponentially local compact Dirac-number charges, each supporting charge-on-site relocalization and
asymptotically local gauge couplings. The constructions are detailed in Appendix~\ref{app:2d_hc_charge_bundle}.

\subsection{Square $\pi$-flux Majoranas: sharp and smooth compact charges}
\label{sec:2d_sq}

\begin{figure}[tb]
    \centering
    \includegraphics[width=0.5\linewidth]{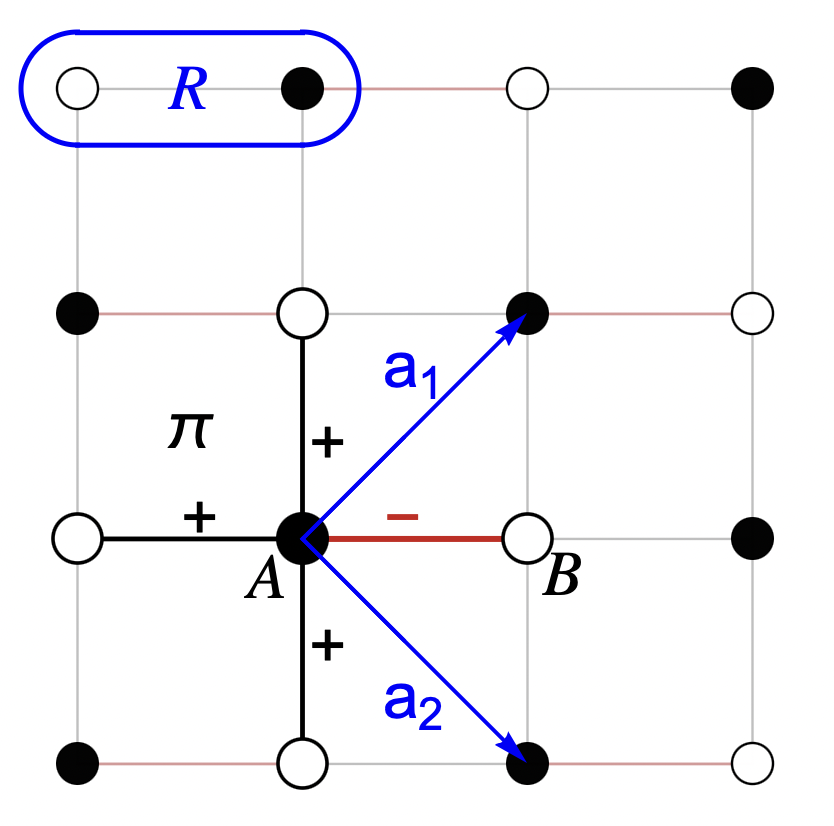}
    \caption{Square-lattice $\pi$-flux Majorana Hamiltonian $H_{\rm sq}$. Filled and open circles denote the $A$ and $B$ sublattices, respectively. Each unit cell, indicated by the capsule, contains one $A$ site and the $B$ site immediately to its left. The Bravais vectors are $\mathbf a_1=\mathbf e_x+\mathbf e_y$ and $\mathbf a_2=\mathbf e_x-\mathbf e_y$. From each $A$ site, the four nearest-neighbor Majorana hoppings have signs $+,+,+,-$, so that the product of hopping signs around every elementary square is $-1$, corresponding to a uniform $\pi$ flux.}
    \label{fig:2d_sq_lattice}
\end{figure}

We now examine these possibilities in the square-lattice $\pi$-flux Majorana model $H_{\rm sq}$, illustrated in
Fig.~\ref{fig:2d_sq_lattice}. Its simple four-bond structure allows us to construct two exact compact
Dirac-number charges: a sharp spectral charge $Q_{s_P}$, analogous to the 1D axial charge $Q_{A,s}$, and an
exponentially local charge $Q_{\rm loc}$. Their common IR action but different microscopic locality will
distinguish the existence of a compact symmetry from the locality of its generator.

With $\mathbf R$ labeling the unit cells, the Hamiltonian is
\begin{equation}
\begin{aligned}
H_{\rm sq}=-\frac{it}{2}\sum_{\mathbf R}
\gamma_{A,\mathbf R}\bigl(&\gamma_{B,\mathbf R}
+\gamma_{B,\mathbf R+\mathbf a_1}\\
&+\gamma_{B,\mathbf R+\mathbf a_2}
-\gamma_{B,\mathbf R+\mathbf a_1+\mathbf a_2}\bigr).
\end{aligned}
\label{eq:2d_sq_micro}
\end{equation}
The negative product of hopping signs around each elementary plaquette realizes a static $\pi$ flux with one
real Majorana per microscopic site. Writing $u=\mathbf k\cdot\mathbf a_1$, $v=\mathbf k\cdot\mathbf a_2$, and
$\gamma_{\mathbf k}=(\gamma_{A,\mathbf k},\gamma_{B,\mathbf k})^T$, we obtain
\begin{equation}
H_{\rm sq}=\frac14\sum_{\mathbf k}
\gamma_{-\mathbf k}^T h_{\rm sq}(u,v)\gamma_{\mathbf k},
\end{equation}
where
\begin{equation}
\begin{aligned}
h_{\rm sq}(u,v)
&=t\left[\operatorname{Im}f_{\rm sq}\,\sigma_x
+\operatorname{Re}f_{\rm sq}\,\sigma_y\right],\\
f_{\rm sq}(u,v)
&=1+e^{iu}+e^{iv}-e^{i(u+v)}.
\end{aligned}
\label{eq:2d_sq_f}
\end{equation}
In physical square-lattice coordinates,
\begin{equation}
k_x=\frac{u+v}{2},
\quad
k_y=\frac{u-v}{2}.
\end{equation}
The structure factor becomes
\begin{equation}
f_{\rm sq}
=
2e^{ik_x}
\left(
\cos k_y-i\sin k_x
\right).
\label{eq:2d_sq_factor}
\end{equation}
A change to the physical orbital-position convention,
$
U(\mathbf k) =
\operatorname{diag}(1,e^{ik_x}),
$
removes the phase $e^{ik_x}$, giving the two-band Majorana Bloch symbol
\begin{equation}
h_{\rm sq}(\mathbf k)
=
2t
\left[
-\sin k_x\,\sigma_x
+\cos k_y\,\sigma_y
\right].
\label{eq:2d_sq_bloch}
\end{equation}
The squared dispersion is
\begin{equation}
E_{\rm sq}^2(\mathbf k)
=
4t^2
\left(
\sin^2k_x+\cos^2k_y
\right),
\label{eq:2d_sq_dispersion}
\end{equation}
whose only zeros in the Brillouin zone of the two-site unit cell are $\mathbf K_{\pm}=(0,\pm\pi/2)$. These
Majorana-reality-related nodes together describe one complex Dirac fermion. Near $\mathbf K_+$,
\begin{equation}
h_{\rm sq}(\mathbf K_++\mathbf q)
=
-2t
\left(
q_x\sigma_x+q_y\sigma_y
\right)
+O(|\mathbf q|^2),
\label{eq:2d_sq_dirac}
\end{equation}
so the Dirac cone is isotropic in the physical square metric without tuning the hopping magnitudes.

The class-D Pfaffian parity follows from the structure factor at the four PHIM of the $(u,v)$ torus:
\begin{equation}
\begin{aligned}
f_{\rm sq}(0,0)&=f_{\rm sq}(\pi,0)=2,\\
f_{\rm sq}(0,\pi)&=2,\quad
f_{\rm sq}(\pi,\pi)=-2.
\end{aligned}
\end{equation}
Hence
\begin{equation}
\mathcal P_D(H_{\rm sq})
=
-1
=
\mathcal P_D(H_K),
\label{eq:2d_sq_pfaffian}
\end{equation}
whereas $\mathcal P_D(H_{\rm hc}^{W})=+1$. Hence $H_{\rm sq}$ shares the odd Pfaffian sector of $H_K$, but, as
we show next, admits compact charge generators with substantially improved locality.

The $\pi$ flux is a fixed background sign pattern, not a dynamical gauge field. Although the same hopping
geometry is familiar from complex-fermion models, here the two nodes are constrained by Majorana reality and
must not be counted as independent complex valleys.

We begin with an exact sharp spectral lift of the IR Dirac number. We choose a sharp spectral sign invariant
under the two bond-centered projective reflections of $H_{\rm sq}$, whose momentum actions are
$(k_x,k_y)\mapsto(-k_x,k_y)$ and $(k_x,k_y)\mapsto(k_x,\pi-k_y)$. A periodic representative on the original
$(u,v)$ Brillouin torus is
\begin{equation}
s_P(u,v)
=\operatorname{sgn}[\sin u-\sin v]
=\operatorname{sgn}[2\cos k_x\sin k_y].
\label{eq:2d_sq_spectral_sign}
\end{equation}
Since $s_P(-u,-v)=-s_P(u,v)$ and $s_P(\mathbf K_{\pm})=\pm1$, this sign is compatible with Majorana reality and
assigns opposite unit charges to the two nodes. Away from its sign seams,
\begin{equation}
J_{s_P}(\mathbf k)
=
-i\,s_P(\mathbf k)\mathbf1,
\ \
J_{s_P}^2=-1,
\ \
[J_{s_P},A_{\rm sq}]=0.
\label{eq:2d_sq_Js}
\end{equation}
As in the honeycomb lattice, we take both linear dimensions $L_1$ and $L_2$ to be even, so that the square
lattice contains $N=L_1L_2\in4\mathbb Z$ two-site $A/B$ unit cells, or $2N$ Majorana operators. For example,
$L_1=L_2=4m$, with PBC along the $u$ direction and APBC along the $v$ direction, gives a grid closed under
momentum inversion and $u\mapsto u+\pi$, while avoiding $\sin u=\sin v$. On such a finite momentum grid, the
centered charge
\begin{equation}
Q_{s_P}
=\frac i4\gamma^T J_{s_P}\gamma
\label{eq:2d_sq_Qs}
\end{equation}
has integer eigenvalues and generates an exact compact $U(1)$. The sharp jumps of $s_P$ produce algebraically
decaying real-space tails. This nonlocality belongs to the particular spectral generator, however, and need not
be a property of every compact charge of the same Hamiltonian.

To obtain an exponentially local compact charge while leaving the gapless finite-range Hamiltonian unchanged, we
exploit an exact pairing of momentum sectors. Folding these sectors exposes additional conserved matrices, which
allow the sharp sign to be replaced by a smooth, everywhere-defined involution without changing its IR action.

Introduce the coordinates $
U=u$ and $V=u+v$.
In these coordinates,
\begin{equation}
f_{\rm sq}(U,V)
=
1+e^{iU}+e^{i(V-U)}-e^{iV},
\end{equation}
and the structure factor satisfies the exact twisted relation
\begin{equation}
f_{\rm sq}(U+\pi,V)
=
-e^{iV}f_{\rm sq}(U,V)^*.
\label{eq:2d_sq_twisted}
\end{equation}
Consequently,
\begin{equation}
E_{\rm sq}(U+\pi,V)
=
E_{\rm sq}(U,V).
\label{eq:2d_sq_spectral_matching}
\end{equation}

In the same coordinates, the reflection-invariant sharp sign is
\begin{equation}
s_P(U,V)=\operatorname{sgn}b_P(U,V),
\end{equation}
where
\begin{equation}
\begin{aligned}
b_P(U,V)&=\frac12[\sin U-\sin(V-U)]\\
&=\cos\frac V2\,\sin\left(U-\frac V2\right).
\end{aligned}
\label{eq:2d_sq_bP}
\end{equation}
The real-analytic function $b_P$ satisfies
\begin{equation}
b_P(U+\pi,V)=-b_P(U,V).
\end{equation}
Thus, away from its zeros, $U\mapsto U+\pi$ exchanges the two sharp-charge sectors while preserving their
spectra.

We combine the two isospectral sectors into the folded Bloch Hamiltonian
\begin{equation}
h_{\rm fold}(U,V)=
\begin{pmatrix}
h_0(U,V)&0\\
0&h_1(U,V)
\end{pmatrix},
\label{eq:2d_sq_hfold}
\end{equation}
where $h_0(U,V)=h_{\rm sq}(U,V)$ and $h_1(U,V)=h_{\rm sq}(U+\pi,V)$. The shift $U\mapsto U+\pi$ exchanges the
two sectors,
\begin{equation}
h_{\rm fold}(U+\pi,V)
=\tau_x h_{\rm fold}(U,V)\tau_x,
\end{equation}
where $\tau_i$ acts on the folded-sector index. This sewing relation arises from the folding convention and does
not represent a physical spectral discontinuity.

The two blocks are related by the orbital matrix
\begin{equation}
\Xi(V)=\cos\frac V2\,\sigma_x-\sin\frac V2\,\sigma_y.
\label{eq:2d_sq_XiV}
\end{equation}
Since $\Xi^\dagger=\Xi$ and $\Xi^2=\mathbf1$, Eq.~\eqref{eq:2d_sq_twisted} gives
\begin{equation}
h_1(U,V)=\Xi(V)h_0(U,V)\Xi(V).
\label{eq:2d_sq_intertwiner}
\end{equation}
Thus $\Xi(V)$ is a Hermitian unitary intertwiner between the paired orbital Hamiltonians.

The pairing supplies three conserved Hermitian matrices:
\begin{equation}
T_1=\tau_z\otimes\mathbf1,
\
T_2=\tau_x\otimes\Xi(V),
\
T_3=\tau_y\otimes\Xi(V).
\label{eq:2d_sq_commuting_triplet}
\end{equation}
Each commutes with $h_{\rm fold}$, and the three anticommute pairwise and satisfy $ T_a^2=\mathbf1.$ Every
Hermitian combination
\begin{equation}
B(U,V)
=
b_1(U,V)T_1
+
b_2(U,V)T_2
+
b_3(U,V)T_3
\end{equation}
therefore commutes with $h_{\rm fold}$, while
$
B^2 =
\left(
b_1^2+b_2^2+b_3^2
\right)\mathbf1.
$

These additional conserved directions allow us to replace the sharp charge by a smooth compact one. We begin
with the unflattened charge matrix $B_P^{(0)}(U,V)=b_P(U,V)T_1$, whose spectral flattening reproduces $s_PT_1$
wherever $b_P\ne0$, but is undefined at the sign seams. Adding suitable components along $T_2$ and $T_3$ removes
these zeros while preserving conservation and the IR charge assignment. An explicit choice is
\begin{equation}
\begin{aligned}
&B_{\rm sq}(U,V)
=
b_P(U,V)\,
\tau_z\otimes\mathbf1
\\
&+
\sin\frac V2\,
\tau_x\otimes\Xi(V)
 -
\cos\left(U+\frac V2\right)
\tau_y\otimes\Xi(V).
\end{aligned}
\label{eq:2d_sq_B}
\end{equation}
The first term retains the original sharp-charge structure, while the two transverse terms vanish at the Dirac
node and eliminate zeros elsewhere. Their relative signs ensure compatibility with the folded sewing and
Majorana-reality conditions.

By Eq.~\eqref{eq:2d_sq_commuting_triplet}, the matrix $B_{\rm sq}$ satisfies
\begin{equation}
[B_{\rm sq},h_{\rm fold}]=0,
\quad
B_{\rm sq}^2=r_{\rm sq}^2\mathbf1,
\end{equation}
where
\begin{equation}
r_{\rm sq}^2(U,V)=b_P(U,V)^2
+\sin^2\frac V2+\cos^2\left(U+\frac V2\right).
\end{equation}
This function is strictly positive on the compact real momentum torus. We can therefore spectrally flatten
$B_{\rm sq}$ to obtain
\begin{equation}
q_{\rm loc}
=
B_{\rm sq}(B_{\rm sq}^2)^{-1/2}
=
\frac{B_{\rm sq}}{r_{\rm sq}},
\quad
J_{\rm loc}=-iq_{\rm loc}.
\label{eq:2d_sq_qloc}
\end{equation}
The resulting matrices satisfy
\begin{equation}
q_{\rm loc}^\dagger=q_{\rm loc},\ q_{\rm loc}^2=1,\
J_{\rm loc}^2=-1,\  [J_{\rm loc},A_{\rm sq}]=0.
\label{eq:2d_sq_qloc_properties}
\end{equation}

At the positive Dirac node $(U,V)=(\pi/2,0)$,
\begin{equation}
b_P=1,\quad
B_{\rm sq}=\tau_z\otimes\mathbf1,\quad
r_{\rm sq}=1,\quad
q_{\rm loc}=\tau_z\otimes\mathbf1.
\label{eq:2d_sq_qloc_IR}
\end{equation}
The paired Majorana sectors therefore carry opposite unit charges, yielding unit Dirac number after
complexification. The smooth and sharp charges consequently have the same IR action:
\begin{equation}
\rho_{\rm IR}(Q_{\rm loc})
=
\rho_{\rm IR}(Q_{s_P})
=
Q_D.
\end{equation}

To establish exponential locality, we must verify that the half-angle factors in Eq.~\eqref{eq:2d_sq_B} are
compatible with the folded momentum sewing and Majorana reality. Define
\begin{equation}
\Gamma_f(U,V)=
\begin{pmatrix}
\gamma(U,V)\\
\gamma(U+\pi,V)
\end{pmatrix}
\end{equation}
and $S=\tau_x\otimes\mathbf1$. Since $\Gamma_f(U+\pi,V)=S\Gamma_f(U,V)$, the charge matrix obeys
\begin{equation}
\begin{aligned}
B_{\rm sq}(U+\pi,V)&=S B_{\rm sq}(U,V)S,\\
B_{\rm sq}(U,V+2\pi)&=B_{\rm sq}(U,V).
\end{aligned}
\label{eq:2d_sq_B_sewing}
\end{equation}
Direct substitution also establishes the Majorana-reality conditions
\begin{equation}
\begin{aligned}
B_{\rm sq}(-U,-V)&=-B_{\rm sq}(U,V)^*,\\
q_{\rm loc}(-U,-V)&=-q_{\rm loc}(U,V)^*.
\end{aligned}
\label{eq:2d_sq_majorana_reality_q}
\end{equation}
Thus $J_{\rm loc}=-iq_{\rm loc}$ defines a real antisymmetric Majorana kernel consistent with the folded sewing.

To demonstrate exponential locality explicitly, introduce $P=2U$ and transform to the periodic doubled-cell
basis\footnote{The even- and odd-site Fourier modes are obtained from the folded sectors as
$\gamma_e(P,V)=[\gamma(U,V)+\gamma(U+\pi,V)]/\sqrt2$ and
$\gamma_o(P,V)=e^{iU}[\gamma(U,V)-\gamma(U+\pi,V)]/\sqrt2$. Thus
$(\gamma_e,\gamma_o)^T=F(U)^\dagger\Gamma_f(U)$. The factor $e^{iU}$ removes the relative intracell phase,
making both components periodic in $P=2U$.} using
\begin{equation}
F(U)=\frac1{\sqrt2}
\begin{pmatrix}
1&e^{-iU}\\
1&-e^{-iU}
\end{pmatrix}\otimes\mathbf1.
\label{eq:2d_sq_F}
\end{equation}
Since $F(U+\pi)=SF(U)$, the matrices
\begin{equation}
\begin{aligned}
\mathcal B(P,V)&=F(U)^\dagger B_{\rm sq}(U,V)F(U),\\
\tilde q(P,V)&=F(U)^\dagger q_{\rm loc}(U,V)F(U)
\end{aligned}
\label{eq:2d_sq_periodic_cell}
\end{equation}
are periodic in $(P,V)$. We reserve $q_{\rm loc}$ for the folded $(U,V)$ frame and $\tilde q$ for the periodic
$(P,V)$ frame. Substitution of Eq.~\eqref{eq:2d_sq_B} shows that $\mathcal B$ is a finite Laurent polynomial in
$e^{iP}$ and $e^{iV}$. Moreover, $\tilde q=\mathcal B/r_{\rm sq}$, with $r_{\rm sq}^2(P,V)$ strictly positive on
the real momentum torus. Consequently, $r_{\rm sq}^{-1}$, $\tilde q$, and $\mathcal J=-i\tilde q$ extend
analytically to a sufficiently narrow complex momentum strip. Shifting the Fourier contours then yields
\begin{equation}
\|\mathcal J(R_1,R_2)\|
\le C_\eta e^{-\eta(|R_1|+|R_2|)}
\label{eq:2d_sq_exp_decay}
\end{equation}
for some $\eta>0$. Hence $Q_{\rm loc}$ is exponentially local, although generally not finite range.

On a finite torus compatible with the doubled cell, an orthogonal Majorana basis transformation brings $J_{\rm
loc}$ to canonical form, yielding $N$ complex fermionic modes $f_j$ and the centered charge
\begin{equation}
Q_{\rm loc}=\frac i4\gamma^T J_{\rm loc}\gamma
=\sum_j\left(f_j^\dagger f_j-\frac12\right).
\label{eq:2d_sq_compactness}
\end{equation}
Since $N$ is even, $Q_{\rm loc}$ has integer eigenvalues and satisfies $e^{2\pi iQ_{\rm loc}}=1$. Thus the
exponentially local charge generates an exact compact $U(1)$ symmetry, although its onsite realization and local
gauging require separate constructions.

The spectral-pairing mechanism extends to more general lattice Hamiltonians. Suppose two Bloch sectors related
by a half-reciprocal-lattice shift $\mathbf g=\mathbf G/2$ admit a unitary intertwiner,
\begin{equation}
h(\mathbf k+\mathbf g)
=\Xi(\mathbf k)h(\mathbf k)\Xi(\mathbf k)^\dagger.
\label{eq:2d_general_spectral_intertwiner_main}
\end{equation}
The folded Hamiltonian then admits three mutually anticommuting conserved Hermitian matrices $T_a$. A
nowhere-vanishing analytic combination $B=\sum_a b_aT_a$, compatible with folded sewing and Majorana reality,
defines a conserved compact involution $q_{\rm loc}=B/\sqrt{B^2}$ with exponentially decaying real-space
couplings. This provides a general sufficient construction of exponentially local compact $U(1)$ charges,
although onsite relocalizability and gauge locality require separate conditions. Explicit identical- and
twisted-sector families are developed in Appendix~\ref{app:2d_gauge_local_families}.

\subsection{Charge-bundle topology, Pfaffian parity, and boundary flow}
\label{sec:2d_sq_CQ}

The sharp charge $Q_{s_P}$ is compact but discontinuous in momentum space, whereas the exponentially local
$Q_{\rm loc}$ defines a smooth, globally sewn charge bundle. Its topology remains well defined at the massless
Dirac point, where the physical class-D Chern number is undefined. We establish its relation to class-D Pfaffian
parity and explore its implications for charge locality, boundary spectral flow, and fermionic lattice
anomalies.

\subsubsection{Charge topology and class-D Pfaffian parity}

In the periodic doubled-cell basis of Eq.~\eqref{eq:2d_sq_periodic_cell}, the smooth Hermitian involution
$\tilde q^2=\mathbf1$ defines a rank-two positive-charge bundle $\mathcal E_+=\operatorname{Im}P_+^Q$, with
$P_+^Q=(\mathbf1+\tilde q)/2$.\footnote{ In the folded-sector basis, the corresponding projector is
$(\mathbf1+q_{\rm loc})/2=FP_+^QF^\dagger$. The sewing relations $F(U+\pi)=SF(U)$ and $q_{\rm
loc}(U+\pi,V)=Sq_{\rm loc}(U,V)S$ ensure that the two representations define the same charge bundle and Chern
number.} Its first Chern number is
\begin{equation}
C_Q=\frac{1}{2\pi i}\int_{T_Q^2}
\operatorname{Tr}\!\left(
P_+^Q\,dP_+^Q\wedge dP_+^Q
\right).
\label{eq:2d_sq_CQ_def}
\end{equation}
Here $T_Q^2$ is the doubled-cell momentum torus with coordinates $(P,V)$, oriented by $dV\wedge dP$, where
$P=2U$. Since $P_+^Q$ is smooth and independent of the occupied-energy subspace, $C_Q$ remains well defined even
at the massless Dirac point.

Majorana reality imposes
$
\tilde q(-\mathbf p)=-\tilde q(\mathbf p)^*.
$
At each PHIM $\Lambda_Q$, the matrix $\mathcal J(\Lambda_Q)=-i\tilde q(\Lambda_Q)$ is therefore real and
antisymmetric. In the canonical periodic doubled-cell frame, its Pfaffian signs are $(+,-,-,-)$ at
$(P,V)=(0,0),(\pi,0),(0,\pi),(\pi,\pi)$, respectively. The charge-Pfaffian identity gives
\begin{equation}
(-1)^{C_Q}
=\prod_{\Lambda_Q}
\operatorname{sgn}\operatorname{Pf}
\mathcal J(\Lambda_Q)
=-1,
\label{eq:2d_sq_q_pfaffians}
\end{equation}
and hence
\begin{equation}
C_Q(H_{\rm sq};Q_{\rm loc})\equiv1\pmod2.
\label{eq:2d_sq_CQ_odd}
\end{equation}
The Pfaffian signs fix only the parity of $C_Q$. The explicit curvature calculation in
Appendix~\ref{app:2d_sq_CQ_integer} gives $C_Q=+1$ with the orientation above.

This charge invariant is related to, but distinct from, the physical class-D topology. For a gapped Majorana
Hamiltonian commuting with a smooth compact charge, Majorana conjugation relates the occupied negative-charge
bundle to the conjugate of the unoccupied positive-charge bundle. Consequently,
\begin{equation}
\nu_{\rm Maj}=2C_{+,\mathrm{occ}}-C_Q,
\label{eq:2d_nu_CQ_relation}
\end{equation}
where $C_{+,\mathrm{occ}}$ is the Chern number of occupied positive-charge states. In particular,
\begin{equation}
\nu_{\rm Maj}\equiv C_Q\pmod2.
\label{eq:2d_nu_CQ_mod2}
\end{equation}
The derivation and the possibility of even shifts in $C_Q$ at fixed $\nu_{\rm Maj}$ are discussed in
Appendix~\ref{app:2d_sq_gapped_characteristic}.

At criticality, $\nu_{\rm Maj}$ is undefined, but the class-D Pfaffian parity $\mathcal P_D(H)$ remains
meaningful when the physical Hamiltonian is nonsingular at the four PHIM. If $H$ commutes with a smooth compact
charge on the same periodic Majorana momentum torus, and the four PHIM lie in one connected nonsingular region,
then
\begin{equation}
\mathcal P_D(H)=(-1)^{C_Q}.
\label{eq:2d_CQ_pfaffian_relation}
\end{equation}

For $H_{\rm sq}$, the physical Pfaffian parity is defined on the primitive-cell torus, where the Dirac nodes lie
away from the PHIM:
\begin{equation}
\mathcal P_D(H_{\rm sq};T_{uv}^2)=-1.
\label{eq:2d_Hsq_original_pfaffian}
\end{equation}
By contrast, the smooth charge is periodic on the doubled-cell torus $T_Q^2$, where folding places the Dirac
nodes at a PHIM and makes the physical Pfaffian singular. The same-torus relation therefore cannot be applied
directly. The canonical-folding analysis in Appendix~\ref{app:2d_sq_gapless_parity} shows that the additional
charge-Pfaffian factor is trivial for $q_{\rm loc}$, yielding the exact massless identity
\begin{equation}
\mathcal P_D(H_{\rm sq})=(-1)^{C_Q}=-1.
\label{eq:2d_sq_direct_gapless_parity}
\end{equation}
No physical mass or gapped descendant is required.

By comparison, the onsite Dirac-number charge of $H_{\rm hc}^{W}$ has a constant projector, and therefore
\begin{equation}
C_Q(H_{\rm hc}^{W};Q_D)=0.
\label{eq:2d_hcstar_CQ_zero}
\end{equation}
Thus $H_{\rm sq}$ and $H_{\rm hc}^{W}$ share the same massless Dirac IR content but differ in charge-bundle
topology. The invariant $C_Q$ characterizes the symmetry-equipped pair $(H,\tilde q)$, not the IR spectrum
alone.

\subsubsection{Locality and charge bulk--edge correspondence}

Nonzero $C_Q$ obstructs a globally smooth periodic basis transformation that makes the charge matrix constant,
although it does not exclude the patched nonlocal relocalization constructed below. It also forbids a strictly
finite-range compact charge: if $\tilde q$ had Laurent-polynomial entries, then $P_+^Q=(\mathbf1+\tilde q)/2$
would define an exactly flat, spectrally isolated band of a finite-range Hamiltonian, contradicting the
flat-band no-go theorem for $C_Q\ne0$~\cite{chenImpossibility2014}. Consequently, the spectral flattening
$q_{\rm loc}=B_{\rm sq}(B_{\rm sq}^2)^{-1/2}$ necessarily has unbounded hopping range, albeit with exponentially
decaying amplitudes.

Topology also imposes a quantitative lower bound on the charge spread. Let $\mathbf k=(k_1,k_2)$ denote periodic
momentum coordinates on $T_Q^2$, and expand the charge matrix in a periodic Bloch basis as $\tilde q(\mathbf
k)=\sum_{\mathbf R}q_{\mathbf R} e^{-i\mathbf k\cdot\mathbf R}$. Define the root-mean-square spread, in
lattice-cell units, by
\begin{equation}
\ell_Q^2=
\frac{\sum_{\mathbf R}|\mathbf R|^2
\operatorname{tr}(q_{\mathbf R}^\dagger q_{\mathbf R})}
{\sum_{\mathbf R}
\operatorname{tr}(q_{\mathbf R}^\dagger q_{\mathbf R})}.
\label{eq:2d_charge_spread_def}
\end{equation}
Then
\begin{equation}
\ell_Q^2\ge\frac{4|C_Q|}{\pi d_{\rm fib}},
\label{eq:2d_charge_spread_bound}
\end{equation}
where $d_{\rm fib}$ is the Bloch-fiber dimension. For $H_{\rm sq}$, the periodic matrix is $\tilde q(P,V)$, with
$(k_1,k_2)=(P,V)$, $d_{\rm fib}=4$, and $C_Q=1$, giving $\ell_Q^2\ge1/\pi$ in doubled-cell units. The derivation
is given in Appendix~\ref{app:2d_charge_spready}.

The same topology enforces boundary charge-spectrum flow. Let $q_{\rm loc}$ now denote the real-space
one-particle charge operator, and let $R$ project onto a half-plane. Although $q_{\rm loc}^2=\mathbf1$, the
compressed operator $q_{\rm cut}=Rq_{\rm loc}R$ can have boundary-localized eigenvalues in $(-1,1)$. A nonzero
$C_Q$ forces the positive-charge hybrid-Wannier centers to wind as the conserved edge momentum traverses its
Brillouin zone. Their passage across the cut produces charge eigenvalues flowing between $-1$ and $+1$. If
$n_\uparrow$ and $n_\downarrow$ count upward and downward zero crossings, respectively, the bulk--edge
correspondence gives
\begin{equation}
n_\uparrow-n_\downarrow=C_Q,
\label{eq:2d_charge_edge_spectral_flow}
\end{equation}
with consistent bulk and boundary orientations
\cite{kellendonkEdge2002,fidkowskiEntanglement2010,
fidkowskiJacksonKlich2011}. Nonzero $C_Q$ also obstructs an exponentially local, exactly compact half-space
charge agreeing with the bulk away from the boundary.

Fig.~\ref{fig:majoranaedge}(d) illustrates this flow for $H_{\rm sq}$. The compressed-charge spectrum depends
only on the charge operator and boundary geometry, not on the Chern mass, and requires no physical energy gap.
In contrast, the physical edge energies depend on that mass; charge-flow states need not coincide with physical
edge excitations. We return to this distinction when discussing the boundary spectra. The bulk--edge
correspondence and compact-termination obstruction are detailed in Appendix~\ref{app:2d_charge_bulk_edge}.

\subsubsection{Majorana-layer anomaly}
\label{sec:2d_sq_majorana_layer}

The odd charge Chern number also reveals an intrinsically fermionic lattice anomaly of the compact Dirac-number
symmetry, with a many-body interpretation extending beyond quadratic fermions.

The quarter-turn of $Q_{\rm loc}$ generates a $\mathbb Z_4^F$ subgroup,
\begin{equation}
G=e^{i\varphi_F}e^{i\pi Q_{\rm loc}/2},
\quad G^2=(-1)^F,
\end{equation}
where $\varphi_F$ fixes the microscopic fermion-parity convention without affecting the action on fermion
operators. Restricting the quadratic generator to a spatial region $A$ gives $G_A$. Whereas an onsite symmetry
satisfies $G_A^2=(-1)^{F_A}$, the restricted non-onsite symmetry leaves a boundary transformation. With the
convention $\Omega_{\partial A}=G_A^{-2}(-1)^{F_A}$, its one-particle action is
\begin{equation}
v_{\partial A}
=\exp(2\pi i RP_+^Q R),
\label{eq:2d_sq_majorana_boundary_walk}
\end{equation}
where $R$ projects onto $A$. Because $P_+^Q$ is exponentially local, its compression deviates from a projector
only through couplings across the cut. Accordingly, $v_{\partial A}$ acts nontrivially only near $\partial A$,
up to exponentially decaying tails.

For a translation-invariant half-plane, a compressed-charge eigenvalue $\lambda_Q(k_\parallel)$ corresponds to
the boundary-unitary eigenvalue $e^{i\pi[1+\lambda_Q(k_\parallel)]}$. Each zero crossing of $\lambda_Q$ thus
produces an eigenphase crossing through $-1$ with the same orientation. By the charge bulk--edge correspondence
of Eq.~\eqref{eq:2d_charge_edge_spectral_flow}, the net boundary-unitary winding is $\chi_{\partial A}=C_Q$.

For Majorana-canonical boundary transformations, this winding measures net Majorana transport. Each transported
Majorana contributes a factor $\sqrt2$ to the fermionic QCA index, as in the 1D shift ${\cal S}$ of
Sec.~\ref{sec:1d_majorana_shift}
\cite{fidkowski2019interacting}. Accounting
for the doubled one-particle representation gives~\cite{kobayashiInamuraShiozaki2026}
\begin{equation}
\operatorname{ind}_{\rm fQCA}(v_{\partial A})^2
=2^{\chi_{\partial A}}=2^{C_Q}.
\end{equation}
For $H_{\rm sq}$, the chosen orientation yields $\operatorname{ind}_{\rm fQCA}(v_{\partial A}) =\sqrt2$. This
irrational index identifies the nontrivial Majorana-layer anomaly
\begin{equation}
[n_2]=C_Q\bmod2=1
\ \ \in H^2(B\mathbb Z_2,\mathbb Z_2).
\label{eq:2d_sq_majorana_layer_index}
\end{equation}

Although computed from a free-fermion charge projector, $[n_2]$ characterizes the many-body symmetry action. Its
interpretation as an interaction-stable lattice anomaly uses the extension of fermionic-QCA indices to
exponentially quasi-local symmetry actions
\cite{kobayashiInamuraShiozaki2026}.
Within this framework, the anomaly is unchanged by local interacting deformations preserving the exact symmetry.
It therefore provides a diagnostic beyond the quadratic charge Chern number and class-D Pfaffian parity. The
obstruction to finite-depth onsite trivialization is nevertheless compatible with the generally nonlocal
canonical relocalization constructed below.

For compatible symmetric gapped invertible phases, the anomaly further requires
\begin{equation}
c_-\in\mathbb Z+\frac12.
\label{eq:2d_sq_majorana_layer_cminus}
\end{equation}
This is consistent with our relation $\nu_{\rm Maj}=2C_{+,\mathrm{occ}}-C_Q$ for gapped charge-preserving
descendants and with the lattice-anomaly analysis of Ref.~\cite{kobayashiInamuraShiozaki2026}. The constraint
does not assign a chiral central charge to the massless $H_{\rm sq}$.

\subsection{Patched relocalization and asymptotic gauge locality}
\label{sec:2d_sq_gauge_locality}

The preceding analysis shows that the nontrivial charge bundle obstructs a globally smooth periodic onsite
trivialization, while the anomaly manifests as a nontrivial fermionic QCA on a 1D boundary. Neither excludes a
generally nonlocal canonical relocalization. We construct such a transformation, making the compact
charge onsite while mapping the massless $H_{\rm sq}$ to a Schwartz-local Hamiltonian in the relocalized
lattice. This provides a compact gauge regulator with controlled finite-range approximations.

\subsubsection{Patched relocalization of compact charges}

\begin{figure}[tb]
\centering
\includegraphics[width=\linewidth]{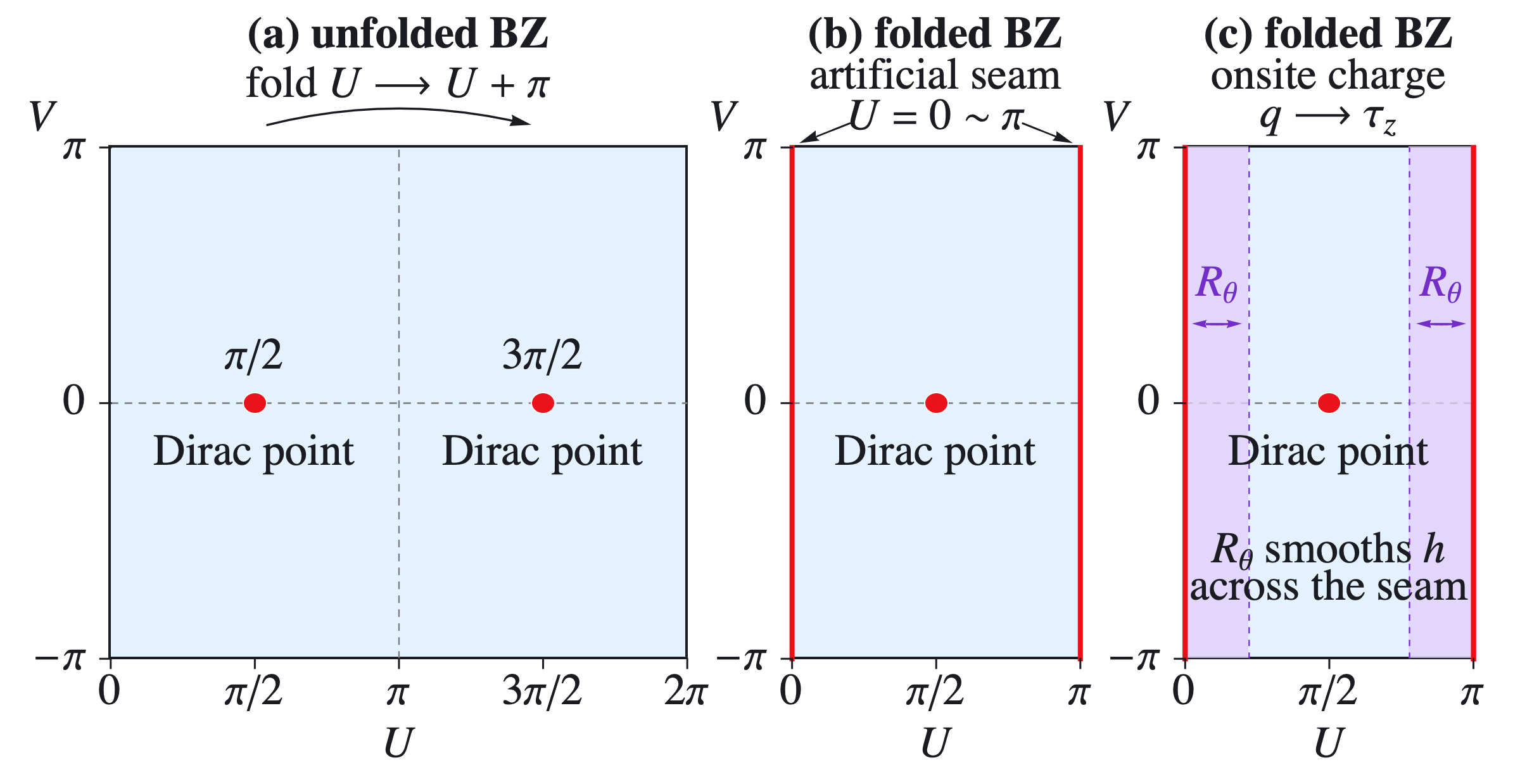}
\caption{Patched charge-on-site relocalization.
(a) The two Majorana-related Dirac nodes at
$(U,V)=(\pi/2,0)$ and $(3\pi/2,0)$ in the
original momentum geometry are related by
$U\mapsto U+\pi$.
(b) Folding and the northern-frame rotation
make the charge constant, leaving an artificial
seam at $U=0\sim\pi$ in a spectrally gapped
region.
(c) The charge-preserving orbital rotation
$R_\theta$, supported near the seam, makes
the matter Hamiltonian smooth and periodic
without modifying the physical Dirac cone.
The resulting hopping amplitudes decay
faster than any power of distance in the
relocalized lattice.}
\label{fig:patched}
\end{figure}

We now construct a canonical relocalization that makes $Q_{\rm loc}$ onsite while preserving Hamiltonian
locality. Although $Q_{\rm loc}$ is exponentially local, its charge densities are not onsite commuting Gauss
generators. The original momentum geometry, with its two Majorana-related Dirac nodes, is illustrated in
Fig.~\ref{fig:patched}(a).

We use the folded Hamiltonian $h_{\rm fold}$ of Eq.~\eqref{eq:2d_sq_hfold}, whose blocks satisfy $h_1=\Xi
h_0\Xi$, with $\Xi(V)$ defined in Eq.~\eqref{eq:2d_sq_XiV}. Conjugation by $\operatorname{diag}(\mathbf1_2,\Xi)$
brings $h_{\rm fold}$ to $\mathbf1_2\otimes h_0$ and the smooth charge to $(\mathbf
n\cdot\boldsymbol\tau)\otimes\mathbf1_2$, where
\begin{equation}
\mathbf n=\frac1{r_{\rm sq}}
\left(\sin\frac V2,\,
-\cos\left(U+\frac V2\right),\,b_P\right).
\end{equation}
The northern-frame rotation
\begin{equation}
R_N(U,V)=
\frac{\mathbf1_2+\tau_z(\mathbf n\cdot\boldsymbol\tau)}
{\sqrt{2(1+n_z)}}
\label{eq:2d_sq_RN_main}
\end{equation}
is smooth on the cut $0\le U<\pi$, $0\le V<2\pi$ and extends smoothly to its boundary. Indeed, $n_z=-1$ would
require $\sin(V/2)=\cos(U+V/2)=0$ and $b_P<0$, whereas their simultaneous zeros on the closed cut have $b_P=1$.
Thus $R_N$ maps the charge to $\tau_z\otimes\mathbf1_2$ without changing $\mathbf1_2\otimes h_0$.

The charge-adapted frame is nevertheless not globally periodic. Its nontrivial charge-bundle topology reappears
as a sewing discontinuity at $U=0\sim\pi$, as illustrated in Fig.~\ref{fig:patched}(b). Crucially, this
artificial UV seam lies away from the physical Dirac node.

The seam is spectrally gapped, with $E_{\rm sq}(0,V)=E_{\rm sq}(\pi,V)=2|t|$. We can therefore smooth the matter
Hamiltonian without altering its spectrum or the constant charge. Write $h_0=d_x\sigma_x+d_y\sigma_y$ and
$e=(d_x^2+d_y^2)^{1/2}$. Choose a smooth collar function $\theta(U)$ equal to $\pi/2$ near both seam edges and
zero near the physical Dirac node. On the support of $\theta$, where $e>0$, define
\begin{equation}
R_\theta(U,V)=
\exp\!\left[
-\frac{i\theta(U)}{2e}
(d_y\sigma_x-d_x\sigma_y)
\right],
\end{equation}
and set $R_\theta=\mathbf1_2$ elsewhere. This charge-preserving rotation, illustrated in
Fig.~\ref{fig:patched}(c), gives
\begin{equation}
\begin{aligned}
h_\sharp^{\rm sq}(U,V)
&=R_\theta h_0R_\theta^\dagger\\
&=\cos\theta\,
(d_x\sigma_x+d_y\sigma_y)
+e\sin\theta\,\sigma_z.
\end{aligned}
\label{eq:2d_sq_hsharp}
\end{equation}
Near the seam, $h_\sharp^{\rm sq}=e\sigma_z$, where
\begin{equation}
e^2=4t^2\left[
1+\frac{\cos(2U-V)-\cos V}{2}
\right].
\end{equation}
Since $e^2$ is smooth and periodic in $(P,V)=(2U,V)$ and strictly positive at the seam, $h_\sharp^{\rm sq}$
extends to a smooth periodic symbol on the relocalized torus $T_X^2$. Moreover, $R_\theta=\mathbf1_2$ near the
Dirac node. The construction is exactly isospectral and introduces no independent UV Hamiltonian deformation.

The combined cut-frame unitary is
\begin{equation}
T=(\mathbf1_2\otimes R_\theta)
(R_N\otimes\mathbf1_2)
\operatorname{diag}(\mathbf1_2,\Xi),
\end{equation}
which satisfies
\begin{equation}
Tq_{\rm loc}T^\dagger
 =\tau_z\otimes\mathbf1_2,\ \
Th_{\rm fold}T^\dagger
 =\mathbf1_2\otimes h_\sharp^{\rm sq}.
\end{equation}
For $t>0$, its transition across the identified $U$-edges is
\begin{equation}
\begin{aligned}
G(V)
&=T(\pi,V)(\tau_x\otimes\mathbf1_2)
T(0,V)^\dagger\\
&=\operatorname{diag}(e^{iV},1,1,e^{-iV}).
\end{aligned}
\label{eq:2d_sq_G_main}
\end{equation}
The positive-charge block has $\det G_+(V)=e^{iV}$ and hence unit winding. Along the other identified edges,
$T(U,V+2\pi)=(\tau_z\otimes\mathbf1_2)T(U,V)$, a constant charge-preserving transition. Thus the charge topology
$C_Q=1$ is encoded in the nonperiodic relocalizing frame, although the transformed Hamiltonian is smooth and
periodic.

It remains to show that this construction defines a canonical transformation of the physical fermions, rather
than only a unitary equivalence of Bloch matrices. Consider a compatible finite momentum grid closed under
folding and inversion, and let $\bar{\mathbf p}$ denote inverse momentum on the reduced torus. Folding combines
the sectors $U$ and $U+\pi$. Under inversion, these sectors are exchanged for $0<U<\pi$, whereas at $U=0$ each
maps to itself. The folded Majorana reality structure is therefore represented by $\Sigma_{\mathbf p}K$, where
$K$ denotes complex conjugation and
\begin{equation}
\Sigma_{\mathbf p}=
\begin{cases}
\mathbf1_4,&U=0,\\
\tau_x\otimes\mathbf1_2,&0<U<\pi.
\end{cases}
\label{eq:2d_sq_finite_Sigma}
\end{equation}
Here $\Sigma_{\bar{\mathbf p}}=\Sigma_{\mathbf p}$ and $\Sigma_{\mathbf p}^2=\mathbf1_4$. The folded Hamiltonian
and charge satisfy
\begin{equation}
\begin{aligned}
h_{\rm fold}(\bar{\mathbf p})
&=-\Sigma_{\mathbf p}
h_{\rm fold}(\mathbf p)^*
\Sigma_{\mathbf p},\\
q_{\rm loc}(\bar{\mathbf p})
&=-\Sigma_{\mathbf p}
q_{\rm loc}(\mathbf p)^*
\Sigma_{\mathbf p}.
\end{aligned}
\end{equation}

Let $E_+=(\mathbf1_2,0)^T$. The positive-charge frame is $V_+(\mathbf p)=T(\mathbf p)^\dagger E_+$. Majorana
conjugation fixes its negative-charge partner:
\begin{equation}
V_-(\mathbf p)
=\Sigma_{\mathbf p}V_+(\bar{\mathbf p})^*.
\end{equation}
Because Majorana reality exchanges the positive- and negative-charge eigenspaces, the combined frame $\mathbb
V=(V_+,V_-)$ is unitary and automatically satisfies
\begin{equation}
\mathbb V(\bar{\mathbf p})
=\Sigma_{\mathbf p}\mathbb V(\mathbf p)^*
(\tau_x\otimes\mathbf1_2).
\label{eq:2d_sq_canonical_sewing}
\end{equation}
It brings the charge and Hamiltonian to the exact Nambu form
\begin{equation}
\begin{aligned}
\mathbb V^\dagger q_{\rm loc}\mathbb V
&=\tau_z\otimes\mathbf1_2,\\
\mathbb V^\dagger h_{\rm fold}\mathbb V
&=
\begin{pmatrix}
h_\sharp^{\rm sq}(\mathbf p)&0\\
0&-h_\sharp^{\rm sq}(\bar{\mathbf p})^T
\end{pmatrix}.
\end{aligned}
\end{equation}

The sewing identity holds also at Majorana-invariant momenta, since $q_{\rm loc}^2=\mathbf1_4$ even at a
physical Dirac zero. Consequently, $\mathbb V$ defines a real orthogonal transformation of the finite Majorana
algebra and hence an exact canonical relocalization $\mathcal W$:
\begin{equation}
\mathcal WQ_{\rm loc}\mathcal W^{-1}
=Q_{\rm onsite},\quad
\mathcal WH_{\rm sq}\mathcal W^{-1}
=H_\sharp^{\rm sq}.
\label{eq:2d_sq_finite_reloc_main}
\end{equation}
The centered charge normalization is preserved. These operator identities establish exact finite-volume
canonical equivalence, rather than merely isospectrality. The relocalizing frame is regular near the physical
Dirac node, where $R_\theta=\mathbf1_2$, and thus acts locally on the continuum fermion field despite its
nonlocal microscopic action.

The resulting Hamiltonian is Schwartz-local: its hopping amplitudes decay faster than any inverse power of
distance. More precisely, writing $h_\sharp^{\rm sq}(\mathbf p) =\sum_{\mathbf R}K_{\mathbf R} e^{i\mathbf
p\cdot\mathbf R}$, smoothness and periodicity imply
\begin{equation}
\|K_{\mathbf R}\|
\le C_M(1+|\mathbf R|)^{-M}
\label{eq:2d_sq_schwartz}
\end{equation}
for every positive integer $M$, with constants $C_M$ independent of $\mathbf R$. Here $\|\cdot\|$ denotes the
matrix operator norm on the two orbital components. This decay is stronger than any fixed power law, although it
need not be exponential. In particular, the Hamiltonian admits finite-range truncations with errors decreasing
faster than any inverse power of the truncation range. The Schwartz-locality result applies to the massless
Hamiltonian and does not automatically extend to its massive descendants under the same relocalization.

Finally, the sharp charge $q_s=s_PT_1$ is related to $q_{\rm loc}$ on seam-free momentum grids by the
Hamiltonian-preserving rotation
\begin{equation}
W_P=
\frac{\mathbf1_4+q_{\rm loc}q_s}
{\sqrt{2(1+c)}},
\quad
c=\frac{|b_P|}{r_{\rm sq}},
\label{eq:2d_sq_WP_main}
\end{equation}
which satisfies $W_Pq_sW_P^\dagger=q_{\rm loc}$ and $W_Ph_{\rm fold}W_P^\dagger=h_{\rm fold}$. Because both
charges satisfy Majorana-odd reality and the folded momentum sewing conditions, $W_P$ defines an exact
Majorana-canonical transformation on these grids. Its composition with $\mathcal W$ maps $Q_{s_P}$ to the same
onsite charge and yields the same Schwartz-local zero-field Hamiltonian. The transported compact gauge
couplings, however, need not coincide in the original lattice geometry.

\subsubsection{Operator-norm dynamical gauge locality}

With the compact charge onsite in the relocalized lattice, the matter Hamiltonian can be coupled to compact
gauge fields while preserving exact gauge invariance at finite cutoff. Restore the lattice spacing $a$ and Dirac
velocity $v$. Let $c_{\mathbf X}$ denote the two-component complex fermion at site $\mathbf X$ of the
relocalized lattice, and let $\mathbf R$ denote a lattice displacement. The hopping matrices $K_{\mathbf R}$ are
the Fourier coefficients of $h_\sharp^{\rm sq}$, defined above. For compact $U(1)$ link variables $U_\ell$, let
$U_{\Gamma_{\mathbf X,\mathbf R}}$ denote the Wilson transporter, given by the product of links along a path
from $\mathbf X$ to $\mathbf X+\mathbf R$. The gauged matter Hamiltonian and its range-$L$ truncation are
\begin{equation}
\begin{aligned}
H_a[U]
&=\frac{v}{a}\sum_{\mathbf X,\mathbf R}
c_{\mathbf X}^\dagger K_{\mathbf R}
U_{\Gamma_{\mathbf X,\mathbf R}}
c_{\mathbf X+\mathbf R},\\
H_{a,L}[U]
&=\frac{v}{a}
\sum_{\substack{\mathbf X,\mathbf R\\|\mathbf R|\le L}}
c_{\mathbf X}^\dagger K_{\mathbf R}
U_{\Gamma_{\mathbf X,\mathbf R}}
c_{\mathbf X+\mathbf R}.
\end{aligned}
\label{eq:2d_sq_gauged_truncation}
\end{equation}
Reverse hoppings use adjoint transporters, ensuring Hermiticity. Under a gauge transformation $c_{\mathbf
X}\mapsto g_{\mathbf X}c_{\mathbf X}$ and $U_{\Gamma_{\mathbf X,\mathbf R}}\mapsto g_{\mathbf
X}U_{\Gamma_{\mathbf X,\mathbf R}} g_{\mathbf X+\mathbf R}^{-1}$, so each dressed hopping is gauge invariant. We
may use shortest paths or weighted averages of a bounded number of symmetry-related shortest paths. Such
averages preserve gauge covariance and have operator norm at most one, although they need not be unitary.
Different path prescriptions may differ by Wilson loops. The estimates require only that the paths have length
$O(1+|\mathbf R|)$; see Appendix~\ref{app:2d_hc_asymptotic_proof}.

The essential question is whether the discarded long-range hoppings become negligible while the physical range
$aL$ shrinks to zero. Keep the momentum-space smoothing profile fixed as $a\to0$. Let $\Phi_a$ denote the
collection of Wilson-line-dressed hopping terms in $H_a[U]$, and $\Phi_{a,L}$ the collection retained in
$H_{a,L}[U]$. Their difference obeys
\begin{equation}
\|\Phi_a-\Phi_{a,L}\|_s
\le C_{s,M}\frac{v}{a}L^{-M},
\label{eq:2d_sq_weighted_tail}
\end{equation}
for every $s\ge0$ and $M>0$. Here $\|\cdot\|_s$ measures the strength and spatial extent of hopping terms near
each matter site or gauge link, using polynomial weights in the standard framework of quasi-local lattice
interactions
\cite{nachtergaele2019quasilocality}.
The constant $C_{s,M}$ is independent of $a$ and $L$; the precise definition and proof are given in
Appendix~\ref{app:2d_hc_asymptotic_proof}. Choosing $L(a)=\lceil a^{-\beta}\rceil$, with $0<\beta<1$, yields,
for any prescribed $p>0$,
\begin{equation}
\|\Phi_a-\Phi_{a,L(a)}\|_s=O(a^p),
\quad aL(a)\to0.
\label{eq:2d_sq_asymptotic_gauge_locality}
\end{equation}
Thus each truncated Hamiltonian is exactly gauge invariant, its physical hopping range vanishes, and the omitted
hopping terms become negligible faster than any prescribed power of $a$. This establishes asymptotic gauge
locality.

More strongly, the same truncation controls the complete many-body Hamiltonian on a fixed physical torus of area
$\mathcal A$, not just the hoppings near an individual site. Define
\begin{equation}
B_{a,L}=H_a[U]-H_{a,L}[U].
\label{eq:2d_sq_BaL}
\end{equation}
The torus contains $O(\mathcal A a^{-2})$ matter cells, while each hopping has energy scale $v/a$. Summing the
omitted hopping terms over the entire torus gives
\begin{equation}
\|B_{a,L}\|
\le C_{\mathcal A,M}\,v\,a^{-3}L^{-M},
\label{eq:2d_sq_fixed_volume}
\end{equation}
with $C_{\mathcal A,M}$ independent of $a$ and $L$. For $L=L(a)$, the bound is $O(a^p)$ for every prescribed
$p>0$. It holds on the full finite-volume matter--gauge Hilbert space, including UV fermionic states and
fluctuating compact gauge links, without assuming smooth gauge backgrounds.

Add the same self-adjoint gauge-field Hamiltonian $H_{g,a}$ to both matter regulators and define $\mathbb
H_a=H_a[U]+H_{g,a}$ and $\mathbb H_{a,L}=H_{a,L}[U]+H_{g,a}$. For any fixed evolution time $t$,
\begin{equation}
\left\|e^{-it \mathbb H_a}
-e^{-it \mathbb H_{a,L}}\right\|
\le |t|\,\|B_{a,L}\|
=O(a^p).
\label{eq:2d_sq_dynamical_convergence}
\end{equation}
The same bound holds on the common Gauss-law subspace. Thus the truncated regulator reproduces the finite-time
dynamics of the complete matter--gauge system in the continuum limit, not merely its low-energy Dirac spectrum.

Unlike the 1D finite-chemical-potential regulator of Sec.~\ref{sec:1d_mu_gauge_locality}, whose algebraic UV
hopping tails prevent comparable full operator-norm estimates, the relocalized massless $H_{\rm sq}$ admits
exactly gauge-invariant finite-range approximations with vanishing physical range and vanishing dynamical error.
This establishes asymptotic gauge locality, without by itself proving the existence of an interacting continuum
fixed point.

\subsubsection{Continuum gauge theories}
\label{sec:2d_wilsonian}

The preceding operator-norm estimates establish asymptotic gauge locality, but do not determine the interacting
continuum limit. The relocalized square model and the Wilson--Dirac representative each contain one charge-one
Dirac fermion in the free IR. For a smooth background $U(1)$ field $A$, their low-energy generating functionals
take the form
\begin{equation}
W_{\rm bg}[A]=-\log\det_{\mathfrak R}(\slashed D_A)+S_{\rm local}[A]+\cdots,
\label{eq:2d_background_effective}
\end{equation}
where $\mathfrak R$ denotes the UV regulator and $S_{\rm local}$ contains local gauge-invariant terms. In
isotropic Dirac coordinates, the quadratic response is
\begin{equation}
\begin{aligned}
\Pi_{\mu\nu}(p)={}&\frac{|p|}{16}\left(\delta_{\mu\nu}-\frac{p_\mu p_\nu}{p^2}\right)\\
&+\frac{ik_{\rm reg}}{2\pi}\epsilon_{\mu\nu\rho}p^\rho+O(p^2/\Lambda),
\end{aligned}
\label{eq:2d_background_polarization}
\end{equation}
with $\Lambda\sim v/a$. The nonanalytic parity-even response is fixed by the massless Dirac fermion, whereas the
parity-odd coefficient $k_{\rm reg}$ depends on the UV completion. The latter will be examined alongside the
competing lattice masses in Sec.~\ref{sec:2d_parity_regulators}.

Promoting $A$ to a dynamical compact gauge connection $a$, the candidate Wilsonian theory is
\begin{equation}
S_{\rm IR}=S_{\rm DM}+S_{\rm reg}+S_{\rm mon}+\cdots,
\label{eq:2d_IR_dirac_maxwell}
\end{equation}
where
\begin{equation}
\begin{aligned}
S_{\rm DM}&=\int d^3x\left[\bar\Psi(\gamma^\mu D_\mu+m_R)\Psi+\frac{f_{\mu\nu}f^{\mu\nu}}{4e_R^2}\right],\\
S_{\rm mon}&=\int d^3x\sum_q\left(\lambda_q\mathcal M_q+\mathrm{H.c.}\right).
\end{aligned}
\end{equation}
Here $D_\mu=\partial_\mu+ia_\mu$, $m_R$ is the renormalized Dirac fermion mass, $e_R$ is the renormalized gauge
coupling, and $f_{\mu\nu}=\partial_\mu a_\nu-\partial_\nu a_\mu$. The term $S_{\rm reg}$ represents the
gauge-invariant regulator completion, while $\mathcal M_q$ denotes a monopole operator with any dressing
required by Gauss's law. The allowed monopole quantum numbers and fugacities depend on the compact UV regulator,
not merely on the free Dirac spectrum.

Unlike in 1D, where the exact identification with the staggered-fermion Schwinger regulator allows us to
establish the massless-fermion continuum limit, the 2D gauge-locality result does not determine whether the
Dirac fermion remains massless after including gauge fluctuations. The 2D compact gauge field also has a
propagating transverse mode and admits monopole events, whose dynamics can affect the IR theory. Thus the
renormalized fermion mass, parity-odd regulator contribution, and monopole couplings require separate analysis.
We next examine the crystalline symmetry constraints through projective reflections and competing mass channels.

\subsection{Crystalline symmetries, masses, and parity anomaly}
\label{sec:2d_sq_gapping_channels}

The charge-on-site relocalization exposes the single-Dirac continuum theory, but the crystalline symmetries and
admissible masses depend on its lattice completion. We first identify the projective reflections, Pfaffian
constraint, and magnetic-translation LSM relation of $H_{\rm sq}$, then construct two competing finite-range
masses. We examine how the charge-preserving mass changes physical edge energies without changing the
compressed-charge spectrum, before distinguishing the electromagnetic parity anomaly from class-D regulator
topology and the response of lattice defects.

\subsubsection{Projective reflections and Pfaffian parity}
\label{sec:2d_sq_parity}

In physical square-lattice coordinates, let $\gamma_{x,y}$ denote the Majorana on the $A$ sublattice for $x+y$
even and on the $B$ sublattice for $x+y$ odd. The $\pi$-flux hopping pattern of Eq.~\eqref{eq:2d_sq_micro} is
invariant under the bond-centered reflections
\begin{equation}
\begin{aligned}
\widetilde{\cal P}_x:\quad\gamma_{x,y}&\longmapsto(-1)^{x+y}\gamma_{-x-1,y},\\
\widetilde{\cal P}_y:\quad\gamma_{x,y}&\longmapsto(-1)^y\gamma_{x,1-y}.
\end{aligned}
\label{eq:2d_sq_projective_parities_real}
\end{equation}
The signs compensate for the background flux, and the reflections square to fermion parity,
\begin{equation}
\widetilde{\cal P}_x^2=\widetilde{\cal P}_y^2=(-1)^F.
\label{eq:2d_sq_projective_parity_square}
\end{equation}
In the physical-orbital Bloch basis $\Gamma_{\mathbf k}=(\gamma_{A,\mathbf k},\gamma_{B,\mathbf k})^T$, their
sewing matrices are
\begin{equation}
\begin{aligned}
\widetilde{\cal P}_x:\quad\Gamma_{k_x,k_y}&\longmapsto
 e^{ik_x}i\sigma_y\,\Gamma_{-k_x,k_y},\\
\widetilde{\cal P}_y:\quad\Gamma_{k_x,k_y}&\longmapsto
 -e^{-ik_y}\sigma_x\,\Gamma_{k_x,\pi-k_y}.
\end{aligned}
\label{eq:2d_sq_projective_parities}
\end{equation}
The bond-center phases are essential for the globally defined lattice action. Both operations fix each
Majorana-reality-related node modulo reciprocal vectors, and preserve its IR Dirac number,
\begin{equation}
\widetilde{\cal P}_i Q_D\widetilde{\cal P}_i^{-1}=Q_D,
\quad i=x,y.
\label{eq:2d_sq_parity_Q_preserve}
\end{equation}
The sharp charge $Q_{s_P}$ is invariant under both bond-centered reflections, since
$s_P=\operatorname{sgn}(2\cos k_x\sin k_y)$ is even under their momentum actions. The smooth charge $Q_{\rm
loc}$, however, has $C_Q=1$ and cannot itself be invariant under a charge-preserving orientation reversal. In
the periodic doubled-cell frame, the physical two-orbital sewings of Eq.~\eqref{eq:2d_sq_projective_parities}
induce four-component matrices $\mathscr S_i(\mathbf p)$, including the action on both folded sectors. With
$\mathbf p=(P,V)$, their momentum maps are $\mathsf p_x(P,V)=(P-2V,-V)$ and $\mathsf p_y(P,V)=(2V-P,V)$ modulo
$2\pi$. They transport the smooth charge to distinct conserved involutions,
\begin{equation}
\begin{aligned}
\tilde q^{(i)}(\mathbf p)
&=\mathscr S_i(\mathbf p)\tilde q(\mathsf p_i\mathbf p)
\mathscr S_i(\mathbf p)^\dagger,\\
C_Q[\tilde q^{(i)}]&=-C_Q[\tilde q]=-1.
\end{aligned}
\label{eq:2d_sq_parity_partner_charge}
\end{equation}
The site-centered reflections instead exchange the Majorana-reality-related nodes and conjugate Dirac number,
${\cal R}_iQ_D{\cal R}_i^{-1}=-Q_D$. Thus a crystalline symmetry of the ungauged Hamiltonian need not preserve a
chosen smooth compact charge or its associated gauge regulator.

The node-fixing reflections imply a more general class-D constraint. Consider a periodic $C^1$ Majorana Bloch
Hamiltonian with exactly one isolated, simple pair of reality-related Dirac nodes $\pm\mathbf K$ away from the
PHIM, and no other zero-energy crossings. If a globally regular, Majorana-compatible orientation-reversing
lattice parity fixes each node and preserves its IR Dirac charge, then
\begin{equation}
\mathcal P_D(H)=-1.
\label{eq:2d_dirac_parity}
\end{equation}
This statement does not require a smooth compact charge. A small parity-odd mass gaps the two nodes; their
Majorana-related Chern jumps agree, whereas parity reverses the Chern number. The resulting gapped descendants
have $\nu_{\rm Maj}=\pm1$, fixing the unchanged PHIM Pfaffian product to $-1$. The precise hypotheses and proof
are in Appendix~\ref{app:2d_dirac_pfaffian}. The bond-centered reflections of $H_{\rm sq}$ meet these
hypotheses, unlike the charge-conjugating site-centered reflections. Equation~\eqref{eq:2d_dirac_parity}
constrains the ungauged class-D realization; it is distinct from the electromagnetic parity anomaly.

Translations impose a complementary constraint. The Bravais vectors $\mathbf a_1=(1,1)$ and $\mathbf a_2=(1,-1)$
carry opposite phases $\pm\pi/2$ at $\mathbf K_+=(0,\pi/2)$, selecting an IR $\mathbb Z_4\subset U(1)_{\rm
Dirac}$. This action admits an exact, generally nonlocal spectral realization through $Q_{s_P}$, analogous to
the 1D axial construction. Independently, the one-site magnetic translations obey
\begin{equation}
\widetilde T_x\widetilde T_y=(-1)^F\widetilde T_y\widetilde T_x.
\label{eq:2d_sq_magnetic_translations}
\end{equation}
With one Majorana per physical site, the magnetic-translation LSM theorem requires an odd class-D Chern number
for a symmetric gapped invertible phase \cite{lu2017lsmSPT}. This agrees with the odd Pfaffian parity $\mathcal
P_D(H_{\rm sq})=(-1)^{C_Q}=-1$: for a gapped charge-preserving Gaussian descendant, $\nu_{\rm
Maj}=2C_{+,\mathrm{occ}}-C_Q$ is necessarily odd. The translation-selected $\mathbb Z_4$ and the magnetic LSM
constraint are nevertheless independent. The staggered pairing mass constructed below enlarges the translation
cell, thereby evading the latter while preserving both bond-centered reflections.

\subsubsection{Competing mass channels}
\label{sec:2d_sq_masses}

Two complementary finite-range masses gap the square Dirac fermion. The first preserves the exact compact charge
but breaks the bond-centered orientation-reversing reflections; the second preserves both reflections while
breaking the compact $U(1)$ to fermion parity. The charge-preserving Chern mass is
\begin{equation}
\delta h_m(U,V)=m\sin U\,\sigma_z.
\label{eq:2d_sq_U1_mass}
\end{equation}
In the real-space convention of Eq.~\eqref{eq:2d_sq_micro}, it is the oriented same-sublattice hopping
\begin{equation}
H_m=\frac{im}{4}\sum_{\mathbf R}\left[
-\gamma_{A,\mathbf R}\gamma_{A,\mathbf R+\mathbf a_1}
+\gamma_{B,\mathbf R}\gamma_{B,\mathbf R+\mathbf a_1}\right].
\label{eq:2d_sq_U1_mass_real}
\end{equation}
The $U\mapsto U+\pi$ charge folding gives
\begin{equation}
\delta h_m^{\rm fold}=m\sin U\,\tau_z\otimes\sigma_z.
\label{eq:2d_sq_U1_mass_folded}
\end{equation}
A direct calculation gives $[\delta h_m^{\rm fold},q_{\rm loc}]=0$, so the mass preserves the smooth compact
charge exactly at finite lattice spacing. Because it anticommutes with the kinetic matrices,
\begin{equation}
E_m^2(U,V)=E_{\rm sq}^2(U,V)+m^2\sin^2U.
\label{eq:2d_sq_U1_mass_spectrum}
\end{equation}
Both nodes are therefore gapped for $m\ne0$. The two signs of $m$ are interchanged by parity, which also
transports a chosen smooth charge completion to its opposite-Chern partner. With the torus orientation used
here, a direct two-band calculation gives
\begin{equation}
\nu_{\rm Maj}(m)=\operatorname{sgn}(m).
\label{eq:2d_sq_U1_mass_chern}
\end{equation}
For $C_Q=1$, the occupied positive-charge Chern number changes from $0$ to $1$ between these phases, consistent
with $\nu_{\rm Maj}=2C_{+,\mathrm{occ}}-C_Q$. Their interface has $\Delta\nu_{\rm Maj}=2$ and supports two
co-propagating chiral Majorana modes, or one chiral complex mode. The Chern sectors are analogous to those of
chiral $p$-wave pairing \cite{readGreen2000}, but here the mass in Eq.~\eqref{eq:2d_sq_U1_mass_real} preserves
an exact compact $Q_{\rm loc}$ at all lattice momenta.

The complementary mass is a reflection-symmetric staggered horizontal-bond perturbation. The two Dirac nodes are
separated by $\mathbf Q=(\pi,\pi)$ in $(u,v)$ coordinates, equivalently $(0,\pi)$ in physical $(k_x,k_y)$
coordinates. With $\mathbf R=n_1\mathbf a_1+n_2\mathbf a_2$, define
\begin{equation}
H_\Delta=-\frac{i\Delta}{4}\sum_{\mathbf R}(-1)^{n_1+n_2}
\gamma_{A,\mathbf R}\bigl(\gamma_{B,\mathbf R}+\gamma_{B,\mathbf R+\mathbf a_1+\mathbf a_2}\bigr).
\label{eq:2d_sq_dimer_mass_real}
\end{equation}
The two bond families connect each $A$ site to its left and right horizontal neighbors. Under the projective
reflections of Eq.~\eqref{eq:2d_sq_projective_parities_real}, $\widetilde{\cal P}_x$ preserves each family,
while $\widetilde{\cal P}_y$ exchanges them, including the staggered signs. Consequently, $H_{\rm sq}+H_\Delta$
preserves both bond-centered reflections. The staggering carries the node-separation momentum and doubles the
translation cell.

In the corresponding node-folded physical-orbital basis, the exact Bloch mass is
\begin{equation}
\delta h_\Delta^{\rm fold}(\mathbf k)
=\Delta\cos k_x\,\tau_x\otimes\sigma_y.
\label{eq:2d_sq_dimer_mass_folded}
\end{equation}
It anticommutes with both kinetic matrices, giving
\begin{equation}
E_\Delta^2(\mathbf k)=4t^2\sin^2k_x+4t^2\cos^2k_y+\Delta^2\cos^2k_x.
\label{eq:2d_sq_dimer_spectrum}
\end{equation}
Thus the spectrum is fully gapped for $t\Delta\ne0$. The mass mixes the two Dirac-charge sectors and acts as a
charge-two pairing mass in the IR. It breaks $Q_{\rm loc}$ but preserves fermion parity; since either preserved
reflection reverses orientation, the gapped phase has $\nu_{\rm Maj}=0$. The enlarged cell may nevertheless
admit weak or crystalline distinctions.

The competition is most transparent in the node-folded low-energy basis,
\begin{equation}
\begin{aligned}
h_{\rm IR}(\mathbf q)&=-vq_x\,\tau_0\otimes\sigma_x
-vq_y\,\tau_z\otimes\sigma_y,\\
 q_D& =\tau_z\otimes\mathbf1,
\end{aligned}
\label{eq:2d_sq_IR_folded}
\end{equation}
where $v=2t$. Because $\cos k_x\to1$ at the Dirac points, the two IR masses are
\begin{equation}
M_{U(1)}=m\,\tau_z\otimes\sigma_z,
\quad M_{\cal P}=\Delta\,\tau_x\otimes\sigma_y.
\label{eq:2d_sq_two_IR_masses}
\end{equation}
Each anticommutes with the kinetic Hamiltonian, but $[M_{U(1)},q_D]=0$ and $\{M_{\cal P},q_D\}=0$. The mass
matrices commute with one another, so
\begin{equation}
E_\pm^2(\mathbf q)=v^2|\mathbf q|^2+(m\pm\Delta)^2.
\label{eq:2d_sq_competing_masses}
\end{equation}
The gap closes at $|m|=|\Delta|$. The two pure-mass limits show that the charge-preserving and
reflection-preserving gaps are distinct, with one or the other symmetry necessarily broken.

\begin{figure*}[tb]
\centering
\includegraphics[width=\linewidth]{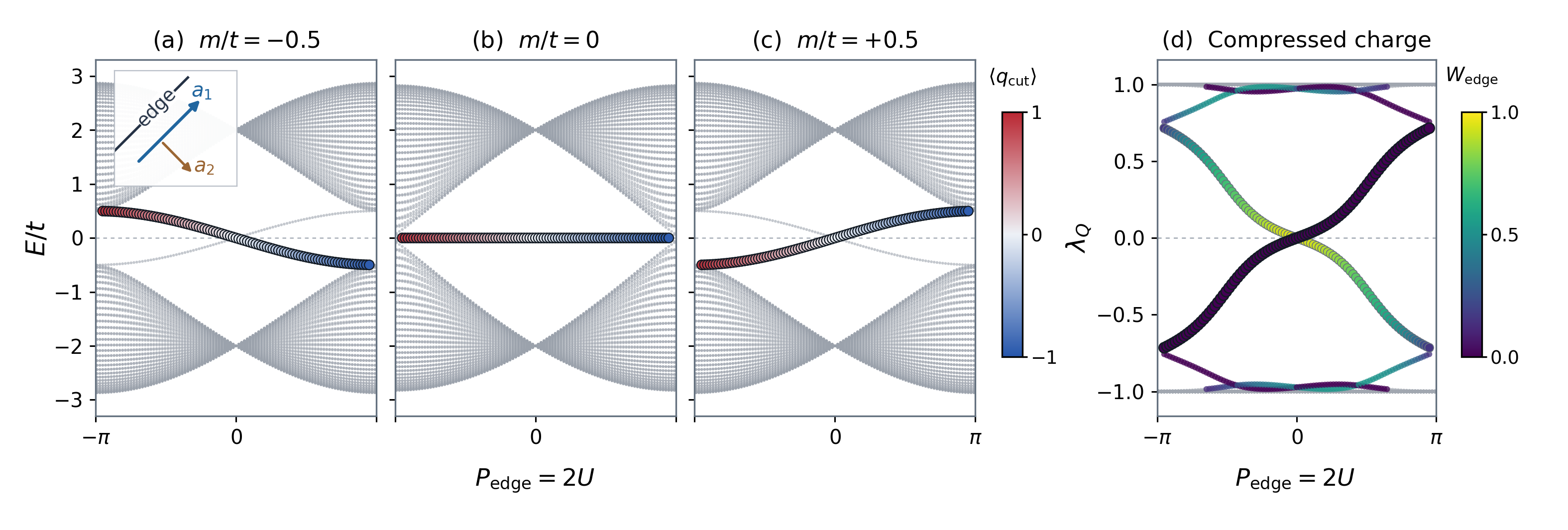}
\caption{Physical edge spectra and compressed Dirac charge
of $H_{\rm sq}+H_m$ on a strip of 30 transverse unit cells,
open along $V$, with conserved momentum
$P_{\rm edge}=2U$ along $\mathbf a_1$
(geometry inset in panel (a)).
(a)--(c) Physical energy spectra for
$m/t=-0.5,0,+0.5$, respectively.
Modes localized at the selected edge are colored by
their charge expectation
$\langle\psi|q_{\rm cut}|\psi\rangle$,
where $q_{\rm cut}=Rq_{\rm loc}R$;
the remaining states are gray.
The edge-state charge expectation varies from approximately
$+1$ to $-1$ with momentum but remains essentially
unchanged as $m$ varies, despite the reversal of
edge-mode chirality.
(d) Mass-independent eigenvalue spectrum $\lambda_Q$
of the compressed charge. Boundary branches associated
with opposite edges exhibit compensating spectral flows
between the bulk charge eigenvalues $\pm1$.
Colors indicate their overlap $W_{\rm edge}$ with the
physical edge-mode subspace
[Eq.~\eqref{eq:2d_sq_edge_charge_overlap}],
while bulk-like charge levels are gray.}
\label{fig:majoranaedge}
\end{figure*}

\subsubsection{Physical edge spectra and charge polarization}
\label{sec:2d_sq_mass_edge}

As established in Sec.~\ref{sec:2d_sq_CQ}, the charge bulk--edge correspondence fixes the spectral flow of the
boundary-compressed charge, independently of whether the physical Hamiltonian is gapped. We now distinguish this
charge flow from physical edge excitations using the charge-preserving Chern mass $H_m$. The smooth charge
$Q_{\rm loc}$ and its Chern number $C_Q=1$ remain unchanged, whereas the physical edge energies depend on $m$.
For $m\ne0$, the class-D Chern number $\nu_{\rm Maj}(m)=\operatorname{sgn}(m)$
[Eq.~\eqref{eq:2d_sq_U1_mass_chern}] requires chiral Majorana edge modes with opposite propagation directions
for the two mass signs. At $m=0$, the fixed-momentum strip is instead a chiral-symmetric SSH-type chain whose
winding supports a flat zero-energy edge band
\cite{suSchriefferHeeger1979,ryuHatsugai2002},
analogous to boundary bands in zigzag graphene
\cite{fujita1996,nakada1996,delplace2011}
and $H_K$ \cite{kitaev2006anyons,thakurathiSenguptaSen2014}. We consider a half-plane open along $V$, with
conserved boundary momentum $P=P_{\rm edge}=2U$ along $\mathbf a_1$. To resolve both boundaries numerically, we
use a finite-width strip with the same orientation, as illustrated in the geometry inset of
Fig.~\ref{fig:majoranaedge}(a).

Let $R$ denote the projection onto the half-plane or finite strip, as appropriate, with the transverse direction
represented by real-space layers. At fixed $P$, define
\begin{equation}
\begin{aligned}
q_{\rm cut}(P) &=Rq_{\rm loc}(P)R,\\
h_{\rm cut}(P;m)
&=R[h_{\rm fold}(P)+\delta h_m^{\rm fold}(P)]R,
\end{aligned}
\end{equation}
where the partially Fourier-transformed operators incorporate the doubled-cell sewing. Although the bulk charge
commutes with the Hamiltonian, their boundary compressions need not commute. We therefore distinguish the charge
eigenstates
\begin{equation}
q_{\rm cut}(P)|\phi_j^Q(P)\rangle
=\lambda_{Q,j}(P)|\phi_j^Q(P)\rangle,
\label{eq:2d_sq_qcut_eigenproblem}
\end{equation}
from physical strip-energy eigenstates $|\psi_n(P;m)\rangle$. The expectation value $
\langle\psi_n(P;m)|q_{\rm cut}(P)|\psi_n(P;m)\rangle
$ 
is not generally a conserved boundary charge quantum number.

To determine whether a charge-flow eigenstate belongs to the physical low-energy edge sector, let $\Pi_{\rm
edge}^H(P;m)$ project onto the physical edge-mode subspace, identified by boundary localization and separation
from the projected bulk spectrum. For the numerical strip, this is the isolated two-edge-mode subspace, treated
as a whole at degeneracies; momenta where it merges with the bulk are excluded. We define
\begin{equation}
W_{\rm edge,j}(P;m)
=\langle\phi_j^Q(P)|
\Pi_{\rm edge}^H(P;m)|\phi_j^Q(P)\rangle.
\label{eq:2d_sq_edge_charge_overlap}
\end{equation}
The charge expectation of a physical edge-energy state and the physical-edge overlap of a charge eigenstate are
thus distinct observables.

Fig.~\ref{fig:majoranaedge} compares the physical edge spectra with the compressed-charge spectral flow. Panels
(a)--(c) show the physical spectra for $m/t=-0.5,0,+0.5$, with modes at the selected edge colored by their
charge expectation $\langle q_{\rm cut}\rangle$. As $P_{\rm edge}$ increases, this expectation changes from
approximately $+1$ to $-1$, with an essentially mass-independent momentum profile. Reversing $m$, however,
reverses the chirality of the physical edge modes.

Panel (d) instead shows the mass-independent eigenvalue flow of $q_{\rm cut}$. The two boundaries contribute
compensating flows, while the oriented crossing number at either half-plane boundary is fixed, up to
orientation, by $C_Q=1$ through Eq.~\eqref{eq:2d_charge_edge_spectral_flow}. These charge-flow eigenstates need
not coincide with physical edge-energy eigenstates because the compressed charge and Hamiltonian generally do
not commute. For the chosen termination, the charge-flow branch at the selected edge has negligible overlap with
the physical edge-mode subspace, whereas the opposite-edge branch reaches $W_{\rm edge}\simeq0.95$ near $P_{\rm
edge}=0$. This asymmetry reflects the mismatch between their boundary wavefunctions and is not determined by
$C_Q$.

The mass independence of the overlap profile follows from the folded Chern mass in
Eq.~\eqref{eq:2d_sq_U1_mass_folded}:
\begin{equation}
h_{\rm cut}(P;m)
=h_{\rm cut}(P;0)+m\sin(P/2)\,\Gamma,
\label{eq:2d_sq_strip_mass_chiral}
\end{equation}
where $\Gamma=\tau_z\otimes\sigma_z$ and $\{h_{\rm cut}(P;0),\Gamma\}=0$. The zero-energy edge subspace of the
massless strip is invariant under $\Gamma$; more generally, an isolated edge cluster containing both chiral
partners has a $\Gamma$-invariant projector. Since the massless Hamiltonian and $\Gamma$ preserve this subspace,
adding the mass leaves its projector unchanged wherever the edge cluster remains isolated from the bulk.
Together with the mass independence of $q_{\rm cut}$, this explains the identical $W_{\rm edge,j}$ profiles for
all three plotted masses. For the chosen termination, localized massless zero modes can be selected as $\Gamma$
eigenstates, so their charge expectations also remain unchanged, apart from finite-width hybridization and
edge--bulk threshold effects. This behavior is specific to the strip geometry and mass, not a consequence of
$C_Q$ alone.

The physical edge-state charge polarization is not topologically quantized. The sharp, spatially nonlocal charge
$Q_{s_P}$ yields qualitatively similar edge-state charge expectations for the open-$V$ strip, showing that
exponential charge locality is not necessary for nonzero edge polarization. For an open-$U$ strip, however,
selected physical edge modes may have nearly vanishing expectation of $Rq_{s_P}R$ even when that of $Rq_{\rm
loc}R$ is nonzero. Thus $C_Q$ fixes the compressed-charge spectral flow, while physical edge energies, charge
polarization, and edge-mode overlaps depend on the Hamiltonian, charge completion, and boundary termination.

\subsubsection{Parity anomaly and regulator topology}
\label{sec:2d_parity_regulators}

The competing mass channels reflect the electromagnetic parity anomaly of a single charged Dirac fermion. For
the charge-preserving Chern mass $H_m$, the standard massive-Dirac Berry-curvature calculation gives a
half-integer IR Hall contribution, while the compact lattice regulator supplies its UV completion. With our
Chern-sign convention, the occupied positive-charge Chern number $C_{+,\mathrm{occ}}(m)$, defined in
Eq.~\eqref{eq:2d_nu_CQ_relation}, satisfies
\begin{equation}
C_{+,\mathrm{occ}}(m>0)-C_{+,\mathrm{occ}}(m<0)=1.
\label{eq:2d_Cplus_jump}
\end{equation}
Each gapped compact regulator has an integer Hall Chern number, equivalently obtained from the many-body Berry
curvature over the torus of boundary holonomies, assuming a unique gapped ground state. The midpoint of the two
massive responses is therefore half-integral, with a regulator-dependent integer part. Formally, the parity-odd
regulator contribution is
\begin{equation}
S_{\rm reg}^{\rm odd}[A]=\frac{i k_{\rm reg}}{4\pi}\int A\wedge dA,
\quad k_{\rm reg}=\frac12\pmod{\mathbb Z}.
\label{eq:2d_background_kreg}
\end{equation}
The half-level term alone is not a well-defined compact gauge action; its global meaning is supplied by the
regulated determinant or $\eta$-invariant
\cite{niemiSemenoff1983,redlich1984}.

The obstruction follows directly from the unit Chern jump. An orientation-reversing symmetry preserving the same
compact charge regulator and interchanging the two mass signs would require
$C_{+,\mathrm{occ}}(m)=-C_{+,\mathrm{occ}}(-m)$. Equation~\eqref{eq:2d_Cplus_jump} would then imply
$2C_{+,\mathrm{occ}}(m>0)=1$, contradicting integer quantization. Thus a standalone compact regulator of one
charged Dirac fermion cannot preserve both the fixed $U(1)_{\rm Dirac}$ and a charge-preserving orientation
reversal. The reflection-symmetric staggered pairing mass avoids this obstruction by breaking the continuous
charge symmetry. The projective reflections of the massless $H_{\rm sq}$ instead transport the chosen smooth
charge $Q_{\rm loc}$ to distinct conserved charges with opposite $C_Q$, rather than preserving the same compact
gauge regulator.

The electromagnetic parity anomaly does not determine the class-D Pfaffian sector. The square model has
$\mathcal P_D(H_{\rm sq})=-1$, whereas the Wilson--Dirac-derived honeycomb model has $\mathcal P_D(H_{\rm
hc}^{W})=+1$, despite their common charged Dirac anomaly. The latter does not satisfy the regular node-fixing,
charge-preserving reflection hypotheses of Eq.~\eqref{eq:2d_dirac_parity}. For fully gapped class-D descendants
preserving their respective Pfaffian sectors, the Majorana Chern numbers therefore differ by an odd integer,
corresponding to a relative odd $p+ip$ sector in the free-fermion classification. This distinction concerns the
neutral ultraviolet topology rather than the common electromagnetic parity anomaly. Neither $C_Q$ nor the
fractional electromagnetic level alone determines compact monopole quantum numbers or the resulting dynamical
gauge theory.

\subsubsection{Vortex defects and the hidden compact charge}

The charge-preserving odd-Chern descendant presents an apparent tension. An ordinary fermion-parity vortex
carries an odd Majorana zero-mode index
\cite{deNittisSchulzBaldesFlux2016}, whereas an exact
compact $U(1)$ symmetry must pair Majorana zero modes into complex fermions. There is no contradiction: although
the hidden $U(1)_{\rm Dirac}$ contains fermion parity, a microscopic parity vortex need not preserve the full
charge. To introduce such vortices, let $z_{ij}=\pm1$ reverse Majorana hopping signs on bonds crossing a branch
cut. A closed path encircling one endpoint acquires an additional minus sign, corresponding to a fermion-parity
$\pi$ flux. The cut can be moved by local Majorana sign redefinitions, while its endpoints identify the
vortices. The resulting Hamiltonian is
\begin{equation}
H_{\rm sq,m}[z]
=\frac{i}{4}\gamma^T A_m[z]\gamma,
\end{equation}
where $A_m[z]$ is real and antisymmetric and $H_{\rm sq,m}[1]=H_{\rm sq}+H_m$. Although fermion parity remains
exact, the vortex generally breaks the hidden compact charge,
\begin{equation}
[A_m[z],J_{\rm loc}]\ne0.
\label{eq:2d_sq_vortex_breaks_charge}
\end{equation}
Thus preserving fermion parity does not guarantee a local extension of the full $U(1)_{\rm Dirac}$ through the
vortex core.

This distinction leads to a stronger locality obstruction. Consider two parity vortices separated by a distance
$d$, each carrying an odd number of localized Majorana zero modes. Globally, these modes can combine into
complex fermions, but at least one Majorana at each vortex must pair with a mode at the distant vortex. Such
pairing becomes nonlocal as $d$ increases. A uniformly local compact symmetry cannot connect the two cores and
must instead mix their zero modes with finite-energy excitations, thereby failing to commute with the defect
Hamiltonian. This physical argument admits a rigorous bound. Suppose the vortex modes are exponentially
localized, their splittings vanish as $d\to\infty$, and a uniform gap $\delta_\pi>0$ separates them from the
remaining spectrum. For any family of uniformly exponentially local complex structures $J_d^T=-J_d$,
$J_d^2=-\mathbf1$, one obtains
$
\liminf_{d\to\infty}
\|[A_m[z_d],J_d]\|\ge\delta_\pi,
$
where $z_d$ denotes the two-vortex configuration. Thus no uniformly exponentially local compact charge can
remain conserved as the vortices separate. This conclusion relies on isolated, gapped vortex modes and does not
extend to gapless or delocalized defect states.

Symmetry relocalization further exposes the distinction between bulk and defect locality. The massless $H_{\rm
sq}$ admits a charge-on-site, Schwartz-local representation. Under this particular canonical relocalization
$\mathcal W$, however, the original odd-Chern mass develops a mismatch of derivatives across the identified UV
seam. The transformed Hamiltonian is continuous but not $C^1$ and therefore lacks Schwartz-decaying hoppings. By
contrast, a genuinely local, gapped onsite-$U(1)$ completion in the symmetry-adapted $X$ coordinate has a
trivial charge bundle, $C_Q^{(X)}=0$. Eq.~\eqref{eq:2d_nu_CQ_relation} then requires $
\nu_{\rm Maj}^{(X)}
=2C_{\rm occ}^{(X)}\in2\mathbb Z$. The same distinction applies to vortices. Transporting an original parity
vortex through $\mathcal W$ preserves its Majorana zero-mode spectrum but generally produces a nonlocal,
charge-breaking defect. Introducing a new local, charge-preserving $\pi$ holonomy directly in the $X$ lattice
instead defines a different defect, whose zero modes, if present, occur in complex pairs. Thus symmetry
relocalization preserves global fermion parity and the IR matter content without necessarily preserving the
locality or zero-mode spectrum of individual defects.

\subsection{Exact AIII and AII bulk completions}
\label{sec:2d_sq_3d_parent}

We embed the single charged Dirac fermion of $H_{\rm sq}$ in gapped 3D lattice Hamiltonians with onsite compact
$U(1)$ symmetry. The aim is an exact realization of the selected boundary pair $(H_{\rm sq},Q_{s_P})$, including
its ultraviolet momentum sewing rather than only the continuum Dirac cone. The direct finite-range parent
$\widetilde{\mathcal H}_\chi$ is in class AIII~\cite{chiu2016classification}. A distinct two-root parent
$\mathcal H_*$ has both chiral and Kramers symmetries; perturbations that annihilate its selected surface states
produce an AIII descendant $\mathcal H_\chi$ and a strong-AII descendant $\mathcal H_\Theta$. These descendants
are connected through $\mathcal H_*$ by a gapped path that leaves the boundary Hamiltonian and compact charge
exactly unchanged.

Exact zero-field surface correspondence does not, however, imply equivalent compact gauge couplings. We first
establish the bulk symmetry protections and canonical boundary identifications, then compare the autonomous 2D
regulator with the independently gauged 3D completions. Magnetic spectral isolation, chiral boundary indices,
and strong-AII axion inflow expose their common parity-anomaly response while keeping their distinct ultraviolet
gauge structures explicit (Appendix~\ref{app:2d_surface_inflow}).

\begin{figure}[tb]
\centering
\includegraphics[width=\linewidth]{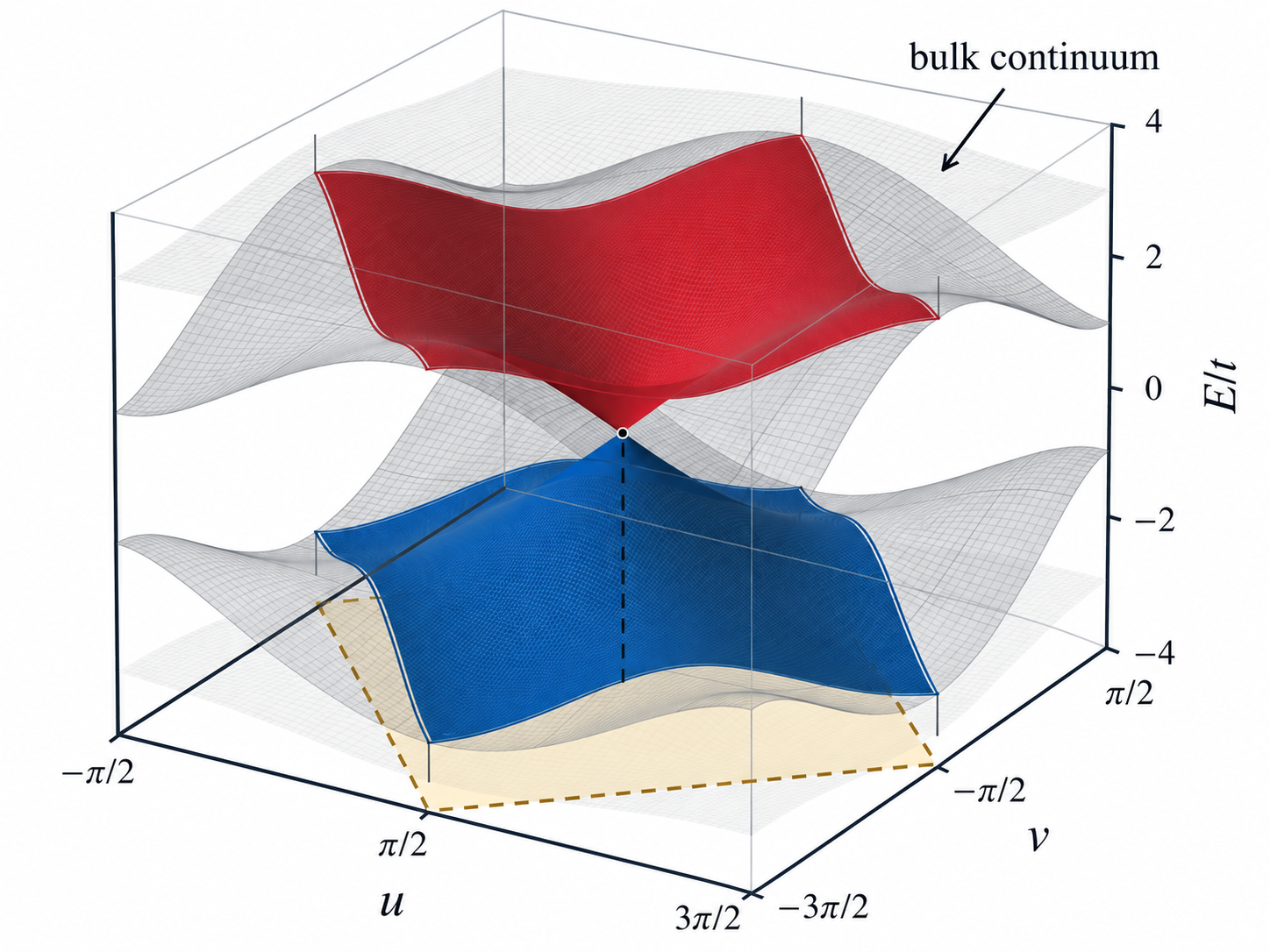}
\caption{Exact bulk--surface band structure of the direct AIII parent $\widetilde{\mathcal H}_\chi$, with $\lambda=1/2$ and $\Lambda=t$. Red and blue denote the exact surface bands $E^\pm(\mathbf k_{uv})
=\pm t|f_{\rm sq}(\mathbf k_{uv})|$,  supported on $D_+=\{(u,v):\sin u-\sin v>0\}$ (pale yellow). They form a single Dirac cone at $(u,v)=(\pi/2,-\pi/2)$. Gray denotes the projected 3D bulk continuum. Along the boundary of $D_+$, the surface localization length diverges and the surface bands merge into the bulk continuum.}
\label{fig:bulkTI}
\end{figure}

\subsubsection{Exact boundary sectors and bulk symmetry protections}

\paragraph{Direct AIII parent: bulk winding and exact surface correspondence.}
In the periodic $(u,v)$ frame of Eq.~\eqref{eq:2d_sq_f}, let $g(u,v)=\sin u-\sin v$ and $M(u,v)=1-\lambda
g(u,v)$, with $0<\lambda<1$. A finite-range 3D AIII parent with transverse momentum $w$ is
\begin{equation}
\begin{aligned}
\widetilde{\mathcal H}_\chi(u,v,w)&
=-\rho_xh_{\rm sq}(u,v)\\
&+\Lambda\bigl[\rho_y\sin w
+\rho_z(M(u,v)+\cos w)\bigr].
\end{aligned}
\label{eq:2d_sq_3d_parent_H}
\end{equation}
The parent conserves onsite particle number and has four complex orbitals per unit cell. Here $\rho_i$ act on an
auxiliary two-component orbital space and $\sigma_i$ on the $A/B$ sublattices of $H_{\rm sq}$. Its chiral
operator $\widetilde S_\chi=\rho_x\sigma_z$ satisfies $\{\widetilde S_\chi,\widetilde{\mathcal H}_\chi\}=0$; the
corresponding antiunitary Fock-space operation exchanges creation and annihilation operators.

The 3D bulk is fully gapped. To compute its AIII winding, write $\widetilde{\mathcal H}_\chi =\mathbf
d\cdot\boldsymbol\Gamma$, where
\begin{equation}
\begin{aligned}
 \boldsymbol\Gamma
=& (\rho_x\sigma_x, \rho_x\sigma_y,\rho_y,\rho_z),\\
\mathbf d
=(-t\operatorname{Im}f_{\rm sq},
& -t\operatorname{Re}f_{\rm sq},
\Lambda\sin w,\Lambda[M+\cos w]).
\end{aligned}
\label{eq:app_2d_3d_d}
\end{equation}
The normalized coefficient vector $\widehat{\mathbf d}=\mathbf d/|\mathbf d|$ defines a map $T^3\to S^3$ with
degree
\begin{equation}
\nu_3=\frac{1}{2\pi^2}\int_{T^3}
\det(\widehat{\mathbf d},\partial_u\widehat{\mathbf d},
\partial_v\widehat{\mathbf d},\partial_w\widehat{\mathbf d})
\,du\,dv\,dw.
\label{eq:app_2d_3d_winding_integral}
\end{equation}
With the stated orientation, the south pole of $S^3$ has a unique preimage,
$
(u,v,w)=(\pi/2,-\pi/2,\pi),
$
with positive Jacobian determinant $2\Lambda t^2>0$. Thus $\nu_3=1$, and the AIII
winding--magnetoelectric relation
\cite{ryu2010tenfold,chiu2016classification} gives
\begin{equation}
\theta=\pi\nu_3=\pi\pmod{2\pi}.
\label{eq:2d_sq_3d_parent_winding}
\end{equation}
The surface correspondence is also exact. For a half-space $z\ge1$, decompose the Hamiltonian into in-plane and
transverse hopping terms:
\begin{equation}
\begin{aligned}
\widetilde{\mathcal H}_\chi
&=\mathcal H_\parallel
+Te^{iw}+T^\dagger e^{-iw},\\
\mathcal H_\parallel
&=\Lambda M\rho_z-\rho_xh_{\rm sq},\\
T&=\frac{\Lambda}{2}(\rho_z-i\rho_y).
\end{aligned}
\label{eq:app_2d_3d_transverse_T}
\end{equation}
The eigenvalue equation is $E\psi_z=\mathcal H_\parallel\psi_z +T\psi_{z+1}+T^\dagger\psi_{z-1}$, with the last
term absent at $z=1$. In the $\rho_x=-1$ sector, take $\psi_z=r^{z-1}u_-\otimes\varphi$. Because $T^\dagger
u_-=0$ and $Tu_-=\Lambda\rho_z u_-$, the transverse terms cancel for $r=-M$, leaving $h_{\rm
sq}\varphi=E\varphi$. This solution also satisfies the open boundary condition exactly.

Normalizability requires $|M|<1$, equivalent to $g>0$ for $0<\lambda<1$. The selected surface therefore occupies
the half-zone $D_+=\{(u,v):g(u,v)>0\}$, with wavefunctions
\begin{equation}
\begin{aligned}
\Psi_{\mathbf k_{uv}}(z)
&=\sqrt{1-M(\mathbf k_{uv})^2}\,
[-M(\mathbf k_{uv})]^{z-1}
u_-\otimes\varphi_{\mathbf k_{uv}},\\
u_-&=\frac1{\sqrt2}(1,-1)^T,
\quad \rho_xu_-=-u_-,\\
& h_{\rm sq}(\mathbf k_{uv})\varphi_{\mathbf k_{uv}}
=E(\mathbf k_{uv})\varphi_{\mathbf k_{uv}}.
\end{aligned}
\label{eq:2d_sq_3d_parent_surface}
\end{equation}
The projected surface Hamiltonian is exactly $h_{\rm sq}$ on $D_+$ (Fig.~\ref{fig:bulkTI}), not merely near its
Dirac node. The penetration length $\xi_z=-1/\log|M|$ diverges as $g\to0^+$, where the state merges with the
projected bulk continuum.

Majorana reality pairs $\mathbf k_{uv}=(u,v)$ with $-\mathbf k_{uv}$. Because $g$ is inversion odd, $D_+$
contains one momentum from each such pair on finite grids avoiding $g=0$. If $a_{\alpha,\mathbf k_{uv}}$
annihilates the selected surface mode in channel $\alpha=A,B$, define
\begin{equation}
\mathcal J_\sharp^\dagger
a_{\alpha,\mathbf k_{uv}}\mathcal J_\sharp
=\frac{\gamma_{\alpha,\mathbf k_{uv}}}{\sqrt2},
\quad \mathbf k_{uv}\in D_+.
\label{eq:2d_AIII_halfzone_canonical}
\end{equation}
The opposite half-zone supplies the adjoints of these independent complex modes. The projected surface
Hamiltonian and the restriction of centered onsite bulk particle number then obey
\begin{equation}
\mathcal J_\sharp^\dagger
H_\partial \mathcal J_\sharp
=H_{\rm sq},\ \
\mathcal J_\sharp^\dagger
Q_\partial \mathcal J_\sharp
 =Q_{s_P}.
\label{eq:2d_AIII_exact_charge_correspondence}
\end{equation}
Thus the compact but spatially nonlocal charge $Q_{s_P}$ is inherited exactly from onsite bulk $U(1)$. The
Hamiltonian-preserving rotation $W_P$ in Eq.~\eqref{eq:2d_sq_WP_main} maps it to the exponentially local $Q_{\rm
loc}$ without changing $H_{\rm sq}$. It also transports the inherited AIII chiral symmetry to a grading
anticommuting with $h_{\rm fold}$ and commuting with $q_{\rm loc}$. Unlike the full charge-onsite relocalization
$\mathcal W$, however, $W_P$ retains the sharp UV seams and does not trivialize $C_Q$.

\begin{table*}[tb]
\centering
\small
\begin{tabular}{@{}lccc@{}}
\hline\hline
Property & AIII $\mathcal H_\chi$ & Common $\mathcal H_*$ & AII $\mathcal H_\Theta$\\
\hline
Bulk AIII chiral symmetry & Yes & Yes & No\\
Bulk Kramers symmetry & No & Yes & Yes\\
Bulk-transported boundary operation & $S_{\rm sq}$ & $S_{\rm sq},\Theta_{\partial}$ & $\Theta_{\partial}$\\
Bulk invariant & $|\nu_3|=1$ & $|\nu_3|=1,\ \nu_0=1$ & $\nu_0=1$\\
\hline\hline
\end{tabular}
\caption{Bulk symmetry protections along the strictly finite-range,
gapped path $\mathcal H_\chi\leftarrow\mathcal H_*\rightarrow
\mathcal H_\Theta$. All three Hamiltonians preserve onsite
$U(1)$ and the same exact selected zero-background boundary
pair $(H_{\rm sq},Q_{s_P})$. The table identifies which
boundary symmetries are transported from each bulk. The transported
$\Theta_{\partial}$ is not claimed to have a local primitive-square realization.}
\label{tab:2d_AIII_AII_boundary_transmutation}
\end{table*}

\paragraph{Two-root parent: distinct bulk symmetries and identical surface.}
A second finite-range parent $\mathcal H_*$ realizes the same selected boundary pair and preserves both AIII
chiral and Kramers symmetries. Introduce the in-plane Dirac Hamiltonian
\begin{equation}
h_{\rm D}(\mathbf p)
=2t(\sin p_x\,\sigma_x+\sin p_y\,\sigma_z),
\label{eq:2d_Dirac_inplane}
\end{equation}
where $\sigma_i$ now act on the parent's Kramers doublet; the physical $A/B$ sublattices of $H_{\rm sq}$ emerge
only after canonical reblocking. With $w=p_z$ and $\zeta=e^{iw}$, define
\begin{equation}
\begin{aligned}
\mathcal H_*(\mathbf p,w)
&=-\rho_xh_{\rm D}(\mathbf p)
+\Lambda\bigl[\rho_y\operatorname{Im}\mathfrak q_*
+\rho_z\operatorname{Re}\mathfrak q_*\bigr],\\
\mathfrak q_*(\mathbf p,\zeta)
&=\zeta^{-1}(\zeta-a)(\zeta-b),\\
a&=1-\cos p_x,\\
b&=1-\frac12\cos p_y
(1+\cos p_x+\cos p_y).
\end{aligned}
\label{eq:2d_AII_bulk_H}
\end{equation}
Here $\Lambda>0$ sets the transverse hopping scale. The two roots $a(\mathbf p)$ and $b(\mathbf p)$ control the
surface-state decay, hence the name
\emph{two-root parent}. Its chiral and Kramers
symmetries are
\begin{equation}
S_*=\rho_x\sigma_y,\ \
\Theta=i\sigma_yK,\ \
S_*^2=1,\ \  \Theta^2=-1,
\end{equation}
They satisfy $\{S_*,\mathcal H_*\}=0$ and $\Theta\mathcal H_*(\mathbf p,w)\Theta^{-1} =\mathcal H_*(-\mathbf
p,-w)$. The parent conserves onsite particle number and has a finite bulk excitation gap $\Delta_*>0$. Its bulk
indices are $|\nu_3|=1$ and $(\nu_0;\nu_x,\nu_y,\nu_z)=(1;0,0,0)$ (Appendix~\ref{app:2d_AII_parent}).

For $z\ge1$, the selected surface state has transverse profile proportional to $(a^z-b^z)/(a-b)$. Both roots
must obey $|a|<1$ and $|b|<1$, which holds precisely on $D=(-\pi/2,\pi/2)^2$. The exact projected Hamiltonian is
\begin{equation}
h_\partial(\mathbf p)=h_{\rm D}(\mathbf p).
\label{eq:2d_AII_surface}
\end{equation}
The two decaying roots do not create two independent surface sectors: the open boundary condition leaves one
transverse profile. Equivalently, the transverse SSH-like winding counts two zeros and one pole, $2-1=1$. The
half-space solution is derived in Appendix~\ref{app:2d_AII_parent}.

As for $H_{\rm sq}$, unfold the selected surface momenta by setting $\mathbf k_{\rm B}=2\mathbf p$. On
compatible finite tori, a canonical Bogoliubov transformation yields the periodic, nearest-neighbor BdG
Hamiltonian
\begin{equation}
\begin{aligned}
h_\partial^{\rm BdG}(\mathbf k_{\rm B})
&=t\bigl[\sin k_{{\rm B},x}\,\kappa_x\sigma_x
+(1-\cos k_{{\rm B},x})\kappa_z\\
&+\sin k_{{\rm B},y}\,\kappa_x\sigma_z
+(1-\cos k_{{\rm B},y})\kappa_x\sigma_y\bigr].
\end{aligned}
\label{eq:2d_AII_edge_BdG}
\end{equation}
Here $\kappa_i$ act on BdG particle--hole space, distinct from the folded-sector matrices $\tau_i$. Its positive
energy is $2t\sqrt{\sin^2(k_{{\rm B},x}/2)+\sin^2(k_{{\rm B},y}/2)}$. A canonical transformation $\mathcal J_*$
identifies the Hamiltonian and centered surface charge exactly with those of $H_{\rm sq}$:
\begin{equation}
\mathcal J_*^\dagger H_{\partial,*}\mathcal J_*
=H_{\rm sq},\quad
\mathcal J_*^\dagger Q_{\partial,*}\mathcal J_*
=Q_{s_P}.
\label{eq:2d_AII_exact_sq_correspondence}
\end{equation}
The explicit transformation is constructed in Appendix~\ref{app:2d_AII_parent}. The direct-parent and two-root
momenta are related by
\begin{equation}
u=\frac\pi2+p_x+p_y,\quad
v=-\frac\pi2+p_x-p_y,
\label{eq:2d_parent_momentum_relation}
\end{equation}
so that $\sin u-\sin v=2\cos p_x\cos p_y>0$ on $D$. In physical square-lattice coordinates, this is simply
$(k_x,k_y)=(p_x,\pi/2+p_y)$.

The canonical correspondence also transports the bulk chiral and Kramers symmetries to the selected boundary
sector of $H_{\rm sq}$. In its physical two-site Bloch frame, the induced chiral symmetry is
\begin{equation}
S_{\rm sq}=\sigma_z,\quad
\{S_{\rm sq},h_{\rm sq}(\mathbf k)\}=0.
\label{eq:2d_sq_induced_chiral}
\end{equation}
The Kramers operation on the square-model boundary is defined by exact canonical transport,
\begin{equation}
\begin{aligned}
\Theta_{\partial}
&:=\mathcal J_*^\dagger\Theta\big|_{\partial,*}\mathcal J_*,\\
\Theta_{\partial}H_{\rm sq}\Theta_{\partial}^{-1}
&=H_{\rm sq},\qquad
\Theta_{\partial}Q_{s_P}\Theta_{\partial}^{-1}=Q_{s_P}.
\end{aligned}
\label{eq:2d_sq_induced_TR}
\end{equation}
Here $\Theta|_{\partial,*}$ is the parent Kramers symmetry restricted to its selected boundary sector. Its
transported one-particle action retains Kramers square $-1$ on the complete compatible finite-volume sector.
However, the intermediate BdG sewing is nonperiodic across the unfolded momentum seams
[Eq.~\eqref{eq:app_2d_AII_TR}]. Canonical transport therefore does not identify $\Theta_{\partial}$ with a
local primitive-square operation having a globally periodic sewing and constant Kramers square. In particular,
locality of $S_{\rm sq}$ does not imply locality of $\Theta_{\partial}$ in the primitive square geometry.

To distinguish the bulk symmetry protections while keeping the selected surface fixed, consider first the
finite-range AII descendant
\begin{equation}
\begin{aligned}
\mathcal H_\Theta(\mathbf p,w)
&=\mathcal H_*(\mathbf p,w)
+\epsilon_\Theta f(\mathbf p)P_{\rho,+},\\
f(\mathbf p)&=2-\cos p_x-\cos p_y,
\end{aligned}
\label{eq:2d_AII_genuine_Htheta}
\end{equation}
where $P_{\rho,+}=\tfrac12(1+\rho_x)$. This perturbation breaks chiral symmetry but preserves Kramers symmetry
and onsite $U(1)$. For $4|\epsilon_\Theta|<\Delta_*$, the bulk remains gapped with strong-AII index $\nu_0=1$.
These symmetries quantize the magnetoelectric angle to $\theta=\pi\pmod{2\pi}$, giving the long-wavelength bulk
response
\begin{equation}
S_\theta[A]=\frac{i\theta}{8\pi^2}\int_{M_4}F\wedge F,
\label{eq:app_2d_AII_theta_response}
\end{equation}
where $F=dA$ is the $U(1)$ field-strength two-form and $M_4=(S^1)^4$ is the Euclidean spacetime torus \cite{qi2008topological, wittenFermion2016}.

Conversely, the AIII descendant
\begin{equation}
\mathcal H_\chi(\mathbf p,w)
=\mathcal H_*(\mathbf p,w)
+\delta P_{\rho,+}\sigma_x,
\quad 0<\delta<\Delta_*,
\label{eq:2d_AIII_chiral_Hchi}
\end{equation}
preserves $S_*$ but breaks Kramers symmetry. Both perturbations act in the complementary $\rho_x=+1$ sector and
therefore annihilate the selected surface states, which have $\rho_x=-1$ at every layer.

The two descendants are joined through $\mathcal H_*$ by a finite-range path. For $-1\le s\le1$, let
$s_-=\max(-s,0)$ and $s_+=\max(s,0)$:
\begin{equation}
\begin{aligned}
\mathcal H_s(\mathbf p,w)
&=\mathcal H_*(\mathbf p,w)
+P_{\rho,+}\bigl[s_-\delta\sigma_x
+s_+\epsilon_\Theta f(\mathbf p)\bigr],\\
\mathcal H_{-1}&=\mathcal H_\chi,\quad
\mathcal H_0=\mathcal H_*,\quad
\mathcal H_{+1}=\mathcal H_\Theta.
\end{aligned}
\label{eq:2d_AIII_AII_fixed_boundary_path}
\end{equation}
Its perturbation strength is at most $\max(\delta,4|\epsilon_\Theta|)<\Delta_*$, so the bulk gap stays open.
Throughout the path, the selected surface wavefunctions, energies, compact charge, and canonical identification
of Eq.~\eqref{eq:2d_AII_exact_sq_correspondence} remain exactly unchanged. The AIII branch retains chiral
symmetry, the AII branch retains Kramers symmetry, and the common parent has both.
Table~\ref{tab:2d_AIII_AII_boundary_transmutation} summarizes these distinct bulk protections.

The direct AIII parent $\widetilde{\mathcal H}_\chi$ and the two-root AIII descendant $\mathcal H_\chi$ have
equal winding-number magnitudes (with signs fixed by orientation conventions), but the path above connects only
descendants of $\mathcal H_*$. Composing either exact surface identification with the full canonical
relocalization $\mathcal W$ gives the same charge-onsite, Schwartz-local $h_\sharp^{\rm sq}$ at zero gauge
field.

\subsubsection{Compact gauge response and parity-anomaly inflow}
\label{sec:2d_surface_magnetic_inflow}

The zero-field boundary correspondences do not imply equivalence of the independently gauged lattice theories.
The autonomous square regulator $h_\sharp^{\rm sq}$ couples through Wilson lines along Schwartz-local hoppings,
whereas the 3D parents have finite-range compact-link couplings and onsite particle number. Moreover, as in the
1D bulk completion of Sec.~\ref{sec:1d_bulk_edge_spt}, the selected surface sector does not admit a separately
gauge-covariant projection. We therefore compare the responses in matched smooth magnetic backgrounds, without
equating ultraviolet spectra or gauge-dependent determinants.

\paragraph{Magnetic correspondence and spectral isolation.}
Thread one flux quantum through the transverse torus, $\int_{T^2}F=2\pi$. At $\lambda=1/2$, the compactly gauged
direct AIII parent and the autonomous regulator approach the same continuum Dirac operator
$D_A=-2t(\sigma_x\Pi_x+\sigma_y\Pi_y)$, where $\Pi_\mu=-i\partial_\mu+A_\mu$. The Dirac index and Riemann--Roch
theorems give precisely one continuum zero mode~\cite{nakahara2018geometry}. For the specified smooth magnetic
backgrounds, Appendix~\ref{app:2d_sq_3d_flux_compactness} establishes lattice spectral isolation and excludes
additional low-energy states. Let $\delta>0$ be the gap above the continuum zero mode and choose $0<c<\delta/2$.
At sufficiently small lattice spacing $a$, the direct AIII half-space has exactly one level in $(-ca,ca)$,
pinned to zero by chiral symmetry:
\begin{equation}
\begin{aligned}
\dim\ker\widetilde{\mathcal H}_{\chi,a}[A]&=1,\\
\operatorname{Spec}\widetilde{\mathcal H}_{\chi,a}[A]
\cap(-ca,ca)&=\{0\}.
\end{aligned}
\label{eq:2d_AIII_mag_isolation_main}
\end{equation}
The autonomous regulator instead has one isolated near-zero level,
\begin{equation}
|E_{{\rm aut},a}|\le C'a^2.
\label{eq:2d_AIII_aut_mag_near_zero_main}
\end{equation}
With matched coordinates, connection, and boundary conditions, the normalized lattice modes converge to the same
continuum zero mode, differing by $O(a)$. Their common magnetic IR content provides a controlled setting in
which to compare the distinct UV mechanisms.

\paragraph{Autonomous UV compensation.}
For the autonomous regulator, the compensating UV contribution can be identified directly. The smooth orbital
rotation defining $h_\sharp^{\rm sq}$ preserves the Dirac cone while aligning the high-energy Hamiltonian with
positive $\sigma_z$ near the momentum-space seams. The nonnegative seam mass $m_W\sigma_z$ can be magnetically
quantized so that its diagonal block satisfies
\begin{equation}
h_{\sharp,a}[A]
=\begin{pmatrix}
M_a&K_a\\
K_a^\dagger&-M_a
\end{pmatrix},\ \ \ M_a\ge-C_Ma^2.
\end{equation}
Shift the Hamiltonian by $s_*\sigma_z$, with $s_*=Ta^2$ and fixed, sufficiently large $T$. The shifted diagonal
block is positive, and the Schur-complement inertia identity gives zero total spectral signature. The isolated
positive-flux Dirac level has continuum polarization $\sigma_z=-1$ and becomes negative under the shift, while
all other levels stay separated from zero by a gap of order $a$. The UV complement consequently has one excess
positive-energy level:
\begin{equation}
\eta_{{\rm UV},a}=+1,\ \
Q_{\rm UV}=-\frac12\eta_{{\rm UV},a}=-\frac12.
\label{eq:2d_AIII_static_eta_UV}
\end{equation}
The finite-lattice proof, including the magnetic positivity bound, is given in
Appendix~\ref{app:2d_sq_3d_static_eta}.

This UV half-charge compensates the sign of the centered IR temporal-holonomy character. Coupling the centered
onsite charge to $\alpha\sim\alpha+2\pi$ gives $Z_a(\alpha)=\prod_n2\cosh[(\beta E_{n,a}-i\alpha)/2]$. In the
scaling regime $\beta a^2\to0$ and $\beta a\to\infty$, with UV thermal corrections suppressed, the isolated
level and its complement yield
\begin{equation}
\begin{aligned}
\frac{Z_{{\rm IR},a}(\alpha)}{Z_{{\rm IR},a}(0)}
&\rightarrow\cos\frac\alpha2,\ \
\frac{Z_{{\rm UV},a}(\alpha)}{Z_{{\rm UV},a}(0)}
\rightarrow e^{-i\alpha/2},\\
\frac{Z_a(\alpha)}{Z_a(0)}
&\rightarrow e^{-i\alpha/2}\cos\frac\alpha2
=\frac{1+e^{-i\alpha}}2.
\end{aligned}
\label{eq:2d_aut_temporal_holonomy_main}
\end{equation}
The centered IR and UV factors are each antiperiodic under $\alpha\mapsto\alpha+2\pi$, but the full
finite-cutoff character is periodic. The UV half-charge therefore cancels the single-Dirac holonomy sign without
requiring spectral flow as $\alpha$ varies.

\paragraph{Direct AIII magnetic index.}
The direct AIII parent realizes the magnetic boundary obstruction through an exact chiral winding. Its
compact-link coupling preserves $\mathcal S_\Phi=\rho_x\epsilon$, where $\epsilon_{x,y}=(-1)^{x+y}$ is the
square-lattice sublattice grading. At transverse flux $\Phi$, its longitudinal chiral block satisfies
$D_\Phi(w)\epsilon=F_\Phi(\cos w)+i\sin w\,\mathbf1$, where $F_\Phi(s)=h_\Phi+\epsilon(M_\Phi+s)$ is Hermitian.
The scalar imaginary part reduces the winding exactly to two endpoint spectral counts:
\begin{equation}
w_1^{(L)}(\Phi)=n_-(F_{\Phi,-})-n_-(F_{\Phi,+}),
\end{equation}
Here $F_{\Phi,\pm}=F_\Phi(\pm1)$, and $n_-(X)$ counts the negative eigenvalues of a Hermitian matrix $X$. The
identity holds at finite volume whenever the endpoints are gapped, without a homotopy or continuum
approximation; see Appendix~\ref{app:2d_sq_3d_flux}.

At zero flux, the endpoints have the same occupied-state count but different Chern numbers, $C_-^{xy}=1$ and
$C_+^{xy}=0$. Under one positive flux quantum, $\Phi=\int_{T^2}F=2\pi$, the lattice St\v{r}eda
relation~\cite{stredaTheory1982,Dana1985} removes one occupied state from the Chern-nontrivial endpoint, leaving the trivial
endpoint unchanged for the Peierls convention $U=e^{+i\int A}$. The weak-field estimate in Appendix~\ref{app:2d_sq_3d_flux} guarantees that both endpoint gaps remain open on sufficiently
large compatible $L\times L$ tori. Hence
\begin{equation}
n_-(F_{2\pi,-}) =\frac{L^2}{2}-1,\ \
n_-(F_{2\pi,+})=\frac{L^2}{2},
\end{equation}
and
\begin{equation}
w_1^{(L)}(0) =0,\ \ w_1^{(L)}(2\pi)=-1.
\end{equation}
More generally, the same argument gives $w_1^{(L)}(2\pi n_\Phi)=-n_\Phi$ for every fixed integer $n_\Phi$ and
sufficiently large compatible tori.

The chiral half-space index fixes the boundary zero-mode parity, $n_0\equiv w_1^{(L)}\pmod2$, while the occupied
determinant Berry holonomy obeys $e^{i\gamma_z}=(-1)^{w_1^{(L)}}$. At unit flux, the boundary therefore carries
an odd number of complex zero modes, and the centered character $Z_0(\alpha)=[2\cos(\alpha/2)]^{n_0}$ changes
sign under $\alpha\mapsto\alpha+2\pi$. For the smooth one-flux backgrounds covered by spectral isolation above,
$n_0=1$, reproducing the normalized IR factor in Eq.~\eqref{eq:2d_aut_temporal_holonomy_main}. Independently,
opposite small charge-preserving masses of $h_\sharp^{\rm sq}$ have a unit Chern-number jump, consistent with
$C_-^{xy}-C_+^{xy}=1$ after matching orientations. This agreement identifies the anomaly class, not the UV
magnetic spectrum.

\paragraph{Two-root magnetic index.}
The common parent $\mathcal H_*$ realizes the same boundary obstruction with a different finite-range
Hamiltonian. Its two longitudinal decay roots select a single zero-field surface profile, but this root count
alone does not determine the magnetic winding. The compact-link coupling preserves $S_*=\rho_x\sigma_y$ at fixed
flux, even though the magnetic field generally breaks Kramers symmetry. In an $S_*$-diagonal basis, the
longitudinal chiral block $D_{*,\Phi}(w)$ defines
\begin{equation}
\nu_z^{(L)}(\Phi)=\frac{1}{2\pi i}
\int_{-\pi}^{\pi}dw\,
\operatorname{tr}\!\left[
D_{*,\Phi}^{-1}\partial_wD_{*,\Phi}
\right]\in\mathbb Z.
\label{eq_2roots_nu}
\end{equation}
Unlike the direct AIII block, its imaginary part is not initially scalar. Appendix~\ref{app:2d_sq_3d_two_root}
establishes a gauge-covariant homotopy that remains gapped and reaches a scalar imaginary part. The same
endpoint-counting identity then gives
\begin{equation}
\nu_z^{(L)}(\Phi)
=n_-(F_{*,\Phi}(-1))-n_-(F_{*,\Phi}(+1)).
\end{equation}
The endpoint Chern numbers are $C_{*,-}^{xy}=0$ and $C_{*,+}^{xy}=1$. Magnetic gap labeling at $\Phi=2\pi$
therefore gives
\begin{equation}
n_-(F_{*,2\pi}(-1))=L^2,\ \
n_-(F_{*,2\pi}(+1))=L^2-1,
\end{equation}
and
\begin{equation}
\nu_z^{(L)}(2\pi)=+1.
\end{equation}
The weak-field bound keeps the entire homotopy and both endpoints gapped on sufficiently large compatible tori.
For any fixed integer $n_\Phi$, the argument likewise yields $\nu_z^{(L)}(2\pi n_\Phi)=+n_\Phi$; no claim is made
for arbitrary finite flux density.

For a chiral-symmetric half-space termination, the bulk--boundary index determines the parity, though not
necessarily the total number, of complex zero modes, $n_0\equiv\nu_z^{(L)}\pmod2$. The occupied determinant
Berry holonomy satisfies $e^{i\gamma_z}=(-1)^{\nu_z^{(L)}}$, while the centered temporal-holonomy character is
$Z_0(\alpha)=[2\cos(\alpha/2)]^{n_0}$. Thus unit flux enforces odd zero-mode parity, Berry holonomy $-1$, and an
antiperiodic centered character. The index permits additional zero-mode pairs of opposite chirality. The direct
and two-root AIII parents therefore realize the same magnetic boundary obstruction through distinct compact-link
Hamiltonians.

The AII descendant $\mathcal H_\Theta$ does not preserve $S_*$, so the chiral magnetic index and exact lattice
zero-mode pinning no longer apply. Nevertheless, its single charged surface Dirac fermion has a continuum zero
mode in unit magnetic flux. Under magnetic spectral isolation, the weak-field lattice limit contains an isolated
level $|E_{\Theta,a}|\le Ca^2$, separated from the other low-energy surface states by a gap of order $a$. Its
normalized centered IR temporal-holonomy character consequently approaches $\cos(\alpha/2)$.

The strong index $\nu_0=1$, together with onsite $U(1)$ and zero-field Kramers symmetry, fixes the
long-wavelength magnetoelectric angle $\theta=\pi\pmod{2\pi}$ [Eq.~\eqref{eq:app_2d_AII_theta_response}]. For
one transverse flux quantum, a specified smooth extension of the temporal holonomy $\alpha$ into the half-space
gives the half-charge factor $e^{-i\alpha/2}$ in the same centered boundary-charge and transverse-flux
orientation as the autonomous calculation. This choice fixes the integer boundary-charge representative;
the strong bulk index alone determines only its half-integer part. Combining it with the centered surface factor yields
\begin{equation}
e^{-i\alpha/2}\cos\frac{\alpha}{2}
=\frac{1+e^{-i\alpha}}{2}.
\label{eq:2d_AII_inflow_holonomy}
\end{equation}
The strong-AII bulk thus provides the same long-wavelength holonomy compensation as the autonomous regulator,
but without a separate finite-cutoff UV spectral-asymmetry identity. The displayed relation concerns universal
factors for the specified half-space extension, not the complete microscopic half-space determinant.

These constructions realize the same IR parity-anomaly response by distinct mechanisms: autonomous UV
half-charge, chiral magnetic boundary indices in the two 3D AIII parents, and axion inflow in the AII
descendant. The full 3D lattice theories remain compact-gauge covariant, but their zero-field canonical
correspondences and common universal response do not equate independently gauged finite-cutoff determinants.
Quantitative matching of determinants and global phases for general backgrounds remains open.

\subsection{Two Majorana flavors: compact $U(2)$ and quaternionic topology}
\label{sec:2d_su2}

Two identical Majorana flavors of $H_{\rm sq}$ describe one fundamental $SU(2)$ Dirac doublet in the IR. The
exponentially local one-flavor complex structure $J_{\rm loc}$ promotes this symmetry to an exact compact $U(2)$
at finite lattice spacing. The resulting $SU(2)$ representation has a quaternionic $\mathbb Z_2$ invariant that
remains well defined at the massless point, although the doubled class-D Pfaffian parity is trivial. We
distinguish this symmetry-bundle topology from the non-Abelian parity anomaly of the IR Dirac doublet and from
the locality of a specified compact gauging.

\subsubsection{Exact compact $U(2)$ from two Majorana flavors}

Let $\boldsymbol\gamma=(\gamma^{(1)},\gamma^{(2)})^T$ collect the two Majorana flavors. The doubled Hamiltonian
is
\begin{equation}
H_2^{\rm sq}=\frac{i}{4}\boldsymbol\gamma^T(\mu_0\otimes A_{\rm sq})\boldsymbol\gamma,
\label{eq:2d_H2}
\end{equation}
where $A_{\rm sq}$ is the real antisymmetric hopping matrix of Eq.~\eqref{eq:2d_sq_micro}. The Pauli matrices
$\mu_i$ act on flavor, with $\mu_0=\mathbf1_2$. Using the inversion-odd sharp sign $s_P$ of
Eq.~\eqref{eq:2d_sq_spectral_sign}, define
\begin{equation}
\begin{aligned}
\mathcal T_0^{(s)}&=-i\mu_0\otimes s_P,& \mathcal T_1^{(s)}&=-i\mu_x\otimes s_P,\\
\mathcal T_2^{(s)}&=-i\mu_y\otimes\mathbf1,& \mathcal T_3^{(s)}&=-i\mu_z\otimes s_P.
\end{aligned}
\label{eq:2d_u2_generators}
\end{equation}
Here $s_P$ acts as the identity on the one-flavor orbital indices. These Majorana-compatible generators commute
with the doubled Hamiltonian and satisfy
\begin{equation}
[\mathcal T_a^{(s)},\mathcal T_b^{(s)}]=2\epsilon_{abc}\mathcal T_c^{(s)},\quad [\mathcal T_0^{(s)},\mathcal T_a^{(s)}]=0,
\label{eq:2d_su2_commutator}
\end{equation}
so they generate an exact compact spectral $U(2)$. Only $\mathcal T_2^{(s)}$ is onsite; the other generators
inherit the algebraic tails of $s_P$.

The smooth one-flavor complex structure satisfies $J_{\rm loc}^2=-\mathbf1$ and $[J_{\rm loc},A_{\rm sq}]=0$. It
yields
\begin{equation}
\begin{aligned}
\mathcal T_0&=\mu_0\otimes J_{\rm loc},&
\mathcal T_1&=\mu_x\otimes J_{\rm loc},\\
\mathcal T_2&=-i\mu_y\otimes\mathbf1,&
\mathcal T_3&=\mu_z\otimes J_{\rm loc}.
\end{aligned}
\label{eq:2d_u2_local_generators}
\end{equation}
All four generators are real antisymmetric, commute with $H_2^{\rm sq}$, and obey
\begin{equation}
[\mathcal T_a,\mathcal T_b]=2\epsilon_{abc}\mathcal T_c,
\quad [\mathcal T_0,\mathcal T_a]=0.
\label{eq:2d_u2_local_algebra}
\end{equation}
They therefore realize an exact compact $U(2)$ with exponentially decaying real-space kernels. The generator
$\mathcal T_2$ remains onsite, whereas the others contain $J_{\rm loc}$.

The sharp and smooth constructions agree on the IR flavor action. On compatible seam-free momentum grids, the
Hamiltonian-preserving rotation $W_P$ of Eq.~\eqref{eq:2d_sq_WP_main}, applied identically to both flavors,
intertwines their generators. This rotation is distinct from the full Majorana-canonical relocalization
$\mathcal W$ considered below. The central $U(1)$ generated by $\mathcal T_0$ assigns equal unit charge to both
Dirac flavors, while the Cartan generator $\mathcal T_3$ assigns opposite charges. By contrast, the onsite
generator $\mathcal T_2=\mathcal T_2^{(s)}$ is the number rotation obtained by pairing the two Majorana flavors
locally, up to orientation. These are distinct compact $U(1)$ subgroups of the same exact $U(2)$.

The doubled model also has conserved finite-range bilinears $Q_{\mathbf r}=(i/2)\sum_{\mathbf x}
\gamma^{(1)}_{\mathbf x}
\gamma^{(2)}_{\mathbf x+\mathbf r}$.
Translating one Majorana flavor makes any chosen $Q_{\mathbf r}$ onsite without changing the Hamiltonian. Their
commutators generate a generalized Onsager algebra whose IR quotient is $\mathfrak{su}(2)_{\rm valley}$. This
quotient describes the low-energy action of the finite-range generators, not an exact microscopic
$\mathfrak{su}(2)$ spanned by them. The exponentially local compact $U(2)$ constructed above instead uses the
orbital-dependent $J_{\rm loc}$ (Appendix~\ref{app:two_copy_onsager_u2}).

\subsubsection{Charge-bundle topology and non-Abelian gauge locality}
\label{sec:2d_su2_bundle}

In the folded one-flavor Bloch frame $(U,V)$ of Eq.~\eqref{eq:2d_sq_qloc}, recall
\begin{equation}
q_{\rm loc}=iJ_{\rm loc},\ \ q_{\rm loc}^2=\mathbf1,
\ \ q_{\rm loc}(-U,-V)=-q_{\rm loc}(U,V)^*.
\label{eq:2d_q_from_J}
\end{equation}
The corresponding periodic doubled-cell charge is $\tilde q(P,V)=F(U)^\dagger q_{\rm loc}(U,V)F(U)$, as in
Eq.~\eqref{eq:2d_sq_periodic_cell}. The fundamental-normalized $SU(2)$ generators on this periodic torus $T_Q^2$
are
\begin{equation}
S_1=\tfrac12\mu_x\otimes \tilde q,\ \
S_2=\tfrac12\mu_y\otimes\mathbf1,\ \
S_3=\tfrac12\mu_z\otimes \tilde q.
\label{eq:2d_su2_Hermitian}
\end{equation}
They satisfy $[S_a,S_b]=i\epsilon_{abc}S_c$ and $S_a(-\mathbf p)=-S_a(\mathbf p)^*$. With $\mathbf p=(P,V)$, let
$\mathscr V$ denote the complexified Majorana Bloch bundle, equipped with Majorana conjugation $\mathcal
C_{\mathbf p}:
\mathscr V_{\mathbf p}\to\mathscr V_{-\mathbf p}$
satisfying $\mathcal C_{-\mathbf p}\mathcal C_{\mathbf p}=1$. Each fiber carries a sum of pseudoreal $SU(2)$
fundamentals $\mathbf2$, with multiplicity bundle
\begin{equation}
\mathcal M_{\mathbf p}=\operatorname{Hom}_{SU(2)}(\mathbf2,\mathscr V_{\mathbf p}).
\label{eq:2d_su2_multiplicity}
\end{equation}
The multiplicity space $\mathcal M_{\mathbf p}$ counts the independent fundamental $SU(2)$ doublets at momentum
$\mathbf p$, so that locally $\mathscr V_{\mathbf p}\simeq
\mathbf2\otimes\mathcal M_{\mathbf p}$.
As momentum varies, the multiplicity spaces form a bundle whose topology need not be trivial, even though
$SU(2)$ acts identically on every doublet. The fundamental representation of $SU(2)$ is pseudoreal: it admits an
antiunitary intertwiner $j=i\mu_yK$, where $K$ denotes complex conjugation, satisfying $jg=gj$ for every $g\in
SU(2)$ and $j^2=-1$. For an intertwiner $\phi\in\mathcal M_{\mathbf p}$, Majorana conjugation induces
\begin{equation}
\Theta_{\mathcal M}\phi
=\mathcal C\circ\phi\circ j^{-1},
\quad \Theta_{\mathcal M}^2=-1,
\label{eq:2d_M_Quaternionic}
\end{equation}
mapping $\mathcal M_{\mathbf p}$ to $\mathcal M_{-\mathbf p}$. Thus $\mathcal M$ is a quaternionic bundle over
the doubled-cell Brillouin torus $T_Q^2$, with $\Theta_{\mathcal M}$ relating fibers at opposite momenta. It
carries the FKMM invariant~\cite{denittisClassification2015}
\begin{equation}
\nu_{SU(2)}:=\nu_{\rm FKMM}(\mathcal M)\in\mathbb Z_2.
\label{eq:2d_nu_SU2_def}
\end{equation}

The central charge splits the one-flavor Bloch bundle into $\mathcal E_\pm=\ker(\tilde q\mp\mathbf1)$. By
Eq.~\eqref{eq:2d_su2_Hermitian}, the $SU(2)$ generators act on $\mathcal E_+$ as $\mu_a/2$, while their action
on $\mathcal E_-$ is related to the same fundamental representation by conjugation with $\mu_y$. Hence $\mathcal
M\simeq\mathcal E_+\oplus\mathcal E_-$. Majorana reality, $\tilde q(-\mathbf p)=-\tilde q(\mathbf p)^*$,
identifies $\mathcal E_-(\mathbf p)$ with the complex conjugate of $\mathcal E_+(-\mathbf p)$, giving
\begin{equation}
\mathcal M_{\mathbf p}\simeq
\mathcal E_+(\mathbf p)\oplus
\overline{\mathcal E_+(-\mathbf p)}.
\label{eq:2d_M_split}
\end{equation}
The antiunitary $\Theta_{\mathcal M}$ exchanges these two sectors. For such a quaternionic double, the FKMM
invariant equals the Chern number of $\mathcal E_+$ modulo two, or equivalently the second Stiefel--Whitney
number of its underlying real bundle, since $w_2((\mathcal E_+)_{\mathbb R})=c_1(\mathcal E_+)\pmod2$. Using the
one-flavor charge Chern number $C_Q=1$, defined on $T_Q^2$ with orientation $dV\wedge dP$, we obtain
\begin{equation}
\nu_{SU(2)}=C_Q\pmod2=1.
\label{eq:2d_nuSU2_CQ}
\end{equation}
This relation is the non-Abelian counterpart of the one-flavor charge-Chern–Pfaffian relation. Although doubling
renders the physical class-D Pfaffian parity even, the $SU(2)$ multiplicity bundle retains a nontrivial
quaternionic $\mathbb Z_2$ invariant.

In the original primitive-cell Bloch convention, the Dirac nodes remain away from the PHIM, and the class-D
Pfaffian parity of the undeformed doubled Hamiltonian is
\begin{equation}
\mathcal P_D(H_2^{\rm sq})=\mathcal P_D(H_{\rm sq})^2=+1.
\label{eq:2d_doubled_PD_trivial}
\end{equation}
There is no contradiction with the one-flavor reflection theorem of Eq.~\eqref{eq:2d_dirac_parity}: doubling
violates its isolated two-band crossing hypothesis. In the periodic doubled-cell frame, the massless nodes
instead fold to a PHIM, where the Bloch Pfaffian vanishes and the corresponding Pfaffian-sign product is
undefined. The primitive-cell sign $+1$ must not be assigned to that massless folded Hamiltonian or to a
perturbation that retains only the doubled translations. The invariant $\nu_{SU(2)}$ detects symmetry topology
independently of either Pfaffian convention.

Unlike a physical occupied-band Chern number, $\nu_{SU(2)}$ remains defined without an energy gap. Although the
central $U(1)$ provides a convenient way to evaluate it, that charge is not required for the invariant to exist.
Appendix~\ref{app:2d_su2_central_breaking} constructs an exponentially local $SU(2)$-preserving perturbation
that breaks the central charge while leaving the Dirac germ unchanged.

The nontrivial $\nu_{SU(2)}$ obstructs a globally smooth, periodic, Majorana-compatible fiberwise change of
basis that makes the entire specified $SU(2)$ momentum independent on the periodic doubled-cell torus $T_Q^2$.
It does not prohibit a nonlocal canonical relocalization. For example, the exponentially local
Hamiltonian-preserving transformation
\begin{equation}
O_{\rm loc}=\exp\!\left(-\frac\pi4\mu_x\otimes J_{\rm loc}\right)
\label{eq:2d_Oloc}
\end{equation}
rotates the individual generator $\mathcal T_3$ to the onsite $\mathcal T_2$, without making the central
generator onsite. Applying the full one-flavor Majorana-canonical relocalization $\mathcal W$ identically to
both flavors instead gives a charge-adapted lattice in which $J_{\rm loc}\mapsto J_{\rm onsite}$, the onsite
complex structure corresponding to $Q_{\rm onsite}$ of Eq.~\eqref{eq:2d_sq_finite_reloc_main}, and
\begin{equation}
\mu_0\otimes J_{\rm onsite},\ \  \mu_x\otimes J_{\rm onsite},\ \
-i\mu_y\otimes\mathbf1,\ \ \mu_z\otimes J_{\rm onsite}
\label{eq:2d_U2_onsite_X}
\end{equation}
are all onsite. The nontrivial bundle topology then resides in the patched relation between the original and
relocalized momentum geometries.

The transformed Hamiltonian $\mu_0\otimes h_\sharp^{\rm sq}$ has Schwartz-decaying hoppings. Since $U(2)$ Wilson
transporters obey the same norm bounds as their $U(1)$ counterparts, the one-flavor gauge-locality argument of
Appendix~\ref{app:2d_hc_asymptotic_proof} extends directly to compact $U(2)$ gauging, yielding exact gauge
covariance and asymptotic dynamical gauge locality. This does not imply finite-range locality at fixed lattice
spacing or parity-invariant gauging.

The construction extends to $n_{\mathrm c}$ identical Majorana flavors, using $J_{\rm loc}$ to generate an
exponentially local compact $U(n_{\mathrm c})$. For $n_{\mathrm c}=2m$, onsite pairing of Majorana flavors
already provides a strictly local $U(m)$ subgroup preserving the original finite-range Hamiltonian. Applying
$\mathcal W$ to every flavor makes the larger $U(n_{\mathrm c})$ onsite in the relocalized geometry and permits
the corresponding Schwartz-local compact gauging.

\subsubsection{$SU(2)$-symmetric descendants and anomaly constraints}
\label{sec:2d_SU2_identification}

Consider a fully gapped quadratic descendant preserving the smooth $SU(2)$ symmetry. Its Bloch fibers locally decompose as $\mathscr V_{\mathbf p}\simeq\mathbf2\otimes\mathcal M_{\mathbf p}$, so Schur's lemma restricts the Hamiltonian to $\mathbf1_2\otimes h_{\mathcal M}$, where $h_{\mathcal M}$ acts on the multiplicity space. Define the occupied multiplicity bundle by
\begin{equation}
F=\operatorname{Im}P_{\rm occ}^{\mathcal M},
\quad
P_{\rm occ}^{\mathcal M}
=\mathbf1_{(-\infty,0)}(h_{\mathcal M}).
\end{equation}
Majorana conjugation relates occupied and unoccupied states at opposite momenta, giving
\begin{equation}
\mathcal M_{\mathbf p}\simeq
F_{\mathbf p}\oplus\overline{F_{-\mathbf p}}.
\label{eq:2d_M_from_F}
\end{equation}
The induced quaternionic structure exchanges these two summands. For such an exchanged pair, the FKMM invariant equals the Chern number of $F$ modulo two~\cite{denittisClassification2015}:
\begin{equation}
\nu_{SU(2)}=c_1(F)[T_Q^2]\pmod2.
\label{eq:2d_nuSU2_Cclass}
\end{equation}

Each occupied multiplicity state forms an $SU(2)$ doublet, so the full occupied Majorana/BdG bundle has twice the Chern number of $F$:
\begin{equation}
C_{\mathcal M}:=c_1(F)[T_Q^2],
\quad
\nu_{\rm full}=2C_{\mathcal M}.
\label{eq:2d_classC_phys}
\end{equation}
Here $C_{\mathcal M}$ is the Chern number of the reduced multiplicity Hamiltonian, while $\nu_{\rm full}$ counts both members of each fundamental $SU(2)$ doublet. This doubling is distinct from the standard 2D class-C even-Chern constraint. For an ordinary onsite fundamental $SU(2)$ representation, the reduced Hamiltonian admits a globally fixed particle--hole operation squaring to $-1$. The class-C classification then requires $C_{\mathcal M}\in2\mathbb Z$~\cite{ryu2010tenfold}, implying $\nu_{\rm full}\in4\mathbb Z$.

In contrast, the present multiplicity space carries a nontrivial quaternionic symmetry bundle that does not admit such a globally fixed class-C particle--hole structure. Odd $C_{\mathcal M}$ is therefore allowed. Combining the two invariants gives $\nu_{\rm full}\equiv2\nu_{SU(2)}\pmod4$; in particular, $\nu_{SU(2)}=1$ requires
\begin{equation}
\nu_{\rm full}\in4\mathbb Z+2.
\label{eq:2d_classC_selection}
\end{equation}
Thus the quaternionic $\mathbb Z_2$ invariant, already defined at the massless point, constrains the Chern number of any gapped descendant preserving the specified smooth $SU(2)$ bundle structure. This selection rule reflects the nontrivial symmetry bundle rather than the ordinary onsite class-C classification.

A finite-range gapped descendant is obtained by adding the same Chern mass $H_m$ to both Majorana flavors. In
the primitive-cell and folded one-flavor Bloch frames, respectively, the mass terms are
\begin{equation}
\begin{aligned}
\delta h_{m,2}(U,V)
&=\mu_0\otimes(m\sin U\,\sigma_z),\\
\delta h_{m,2}^{\rm fold}(U,V)
&=\mu_0\otimes(m\sin U\,\tau_z\otimes\sigma_z).
\end{aligned}
\label{eq:2d_sq_U2_mass}
\end{equation}
They preserve the exact compact $U(2)$ symmetry and the relative-translation Onsager charges. From the
one-flavor Chern number in Eq.~\eqref{eq:2d_sq_U1_mass_chern}, we obtain
\begin{equation}
\begin{aligned}
\nu_{\rm full}(m)
&=2\nu_{\rm Maj}(m)=2\operatorname{sgn}(m),\\
C_{\mathcal M}(m)&=\operatorname{sgn}(m).
\end{aligned}
\label{eq:2d_enriched_classC_minimal}
\end{equation}
Thus the two massive descendants realize
\begin{equation}
(\nu_{\rm full},\nu_{SU(2)})=(\pm2,1).
\label{eq:2d_enriched_descendants}
\end{equation}

These descendants also satisfy the magnetic-translation LSM constraint of the onsite complex $\pi$-flux model.
The relevant microscopic particle-number symmetry is generated by the onsite $\mathcal T_2$, rather than the
central $\mathcal T_0$. At flux $\phi=1/2$ and filling $\bar\rho=1/2$, a unique, gapped, short-range-entangled
phase preserving onsite particle number and magnetic translations satisfies the magnetic LSM constraint
$\phi\sigma_{xy}=\bar\rho\pmod1$~\cite{luFillingenforced2020}. It consequently requires an odd Hall
conductivity.
The two signs of the Chern mass realize $\sigma_{xy}=\pm1$.

The magnetic LSM constraint is distinct from the non-Abelian parity anomaly. A single fundamental
$SU(2)$ Dirac doublet carries a fractional Chern--Simons level $k_{\rm frac}=1/2\pmod{\mathbb Z}$
in fundamental-trace normalization. Its mod-two anomaly index agrees with $\nu_{SU(2)}=1$ for the explicit mass
family. Extending this correspondence to general regulators requires compatible mass-inversion sewing and
control of additional UV gap closings; it does not establish a universal lattice gauge-Pfaffian holonomy
identity.

The bond-centered reflections $\widetilde{\mathcal P}_{x,y}$ of Eq.~\eqref{eq:2d_sq_projective_parities}
preserve the sharp spectral $U(2)$ but transport the specified smooth $SU(2)$ to a distinct realization. Their
action involves the full four-component folded sewing, rather than only the two-orbital matrices of
Eq.~\eqref{eq:2d_sq_projective_parities}. The reflected charge cannot equal $\tilde q$, because its Chern number
has the opposite sign, or $-\tilde q$, because both charges have the same Dirac-number action at the nodes.
Since a flavor-diagonal reflection normalizes the specified $SU(2)$ only if it maps $\tilde q$ to $\pm\tilde q$,
neither bond-centered reflection does so. Appendix~\ref{app:2d_su2_dirac_inversion} establishes these results
and the separate conditional gap obstruction for an orientation-reversing symmetry that preserves the smooth
$SU(2)$.

Finally, consider the flavor-opposed mass patterns $(+m,-m)$ and $(-m,+m)$. Their interface breaks the
mixing $SU(2)$ but preserves two commuting, exponentially local $U(1)$ charges. After an exponentially localized
correction ensuring exact charge conservation, the positive-charge occupied sectors have relative Chern numbers
$-1$ and $+1$. These protect counterpropagating charge-resolved spectral flow. The resulting mixed IR
$U(1)_V$--$U(1)_A$ anomaly has $\mathcal A_{VA}=\mathcal A_{AV}=\pm2$ and vanishing diagonal entries,
despite cancellation of the total class-D Chern-number difference. The relative indices determine the protected
spectral flow and its anomaly coefficients independently of the detailed interface dispersion
(Appendix~\ref{app:2d_su2_interfaces}).

\section{3D: Compact symmetries, charge topology, and anomalies}
\label{sec:3d_vectorlike}

We study compact symmetries and gauge locality of a 3D lattice Dirac fermion. Unlike in 1D, the axial $U(1)_A$
has a pure perturbative gauge anomaly, in addition to the mixed $AVV$ anomaly diagnosed through vector-flux
insertion. We therefore focus on the anomaly-free vector $U(1)_V$, constructing an exponentially local compact
charge with strong quaternionic topology and an exact nonlocal relocalization preserving Hamiltonian locality.
Finally, we construct an exact compact $SU(2)$ with one independent Weyl doublet in the IR, exhibiting the
Witten global anomaly, and establish a general no-go theorem for onsite relocalization compatible with
Hamiltonian locality.

\subsection{Weyl-semimetal realization, relocalization, and anomalies}
\label{sec:3d_dirac}
Our benchmark is a finite-range two-band Weyl semimetal. It has ordinary onsite vector charge and a separately
conserved spectral axial charge, so the failure of local axial gauging can be traced without first assuming any
particular UV gauge coupling. Consider the Bloch Hamiltonian
\begin{equation}
\begin{aligned}
h_{\rm WSM}(\mathbf k)=& t\bigl[\sin k_x\,\sigma_x
+\sin k_y\,\sigma_y\\
&+(2-\cos k_x-\cos k_y-\cos k_z)\sigma_z\bigr].
\end{aligned}
\label{eq:3d_dirac_wsm}
\end{equation}
Its only gapless points are $\mathbf K_\pm=(0,0,\pm\pi/2)$, where
\begin{equation}
h_{\rm WSM}(\mathbf K_\pm+\mathbf p)
=t(p_x\sigma_x+p_y\sigma_y\pm p_z\sigma_z)
+O(|\mathbf p|^2).
\label{eq:3d_dirac_node_expansion}
\end{equation}
For a Weyl cone $\sum_{ij}v_{ij}p_j\sigma_i$, $\chi=\operatorname{sgn}\det(v_{ij})$ is its chirality: the two
nodes have $\chi_\pm=\pm1$ and form a single four-component massless Dirac fermion in the IR. The onsite vector
charge is ordinary centered particle number,
\begin{equation}
Q_V=\sum_{\mathbf x}(c_{\mathbf x}^\dagger c_{\mathbf x}-1),
\label{eq:3d_QV}
\end{equation}
where the subtraction is for the two orbitals per site. The same free Hamiltonian admits an exact compact
spectral axial charge
\begin{equation}
Q_A=\sum_{\mathbf k}s(k_z)
(c_{\mathbf k}^\dagger c_{\mathbf k}-1),
\quad s(k_z)=\operatorname{sgn}(\sin k_z).
\label{eq:3d_QA}
\end{equation}
We choose finite momentum grids avoiding the seams $k_z=0,\pi$. Both charges commute with the free Hamiltonian
and assign integer charges to the fermionic modes, but they have different locality: $Q_V$ is onsite, whereas
the step function in $Q_A$ gives algebraic real-space tails along $z$. Near the two nodes the charges are
$(q_V^\pm,q_A^\pm)=(1,\pm1)$, precisely the vector and axial assignments of the continuum Dirac theory.

The perturbative 't Hooft anomalies of the IR Dirac theory are well known and are determined by the Weyl-node
charges and chiralities. Label the vector and axial symmetries by $a,b,c\in\{V,A\}$. For each Weyl node
$\eta=\pm$, let $q_a^\eta$ denote its charge under $U(1)_a$ and $\chi_\eta$ its chirality. The anomaly
coefficients are
\begin{equation}
d_{abc}=\sum_{\eta=\pm}\chi_\eta
q_a^\eta q_b^\eta q_c^\eta,
\quad
b_a=\sum_{\eta=\pm}\chi_\eta q_a^\eta.
\label{eq:3d_node_anomaly_tensor}
\end{equation}
The symmetric cubic coefficients $d_{abc}$ characterize perturbative gauge anomalies, while $b_a$ characterize
mixed gauge--gravitational anomalies associated with spacetime curvature. Nonzero coefficients obstruct gauging
the corresponding symmetries without additional anomaly cancellation. For the charges above, one obtains
\begin{equation}
\begin{gathered}
d_{VVV}=d_{VAA}=0,\quad
d_{AVV}=d_{AAA}=2,\\
b_V=0,\quad b_A=2.
\end{gathered}
\label{eq:3d_Dirac_anomaly_tensor}
\end{equation}
Thus the vector $U(1)_V$ is perturbatively anomaly free when gauged alone, whereas the axial $U(1)_A$ has both
pure and mixed anomalies. In particular, $\chi_+q_A^+=\chi_-q_A^-=1$, so the two Weyl nodes contribute with the
same sign to the mixed $AVV$ anomaly. In a purely vector electromagnetic background, this gives the familiar
axial-current response
\begin{equation}
\partial_\mu j_A^\mu
=\frac{1}{2\pi^2}\mathbf E_V\cdot\mathbf B_V,
\label{eq:3d_axial_EB}
\end{equation}
in the convention $U=e^{+i\int A}$, with $\mathbf E_V=-\partial_t\mathbf A_V$ in temporal gauge.

The obstruction to local axial gauging can be seen by relocalizing the exact spectral charge. As in 1D, we first
use a Gaussian particle--hole map on the negative-$k_z$ branch:
\begin{equation}
\eta_{\mathbf k}=\mathcal Wc_{\mathbf k}\mathcal W^{-1}
=\begin{cases}
c_{\mathbf k},&k_z>0,\\
-i\sigma_y(c_{\mathbf k}^\dagger)^T,&k_z<0.
\end{cases}
\label{eq:3d_W_axial_action}
\end{equation}
Because $h_{\rm WSM}$ is traceless, $\sigma_y h_{\rm WSM}(\mathbf k)^T\sigma_y =-h_{\rm WSM}(\mathbf k)$.
Consequently,
\begin{equation}
\begin{aligned}
\mathcal WH_{\rm WSM}\mathcal W^{-1}&=H_{\rm WSM},\\
\mathcal WQ_V\mathcal W^{-1}&=Q_A,\\
\mathcal WQ_A\mathcal W^{-1}& =Q_V.
\end{aligned}
\label{eq:3d_W_axial_identities}
\end{equation}
This looks like a complete solution: the new charge is onsite and the free Hamiltonian remains finite range. But
the map conjugates the left-handed field at the same momentum. Fourier transformation then gives
\begin{equation}
\eta_L(\mathbf x)=-i\sigma_y
[\psi_L^\dagger(-\mathbf x)]^T,
\quad \eta_R(\mathbf x)=\psi_R(\mathbf x).
\label{eq:3d_W_IR_inversion}
\end{equation}
A spatially varying onsite phase $\alpha(\mathbf x)$ for $\eta_L$ therefore induces the parameter
$-\alpha(-\mathbf x)$ on the original left Weyl field, not the desired axial rotation by $-\alpha(\mathbf x)$.
Preservation of a translation-invariant Hamiltonian and of the global charge is not enough to reproduce the
physical local action of a symmetry.

One can repair this IR spatial mismatch by reflecting momentum about the left Weyl node before conjugation as in
1D:
\begin{equation}
\eta'_{\mathbf k}=\begin{cases}
c_{\mathbf k},&k_z>0,\\
-i\sigma_y(c_{\bar{\mathbf k}}^\dagger)^T,&k_z<0,
\end{cases}
\quad
\bar{\mathbf k}=(-k_x,-k_y,-\pi-k_z).
\label{eq:3d_W_reflected_action}
\end{equation}
Now $\eta'_L(\mathbf x)=-i\sigma_y [\psi_L^\dagger(\mathbf x)]^T$, so onsite rotations of $\eta'$ induce the
correct local axial transformation. The price is UV nonlocality. Indeed, writing $m_\perp=2-\cos k_x-\cos k_y$,
the transformed one-particle symbol is
\begin{equation}
\begin{aligned}
h_{\eta'}(\mathbf k)=t\bigl[&s(k_z)
(\sin k_x\,\sigma_x+\sin k_y\,\sigma_y)\\
&+(m_\perp-s(k_z)\cos k_z)\sigma_z\bigr].
\end{aligned}
\label{eq:3d_W_reflected_symbol}
\end{equation}
It is local at either Weyl node, but the sign function jumps across $k_z=0,\pi$, producing hopping with
algebraic tails. The failure is not just a poorly chosen phase convention: the reflected branch generically
changes the squared microscopic energy by
\begin{equation}
E^2(\bar{\mathbf k})-E^2(\mathbf k)
=4t^2m_\perp\cos k_z,
\label{eq:3d_W_reflection_obstruction}
\end{equation}
although the mismatch vanishes to leading order at the left Weyl node. Correspondingly, the two relocalized Weyl
cones have the same chirality: conjugating the left Weyl fermion changes its chirality, as required for an
onsite realization of its opposite axial charge. A smooth periodic local two-band Hamiltonian cannot contain
only two same-chirality Weyl nodes without compensating UV structure
(Appendix~\ref{app:friedan_charge_doubling}).

As in the 1D vector-flux spectral-flow argument, the mixed anomaly coefficient $d_{AVV}=2$ has an exact lattice
manifestation corresponding to the Adler--Bell--Jackiw response in Eq.~\eqref{eq:3d_axial_EB}. Thread $n_\Phi$
units of vector magnetic flux through the $xy$ torus and vary the longitudinal holonomy
$\Phi_V=\oint_z A_V$ from $0$ to $2\pi$. With $E_{V,z}=-\partial_t A_{V,z}$, this increasing holonomy
has negative physical electric field. Each Weyl node supports a chiral zeroth Landau level with degeneracy
$|n_\Phi|$ for nonzero flux. The two levels propagate
in opposite directions, but their contributions to the axial charge have the same sign. For positive magnetic
flux, their longitudinal velocities are opposite to those of the corresponding 1D chiral branches, so the same
positive holonomy insertion reverses the axial spectral-flow orientation. Consequently, $\Delta Q_A=-2n_\Phi$.
See Appendix~\ref{app:3D_axial} for details.

For APBC along $z$ with $L_z\in4\mathbb Z$, let $\delta k=2\pi/L_z$ and define the centered vector charge of
each transverse momentum slice $k_z=k$ by
\begin{equation}
Q_{V,k}=\sum_{\mathbf x_\perp,\alpha}
\left(
c_{\mathbf x_\perp,k,\alpha}^\dagger
c_{\mathbf x_\perp,k,\alpha}-\frac12
\right).
\end{equation}
For this positive longitudinal holonomy, use the same large vector gauge transformation as in 1D. Its momentum action is
$
\mathscr U_{V}c_k\mathscr U_{V}^{-1}=c_{k+\delta k}.
$ 
It maps the $\Phi_V=2\pi$ Hamiltonian back to $\Phi_V=0$. The corresponding inverse adjoint action on the
axial charge gives the exact operator identity
\begin{equation}
\begin{aligned}
&\mathscr U_{V}^{-1}Q_A
\mathscr U_{V}-Q_A\\
&\quad=2\left(
Q_{V,-\delta k/2}
-Q_{V,\pi-\delta k/2}
\right).
\end{aligned}
\label{eq:3d_main_AVV_operator_identity}
\end{equation}
The slices at $k=-\delta k/2$ and $k=\pi-\delta k/2$ have occupied-band Chern numbers $1$ and $0$,
respectively, and finite zero-field gaps. Choose a
magnetic torus with quantized flux $\Phi_{xy}=2\pi n_\Phi$ and compatible Peierls boundary conditions. Provided
both slice gaps remain open as the flux is introduced, magnetic gap labeling gives $\langle
Q_{V,k}\rangle=-C(k)n_\Phi$ in this flux convention and hence
\begin{equation}
\left\langle
\mathscr U_{V}^{-1}Q_A
\mathscr U_{V}-Q_A
\right\rangle_{\mathrm{FS}(B_V)}
=-2n_\Phi,
\label{eq:3d_main_AVV_result}
\end{equation}
for the stated flux and holonomy orientations. Thus the IR coefficient $d_{AVV}=2$ follows from an
exact finite-lattice comparison of gauge-equivalent endpoints, providing a lattice counterpart of Eq.~\eqref{eq:3d_axial_EB}. Although $Q_A$
remains conserved throughout the flux insertion in a fixed gauge, the large vector gauge transformation acts
nontrivially on this nonlocal charge.

The two lattice diagnostics illuminate different aspects of the anomaly. Relocalization identifies the locality
obstruction: the Hamiltonian-preserving map gives the wrong local spacetime action, whereas the map with the
correct axial action produces a UV seam and long-range hopping. Vector-flux insertion independently detects
$d_{AVV}=2$ through the monodromy of the spectral charge. The pure-axial and mixed gauge--gravitational
coefficients in Eq.~\eqref{eq:3d_Dirac_anomaly_tensor} follow from the Weyl-node data without matching a
nonlocal axial-gauging prescription to continuum current definitions. Likewise, a single unit-charge Weyl
fermion has $d_{AAA}=b_A=1$ and cannot admit standalone local gauging of its chiral $U(1)$.

\subsection{Quaternionic charge topology and relocalization}
\label{sec:3d_strong_charge}

\subsubsection{Charge-bundle topology}

In the preceding Weyl-semimetal model, the vector charge acts as the identity on each Bloch fiber, $q_V^{\rm
WSM}(\mathbf k)=\mathbf1_2$, and its positive-charge projector is trivial. For comparison, identify the oriented
transverse torus $(k_x,k_y)$ with the square-model periodic torus $(V,P)$, so that $dk_x\wedge dk_y=dV\wedge
dP$.  Pulling back the square positive-charge projector $P_+^Q$ and tensoring with a trivial rank-two factor gives a weak 3D charge bundle~\cite{ryu2010tenfold, chiu2016classification}
\begin{equation}
P_{+,\mathrm{weak}}(\mathbf k)=P_+^Q(k_x,k_y)\otimes\mathbf1_2.
\end{equation}
Its first Chern numbers are
\begin{equation}
C_{Q,xy}=2C_Q(H_{\rm sq};Q_{\rm loc})=2,\ 
C_{Q,yz}=C_{Q,zx}=0.
\label{eq:3d_square_dirac_chern}
\end{equation}
These invariants characterize the charge projector rather than the physical Hamiltonian. By the charge bulk--edge correspondence of Eq.~\eqref{eq:2d_charge_edge_spectral_flow}, the weak topology enforces spectral flow of the compressed charge on side boundaries normal to $x$ or $y$, but not on a surface normal to $z$; it does not imply physical surface-energy modes. We now construct a massless 3D Dirac fermion with an exact compact vector charge carrying strong, rather than weak, topology.

Let $\mu_i$, $\rho_i$, and $\sigma_i$ act on flavor, orbital, and Kramers spaces, respectively, with
$\mu_0=\rho_0=\sigma_0=\mathbf1_2$. Define
\begin{equation}
\begin{aligned}
S(\mathbf k)&=\sum_{j=x,y,z}\sin k_j\,\sigma_j,
\quad Z=\rho_z\otimes\sigma_0,\\
M_s(\mathbf k)&=-2+\sum_j\cos k_j,
\end{aligned}
\label{eq:3d_strong_wilson_data}
\end{equation}
and the auxiliary Wilson--Dirac Hamiltonian
\begin{equation}
\begin{aligned}
D_s(\mathbf k)
&=\rho_x\otimes S(\mathbf k)+M_s(\mathbf k)Z,\\
E_s(\mathbf k)
&=\sqrt{M_s(\mathbf k)^2+\sum_j\sin^2k_j}.
\end{aligned}
\end{equation}
The gapped four-band $D_s$ realizes a class-AII strong topological insulator \cite{fu2007topological}. Its flattened involution $q_{\rm
AII}=D_s/E_s$ satisfies $q_{\rm AII}^2=\mathbf1_4$ and is exponentially local. We use it to define the compact
vector charge of a distinct Majorana-compatible system:
\begin{equation}
q_V(\mathbf k)=q_{\rm AII}(\mathbf k)\oplus Z,
\quad q_V^2=\mathbf1_8.
\label{eq:3d_strong_charge}
\end{equation}
With $m_\delta(\mathbf k)=3-\sum_j\cos k_j+\delta$, construct
\begin{equation}
\begin{aligned}
X_\delta(\mathbf k)
&=\mu_x\otimes\rho_0\otimes S(\mathbf k)
-m_\delta(\mathbf k)\mu_z\otimes\mathbf1_4,\\
h_\delta(\mathbf k)
&=\tfrac12\{q_V(\mathbf k),X_\delta(\mathbf k)\}.
\end{aligned}
\label{eq:3d_strong_dirac_h}
\end{equation}
Since $[q,\{q,X\}]=[q^2,X]$, it is easy to see that $[h_\delta,q_V]=0$. Both operators are exponentially local,
although generally not strictly finite range.

The  Kramers and Majorana operations are
\begin{equation}
\Theta=\mu_0\otimes\rho_0\otimes i\sigma_yK,
\quad
\mathcal C=\mu_0\otimes\rho_y\otimes\sigma_yK,
\label{eq:3d_strong_antiunitary}
\end{equation}
with $\Theta^2=-1$, $\mathcal C^2=+1$, and
\begin{equation}
\begin{aligned}
\Theta h_\delta(\mathbf k)\Theta^{-1}
&=h_\delta(-\mathbf k),\ \ \mathcal C h_\delta(\mathbf k)\mathcal C^{-1} =-h_\delta(-\mathbf k),\\
\Theta q_V(\mathbf k)\Theta^{-1}
&=q_V(-\mathbf k), \ \ 
\mathcal C q_V(\mathbf k)\mathcal C^{-1}
=-q_V(-\mathbf k).
\end{aligned}
\label{eq:3d_strong_symmetries}
\end{equation}
The chiral grading $\Sigma=\mu_0\otimes\rho_y\otimes\sigma_0$ anticommutes with $h_\delta$, placing its gapped
phases in class DIII. Direct multiplication gives
\begin{equation}
\begin{aligned}
h_\delta(\mathbf k)^2
&=\varepsilon_\delta(\mathbf k)^2\mathbf1_8,\\
\varepsilon_\delta(\mathbf k)^2
&=m_\delta(\mathbf k)^2
+\frac12\left(1+\frac{M_s(\mathbf k)}{E_s(\mathbf k)}\right)
\sum_j\sin^2k_j.
\end{aligned}
\label{eq:3d_strong_dirac_spectrum}
\end{equation}
At $\delta=0$, the only zero is $\mathbf k=\mathbf0$. On the four-dimensional positive-charge fiber, the IR
Hamiltonian is
\begin{equation}
h_{V,+}^{\rm IR}(\mathbf p)
=\mu_x\otimes(\mathbf p\cdot\boldsymbol\sigma)
+O(|\mathbf p|^2),
\quad q_V^{\rm IR}=\mathbf1_4.
\label{eq:3d_strong_dirac_IR}
\end{equation}
The two opposite-chirality Weyl components thus form one independent massless 3D Dirac fermion, with $Q_V^{\rm
IR}=\int d^3x\, :\!\Psi_D^\dagger\Psi_D\!:$; the negative-charge sector is its Majorana conjugate. A uniform
$\delta$ adds the symmetry-preserving mass $-\delta\mu_zq_V$. The Dirac point is therefore fine-tuned,
consistent with the single-Dirac symmetry-protection no-go theorem of Ref.~\cite{gioiaBurkovHughes2026}.

Let $P_s=(\mathbf1_4+q_{\rm AII})/2$ and $P_0=(\mathbf1_4+Z)/2$. The positive-charge projector is
\begin{equation}
P_{V,+}(\mathbf k)
=\frac{\mathbf1_8+q_V(\mathbf k)}2
=P_s(\mathbf k)\oplus P_0.
\label{eq:3d_strong_charge_projector}
\end{equation}
It is a quaternionic bundle: $\Theta$ maps the fiber at $\mathbf k$ antiunitarily to that at $-\mathbf k$ and
squares to $-1$, whereas $\mathcal C$ exchanges the positive- and negative-charge fibers. The rank-two $P_s$
summand carries the nontrivial strong topology, while $P_0$ is trivial. Indeed, inversion acts as $Z$ on $q_{\rm
AII}$, and at time-reversal-invariant momenta $\boldsymbol\Lambda$, $q_{\rm AII}(\boldsymbol\Lambda)
=\operatorname{sgn}M_s(\boldsymbol\Lambda)Z$. Since $M_s$ is positive at $\mathbf k=\mathbf0$ and negative at
the other seven momenta, the Fu--Kane inversion criterion \cite{fuKane2007inversion} yields
\begin{equation}
(\nu_0;\nu_1\nu_2\nu_3)=(1;000),\quad
C_{Q,xy}=C_{Q,yz}=C_{Q,zx}=0.
\label{eq:3d_strong_charge_index}
\end{equation}
This index characterizes the charge eigenspace, not the occupied-energy bundle, and remains defined at the
massless point and throughout the $\delta$ interpolation. On the $k_z=0$ plane, the two time-reversed sectors of
$J=\rho_z\otimes\sigma_z$ have Chern numbers $\pm1$, whereas their $k_z=\pi$ counterparts are trivial. Thus a
vectorlike Dirac fermion can carry nontrivial microscopic charge-bundle topology.

\subsubsection{Boundary spectra and mass interfaces}

The distinction between charge and energy topology is visible at a boundary. For a half-space projector $R$,
define $q_{V,\rm cut}=Rq_VR$ and $h_{\rm cut}=Rh_\delta R$. In general,
\begin{equation}
q_{V,\rm cut}^2
=R-Rq_V(1-R)q_VR,
\ \  [h_{\rm cut},q_{V,\rm cut}]\ne0.
\label{eq:3d_strong_boundary_charge_algebra}
\end{equation}
The boundary eigenstates of the charge and Hamiltonian therefore need not coincide. At $k_x=k_y=0$, the
nontrivial charge block for a $z$ boundary reduces to
\begin{equation}
q_{\rm AII}(0,0,k_z)=\rho_z\cos k_z
+\rho_x\otimes\sigma_z\sin k_z.
\label{eq:3d_strong_charge_boundary_reduction}
\end{equation}
Its half-space compression has an exact zero-eigenvalue Kramers pair supported on the first layer, characterized
for the chosen termination by $\rho_y\varphi_s=\sigma_z\varphi_s=s\varphi_s$ ($s=\pm1$). The in-plane charge
spectrum is
\begin{equation}
\lambda_\pm(k_x,k_y)
=\pm\sqrt{k_x^2+k_y^2}+O(k_x^2+k_y^2).
\label{eq:3d_strong_charge_surface_cone}
\end{equation}
This Dirac cone belongs to the spectrum of the compressed charge, not to physical energies. It reflects the
strong charge topology and is independent of $\delta$.

Physical surface-energy modes can be analyzed separately. At $k_x=k_y=0$, the chiral block in the $\sigma_z=s$
sector has determinant
\begin{equation}
\begin{aligned}
&\det\mathscr D_s(k_z)
=\\
& -e^{-isk_z}\left[
(\delta+1-\cos k_z)^2
+\sin^2k_z\cos^2\frac{k_z}{2}\right].
\end{aligned}
\label{eq:3d_strong_boundary_det}
\end{equation}
For $\delta\ne0,-2$, the bracket is positive on the real momentum circle, giving winding $w_s=-s$ and a physical
zero-energy surface Kramers pair for the standard half-space termination. Complex roots determine the
penetration lengths
\begin{equation}
\xi_z\sim|\delta|^{-1}\ \ (\delta\to0),\quad
\xi_z\sim|\delta+2|^{-1/2}\ \ (\delta\to-2).
\label{eq:3d_strong_edge_lengths}
\end{equation}
The physical surface pair delocalizes at these bulk critical points, whereas the boundary charge cone is
unchanged. Additional bulk closings occur at $\delta=-4,-6$ at other surface momenta. The half-space
calculations are detailed in Appendix~\ref{app:3d_strong_boundary}.

The mass interpolation also admits an exactly charge-preserving interface. Let $q_V$ be the real-space
single-particle charge involution, $h_{\delta=0}$ the massless one-particle Hamiltonian, and $M$ a bounded, real
onsite mass profile with asymptotic values $\delta_-\in(-2,0)$ and $\delta_+>0$ on opposite sides. Define
\begin{equation}
h_{\rm int}=h_{\delta=0}-\frac12\mu_z\{M,q_V\}.
\label{eq:3d_strong_interface}
\end{equation}
It is easy to see that $[h_{\rm int},q_V]=0$, using $q_V^2=1$ and $[\mu_z,q_V]=0$; a real profile also preserves
Kramers symmetry. For uniform $M=\delta$, this reduces to $h_\delta$. When the profile is constant outside a
finite-width wall, the correction to naive multiplication by $M$ is exponentially localized near the wall. The
IR positive-charge Hamiltonian is
\begin{equation}
h_{V,+}^{\rm int}
=-i\mu_x\sigma_z\partial_z
+\mu_x(k_x\sigma_x+k_y\sigma_y)
-\delta(z)\mu_z,
\label{eq:3d_strong_interface_IR}
\end{equation}
which supports the familiar 3D topological-insulator mass-domain-wall Dirac mode. This physical interface mode
is distinct from the boundary charge-spectrum cone: $q_V$ is unchanged across the mass profile, so no separate charge-spectrum Dirac
cone is required at the wall.

\subsubsection{Nonlocal symmetry relocalization and compact gauging}
\label{sec:3d_strong_relocalization}

The positive-charge bundle has vanishing first Chern classes and is trivial as an ordinary complex bundle, but its strong quaternionic index is nonzero. A smooth charge-adapted frame can make $U(1)_V$ onsite while retaining Schwartz locality, but cannot simultaneously keep Kramers time reversal onsite. This obstruction does not exclude a nonlocal transformation that makes both symmetries onsite while preserving exponential locality of a particular Hamiltonian.

Such a transformation exists explicitly for the uniform massless model. Let $r^2=\sum_j\sin^2k_j$, $m=3-\sum_j\cos k_j$, $\widetilde m=m-r^2/2$, and $\alpha=(1+M_s/E_s)/2$, and define the positive analytic function $v(\mathbf k)$ through $v^2=\alpha+(m+\widetilde m)/2$. A momentum-dependent Gaussian canonical transformation $\mathcal W$ satisfies
\begin{equation}
\mathcal W q_V\mathcal W^\dagger=q_0,\ 
\mathcal W h_0\mathcal W^\dagger=h_{\rm reloc},
\label{eq:3d_strong_reloc_exact}
\end{equation}
where $q_0=\mu_0\otimes Z$ and
\begin{equation}
h_{\rm reloc}(\mathbf k)
=q_0\bigl[
v(\mathbf k)\,\mu_x\otimes\rho_0\otimes S(\mathbf k)-\widetilde m(\mathbf k)\,
\mu_z\otimes\mathbf1_4\bigr].
\label{eq:3d_strong_reloc_h}
\end{equation}
The transformation first aligns the charge eigenspaces and then rotates the kinetic and mass matrices to obtain an analytic Hamiltonian. Both the compact vector charge and Kramers time reversal are onsite in the new representation, while $h_{\rm reloc}$ is exponentially local. Near the physical Dirac point, $\mathcal W$ is smooth and approaches the identity, so the original and relocalized continuum Dirac fields agree at leading order.

The transformation itself is nonlocal: its Bloch representative is discontinuous at the seven time-reversal-invariant momenta where $M_s<0$. It therefore does not provide a smooth, Kramers-compatible trivialization of the strong charge bundle or a smooth homotopy between local Bloch Hamiltonians. In particular, band-topological invariants cannot generally be transported through this transformation. The two-step construction and symmetry checks are given in Appendix~\ref{app:3d_strong_reloc}.

The construction extends to each uniform $\delta>-1/2$, including gapped descendants, but the resulting transformation $\mathcal W_\delta$ generally depends on $\delta$. It therefore does not establish a common relocalization preserving Hamiltonian locality for spatially varying mass interfaces. In each uniform relocalized model, the onsite compact $U(1)_V$ can be gauged through Wilson lines while retaining exponentially local couplings, although transforming the gauge coupling back to the original variables need not preserve locality.

\subsection{$SU(2)$ Weyl doublet: global anomaly and locality obstruction}
\label{sec:3d_weyl_doublet}

In the continuum, one massless Dirac fermion can be reorganized into a left-handed Weyl doublet. Writing its
opposite-chirality components as $\psi_L$ and $\psi_R$, define
\begin{equation}
\chi_L=
\begin{pmatrix}
\psi_L\\
\psi_R^c
\end{pmatrix},
\quad
\psi_R^c=i\sigma_y\psi_R^*.
\label{eq:3d_SU2_dirac_reorganization}
\end{equation}
The two components have the same chirality, and the massless kinetic term admits a flavor $SU(2)$ acting as
$\chi_L\mapsto(U\otimes\mathbf1_2)\chi_L$. The original vector $U(1)_V$ is its Cartan subgroup
$U=e^{i\alpha\mu_z}$, since charge conjugation reverses the vector charge of the second component. This
reorganization does not produce an exact $SU(2)$ symmetry of the preceding microscopic Dirac model; in
particular, its vector-preserving mass breaks the enlarged symmetry.

A single fundamental $SU(2)$ Weyl doublet has vanishing perturbative cubic gauge anomaly, $\operatorname{tr}(t_a\{t_b,t_c\})=0$, where $t_a=\mu_a/2$, but its continuum theory carries the Witten global anomaly: a gauge transformation representing the nontrivial element of $\pi_4(SU(2))=\mathbb Z_2$ reverses the sign of the fermionic Pfaffian~\cite{witten1982su2}. We construct a lattice Hamiltonian with an exact compact global $SU(2)$ whose IR spectrum contains one independent Weyl doublet and therefore has this anomalous continuum content, without assuming a microscopic gauge coupling.

We reuse the class-AII involution $q_{\rm AII}=D_s/E_s$ from the preceding subsection and denote its
positive-eigenspace projector by
\begin{equation}
P_s(\mathbf k)=\frac12
\left(\mathbf1_4+q_{\rm AII}(\mathbf k)\right),
\quad
\mathcal M=\operatorname{Ran}P_s.
\label{eq:3d_SU2_parent_projector}
\end{equation}
Introduce a two-dimensional flavor space with Pauli matrices $\mu_a$ and define
\begin{equation}
T_a(\mathbf k)
=\frac12\mu_a\otimes P_s(\mathbf k),
\quad a=x,y,z.
\label{eq:3d_SU2_generators}
\end{equation}
It is easy to see that $[T_a,T_b]=i\epsilon_{abc}T_c$. The generators exponentiate to
\begin{equation}
\mathscr U_g
=g\otimes P_s+\mu_0\otimes(\mathbf1_4-P_s),
\quad g\in SU(2),
\end{equation}
with $\mathscr U_g\mathscr U_{g'}=\mathscr U_{gg'}$. Thus the symmetry acts as a fundamental doublet on $\mathbb C_\mu^2\otimes\mathcal M$ and trivially on its complement. Since $P_s$ is analytic, the generators are exponentially local. The multiplicity bundle $\mathcal M=\operatorname{Ran}P_s$ is the nontrivial quaternionic summand of the positive-charge bundle in the preceding subsection. It has strong index $\nu_0(\mathcal M)=1$, obstructing a globally smooth, periodic, Kramers-compatible trivialization.

To obtain the Weyl dispersion, introduce the Kramers-odd Hermitian matrix $s_z=\rho_z\otimes\sigma_z$ and define
$h_W=P_ss_zP_s$. Setting $B=\{D_s,s_z\}/2$ gives
\begin{equation}
\begin{aligned}
&B=M_s\,\rho_0\otimes\sigma_z
-\sin k_x\,\rho_y\otimes\sigma_y
+\sin k_y\,\rho_y\otimes\sigma_x,\\
&[D_s,B]=0,\ \ B^2=(M_s^2+\sin^2k_x+\sin^2k_y)\mathbf1_4.
\end{aligned}
\end{equation}
Since $D_sP_s=E_sP_s$, it follows that $h_W=P_sB/E_s$, and hence
\begin{equation}
\operatorname{Spec}(h_W|_{\mathcal M})
=
\left\{
\pm\frac{\sqrt{
M_s(\mathbf k)^2+\sin^2k_x+\sin^2k_y
}}{E_s(\mathbf k)}
\right\}.
\label{eq:3d_projected_weyl_spectrum}
\end{equation}
The only nodes occur at $\mathbf K_\pm=(0,0,\pm\pi/2)$. Near them, $M_s(\mathbf K_\pm+\mathbf p) =\mp
p_z+O(|\mathbf p|^2)$ and $E_s=1+O(|\mathbf p|^2)$, yielding two linear Weyl cones with opposite chiralities.

On the enlarged one-particle space, define
\begin{equation}
\mathcal H_{\rm Wd}(\mathbf k)
=\mu_0\otimes P_s(\mathbf k)s_zP_s(\mathbf k)
+\Delta\mu_z\otimes
[\mathbf1_4-P_s(\mathbf k)],
\label{eq:3d_SU2_physical_H}
\end{equation}
where $\Delta\ne0$. The Weyl term is flavor independent, while the complementary states are $SU(2)$ singlets
with gap $|\Delta|$. Consequently, $[\mathcal H_{\rm Wd},T_a]=0$ for all $a$, and the Hamiltonian is
exponentially local.

The Majorana particle--hole operation is
\begin{equation}
\mathcal C_{\rm Wd}
=(i\mu_y)\otimes\rho_0\otimes(i\sigma_y)K,
\quad
\mathcal C_{\rm Wd}^2=+1.
\label{eq:3d_SU2_particle_hole}
\end{equation}
Using the Kramers covariance of $P_s$, the oddness of $s_z$, and $(i\mu_y)\mu_a^*(i\mu_y)^{-1}=-\mu_a$, it can
be shown that
\begin{equation}
\begin{aligned}
\mathcal C_{\rm Wd}
\mathcal H_{\rm Wd}(\mathbf k)
\mathcal C_{\rm Wd}^{-1}
&=-\mathcal H_{\rm Wd}(-\mathbf k),\\
\mathcal C_{\rm Wd}
T_a(\mathbf k)\mathcal C_{\rm Wd}^{-1}
&=-T_a(-\mathbf k).
\end{aligned}
\end{equation}
The latter relation is consistent with the finite $SU(2)$ action because $\mathcal C_{\rm Wd}$ is antiunitary.
Majorana reality identifies the opposite-node representatives, leaving one independent fundamental $SU(2)$ Weyl
doublet in the IR.

A general no-go theorem excludes translation-preserving Majorana-canonical Gaussian relocalizations that make
the full compact $SU(2)$ onsite while retaining the original oriented Weyl content---in particular, odd net Weyl
charge in an open half Brillouin zone with gapped boundary planes---and a periodic $C^1$ transformed
Hamiltonian. The transformation itself need not be local or globally continuous; see
Appendix~\ref{app:3d_SU2_no_go}. Neither this locality obstruction nor the quaternionic bundle topology
establishes the sign holonomy of a background-gauged lattice Pfaffian, which requires an additional
gauge-coupling construction. Thus the exact compact global $SU(2)$ does not by itself provide a standalone local
gauge regulator.

\section{Discussion and outlook}
\label{sec:discussion}

\paragraph{Nonlocal canonical transformations and IR locality.}
The central insight of this work is that a compact symmetry can be spatially nonlocal in the UV yet admit local gauge dynamics in the IR. A generally nonlocal canonical transformation can make such an exact symmetry onsite while retaining sufficient Hamiltonian locality for compact gauging. Although the resulting gauge couplings are generally nonlocal in the original lattice variables, our successful relocalizations preserve the local spatial structure of the continuum fermion fields, despite their microscopic nonlocality. Together with the established gauge-locality properties of the relocalized Hamiltonians, this provides a mechanism for local IR gauge dynamics in the original continuum description. This construction goes beyond the emergence of an IR symmetry algebra, which may arise as a quotient of a larger microscopic algebra without furnishing an exact compact symmetry suitable for gauging. It also demonstrates that the same low-energy fermions can admit distinct microscopic realizations of symmetry and gauge locality without introducing additional fermionic degrees of freedom.

This perspective extends beyond locality-preserving QCAs, which provide an important framework for classifying topological phases. Our transformations need not map arbitrary local operators to local operators and therefore do not generally define equivalences under conventional locality-preserving classifications. For the more limited purpose of obtaining local IR dynamics and gauge couplings, preserving all microscopic operator localities is unnecessary. A natural mathematical problem is to characterize canonical transformations that are nonlocal in the UV but regular near the gapless modes, and to determine when they connect local or quasilocal Hamiltonians. The canonical centralizer provides a useful starting point, while the classification of these transformations, their stability, and their possible topological invariants remain largely unexplored.

\paragraph{Emergent gauge locality and UV seams.}
Gauge locality can arise through several mechanisms. Exact local-to-local relocalization permits conventional
compact gauging at finite spacing. In 2D, Schwartz-local Hamiltonians admit finite-range, gauge-invariant
approximants with operator-norm and dynamical control as the lattice spacing vanishes. In 3D, the vectorlike
construction retains exponentially local hopping and permits compact gauging with quasilocal couplings. The 1D
finite-chemical-potential construction is different: its gauged Hamiltonian remains genuinely nonlocal at each
finite cutoff, yet converges in norm resolvent to a strictly local dynamical gauge regulator. Thus, once the
compact symmetry has been made onsite, finite-cutoff Hamiltonian locality permits conventional local gauging,
but is not necessary for gauge locality to emerge in a controlled continuum limit.

UV seams likewise do not constitute an obstruction by themselves. Certain seam prescriptions produce singular
gauge vertices or nonlocal contributions that survive the continuum limit, but the relevant test is a controlled
comparison of the complete matter--gauge theories. In the finite-chemical-potential example, this comparison
includes dynamical gauge fields, Gauss-law constraints, and holonomy even though the Hamiltonians do not
converge in operator norm. Characterizing which seam structures are harmless, and for which gauge-coupling
prescriptions, remains an open problem.

\paragraph{The unresolved gauge locality of the Kitaev--Majorana model.}
This distinction is particularly relevant to the original Kitaev--Majorana honeycomb Hamiltonian $H_K$, which
motivated our investigation. Although it admits an exact compact spectral Dirac-number symmetry, its natural
onsite relocalization encounters UV seams. We establish restrictions on smooth compact quadratic completions
within the specified lattice classes, while a particular gauging prescription exhibits a divergent holonomy
response. These results constrain the constructions examined here; they neither prove nor disprove gauge
locality of the undeformed $H_K$ in general. More general canonical transformations or gauge-coupling
prescriptions may lead to different conclusions. Determining whether its exact Dirac-number symmetry admits any
gauge-local realization, possibly only in a controlled continuum limit, remains an important open problem.

\paragraph{Charge topology, anomalies, and boundary locality.}

Charge-bundle topology and 't Hooft anomalies impose distinct constraints. Nontrivial Chern and quaternionic invariants can obstruct smooth, periodic onsite trivializations without excluding more general nonlocal canonical transformations. By contrast, an uncanceled 't Hooft anomaly obstructs autonomous local gauging of the full anomalous symmetry structure, irrespective of the microscopic regulator. For the 3D fundamental $SU(2)$ Weyl doublet, the Witten global anomaly is consistent with our more specific Gaussian no-go theorem under its stated Hamiltonian-locality assumptions. Although the quaternionic charge indices in our 2D and 3D constructions accompany corresponding mod-two continuum anomaly data, this does not establish that they compute 't Hooft anomaly indices or gauge-Pfaffian holonomy.

A broader question concerns how charge-bundle topology constrains symmetry locality at boundaries. Our 2D and 3D constructions exhibit protected boundary spectra of compressed charge operators, even when these are distinct from the physical boundary-energy spectra. These charge-boundary spectra fall within the established bulk--boundary correspondence for spectrally flattened projectors and associated free-fermion entanglement spectra \cite{fidkowskiEntanglement2010,fidkowskiJacksonKlich2011}. Under a nonlocal canonical relocalization, the charge can become onsite while the original spatial boundary is transformed into a generally nonlocal cut. The original boundary charge spectrum is preserved when the spatial cut is transformed together with the charge, but need not survive an ordinary local cut in the relocalized representation. These observations motivate a systematic study of how charge-bundle topology constrains local compact symmetries at boundaries, how boundary charge spectra relate to physical edge dynamics, and which properties survive canonical transformations that are nonlocal in the UV but preserve spatial locality in the IR.

\paragraph{Bulk completion and anomaly inflow.}

Higher-dimensional bulk completions provide a complementary route to understanding anomalous lower-dimensional matter. Our exact bulk--boundary correspondences show that a non-onsite boundary symmetry may originate from an onsite bulk symmetry, and that distinct bulk symmetry structures can support the same selected boundary theory. The autonomous theory and its boundary realization share the same IR 't Hooft anomaly, but their microscopic gauge responses and dynamical gauge theories need not agree. In the bulk-induced construction, the anomalous boundary contribution is compensated by bulk inflow, so the complete bulk--boundary theory remains gauge invariant even when the selected boundary sector is not independently gauge covariant. An important open problem is to establish quantitative IR factorization of the full gauge-invariant partition function into anomalous boundary and compensating bulk amplitudes, including controlled errors, global anomaly phases, and symmetry-twisted backgrounds. Such a result would clarify how the common IR anomaly is realized within the higher-dimensional microscopic theory.

\paragraph{Gauge locality versus interacting IR dynamics.}
Gauge locality does not determine the resulting interacting IR physics. Our operator-norm and norm-resolvent
estimates compare specified dynamical gauge regulators; they do not establish that the regulators flow to a
prescribed interacting QFT. For example, reaching massless one-flavor QED$_3$ generally requires tuning the
fermion mass, while monopole perturbations, Chern--Simons levels, and strong gauge fluctuations may further
influence its IR behavior. Likewise, the exact matter correspondence underlying fermionic particle--vortex
duality does not establish the required monopole quantum numbers, global gauge structure, or interacting
duality. The existence of an exact compact symmetry, the construction of a gauge-local regulator, and control of
the interacting continuum limit are therefore three distinct problems. Our work primarily makes progress on the
first two.

\paragraph{Beyond free fermions.}

Our locality theorems concern Gaussian transformations of free fermions, but the broader idea need not be limited to quadratic systems. In Schulz--Shastry-type correlated-hopping models, nonlocal density-string transformations can connect local interacting Hamiltonians while transporting onsite symmetries into non-onsite ones \cite{schulzShastry1998}. Such constructions rely on special cancellations and do not establish general interacting gauge locality. A natural next step is to identify nonlocal operator-algebra automorphisms that preserve the intended IR spacetime and symmetry actions while retaining sufficient locality of interacting dynamics, possibly only in a controlled continuum limit. Such extensions could clarify when nonlocal transformations connect local interacting Hamiltonians and whether they reveal relations between phases beyond conventional locality-preserving equivalences.

\paragraph{Discrete and generalized symmetries.}

Although our constructions begin with fermionic canonical transformations, their implications extend beyond fermionic systems. The 1D vector--axial transformation provides a concrete example: through the Jordan--Wigner correspondence and bosonization, $\mathcal W$ realizes momentum--winding T-duality of the compact boson, while parity projection on the XX spin chain produces an exact noninvertible spectral operator $\mathcal D$ satisfying $\mathcal D^2=1+U$. In the IR, $\mathcal D$ flows to a noninvertible T-duality topological defect of the compact boson, while the corresponding projected operator in the isolated Ising sector flows to the conventional Kramers--Wannier defect. Thus noninvertible topological defects can emerge from algebraically nonlocal microscopic transformations, even without manifestly local lattice realizations. We will explore the broader connection between nonlocal dualities and generalized symmetries in forthcoming work. Extending this perspective to higher-form symmetries would require a broader notion of relocalization beyond ordinary onsite group representations.

\paragraph{Chiral lattice gauge theories.}
A further direction is to extend symmetry relocalization to chiral lattice gauge theories. Our autonomous compact-gauging constructions primarily concern symmetry sectors that can be gauged individually without an intrinsic anomaly obstruction, while the Weyl constructions expose additional constraints from Hamiltonian locality and topology. For genuinely chiral representations, an exact compact spectral symmetry and its onsite realization are insufficient: a standalone gauge-invariant regulator must also have local matter--gauge couplings, satisfy the relevant anomaly-cancellation conditions, and admit a consistent fermion measure. For anomaly-free chiral representations, one possible route combines symmetry relocalization with a local mirror-fermion completion, using symmetric mass generation to gap the unwanted mirror modes while preserving the gauge symmetry \cite{wangYou2022SMG}. However, the required interactions need not remain local under relocalization, and anomaly cancellation alone does not guarantee successful mirror-sector decoupling.

Lüscher's nonperturbative construction of anomaly-free Abelian chiral gauge theories with exact lattice gauge invariance, and its extension to non-Abelian theories to all orders in perturbation theory, provide important precedents \cite{luscherAbelian1999a,luscherLattice2000}. Recent proposals pursue nonperturbative formulations of anomaly-free non-Abelian chiral gauge theories through overlap-based fermion measures or continuum Weyl fermions coupled to lattice-interpolated gauge fields \cite{craiggaugeinvariant2026,hsuStandard2026,wittenNote2026}. A distinct Hamiltonian approach employs locality-preserving symmetry disentanglers and lattice rotor models \cite{thorngrenChiral2026a}. Our approach instead permits spatially nonlocal canonical transformations while retaining local or asymptotically local dynamics. Whether this additional flexibility can provide complementary Hamiltonian constructions of anomaly-free chiral gauge theories remains an open question.

\paragraph{Anomaly cancellation and onsiteability.}
Our constructions raise the question of when an exact compact symmetry of a local Hamiltonian can be made onsite by a generally nonlocal canonical transformation while preserving the same symmetry-enriched IR theory and sufficient Hamiltonian and gauge locality. As reviewed in Appendix~\ref{app:friedan_charge_doubling}, Friedan's charge-resolved doubling theorem requires chirality cancellation separately in globally regular charge sectors, even when their bundles are topologically nontrivial \cite{friedanproof1982a}. This necessary constraint neither guarantees onsite relocalization nor establishes gauge locality. A distinct, less restrictive question is whether another local Hamiltonian, possibly involving interactions or ancillary degrees of freedom, can realize the same IR theory with exact onsite symmetry and consistent local gauging, provided the relevant 't Hooft anomalies cancel. Our 2D constructions illustrate this distinction: although gauge locality remains unresolved for the original Kitaev--Majorana Hamiltonian $H_K$, the square-Majorana Hamiltonian $H_{\rm sq}$ realizes the same single-Dirac IR content and admits asymptotically local compact $U(1)$ gauging, without eliminating the associated parity-anomaly constraints. Thus the existence of a suitable regulator does not imply the relocalizability of a particular microscopic Hamiltonian.

Recent studies have clarified the distinctions among lattice anomalies, onsiteability, and gaugeability \cite{seifnashriShirley2026,kapustinHigher2025,kawagoeShirley2025,tuLongElse2026}. In particular, certain microscopic symmetry actions admit trivial symmetric gapped states yet remain non-onsiteable and nongaugeable within locality-preserving frameworks \cite{shirleyZhangJiLevin2026}. Such obstructions do not necessarily exclude generally nonlocal canonical relocalizations or alternative local regulators, whereas an uncanceled 't Hooft anomaly remains an obstruction to autonomous local gauging of the full anomalous symmetry structure. Our constructions demonstrate that nonlocal transformations can change microscopic symmetry locality while preserving the intended IR fermion theory and its established anomaly responses. Determining the conditions for simultaneous onsite symmetry, local dynamics, and consistent gauging remains an open problem.

Taken together, our results show that the microscopic nonlocality of an exact compact symmetry need not obstruct local gauge dynamics in the IR. Nonlocal canonical transformations can make such symmetries onsite while preserving the local spatial structure of the continuum fermion fields and sufficient Hamiltonian locality for compact gauging. Establishing general principles governing these relocalizations, their stability, and their extension beyond quadratic fermions remains an important challenge. Developing these principles would lay the foundations for a general theory of \emph{Locality from Nonlocality}.

\begin{acknowledgements}  The author thanks  Andriy Nevidomskyy, Dmitriy Pavshinkin, Alexei Tsvelik, and Meng Zeng for useful discussions. This work was supported by the U.S. Department of Energy (DOE) Office of Basic Energy Sciences (BES), Division of Materials Sciences and Engineering under contract No. DE-SC0012704 with Brookhaven National Laboratory. ChatGPT was used to assist with derivations, proofs, and manuscript preparation. 
\end{acknowledgements}

\appendix

% Appendices are ordered from general background through 1D, 2D, and 3D.
% All cross-references use stable labels, so their letters update automatically.
\section{Charge-resolved doubling: Friedan's theorem and Majorana extension}
\label{app:friedan_charge_doubling}

We review Friedan's charge-resolved doubling theorem and its geometric proof, emphasizing that smooth,
periodic compact symmetries can define topologically nontrivial charge bundles without invalidating chirality
cancellation. We then explain its application to conserved Majorana complex structures, distinguish independent
complex fermions from BdG redundancy, and relate charge-bundle topology to symmetry relocalization and gauge
locality.

Friedan's refinement of the Nielsen--Ninomiya theorem establishes chirality cancellation separately in each
sufficiently regular irreducible representation of an internal symmetry group
\cite{friedanproof1982a}. No globally momentum-independent symmetry frame is assumed. Let the spatial
dimension be odd, $d=2\ell+1$, and consider a translation-invariant, number-conserving free-fermion Hamiltonian
acting on independent complex fermions,
\begin{equation}
H=\int_{T^d}d^dk\,c^\dagger(\mathbf k)h(\mathbf k)c(\mathbf k),
\label{eq:friedan_general_H}
\end{equation}
where $h(\mathbf k)$ is finite dimensional, periodic, and smooth for the differential-form argument below. Let
$G$ be a compact internal symmetry commuting with $h$, and let $P_Q(\mathbf k)$ project onto an irreducible
symmetry sector $Q$. We initially assume that $h$ and these projectors are smooth and periodic; continuity will
suffice for the topological conclusion.

For $G=U(1)^r$, let $q_a(\mathbf k)$, $a=1,\ldots,r$, be mutually commuting Hermitian generators with integer
one-particle weights, $e^{2\pi i q_a(\mathbf k)}=1$, and
\begin{equation}
[q_a(\mathbf k),q_b(\mathbf k)]=0,\ \
[q_a(\mathbf k),h(\mathbf k)]=0.
\label{eq:friedan_compact_generators}
\end{equation}
For a joint charge vector $\mathbf n=(n_1,\ldots,n_r)$, define
\begin{equation}
P_{\mathbf n}(\mathbf k)=\prod_{a=1}^{r} P^{(a)}_{n_a}(\mathbf k),
\label{eq:friedan_joint_projector}
\end{equation}
where
\begin{equation}
P^{(a)}_{n_a}(\mathbf k)=\frac{1}{2\pi i}\oint_{\Gamma_{n_a}}dz\,
\left[z\mathbf 1-q_a(\mathbf k)\right]^{-1},
\label{eq:friedan_spectral_projector}
\end{equation}
and $\Gamma_{n_a}$ encloses only the eigenvalue $n_a$. They obey
\begin{equation}
q_aP_{\mathbf n}=n_aP_{\mathbf n},\ \
P_{\mathbf n}P_{\mathbf n'}=\delta_{\mathbf n,\mathbf n'}P_{\mathbf n},\ \
\sum_{\mathbf n}P_{\mathbf n}=\mathbf1.
\label{eq:friedan_projector_relations}
\end{equation}
Smooth, periodic compact generators have smooth, periodic projectors of constant rank because their eigenvalues
lie in a discrete set. The theorem requires regularity of the projectors, however, not a globally defined
symmetry frame; compactness and commutation alone do not ensure this regularity. The associated bundles
\begin{equation}
\mathcal E_{\mathbf n}=\operatorname{Ran}P_{\mathbf n}\rightarrow T^d
\label{eq:friedan_charge_bundle}
\end{equation}
may be topologically nontrivial even when the compact generators and their projectors are smooth and periodic.
Thus a regular charge-sector decomposition need not admit a globally smooth periodic onsite frame. Since
$[P_{\mathbf n},h]=0$, the Hamiltonian acts within each bundle.

Assume that zero energy occurs only at finitely many isolated Fermi points $\mathbf K_j$, with relativistic
Clifford low-energy blocks. If $n_{\mathrm R}(\mathbf n)$ and $n_{\mathrm L}(\mathbf n)$ count the right- and
left-handed massless fermions in sector $\mathbf n$, Friedan's theorem states
\begin{equation}
I_{\mathbf n}:=n_{\mathrm R}(\mathbf n)-n_{\mathrm L}(\mathbf n)=0
\quad\text{for every }\mathbf n.
\label{eq:friedan_charge_resolved_index}
\end{equation}
More generally, the topological index is defined without the relativistic assumption; the latter identifies it
with net fermion chirality.

For the geometric proof, remove disjoint small balls $B_j$ around the Fermi points:
\begin{equation}
\mathcal B^\circ=T^d\setminus\bigcup_j\operatorname{int}B_j.
\label{eq:friedan_punctured_BZ}
\end{equation}
At energy $E$, define the charge-resolved spectral projector
\begin{equation}
S_{\mathbf n}(E,\mathbf k)
=P_{\mathbf n}(\mathbf k)\,\mathbf 1_{(E,\infty)}\!\left(h(\mathbf k)\right).
\label{eq:friedan_positive_projector}
\end{equation}
On $\mathcal B^\circ$, $S_{\mathbf n}:=S_{\mathbf n}(0,\mathbf k)$ is smooth and determines the positive-energy
bundle $\mathcal F_{\mathbf n}=\operatorname{Ran}S_{\mathbf n}\to\mathcal B^\circ$. The projected connection
$\nabla_{\mathbf n}=S_{\mathbf n}d_{\mathbf k}$ has curvature $F_{\mathbf n}=S_{\mathbf n}(d_{\mathbf
k}S_{\mathbf n})^2S_{\mathbf n}$, with components $(F_{\mathbf n})_{ij}=S_{\mathbf n}[\partial_iS_{\mathbf
n},\partial_jS_{\mathbf n}]S_{\mathbf n}$. Its degree-$2\ell$ Chern character is closed:
\begin{equation}
\operatorname{ch}_\ell(\mathcal F_{\mathbf n})
=\frac{1}{\ell!(2\pi i)^\ell}\operatorname{tr}\!\left(F_{\mathbf n}^{\,\ell}\right),
\ \ d\,\operatorname{ch}_\ell(\mathcal F_{\mathbf n})=0.
\label{eq:friedan_chern_character}
\end{equation}
With the flat metric on the Brillouin torus, this also defines a conserved momentum-space topological current,
\begin{equation}
j_{\mathbf n}^{\rm top}=\star\operatorname{ch}_\ell(\mathcal F_{\mathbf n}),
\quad\delta j_{\mathbf n}^{\rm top}=0\quad\text{on }\mathcal B^\circ,
\label{eq:friedan_chiral_current}
\end{equation}
which is not a spacetime axial Noether current. The flux surrounding a Fermi point measures its chirality, up to
the orientation convention:
\begin{equation}
\nu_{\mathbf n,j}=\int_{\partial B_j}\operatorname{ch}_\ell(\mathcal F_{\mathbf n}).
\label{eq:friedan_local_chirality}
\end{equation}
Since $\partial\mathcal B^\circ$ has the opposite orientation to the boundaries of the removed balls, Stokes'
theorem gives
\begin{equation}
\begin{aligned}
I_{\mathbf n}
&=\sum_j\nu_{\mathbf n,j}
=-\int_{\partial\mathcal B^\circ}\operatorname{ch}_\ell(\mathcal F_{\mathbf n})\\
&=-\int_{\mathcal B^\circ} d\,\operatorname{ch}_\ell(\mathcal F_{\mathbf n})=0.
\end{aligned}
\label{eq:friedan_stokes}
\end{equation}
No global frame for $\mathcal E_{\mathbf n}$ or $\mathcal F_{\mathbf n}$ is required. In particular,
nontrivial charge-bundle Chern classes do not invalidate charge-resolved chirality cancellation.

The geometric argument extends to continuous, periodic Hamiltonians and charge-sector projectors,
using vector-bundle topology rather than differential forms. If both $h$ and $P_{\mathbf n}$ are
continuous, the spectral gap on the compact set $\mathcal B^\circ$ makes
$S_{\mathbf n}(0,\mathbf k)$ continuous there, so $\mathcal F_{\mathbf n}\to\mathcal B^\circ$
remains a complex vector bundle. Its Chern character is then understood as a cohomology class.
With $\iota:\partial\mathcal B^\circ\hookrightarrow\mathcal B^\circ$,
\begin{equation}
I_{\mathbf n}
=-\left\langle\iota^*\operatorname{ch}_\ell(\mathcal F_{\mathbf n}),[\partial\mathcal B^\circ]\right\rangle=0,
\label{eq:friedan_continuous_extension}
\end{equation}
because $\iota_*[\partial\mathcal B^\circ]=0$ in $H_{2\ell}(\mathcal B^\circ;\mathbb Z)$. Thus continuity
suffices for topological chirality cancellation, whereas a relativistic interpretation of the local charges
still requires appropriate low-energy regularity. Continuity does not imply exponential locality or gauge
locality.

For a relativistic low-energy theory in even spacetime dimension $D=d+1=2\ell+2$, charge-resolved doubling also
forces the associated perturbative Abelian anomaly coefficients to cancel. All fermions in a fixed sector
$\mathbf n$ have identical charges, so their anomaly-polynomial contribution is proportional to $I_{\mathbf n}$:
\begin{equation}
\mathcal I_{D+2}^{(\mathbf n)}
=I_{\mathbf n}\left[\widehat A(T)\exp\!\left(
\sum_a n_a\frac{F_a}{2\pi}\right)\right]_{D+2},
\label{eq:friedan_anomaly_polynomial}
\end{equation}
up to conventional factors of $i$. In $(3{+}1)$D, in particular,
\begin{equation}
\mathcal A_{abc}\propto\sum_{\mathbf n}n_an_bn_c I_{\mathbf n}=0,
\ \
\mathcal A_a^{\rm grav}\propto\sum_{\mathbf n}n_a I_{\mathbf n}=0.
\label{eq:friedan_4d_anomaly_corollary}
\end{equation}
In $(1{+}1)$D, $\ell=0$ and $\operatorname{ch}_0$ is the rank, reducing the argument to cancellation of signed
Fermi crossings in each continuous charge sector. These are perturbative anomaly statements for independent
complex fermions; they neither construct a gauge-local background-field coupling nor exclude global or torsion
anomalies, such as the $SU(2)$ Witten anomaly. They do not address the odd-spacetime-dimensional parity anomaly.

For Majorana fermions, a conserved orthogonal complex structure generates an exact compact $U(1)$ symmetry and
provides the complex one-particle bundle needed to apply Friedan's theorem to independent physical fermions. On
the real-space Majorana one-particle space, write
\begin{equation}
H_{\rm Maj}=\frac{i}{4}\gamma^T A\gamma,
\quad A^T=-A.
\end{equation}
Suppose a translation-covariant real complex structure $J$ has a continuous, periodic Fourier symbol and
satisfies, as a real-space operator,
\begin{equation}
J^T=-J,\quad J^2=-\mathbf1,\quad[J,A]=0.
\label{eq:friedan_majorana_J}
\end{equation}
In a fixed real Majorana basis, its momentum-space sewing is $J(-\mathbf k)=J(\mathbf k)^*$ and $J(\mathbf
k)^\dagger=-J(\mathbf k)$; the symbol need not be pointwise real or antisymmetric under ordinary matrix
transpose. In the Hermitian charge convention used throughout this work, $q=iJ$ and $q^2=\mathbf1$. The complex
structure acts as multiplication by $i$ on the underlying real one-particle bundle, making $A$ complex linear.
Its associated Hermitian one-particle Hamiltonian is
\begin{equation}
h_J(\mathbf k)=J(\mathbf k)A(\mathbf k),
\label{eq:friedan_majorana_hJ}
\end{equation}
which commutes with $J(\mathbf k)$; $JA$ is real symmetric on the original real-space Majorana space. In the
complexification, the two complementary spectral projectors are
\begin{equation}
P_\pm^{(q)}(\mathbf k)=\frac12\left(\mathbf1\pm iJ(\mathbf k)\right).
\label{eq:friedan_majorana_projectors}
\end{equation}
Here the signs denote eigenvalues $q=\pm1$. These projectors specify a complex bundle and its conjugate, not two
independent sets of physical fermions. Choosing the complex structure identifies the independent complex bundle,
on which $h_J$ defines a number-conserving Hamiltonian. When the corresponding projector is globally continuous
and periodic and the zero-energy crossings satisfy the hypotheses above, Friedan's charge-resolved cancellation
applies to this physical complex bundle.

The distinction is essential in a redundant BdG/Nambu representation: its doubled bands cannot be counted as
independent chiral fermions. Moreover, continuity of $J$ need not provide a global onsite Majorana pairing or a
trivial charge frame. If $J(\mathbf k)$ is analytic and periodic, its Fourier coefficients decay exponentially,
yet the associated bundle may remain nontrivial. Without a compact complex structure, a purely real Majorana
problem is not covered by Friedan's complex-bundle argument and requires the appropriate real or quaternionic
analysis.

This formulation also clarifies our 1D vector--axial relocalization. In the relocalized frame the axial charge
is onsite, so its regular charge sectors satisfy the doubling constraint. The mixed $U(1)_V\times U(1)_A$
anomaly nevertheless persists: the transported vector charge has an endpoint spectral seam and does not define a
simultaneous continuous periodic charge decomposition in that frame. Friedan's theorem constrains each available
regular charge decomposition, but does not assert simultaneous regularity, onsite relocalizability, or gauge
locality for different microscopic symmetries.

The geometric role of charge-bundle topology is illustrated by our higher-dimensional constructions. In 2D,
the square-Majorana model has an exponentially local compact charge with $C_Q=1$. Its smooth positive-charge
projector defines a nontrivial complex bundle and obstructs a globally smooth periodic onsite trivialization,
although a generally nonlocal onsite relocalization exists. In 3D, the compact vector charge has vanishing first
Chern numbers but a nontrivial strong quaternionic index $\nu_0=1$, which obstructs a globally smooth periodic
Kramers-compatible charge frame. These statements are compatible with Friedan's theorem: the 2D example is
outside its odd-spatial-dimensional setting, while the 3D strong quaternionic index does not measure net Weyl
chirality. Thus charge-bundle topology can remain nontrivial for a regular compact symmetry and provides
additional information relevant to symmetry relocalization, without by itself deciding gauge locality.

\section{1D: Bosonization, spectral winding symmetry, and T-duality}
\label{app:bosonization_tduality}

This appendix relates the exact vector--axial relocalization $\mathcal W$ to bosonic T-duality and noninvertible symmetries. In the XX chain, we identify the spectral winding charge and its fermion-parity and boundary-sector dependence. Projecting $\mathcal W$ onto a physical spin sector produces a noninvertible lattice operator whose IR limit realizes T-duality at the XX radius. We then show that $\mathcal W$ exchanges SSH dimerization and Kitaev pairing while acting nontrivially on only one Majorana species. At the self-dual critical point, this yields a standalone Ising construction whose parity-projected operator flows to the Kramers--Wannier topological defect.

\subsection{Spectral winding charge and T-duality}
\label{app:XX_spectral_charges}

Consider the APBC chain of Eq.~\eqref{eq:1d_H0} with $N\in4\mathbb Z$. Its Jordan--Wigner (JW) map to XX spins $X_j,Y_j,Z_j$ reads
\begin{equation}
c_j=J_j\frac{X_j-iY_j}{2},\quad
Z_j=2c_j^\dagger c_j-1,
\label{eq:app_XX_JW}
\end{equation}
where $J_j=\prod_{r<j}(-Z_r)$. For even $N$, fermion parity $P_F=(-1)^F$ becomes $U=\prod_j(-Z_j)=\prod_j Z_j$.
The periodic and antiperiodic XX Hamiltonians are
\begin{equation}
H_{\rm XX}^{\rm P,A}
=-\frac t2\sum_{j=0}^{N-2}
(X_jX_{j+1}+Y_jY_{j+1})\mp\frac t2 B,
\end{equation}
where $B=X_{N-1}X_0+Y_{N-1}Y_0$. The APBC fermion Hamiltonian corresponds to
\begin{equation}
H_{\rm XX}^{\rm PA}
=\Pi_+H_{\rm XX}^{\rm P}
+\Pi_-H_{\rm XX}^{\rm A},
\label{eq:app_XX_JW_boundary}
\end{equation}
with $\Pi_\pm=(1\pm U)/2$. Thus the APBC fermion chain combines the even-parity periodic and odd-parity antiperiodic XX spin sectors.

Define the momentum charge and spectral winding charge as
\begin{equation}
\begin{aligned}
Q_M&=\frac12\sum_j Z_j=Q_V,\\
Q_W&=\frac12Q_{A,0}
=\frac12\sum_k s_0(k)(n_k-\tfrac12).
\end{aligned}
\label{eq:app_XX_MW_definitions}
\end{equation}
For $N\in4\mathbb Z$, the APBC grid excludes $k=0,\pi,\pm\pi/2$. Both charges are diagonal in momentum
occupation numbers, hence
\begin{equation} [H_{\rm XX}^{\rm PA},Q_M]
=[H_{\rm XX}^{\rm PA},Q_W]=0,\ \ [Q_M,Q_W]=0.
\label{eq:app_XX_MW_commuting}
\end{equation}
Their parity dependence is fixed by
\begin{equation}
e^{i\pi Q_M}=e^{2\pi iQ_W}=U.
\label{eq:app_XX_MW_parity}
\end{equation}
Thus $Q_W$ is integer valued in the even-parity sector and half-integer valued in the odd-parity sector. Since
$2Q_W=Q_{A,0}$ is integral on the full APBC Fock space, $e^{i\lambda Q_W}$ is $4\pi$-periodic, and its
$2\pi$ rotation equals $U$.

Let $S_{j\ell}$ denote the Fourier kernel of $s_0(k)=\operatorname{sgn}(\sin k)$ defined in
Eq.~\eqref{eq:1d_axial_charge}. Then
\begin{equation}
Q_W=\frac12\sum_{j,\ell}c_j^\dagger S_{j\ell}c_\ell,
\end{equation}
and
\begin{equation}
S^\dagger=S,\quad S^T=-S,\quad S^2=I.
\label{eq:app_XX_S_involution}
\end{equation}
With $\mathcal J_{j\ell}=\prod_{j<r<\ell}(-Z_r)$ for $j<\ell$, antisymmetry of $S$ and the JW map yield
\begin{equation}
Q_W=-\frac{i}{4}\sum_{0\le j<\ell<N}S_{j\ell}
\bigl[X_j\mathcal J_{j\ell}Y_\ell
-Y_j\mathcal J_{j\ell}X_\ell\bigr].
\label{eq:app_XX_QW_spin_strings}
\end{equation}
Thus $Q_W$ consists of algebraically decaying spin-current strings and commutes with $Q_M$. By contrast, the
finite-range winding charge on a periodic spin chain,
\begin{equation}
Q_W^{\rm loc}=\frac14\sum_r
(X_{2r}Y_{2r+1}-Y_{2r+1}X_{2r+2}),
\label{eq:app_XX_local_winding_other}
\end{equation}
uses periodic site labels and does not commute with $Q_M$; together they belong to the Onsager
algebra~\cite{pace2024lattice, chatterjee2025quantized}.

To represent $\mathcal W$ in spin variables, introduce the onsite Majoranas
\begin{equation} a_j=c_j+c_j^\dagger=J_jX_j,\ \ b_j=\frac{c_j-c_j^\dagger}{i}=-J_jY_j.
\label{eq:app_XX_majoranas}
\end{equation}
These obey $ia_jb_j=Z_j$. Let $e,o$ label even and odd sites. Since $S$ connects opposite sublattices
and $S^2=I$, $-\sum_o
S_{eo}S_{e'o}=\delta_{ee'}$; hence $-iS_{eo}$ is a real orthogonal matrix. In this basis, the microscopic
transformation of Eq.~\eqref{eq:1d_eta_microscopic_raw} becomes
\begin{equation}
\begin{aligned}
\mathcal W:\quad &a_e\mapsto a_e,
\quad b_o\mapsto b_o,\\
&b_e\mapsto-i\sum_oS_{eo}a_o,\quad
a_o\mapsto-i\sum_eS_{eo}b_e.
\end{aligned}
\label{eq:app_XX_W_Majorana_action}
\end{equation}
The staggered rotation
\begin{equation}
f_j=e^{-i\pi j/2}c_j=\frac{\chi'_j+i\chi_j}{2}
\label{eq:app_XX_SSH_shifted_Majoranas}
\end{equation}
separates the chain into two APBC Majorana species $\chi$ and $\chi'$. With $g_j=(-1)^{\lfloor j/2\rfloor}$,
\begin{equation}
\chi_j=g_j\begin{cases}
b_j,&j\ \mathrm{even},\\
-a_j,&j\ \mathrm{odd},
\end{cases}\ \
\chi'_j=g_j\begin{cases}
a_j,&j\ \mathrm{even},\\
b_j,&j\ \mathrm{odd}.
\end{cases}
\label{eq:app_XX_SSH_AB_from_alpha_beta}
\end{equation}
Equation~\eqref{eq:app_XX_W_Majorana_action} reduces to
\begin{equation}
\mathcal W:\quad\chi_q \mapsto\operatorname{sgn}(\cos q)\chi_q,\ \
\chi'_q \mapsto\chi'_q.
\label{eq:app_XX_SSH_W_AB_momentum}
\end{equation}
A staggered $\chi$ bilinear couples $q$ to $\pi-q$. Because
$\operatorname{sgn}(\cos q)\operatorname{sgn}[\cos(\pi-q)]=-1$, $\mathcal W$ reverses its dimerization
mass while leaving the $\chi'$ mass unchanged.

An explicit Gaussian implementer follows by defining
\begin{equation}
\xi_e=\frac1{\sqrt2}
\left(b_e+i\sum_oS_{eo}a_o\right),
\label{eq:app_XX_xi_Majoranas}
\end{equation}
with $\{\xi_e,\xi_{e'}\}=2\delta_{ee'}$. The $\xi_e$ span the $-1$ eigenspace of
Eq.~\eqref{eq:app_XX_W_Majorana_action}. Because $N/2$ is even, they can be paired to give
\begin{equation}
\mathcal W
=\zeta_N\prod_{r=0}^{N/4-1}i\xi_{4r}\xi_{4r+2},
\quad \zeta_N=\pm1.
\label{eq:app_XX_W_xi_implementer}
\end{equation}
Fix $\zeta_N$ by requiring $\mathcal W|\Omega_N\rangle=|\Omega_N\rangle$ for the unique critical APBC ground
state. The factors are commuting Hermitian involutions. Since $b_e=g_e\chi_e$ and $a_o=-g_o\chi_o$, $\mathcal
W$ belongs entirely to the $\chi$ Majorana algebra, and
\begin{equation}
\mathcal W^\dagger=\mathcal W,\quad
\mathcal W^2=1,\quad
[\mathcal W,P_F]=0.
\label{eq:app_XX_W_properties}
\end{equation}
In XX spin variables, set
\begin{equation}
\Lambda_e=J_eY_e
-i\sum_{o\ \mathrm{odd}}S_{eo}J_oX_o=-\sqrt2\,\xi_e.
\label{eq:app_XX_Lambda_spin}
\end{equation}
Then
\begin{equation}
\mathcal W
=\zeta_N\prod_{r=0}^{N/4-1}
\left(\frac{i}{2}\Lambda_{4r}\Lambda_{4r+2}\right).
\label{eq:app_XX_W_explicit_spin}
\end{equation}
Each factor is a product of algebraically extended Pauli strings, and the factors commute.\footnote{For $N=4m+2$, the APBC grid
contains the zero modes $k=\pm\pi/2$ and $\mathcal W$ conjugates an odd number $N/2$ of complex modes. Thus
$\mathcal WP_F\mathcal W^{-1}=-P_F$, and the parity-even unitary of Eq.~\eqref{eq:app_XX_W_xi_implementer} must
be replaced by a parity-odd one. The kernel $S$ and exact charge exchange remain valid, while the rotated
Majoranas obey PBC since $f_{j+N}=f_j$.}

Under $\mathcal W$,
\begin{equation}
\mathcal W Z_e\mathcal W^{-1} =\sum_o S_{eo}a_ea_o,\ \
\mathcal W Z_o\mathcal W^{-1} =\sum_e S_{eo}b_eb_o.
\label{eq:app_XX_W_Z_image}
\end{equation}
For $j<\ell$, the relevant bilinears are JW strings:
\begin{equation}
ia_ja_\ell =-Y_j\mathcal J_{j\ell}X_\ell,\quad
ib_jb_\ell = X_j\mathcal J_{j\ell}Y_\ell.
\label{eq:app_XX_Majorana_bilinears_spin}
\end{equation}
Summing the transformed densities using these JW strings reproduces Eq.~\eqref{eq:app_XX_QW_spin_strings} and
gives
\begin{equation} 
\mathcal WQ_M\mathcal W^{-1}
=2Q_W,\ \
\mathcal WQ_W\mathcal W^{-1}
=\frac12Q_M. 
\label{eq:app_XX_exact_MW_exchange}
\end{equation}
These identities hold exactly at finite $N$ with APBC, and $\mathcal W$ preserves the corresponding XX Hamiltonian.

\subsection{Spin sectors and noninvertible spectral duality}
\label{app:XX_twists_gauging}

Following Ref.~\cite{li2023non}, write fermion parity as $U=(-1)^u$ and label periodic and antiperiodic spin boundary conditions by $t=0,1$, respectively. The JW fermions obey APBC when $u=t$ and PBC otherwise:
\begin{equation}
\begin{array}{c|c|c}
(u,t)&\text{spin sector}&\text{JW fermions}\\\hline
(0,0)&\text{periodic, even}&\text{APBC}\\
(1,1)&\text{antiperiodic, odd}&\text{APBC}\\
(1,0)&\text{periodic, odd}&\text{PBC}\\
(0,1)&\text{antiperiodic, even}&\text{PBC}
\end{array}
\label{eq:app_XX_sector_table}
\end{equation}
In the APBC sectors, Eq.~\eqref{eq:app_XX_MW_parity} gives
\begin{equation}
Q_M\equiv 2Q_W\equiv u=t\pmod2.
\label{eq:app_XX_sector_quantization}
\end{equation}
Since $[\mathcal W,U]=0$, the APBC transformation preserves both sectors.\footnote{For $N=4m+2$, the APBC grid contains the fixed mode $k_*=-\pi/2$. The explicit parity-odd implementer of Eq.~\eqref{eq:1d_W_fixed_mode} gives $\mathcal W U\mathcal W^{-1}=-U$. Since $Q_M\equiv u+1$ and $2Q_W\equiv u=t\pmod2$, the transformation exchanges the APBC-fermion sectors $(0,0)\leftrightarrow(1,1)$ under the JW correspondence.} Extending the transformation to PBC requires a prescription for the modes at $k=0,\pi$.

On the periodic XX ring, retain $\mathcal W$ in the even-parity APBC block and set it to zero in the odd-parity PBC block. The resulting noninvertible operator, with conventional fusion normalization, is
\begin{equation}
\mathcal D
=\sqrt2\,\Pi_+\mathcal W\Pi_+.
\label{eq:app_XX_Dspec_def}
\end{equation}
Since $\mathcal W^2=1$ and $[\mathcal W,U]=0$, the factor $\sqrt2$ gives the fusion relation
\begin{equation}
\mathcal D^{\,2} =1+U,\quad
U\mathcal D
=\mathcal DU
=\mathcal D.
\label{eq:app_XX_Dspec_fusion}
\end{equation}
Moreover, $\mathcal D$ commutes with $H_{\rm XX}^{\rm P}$. The exact charge exchange~\eqref{eq:app_XX_exact_MW_exchange} yields
\begin{equation}
\mathcal DQ_M
=2Q_W\mathcal D,\ \ 
\mathcal DQ_W
=\frac12Q_M\mathcal D. 
\label{eq:app_XX_Dspec_charge_intertwine}
\end{equation}
These relations hold on the even periodic-spin sector and extend to the full spin Hilbert space under any parity-preserving extension of $Q_W$ that agrees on this sector.

The action on a local spin density makes the microscopic nonlocality of $\mathcal D$ explicit. For even $e$,
\begin{equation}
\mathcal DZ_e
=\left(\sum_{o\ {\rm odd}}S_{eo}a_ea_o\right)
\mathcal D.
\label{eq:app_XX_Dspec_Z_intertwine}
\end{equation}
Write $A_e=\sum_o S_{eo}a_ea_o$. For any odd $o$, $[A_e,Z_o]\Pi_+=2iS_{eo}a_eb_o\Pi_+$, so $\|[A_e,Z_o]\Pi_+\|=2|S_{eo}|$. The operator $ia_eb_o$ is parity preserving and unitary. If a bounded-support spin operator $B_e$ satisfied $\mathcal DZ_e=B_e\mathcal D$, surjectivity of $\mathcal D$ onto the even sector would imply $B_e\Pi_+=A_e\Pi_+$. However, $B_e$ commutes with sufficiently distant $Z_o$, contradicting the nonzero commutator above. Thus parity projection does not make this density's image uniformly finite range in $N$. This differs from the local bond intertwiners obtained by $\mathbb Z_2$ gauging and from the local XX duality of Ref.~\cite{pace2024lattice, su2025z2}, whose square also involves translation and a momentum-charge rotation.

To intertwine different spin twists, extend $\mathcal W$ to all four sectors. On the PBC momentum grid, use $c_k\mapsto c_{-\pi-k}^\dagger$ for $-\pi<k<0$, leaving the remaining modes, including $k=0,\pi$, unchanged. For $N\in4\mathbb Z$, the negative branch contains $N/2-1$ modes, an odd number. The fixed mode $k_*=-\pi/2$ maps to its own adjoint, while the remaining negative momenta pair at opposite energies. Thus the PBC implementer $\mathcal W_{\rm P}$ preserves $H_0$ but reverses fermion parity.\footnote{For $N=4m+2$, the PBC grid excludes $k_*=-\pi/2$, and the negative branch contains $N/2-1=2m$ modes. The PBC branch rule inherited from Eq.~\eqref{eq:1d_W_explicit} still preserves $H_0$, but its implementer is parity even. Together with the parity-odd APBC implementer of Eq.~\eqref{eq:1d_W_fixed_mode}, it gives the sector permutation $(u,t)\mapsto(t+1,u+1)$ modulo $2$.} Choose its phase so that $\mathcal W_{\rm P}^2=1$. Together with the APBC implementer, it defines
\begin{equation}
\mathscr W:\mathcal H^{(u,t)}
\longrightarrow\mathcal H^{(t,u)},
\quad \mathscr W^2=1.
\label{eq:app_XX_completed_W}
\end{equation}

The completed transformation defines the winding charge in all four sectors through $2Q_W=\mathscr W Q_M\mathscr W^{-1}$. On the PBC grid, the corresponding spectral sign $s_{\rm PBC}(k)$ agrees with $\operatorname{sgn}(\sin k)$ on the open branches and equals $+1$ at $k=0,\pi$, since these modes are unchanged. Unlike the balanced opposite-sign prescription mentioned in the main text, these endpoint values are fixed by the canonical sector exchange. Indeed, an inversion-odd sign taking values $\pm1$ cannot be assigned at these inversion-invariant momenta. Consequently, $\sum_k s_{\rm PBC}(k)=2$, giving $2Q_W=\sum_k s_{\rm PBC}(k)(n_k-\tfrac12)=\sum_k s_{\rm PBC}(k)n_k-1$. The two prescriptions agree near the Fermi points but differ in their finite-size winding-charge quantization. Since $e^{i\pi Q_M}=(-1)^u$, Eq.~\eqref{eq:app_XX_completed_W} implies
\begin{equation}
e^{2\pi iQ_W}=(-1)^t\quad\text{on }\mathcal H^{(u,t)}.
\label{eq:app_XX_completed_winding_period}
\end{equation}
Thus $Q_W$ is integral throughout the periodic spin chain. Since $2Q_W$ has only even eigenvalues in this sector, the intertwining relation $\mathcal DQ_M=2Q_W\mathcal D$ forces $\mathcal D$ to annihilate all odd-$Q_M$ states. Conversely, $\mathcal D$ is invertible on the even block, so these states constitute its entire kernel.

To describe the duality between different spin boundary conditions, let $\mathsf P_t$ project onto spin twist $t$ and define
\begin{equation}
\mathcal N_{t'\leftarrow t}
=\sqrt2\,\mathsf P_{t'}\mathscr W\mathsf P_t.
\label{eq:app_XX_twist_intertwiner}
\end{equation}
The sector permutation gives
\begin{equation}
\begin{aligned}
\mathcal N_{t'\leftarrow t}^\dagger
\mathcal N_{t'\leftarrow t}
&=1+(-1)^{t'}U,\\
\mathcal N_{t'\leftarrow t}
\mathcal N_{t'\leftarrow t}^\dagger
&=1+(-1)^tU,
\end{aligned}
\label{eq:app_XX_twist_fusion}
\end{equation}
on the input and output twist sectors, respectively. In particular, $\mathcal N_{0\leftarrow0}=\mathcal D$. These identities show that the projected operators are noninvertible because they retain only selected parity sectors of the invertible transformation $\mathscr W$. Their products depend only on the sector permutation, not on the microscopic details of its completion.

\subsection{IR winding charge and noninvertible T-duality}
\label{app:XX_bosonization_winding}

To identify the IR winding charge, split the APBC momentum grid at $k=0$ and define the centered chiral
charges
\begin{equation}
Q_R=\sum_{0<k<\pi}(n_k-\tfrac12),\quad
Q_L=\sum_{-\pi<k<0}(n_k-\tfrac12).
\end{equation}
They satisfy the exact identities
\begin{equation}
Q_M=Q_R+Q_L,\quad 2Q_W=Q_R-Q_L.
\label{eq:app_XX_chiral_MW}
\end{equation}
The filled Fermi sea obeys $Q_R=Q_L=0$. Linearizing near $k_F=\pm\pi/2$ at fixed conformal excitation level gives
\begin{equation}
E-E_0=\frac{2\pi v_F}{N}
\left[\frac{Q_R^2+Q_L^2}{2}+N_R+N_L\right]
+o(N^{-1}),
\label{eq:app_XX_chiral_energy}
\end{equation}
Here $v_F=2t$ for $t>0$ and unit lattice spacing, and $N_R,N_L\in\mathbb Z_{\geq0}$ are oscillator levels.
Equation~\eqref{eq:app_XX_chiral_MW} gives
\begin{equation}
E-E_0=\frac{2\pi v_F}{N}
\left[\frac{Q_M^2}{4}+Q_W^2+N_R+N_L\right]
+o(N^{-1}).
\label{eq:app_XX_MW_energy}
\end{equation}
This is the sector-resolved $c=1$ spectrum at the XX radius $R=\sqrt2$. The bosonic momentum and winding are
$m=Q_M$ and $w=Q_W$, with primary dimension $\Delta_{m,w}=m^2/4+w^2$. The APBC fermions supply even $m$,
integral $w$ for periodic spins, and odd $m$, half-integral $w$ for antiperiodic spins. Periodic-spin states
with odd $m$ come from PBC fermions, whose winding charge is defined by the completion in Eq.~\eqref{eq:app_XX_completed_W}.
A $2\pi$ vector-flux insertion transfers one fermion between the Fermi
points, changing $Q_{A,0}=2Q_W$ by two with the flux orientation of the main text.

Using the symmetry action $O\mapsto e^{i\lambda Q}Oe^{-i\lambda Q}$ and suppressing Klein factors, write the chiral
creation fields and bosonic vertices as
\begin{equation}
\begin{gathered}
\psi^\dagger_{R,L}\sim e^{i(\theta\pm\phi)},\quad
V_{n,m}=e^{in\phi+im\theta},\\
\Delta_{n,m}(K)=\frac14\left(n^2K+\frac{m^2}{K}\right).
\end{gathered}
\label{eq:app_XX_bosonization_vertices}
\end{equation}
The free Dirac point corresponds to $K=1$. The charges $Q_M$ and $Q_{A,0}$ generate shifts of $\theta$ and
$\phi$, respectively, so $V_{n,m}$ carries momentum $m$ and winding $w=n/2$:
\begin{equation}
[Q_M,V_{n,m}]=m V_{n,m},\quad [Q_{A,0},V_{n,m}]=n V_{n,m}.
\label{eq:app_vector_axial_momentum_winding}
\end{equation}
In particular, $e^{i\lambda Q_W}$ shifts $\phi$ by $\lambda/2$.

On the IR chiral fields, $\mathcal W$ acts as
\begin{equation}
\psi_R\mapsto\psi_R,\quad
\psi_L\mapsto\psi_L^\dagger.
\end{equation}
Consequently $J_R\mapsto J_R$, $J_L\mapsto-J_L$, and
\begin{equation}
\begin{gathered}
\phi\leftrightarrow\theta,\quad
(n,m,K)\mapsto(m,n,K^{-1}),\\
\Delta_{n,m}(K)=\Delta_{m,n}(K^{-1}).
\end{gathered}
\label{eq:app_XX_fermion_T_duality}
\end{equation}
The current--current interaction $gJ_RJ_L$ changes sign, consistent with $K\leftrightarrow K^{-1}$. This
continuum transformation does not imply a local duality of interacting lattice Hamiltonians: $\mathcal W$ need
not preserve the locality of lattice interactions.

The factor of two comes from the local XX spin vertex algebra, where $n\in2\mathbb Z$ and $m\in\mathbb Z$. Its
compactification is
\begin{equation}
\phi\sim\phi+\pi,\quad\theta\sim\theta+2\pi.
\label{eq:app_XX_bosonic_periods}
\end{equation}
These periods characterize the local bosonic operator algebra, not the full fermionic theory. In terms of
$2\pi$-periodic fields $\Phi=2\phi$ and $\Theta=\theta$, the duality becomes
\begin{equation}
(\Phi,\Theta)\mapsto(2\Theta,\Phi/2),
\quad
(m,w)\mapsto(2w,m/2).
\label{eq:app_XX_bosonic_T_duality}
\end{equation}
For odd $m$, the image has half-integer winding and belongs to a twisted sector. On the Neveu--Schwarz (NS)
fermionic vertex lattice, $n\equiv m\pmod2$, and chiral reflection preserves this lattice. Restricting to the
same periodic-spin sector leaves states with $(m,w)\in2\mathbb Z\times\mathbb Z$. At $R=\sqrt2$, gauging momentum $\mathbb Z_2$
followed by ordinary T-duality gives $R\mapsto R/2\mapsto2/R$, returning to the same bosonic
CFT~\cite{pace2024lattice,su2025z2}. A fermionic Villain realization instead uses an enlarged constrained Hilbert space
and Majorana translation~\cite{luFermionic2026}.

Equation~\eqref{eq:app_XX_SSH_W_AB_momentum} determines the IR action of $\mathcal D$. On the periodic XX
ring, $\mathcal H_{\rm XX}^{\rm P}=\mathcal H_{\rm NS}^{+}\oplus\mathcal H_{\rm R}^{-}$. By its fusion-normalized
definition, $\mathcal D$ acts as $\sqrt2\mathcal W$ on the even Neveu--Schwarz (NS) sector and vanishes on the
odd Ramond (R) sector. Near the critical $\chi$ nodes $q=0,\pi$, the spectral signs are $+1,-1$. With the vacuum normalization fixed in
Eq.~\eqref{eq:app_XX_W_xi_implementer}, $\mathcal W$ therefore becomes the chiral reflection
$C_L=(-1)^{F_L^{\chi}}$. Because the signs are constant near the nodes, this identification holds on each fixed
low-energy Fock subspace for sufficiently large $N$. Hence
\begin{equation}
\mathcal D_{\rm IR}
=\left.\sqrt2\,C_L\right|_{\mathcal H_{\rm NS}^{+}}
\oplus 0_{\mathcal H_{\rm R}^{-}},
\quad
\mathcal D_{\rm IR}|0\rangle=\sqrt2|0\rangle.
\label{eq:app_XX_D_IR_operator}
\end{equation}
The chiral reflection preserves both Virasoro algebras. Together with the exact charge
intertwiners~\eqref{eq:app_XX_Dspec_charge_intertwine}, it exchanges $(m,w)\mapsto(2w,m/2)$ on even-$m$ primary
sectors, reflects the left-moving oscillators, and annihilates odd-$m$ states. The resulting topological defect
realizes noninvertible T-duality at $R=\sqrt2$, up to symmetry dressing~\cite{pace2024lattice}. The
Gaussian transformation fixes phases in the fermionic Fock basis, while specifying a bosonic representative
also requires a consistent vertex-operator cocycle convention.

\subsection{$U(1)$-breaking SSH--Kitaev duality and Ising criticality}
\label{app:XX_SSH_Kitaev_duality}

\subsubsection{SSH--Kitaev exchange and self-dual criticality}

The spectral reflection~\eqref{eq:app_XX_SSH_W_AB_momentum} exchanges two local perturbations, each preserving
a different continuous charge. On the APBC chain, let
\begin{equation}
\begin{aligned}
P_{\rm dimer}&=\sum_j(-1)^j\bigl(c_j^\dagger c_{j+1}+\mathrm{H.c.}\bigr),\\
P_{\rm sc}&=\sum_j\bigl(c_jc_{j+1}+\mathrm{H.c.}\bigr),\\
H(\delta,\Delta)&=H_0+\delta P_{\rm dimer}+\Delta P_{\rm sc},
\end{aligned}
\label{eq:app_XX_SSH_Kitaev_family}
\end{equation}
where $\delta,\Delta\in\mathbb R$, with APBC also imposed on boundary-crossing terms. The perturbations
represent SSH hopping dimerization and nearest-neighbor spinless $p$-wave pairing. Although $\mathcal W$ has
algebraic tails on generic operators, it exchanges these two local terms exactly:
\begin{equation}
\begin{aligned}
\mathcal W P_{\rm dimer}\mathcal W^{-1}&=P_{\rm sc},\\
\mathcal W P_{\rm sc}\mathcal W^{-1}&=P_{\rm dimer},\\
\mathcal W H(\delta,\Delta)\mathcal W^{-1}&=H(\Delta,\delta).
\end{aligned}
\label{eq:app_XX_SSH_pairing_exchange}
\end{equation}
For $k>0$, staggered hopping couples $k$ to $k-\pi$ with a $\sin k$ form factor, while $\mathcal
Wc_{k-\pi}\mathcal W^{-1}=c_{-k}^\dagger$. The image is nearest-neighbor pairing, with no nonlocal
terms; the reverse relation follows from $\mathcal W^2=1$.

In the Majorana basis~\eqref{eq:app_XX_SSH_shifted_Majoranas}, the same family is
\begin{equation}
\begin{aligned}
H(\delta,\Delta)=-\frac{i}{2}\sum_j\bigl[&
(t-m_\chi(-1)^j)\chi_j\chi_{j+1}\\
&+(t-m_{\chi'}(-1)^j)\chi'_j\chi'_{j+1}\bigr].
\end{aligned}
\label{eq:app_XX_SSH_Kitaev_Majorana_H}
\end{equation}
The two dimerization masses are $m_\chi=\delta-\Delta$ and $m_{\chi'}=\delta+\Delta$. The transformation leaves
the $\chi'$ chain unchanged and reverses $m_\chi$. The self-dual line is $\delta=\Delta$, where
\begin{equation}
H_{\rm sd}
=-\frac{it}{2}\sum_j\chi_j\chi_{j+1}
-\frac{i}{2}\sum_j[t-2\delta(-1)^j]\chi'_j\chi'_{j+1}.
\label{eq:app_XX_SSH_self_dual_H}
\end{equation}
For $\delta\ne0$, $\chi'$ is gapped and $\chi$ is a uniform critical Majorana chain. The self-dual Hamiltonian
therefore has one massless Majorana species ($c=1/2$), with an APBC finite-size gap of order $N^{-1}$. The
spectral transformation acts on the critical chain independently of its gapped spectator, as demonstrated by
the Ising construction below. The other Ising critical line, $\delta=-\Delta$, has $m_{\chi'}=0$ but is
not self-dual except at $\delta=\Delta=0$.

Self-dual criticality persists even though both continuous charges are broken:
\begin{equation}
\begin{aligned}
&[Q_V,P_{\rm dimer}]=0,\quad [Q_{A,0},P_{\rm sc}]=0,\\
&[Q_V,P_{\rm sc}]\ne0,\quad [Q_{A,0},P_{\rm dimer}]\ne0.
\end{aligned}
\label{eq:app_XX_SSH_broken_U1}
\end{equation}
At $\delta=\Delta\ne0$, both $U(1)_V$ and $U(1)_{A,0}$ are broken, but the discrete $\mathcal W$ symmetry
remains exact. On the coordinate axes, SSH dimerization preserves $U(1)_V$, while pairing preserves
$U(1)_{A,0}$ and is number conserving in the charge-adapted variables. These identities
concern the APBC ring; open boundaries require separate treatment.

More generally, $\mathcal W$ protects the gaplessness of $\chi$ within the class of uniformly finite-range,
two-site-periodic quadratic Hamiltonians. Mixed $\chi$--$\chi'$ bilinears are odd under $\mathcal W$ on an open momentum
interval,
while staggered $\chi$ bilinears are odd at every allowed momentum because $\operatorname{sgn}(\cos
q)\operatorname{sgn}[\cos(q+\pi)]=-1$. Finite range makes their coefficients trigonometric polynomials of degree
bounded independently of $N$. Invariance forces these polynomials to vanish at arbitrarily many APBC momenta,
and hence identically for sufficiently large $N$. The remaining $\chi$ bilinears are one-site translation
invariant, with an odd Majorana dispersion vanishing at $q=0,\pi$; they cannot open a thermodynamic gap. This
argument does not cover interactions.

\subsubsection{Transverse-field Ising duality}
\label{app:XX_standalone_Ising}

Because $\mathcal W$ acts nontrivially only on $\chi$, the construction applies directly to a single
Majorana chain. Parity projection then produces a noninvertible symmetry of the periodic
transverse-field Ising model. To keep track of boundary conditions, set $M=N/2$ with even $M$, and use
$X_j,Y_j,Z_j$ for Pauli operators on the $M$-site Ising chain, distinct from the XX spins above. The
Ising parity is $U_{\rm I}=\prod_{j=0}^{M-1}(-Z_j)=\prod_{j=0}^{M-1}Z_j$. Choose Ising axes such that
\begin{equation}
\chi_{2j}=\left(\prod_{\ell<j}Z_\ell\right)X_j,\ \ 
\chi_{2j+1}=-\left(\prod_{\ell<j}Z_\ell\right)Y_j.
\label{eq:app_XX_Ising_JW}
\end{equation}
The relative sign specifies the Ising-axis convention. On interior bonds,
$i\chi_{2j}\chi_{2j+1}=Z_j$ and $i\chi_{2j+1}\chi_{2j+2}=X_jX_{j+1}$, while the APBC Majorana boundary
bond gives $U_{\rm I}X_{M-1}X_0$. Defining
\begin{equation}
\begin{aligned}
H_{\rm I,1}&=\sum_{j=0}^{M-1}Z_j,\\
H_{\rm I,2}&=\sum_{j=0}^{M-2}X_jX_{j+1}
+U_{\rm I}X_{M-1}X_0,
\end{aligned}
\label{eq:app_XX_Ising_FG}
\end{equation}
the single-Majorana Hamiltonian takes the form
\begin{equation}
H_{\chi}(m_\chi)
=-\frac12\bigl[(t-m_\chi)H_{\rm I,1}
+(t+m_\chi)H_{\rm I,2}\bigr].
\label{eq:app_XX_Ising_H}
\end{equation}
The field and bond couplings are $h=(t-m_\chi)/2$ and $J=(t+m_\chi)/2$. The boundary bond in $H_{\rm I,2}$
is periodic for $U_{\rm I}=+1$ and antiperiodic for $U_{\rm I}=-1$. At $m_\chi=0$, $h=J$, and the
chain is critical in the thermodynamic limit.

Let $T_{\chi}$ denote one-site translation on the APBC Majorana chain. Its spectral reflection is
the real orthogonal transformation
\begin{equation}
R_{\rm I}=\operatorname{sgn}\!\left[\frac12
(T_{\chi}+T_{\chi}^{-1})\right],\ \ 
R_{\rm I}(q)=\operatorname{sgn}(\cos q).
\label{eq:app_XX_Ising_spectral_reflection}
\end{equation}
Equation~\eqref{eq:app_XX_Ising_spectral_reflection} is precisely the action of $\mathcal W$ on $\chi$ in
Eq.~\eqref{eq:app_XX_SSH_W_AB_momentum}. Since $M$ is even, the APBC momentum grid excludes $q=\pm\pi/2$
and the negative eigenspace of $R_{\rm I}$ has even dimension. The reflection therefore admits a parity-even
Gaussian implementer, again denoted by $\mathcal W$. Fix its phase so that $\mathcal W^2=1$ and it leaves
the critical APBC vacuum invariant. The single-chain implementation preserves
$U_{\rm I}$ without requiring a spectator Majorana. Since it preserves uniform Majorana hopping and reverses
the staggered bilinear, it exchanges the field and boundary-twisted bond terms exactly:
\begin{equation} 
\mathcal WH_{\rm I,1}\mathcal W^{-1} =H_{\rm I,2},\ \
\mathcal WH_{\rm I,2}\mathcal W^{-1}=H_{\rm I,1}. 
\label{eq:app_XX_Ising_coupling_exchange}
\end{equation}
Consequently, $\mathcal WH_{\chi}(m_\chi)\mathcal W^{-1}=H_{\chi}(-m_\chi)$, including the
parity-dependent boundary term. The identity holds directly in the standalone Ising representation.

To define an operator on the ordinary periodic Ising chain, consider the physical spin Hamiltonian
\begin{equation}
H_{\rm I}^{\rm P}(h,J)
=-h\sum_{j=0}^{M-1}Z_j-J\sum_{j=0}^{M-1}X_jX_{j+1},
\label{eq:app_XX_Ising_periodic_H}
\end{equation} 
This differs from the parity-twisted Hamiltonian above in the odd-parity sector. In the even sector,
$U_{\rm I}=+1$, they agree and the Majorana boundary condition is antiperiodic. With
$P_{\rm I,+}=(1+U_{\rm I})/2$, define the spectral defect operator
\begin{equation}
\mathcal D_{\rm I}=\sqrt2\,P_{\rm I,+}\mathcal W.
\label{eq:app_XX_Ising_standalone_defect}
\end{equation}
Because $[\mathcal W,U_{\rm I}]=0$ and $\mathcal W^2=1$, the defect obeys the exact finite-size
relations
\begin{equation}
\begin{aligned}
\mathcal D_{\rm I}^2&=1+U_{\rm I},\\
\mathcal D_{\rm I}H_{\rm I}^{\rm P}(h,J)
&=H_{\rm I}^{\rm P}(J,h)\mathcal D_{\rm I}.
\end{aligned}
\label{eq:app_XX_Ising_standalone_intertwining}
\end{equation}
Thus $\mathcal D_{\rm I}$ is an exact, algebraically nonlocal, noninvertible operator at finite $M$ and
commutes with the periodic Ising Hamiltonian at $h=J$. It acts as $\sqrt2\mathcal W$ on the even-parity
APBC-Majorana sector and annihilates the odd-parity sector, whose physical Majorana boundary condition is
periodic. The projection is essential: the unprojected coupling exchange applies to the parity-twisted
Hamiltonian, not to the physical periodic spin Hamiltonian in both sectors. The spectral transformation also
differs from the local Ising order--disorder map: $\mathcal W$ preserves parity and cannot conjugate
the $U_{\rm I}$-odd order field $X_j$ into the $U_{\rm I}$-even disorder string
$\prod_{\ell\le j}Z_\ell$.

Seiberg and Shao related the IR chiral Majorana parity $(-1)^{F_L}$, arising from lattice translation, to
the Kramers--Wannier topological defect of the critical Ising CFT~\cite{seiberg2024majorana}. The
spectral reflection has eigenvalues $+1$ and $-1$ near $q=0$ and $q=\pi$, respectively. The vacuum-normalized
$\mathcal W$ therefore agrees with $(-1)^{F_L}$ on every fixed low-energy Fock subspace for sufficiently
large $M$. The IR eigenvalues of $\mathcal D_{\rm I}$ on the identity, energy, and spin
conformal families are $(\sqrt2,-\sqrt2,0)$, giving
\begin{equation}
(\mathcal D_{\rm I})_{\rm IR}
=\sqrt2(P_{\mathbf1}-P_\varepsilon)=D_{\rm KW},
\label{eq:app_XX_Ising_D_IR}
\end{equation}
where $P_{\mathbf1}$ and $P_\varepsilon$ project onto the corresponding conformal families. In the IR, the
exact field--bond intertwining exchanges the leading field and bond scaling operators. Their Ising-energy
components have opposite signs, consistent with $D_{\rm KW}\varepsilon=-\varepsilon D_{\rm KW}$.

\section{1D: Symmetry relocalization and gauge locality}
\label{app:1d_gauge_locality}

This appendix provides technical details of the 1D constructions. We establish the exact equivalence of separately vector- and axial-gauged compact theories, extend symmetry relocalization and gauging to general spectral charges and finite-range dispersions, and examine constraints on Gaussian locality. Finally, at fixed physical chemical potential, we establish the low-energy equivalence of nonlocal and local gauge regulators through a norm-resolvent estimate that includes dynamical gauge fields and holonomy.

\subsection{Exact equivalence of vector and axial compact gaugings}
\label{app:1d_tduality_gauged_intertwiner}

The canonical relocalization $\mathcal W$ of Eq.~\eqref{eq:1d_W_explicit} intertwines the two compact gauge theories introduced in Sec.~\ref{sec:1d_exact_finite_spacing_gauge_locality}. We work on the APBC ring with $N\in4\mathbb Z$ and $\eta_j=\mathcal W c_j\mathcal W^{-1}$. As in Eqs.~\eqref{eq:1d_vector_density} and \eqref{eq:1d_eta_density}, use the same staggered background $\nu_j=[1-(-1)^j]/2$ in the two onsite densities,
\begin{equation}
q_{V,j}=c_j^\dagger c_j-\nu_j,\qquad
q_{A,j}=\eta_j^\dagger\eta_j-\nu_j.
\end{equation}
Their sums are $Q_V$ and $Q_{A,0}$, respectively. The two gauge-link Hilbert spaces are initially separate, with $[E_j^I,U_l^I]=\delta_{jl}U_l^I$ for $I=V,A$. In temporal gauge, in lattice units, and with equal couplings $g_V=g_A=g$, the vector theory is
\begin{equation}
\begin{aligned}
H_V[U^V,E^V]
&=-t\sum_j(c_j^\dagger U_j^Vc_{j+1}+\mathrm{H.c.})
+\frac{g^2}{2}\sum_j(E_j^V)^2,\\
\mathcal G_{V,j}&=E_j^V-E_{j-1}^V-q_{V,j},
\end{aligned}
\label{eq:app_1d_tduality_HV_G}
\end{equation}
while the axial theory is
\begin{equation}
\begin{aligned}
H_A[U^A,E^A]
&=-t\sum_j(\eta_j^\dagger U_j^A\eta_{j+1}+\mathrm{H.c.})
+\frac{g^2}{2}\sum_j(E_j^A)^2,\\
\mathcal G_{A,j}&=E_j^A-E_{j-1}^A-q_{A,j}.
\end{aligned}
\label{eq:app_1d_tduality_HA_G}
\end{equation}
Both hopping terms obey the same APBC prescription. The vector theory gauges onsite number in the original $c$ variables; the axial theory gauges the compact spectral symmetry after it has been made onsite in the $\eta$ variables.

Identify the link-rotor Hilbert spaces, $U_j^V\leftrightarrow U_j^A$ and $E_j^V\leftrightarrow E_j^A$, and extend the matter transformation as $\mathbb W=\mathcal W\otimes\mathbf1_{\rm gauge}$. Then
\begin{equation}
\begin{aligned}
\mathbb W H_V[U^V,E^V]\mathbb W^{-1}
&=H_A[U^A,E^A],\\
\mathbb W\mathcal G_{V,j}\mathbb W^{-1}
&=\mathcal G_{A,j}.
\end{aligned}
\label{eq:app_1d_tduality_exact_intertwiner}
\end{equation}
These are exact finite-volume operator identities for arbitrary compact links and electric fields. In particular, $\mathbb W\mathcal H_{\rm phys}^V=\mathcal H_{\rm phys}^A$, where $\mathcal H_{\rm phys}^I$ is the physical space. The two constrained theories therefore have identical spectra and matrix elements of correspondingly transported gauge-invariant observables. This is an equivalence between symmetry-adapted gauge regularizations, not an assertion that arbitrary deformations in the original $c$ coordinates remain local after relocalization.

Local gauge transformations are intertwined together with the Gauss laws. For arbitrary site angles $\alpha_j$,
\begin{equation}
\mathbb W\exp\!\left(i\sum_j\alpha_j\mathcal G_{V,j}\right)
\mathbb W^{-1}
=\exp\!\left(i\sum_j\alpha_j\mathcal G_{A,j}\right).
\label{eq:app_1d_tduality_large_gauge}
\end{equation}
For a uniform background holonomy $U_j^I=e^{i\Phi_I/N}$, use the same matter-sector large-gauge convention as in Sec.~\ref{sec:1d_anomaly}:
\begin{equation}
\begin{aligned}
&\mathscr U_I
=\exp\!\left[\frac{2\pi i}{N}\sum_{j=0}^{N-1}j\,q_{I,j}\right],\\
& \mathscr U_I
H_{I,\rm bg}(\Phi_I+2\pi)
\mathscr U_I^{-1}
=H_{I,\rm bg}(\Phi_I),
\end{aligned}
\label{eq:app_1d_tduality_flux_endpoint}
\end{equation}
where $H_{I,\rm bg}$ denotes the matter Hamiltonian with prescribed links, not the dynamical gauge Hamiltonian. Since $\mathbb W\mathscr U_V\mathbb W^{-1}=\mathscr U_A$, the large-gauge endpoint identification is transported exactly. 

The intertwiner preserves the electric fields and the compact holonomy $\Omega_I=\prod_jU_j^I=e^{i\Phi_I}$ under this identification. Writing $Q_{I,j}=\sum_{l=0}^j q_{I,l}$, the Gauss laws imply
\begin{equation}
E_j^I=m_I+Q_{I,j},\qquad m_I=E_{N-1}^I\in\mathbb Z,
\quad I=V,A.
\label{eq:app_1d_tduality_Gauss_solved}
\end{equation}
Periodicity requires $Q_{I,N-1}=\sum_jq_{I,j}=0$, so both physical matter sectors are half filled. The unitary $\mathbb W$ maps their accumulated charges site by site while leaving the global electric-flux variable unchanged.

The complementary spectral charges are also transported. With the covariant shifts $(T_I\psi)_j=U_j^I\psi_{j+1}$ (including the APBC sign) and $D_I=(T_I-T_I^\dagger)/(2i)$, use the same definitions as in Sec.~\ref{sec:1d_anomaly}:
\begin{equation}
\begin{aligned}
&\mathbb W Q_{A,0}^{\mathrm{cov}}[U^V]\mathbb W^{-1}
=Q_V^{\mathrm{cov}}[U^A],\\
&Q_{A,0}^{\mathrm{cov}}[U^V]
=c^\dagger\operatorname{sgn}(D_V)c,\\
& Q_V^{\mathrm{cov}}[U^A]
=\eta^\dagger\operatorname{sgn}(D_A)\eta.
\end{aligned}
\label{eq:app_1d_tduality_spectator_covariant}
\end{equation}
Away from zero crossings, these spectral charges remain compact. Replacing $\operatorname{sgn}(D_I)$ by $D_I$ similarly transports the local bond representatives, and the background-field Ward identities and integrated spectral-flow responses match under $\mathbb W$. The complementary charges are not generally conserved once the corresponding electric fields become dynamical.

The exact equivalence does not smooth the spectral sign: during a uniform $2\pi$-flux cycle, two eigenvalues of $D_I$ cross zero at $\Phi_I=\pi$, producing the rank-two jump described in Sec.~\ref{sec:1d_anomaly}. The fixed transformation $\mathbb W$ identifies the two descriptions without eliminating this discontinuity. It leaves the gauge links and electric fields unchanged and is therefore not an electromagnetic duality exchanging them. Nor does the equivalence imply anomaly-free simultaneous gauging of $U(1)_V\times U(1)_A$ within a standalone 1D theory.

\subsection{Generality and limits of 1D symmetry relocalization}
\label{app:1d_general_s_details}

We give the finite-volume constructions and locality constraints for the extensions introduced in Sec.~\ref{sec:1d_general_s_gauge_locality}.

\subsubsection{Spectral axial-charge families and gauge locality}

Consider $N=2M\in4\mathbb Z$ sites with APBC. The momentum grid is invariant under $k\mapsto-k$ and $k\mapsto\pi-k$ modulo $2\pi$, and excludes $0,\pi,\pm\pi/2$. For an inversion-odd sign on this grid,
\begin{equation}
s(k)\in\{\pm1\},\qquad s(-k)=-s(k),
\label{eq:1d_s_general_hypotheses}
\end{equation}
To reproduce the standard continuum axial symmetry and mixed anomaly, require $s(k)=s_0(k)$ in neighborhoods of both Fermi points $\pm k_F$. Define the conserved compact charge
\begin{equation}
Q_{A,s}=\sum_k s(k)\left(n_k-\frac12\right).
\label{eq:1d_s_general_charge}
\end{equation}
Because $\sum_k s(k)=0$, the charge has integer spectrum. The momentum-swap unitary $\mathcal U_s$ defined in Eqs.~\eqref{eq:1d_s_U_swap} and \eqref{eq:1d_s_U_action} exchanges selected degenerate pairs $(k,-k)$, giving
\begin{equation}
\begin{aligned}
[\mathcal U_s,H_0]&=0,\\
[\mathcal U_s,Q_V]&=0,\\
\mathcal U_sQ_{A,0}\mathcal U_s^{-1}&=Q_{A,s}.
\end{aligned}
\label{eq:1d_s_U_identities}
\end{equation}
The generalized two-way vector--axial exchange is the conjugated transformation
\begin{equation}
\mathcal W_s=\mathcal U_s\mathcal W\mathcal U_s^{-1},
\label{eq:1d_s_W_exchange}
\end{equation}
which obeys
\begin{equation}
\begin{aligned}
\mathcal W_sH_0\mathcal W_s^{-1}&=H_0,\\
\mathcal W_sQ_V\mathcal W_s^{-1}&=Q_{A,s},\\
\mathcal W_sQ_{A,s}\mathcal W_s^{-1}&=Q_V.
\end{aligned}
\label{eq:1d_s_W_identities}
\end{equation}
These identities hold at finite volume for every inversion-odd sign, without requiring regularity in momentum space.

For the local representation, instead transform the reference adapted fermions $\eta_j=\mathcal Wc_j\mathcal W^{-1}$:
\begin{equation}
\xi_j^{(s)}=\mathcal U_s\eta_j\mathcal U_s^{-1}
=(\mathcal U_s\mathcal W)c_j
(\mathcal U_s\mathcal W)^{-1}.
\label{eq:1d_s_xi_def}
\end{equation}
Thus $\mathcal W_s$ generates the two-way charge exchange, whereas $\mathcal U_s\mathcal W$ defines the adapted fermions. Using Eq.~\eqref{eq:1d_s_U_identities} and the reference vector--axial identities gives
\begin{equation}
\begin{aligned}
H_0[c]&=H_0[\xi^{(s)}],\\
Q_{A,s}[c]&=\sum_j\left(\xi_j^{(s)\dagger}\xi_j^{(s)}-\frac12\right),\\
Q_V[c]&=Q_{A,0}[\xi^{(s)}].
\end{aligned}
\label{eq:1d_s_localization_identity}
\end{equation}
The kinetic Hamiltonian is therefore nearest-neighbor in $\xi^{(s)}$, while $Q_{A,s}$ is onsite.

Its compact gauging follows from Eqs.~\eqref{eq:1d_local_axial_gauge} and \eqref{eq:1d_local_Gauss} by replacing $\eta_j$ with $\xi_j^{(s)}$. Here $U_j$ and $E_j$ abbreviate the axial gauge-link operators $U_j^A$ and $E_j^A$. With $\nu_j=[1-(-1)^j]/2$, the Gauss operators are
\begin{equation}
\mathcal G_{j,s}=E_j-E_{j-1}
-\left(\xi_j^{(s)\dagger}\xi_j^{(s)}-\nu_j\right).
\label{eq:1d_s_Gauss}
\end{equation}
Denote the corresponding local gauge Hamiltonian by $H_s^A[U,E]$, in lattice units. For any $s,s'$, the unitary $\mathcal V_{s\to s'}=\mathcal U_{s'}\mathcal U_s^{-1}$ acts trivially on the link Hilbert space and satisfies
\begin{equation}
\begin{aligned}
\mathcal V_{s\to s'}H_s^A[U,E]
\mathcal V_{s\to s'}^{-1}&=H_{s'}^A[U,E],\\
\mathcal V_{s\to s'}\mathcal G_{j,s}
\mathcal V_{s\to s'}^{-1}&=\mathcal G_{j,s'}.
\end{aligned}
\label{eq:1d_s_gauge_intertwining}
\end{equation}
The complete gauge theories, including their constrained physical Hilbert spaces and transported gauge-invariant observables, are thus unitarily equivalent. Their embeddings in the original $c$ variables are nevertheless generally nonlocal.

For example, consider
\begin{equation}
s_\lambda(k)=\operatorname{sgn}(\sin k)
\operatorname{sgn}(1+2\lambda\cos k),
\label{eq:1d_s_lambda_family}
\end{equation}
for $\lambda\ge0$.
If the second factor vanishes on the finite momentum grid, choose its values at $k$ and $-k$ equal to preserve inversion oddness. For $0\le\lambda\le1/2$, $s_\lambda=s_0$. For $\lambda>1/2$, additional sign changes occur at $\pm k_c(\lambda)$, with
\begin{equation}
\begin{aligned}
&k_c(\lambda)=\arccos\!\left(-\frac1{2\lambda}\right),\\
&\delta_\lambda =k_c(\lambda)-\frac\pi2
=\arcsin\!\left(\frac1{2\lambda}\right).
\end{aligned}
\label{eq:1d_s_lambda_window}
\end{equation}
Every fixed finite $\lambda$ preserves neighborhoods of $\pm k_F$. If $\lambda$ varies as the lattice spacing $a\to0$, matching over every fixed physical momentum window requires $\delta_{\lambda(a)}/a\to\infty$, equivalent to $a\lambda(a)\to0$ when $\lambda(a)\to\infty$.

Since $\epsilon(k)=-2t\cos k$ is one-to-one on $(0,\pi)$, any inversion-odd sign on the grid admits the representation
\begin{equation}
s(k)=s_0(k)F(\epsilon(k)),\qquad
F(E)\in\{\pm1\},
\label{eq:1d_s_energy_sign_family}
\end{equation}
where $F$ need not be continuous. This representation need not hold for nonmonotone dispersions. More generally, matching $s_0$ within $|k\mp k_F|<\delta_a$ with $\delta_a/a\to\infty$ covers every fixed physical IR momentum window. At finite volume, the sign choices are discrete, and their relocalizations need not vary continuously in operator norm.

\subsubsection{Finite-range deformations and Gaussian locality constraints}
\label{app:1d_extended_dispersions}

Consider the dispersion family $\epsilon_\lambda(k)$ of Eq.~\eqref{eq:1d_general_dispersion_main}, with $H_\lambda=\sum_k\epsilon_\lambda(k)n_k$. Positivity of $P_\lambda$ ensures that the only zeros are the simple Fermi points $\pm\pi/2$, while $P_\lambda(0)=1$ fixes their Fermi velocity to $2t$. Since $\epsilon_\lambda(k)$ is inversion even and $\pi$-antiperiodic, it contains only odd cosine harmonics, with maximum hopping range $2d+1$ for $d=\deg P_\lambda$.

The reference transformation $\mathcal W$ is independent of $\lambda$. Because $\epsilon_\lambda(-\pi-k)=-\epsilon_\lambda(k)$, it obeys
\begin{equation}
\mathcal WH_\lambda\mathcal W^{-1}=H_\lambda,
\qquad
\mathcal WQ_{A,0}\mathcal W^{-1}=Q_V.
\label{eq:1d_W_dispersion_exchange}
\end{equation}
In the adapted $\eta$ variables, $H_\lambda$ has finite-range odd-distance hopping,
\begin{equation}
H_\lambda=
\sum_{\substack{r\ge1\\r\ \mathrm{odd}}}\sum_j
t_{\lambda,r}\left(\eta_j^\dagger\eta_{j+r}+\mathrm{H.c.}\right),
\label{eq:1d_W_dispersion_odd_hopping}
\end{equation}
where $t_{\lambda,r}=0$ for $r>2d+1$, and the axial charge remains onsite. In the periodic two-orbital frame of Eq.~\eqref{eq:1d_Rp}, the corresponding Bloch Hamiltonian is
\begin{equation}
\begin{aligned}
h_{{\rm loc},\lambda}(p)
={}&P_\lambda\!\left(\frac{1-\cos p}{2}\right)\\
&\times\left[-t(1-\cos p)\sigma_x
+t\sin p\,\sigma_y\right].
\end{aligned}
\label{eq:1d_localizable_hloc}
\end{equation}
Its finite Laurent-polynomial dependence on $e^{ip}$ makes strict finite-range locality explicit. Conversely, every real, inversion-even, $\pi$-antiperiodic finite-range dispersion is a finite sum of odd cosine harmonics and can be written as $\cos k$ times a polynomial in $\cos^2 k$, without necessarily satisfying the positivity or normalization conditions.

The spectral-sign and dispersion deformations can be combined because $[\mathcal U_s,H_\lambda]=0$. In the $\xi^{(s)}$ variables, the deformed Hamiltonian retains the same odd-distance hopping, while $Q_{A,s}$ becomes onsite. Dressing each hopping term with the corresponding oriented Wilson product gives, in lattice units,
\begin{align}
H_{s,\lambda}^A[U,E]
={}&\sum_{\substack{r\ge1\\r\ \mathrm{odd}}}\sum_j
t_{\lambda,r}\bigl(
\xi_j^{(s)\dagger}
U_jU_{j+1}\cdots U_{j+r-1}
\xi_{j+r}^{(s)}\nonumber\\
&\hspace{3em}+\mathrm{H.c.}\bigr)
+\frac{g_A^2}{2}\sum_jE_j^2.
\label{eq:1d_s_lambda_gauging}
\end{align}
The APBC convention applies to hoppings crossing the boundary. This Hamiltonian commutes with every Gauss operator $\mathcal G_{j,s}$ of Eq.~\eqref{eq:1d_s_Gauss} and has hopping range bounded independently of $N$ for fixed $P_\lambda$. Because all hopping connects opposite sublattices, the Gauss-law derivation of the discrete-chiral mass reflection extends from the nearest-neighbor regulator. Restoring physical units gives the same symmetry-centered bare mass $m_{\rm lat}^*=-g_A^2a/8$~\cite{dempsey2022discrete}.

The $\pi$-antiperiodicity condition is also necessary and sufficient for Hamiltonian preservation by the specified reference transformation $\mathcal W$ for an inversion-even dispersion:
\begin{equation}
\mathcal WH_\epsilon\mathcal W^{-1}=H_\epsilon
\quad\Leftrightarrow\quad
\epsilon(k+\pi)=-\epsilon(k).
\label{eq:1d_W_dispersion_criterion}
\end{equation}
This criterion concerns the given transformation, not every possible canonical relocalization.

A more general constraint holds for Gaussian rotations acting within each fixed reduced-momentum doublet. For a real-analytic, inversion-even dispersion, the particle--hole folded Hamiltonian is, up to a constant, $h_{{\rm br},\epsilon}(p)=\epsilon(k_F+p/2)\sigma_z$. Each fiberwise unitary rotation preserves the spectral invariant
\begin{equation}
\operatorname{tr}\,h_{{\rm br},\epsilon}(p)^2
=2\epsilon\!\left(k_F+\frac p2\right)^2.
\label{eq:1d_gaussian_spectral_invariant}
\end{equation}
For a smooth periodic rotated Hamiltonian, all derivatives of this analytic invariant must agree across the seam $p=\pm\pi$. Analytic continuation then requires
\begin{equation}
\epsilon(k+\pi)^2=\epsilon(k)^2.
\label{eq:1d_gaussian_locality_condition}
\end{equation}
Real analyticity implies either $\epsilon(k+\pi)=\epsilon(k)$ or $\epsilon(k+\pi)=-\epsilon(k)$ globally. When the only zeros are the two simple Fermi points $\pm\pi/2$, the first possibility is excluded: $\pi$ periodicity would give equal slopes at the two Fermi points, whereas inversion symmetry gives opposite slopes. Thus smooth periodic Hamiltonian locality within this fixed-reduced-momentum Gaussian class requires $\pi$-antiperiodicity. Momentum-rearranging or non-Gaussian canonical relocalizations are not excluded.

\subsubsection{Chemical potential and finite-spacing locality}
\label{app:1d_s_mu_locality}

For $H_0-\mu Q_V$, every $Q_{A,s}$ remains conserved and $[\mathcal U_s,H_0-\mu Q_V]=0$. In the adapted variables, however,
\begin{equation}
H_0[c]-\mu Q_V[c]
=H_0[\xi^{(s)}]-\mu Q_{A,0}[\xi^{(s)}],
\label{eq:1d_s_mu_universal_form}
\end{equation}
independently of $s$. The discontinuities of $s_0(k)=\operatorname{sgn}(\sin k)$ produce algebraically decaying hopping for $\mu\ne0$, so the strictly local finite-spacing gauge construction does not extend simply by dressing the transformed hopping with Wilson products.

A Hamiltonian-preserving exchange between $Q_V$ and $Q_{A,s}$ is also impossible at $\mu\ne0$. Indeed, the normalized full-Fock-space trace gives
\begin{equation}
\begin{aligned}
2^{-N}\operatorname{Tr}\!\left[(H_0-\mu Q_V)Q_V\right]
&=-\frac{\mu N}{4},\\
2^{-N}\operatorname{Tr}\!\left[(H_0-\mu Q_V)Q_{A,s}\right]
&=0.
\end{aligned}
\label{eq:1d_s_mu_trace_obstruction}
\end{equation}
A unitary preserving $H_0-\mu Q_V$ and mapping $Q_V$ to $Q_{A,s}$ would have to preserve both trace pairings, contradicting these identities.

At fixed physical $\mu$, the Fermi-point displacement is $O(a)$ and stays within any fixed node-matching neighborhood. For cutoff-dependent signs, the stronger condition $\delta_a/a\to\infty$ likewise retains the fixed physical IR windows. Transporting a fixed shortest-path compact-gauging prescription through the exact sign-dependent unitary identification makes the norm-resolvent comparison of Appendix~\ref{app:1d_mu_stability} uniform over the inversion-odd sign family. This result is established for nearest-neighbor hopping; deformed dispersions require separate estimates.

\subsection{Dynamical gauge locality at fixed physical chemical potential}
\label{app:1d_mu_stability}

We compare the nonlocal compact gauging of the finite-chemical-potential chain with a strictly local regulator.
The comparison controls the full matter--gauge dynamics, not merely the free-fermion dispersion.

\paragraph{Physical setting and the two lattice regulators.}
At zero chemical potential, the Hamiltonian-preserving vector--axial exchange yields a local Hamiltonian in the relocalized variables. At nonzero $\mu$, the term $-\mu Q_V$ becomes hopping with spectral symbol $-\mu s_0(k)$, where $s_0(k)=\operatorname{sgn}(\sin k)$; see Eq.~\eqref{eq:1d_exact_H_preserving_VA_exchange}. The jumps in $s_0$ produce algebraically decaying hopping, whose shortest-path compact gauging is nonlocal at every finite lattice spacing. We ask whether this UV nonlocality affects the fully dynamical low-energy gauge theory.

We take the continuum limit at fixed physical circumference $\ell$, Fermi velocity $v_F$, gauge coupling
$g_A>0$, and chemical potential $\mu\in\mathbb R$, with
\begin{equation}
N=2M\in4\mathbb Z,\quad a=\frac{\ell}{N},\quad
t_a=\frac{v_F}{2a}.
\label{eq:app_1d_mu_scaling}
\end{equation}
Constants below may depend on these fixed parameters but not on $N$. For $N\in4\mathbb Z$, the APBC grid avoids both Fermi points $\pm\pi/2$ and the discontinuities of $s_0$ at $0,\pi$.
In the $\eta$ variables, the ungauged Hamiltonian is, up to the same additive scalar in both
regulators,
\begin{equation}
H_{a,\mu}=\sum_k[-2t_a\cos k-\mu s_0(k)]n_{\eta,k},
\quad n_{\eta,k}=\eta_k^\dagger\eta_k.
\label{eq:app_1d_mu_ungauged}
\end{equation}
The axial $U(1)$ acts onsite on the $\eta$ fermions. Its charge density and Gauss-law constraint are
\begin{equation}
\begin{aligned}
q_{A,j}&=\eta_j^\dagger\eta_j-\nu_j,\quad
\nu_j=\frac{1-(-1)^j}{2},\\
\mathcal G_j&=E_j-E_{j-1}-q_{A,j}=0.
\end{aligned}
\label{eq:app_1d_mu_gauss}
\end{equation}
Here $U_j=e^{i\vartheta_j}$ are compact gauge links, $E_j$ are integer electric fields, and
$[E_j,U_l]=\delta_{jl}U_l$. Summing Gauss's law around the ring gives $\sum_jq_{A,j}=0$, so physical states contain exactly $M$ relocalized fermions. This fixes the axial charge but does not constrain the original vector charge coupled to $\mu$.

The compact gauging of the algebraic hopping must retain its finite-volume Fourier coefficients. Define the trigonometric interpolant
\begin{equation}
s_N(x)=\frac4N\sum_{\substack{1\le r<M\\r\,{\rm odd}}}
\frac{\sin(rx)}{\sin(\pi r/N)},
\label{eq:app_1d_mu_sN}
\end{equation}
On the allowed momentum grid, discrete Fourier inversion gives the exact identity
\begin{equation}
s_N(k_j)=s_0(k_j).
\label{eq:app_1d_mu_interpolation}
\end{equation}
Away from the grid, $s_N$ generally differs from $s_0$, a distinction that matters when the dynamical holonomy shifts the momenta. Introduce the covariant forward shift
$(T_U\psi)_j=U_j\psi_{j+1}$, with $\psi_N=-\psi_0$, and the number-conserving second quantization
$d\Gamma(B)=\sum_{jl}\eta_j^\dagger B_{jl}\eta_l$. Shortest-path Wilson dressing of the finite-volume hopping gives
\begin{equation}
s_N(T_U)=\frac4N\sum_{\substack{1\le r<M\\r\,{\rm odd}}}
\frac{T_U^r-T_U^{-r}}{2i\sin(\pi r/N)}.
\label{eq:app_1d_mu_sNT}
\end{equation}
Only odd distances $r<M=N/2$ occur, avoiding the antipodal link ambiguity. The gauging retains the original Fourier coefficients; in particular, it does not recompute the spectral sign in the gauge background.

The resulting nonlocal compact gauge Hamiltonian is
\begin{equation}
\begin{aligned}
H_a^{\rm nl}&=-t_a d\Gamma(T_U+T_U^\dagger)
-\mu d\Gamma(s_N(T_U))+H_E,\\
H_E&=\frac{g_A^2a}{2}\sum_jE_j^2.
\end{aligned}
\label{eq:app_1d_mu_Hnl}
\end{equation}
For comparison, replace the spectral-sign hopping by the nearest-neighbor current term
\begin{equation}
H_a^{\rm loc}=-t_a d\Gamma(T_U+T_U^\dagger)
-\mu d\Gamma\!\left(\frac{T_U-T_U^\dagger}{2i}\right)+H_E.
\label{eq:app_1d_mu_Hloc}
\end{equation}
At zero gauge field this replaces $s_0(k)$ by $\sin k$, which agrees at the two Fermi points and has the same linearized Dirac limit. Unlike the sharp sign, $\sin k$ admits nearest-neighbor Peierls gauging. The comparison below controls the complete momentum grid and all dynamical holonomies.

The local regulator is a nearest-neighbor chain with complex forward-hopping amplitude
$t_a-i\mu/2=t_a'e^{-i\alpha_a}$, where
\begin{equation}
t_a'=\sqrt{t_a^2+\frac{\mu^2}{4}},\quad
\alpha_a=\arctan\frac{\mu}{2t_a},
\end{equation}
so its matter hopping can be written as
\begin{equation}
H_{{\rm m},a}^{\rm loc}=-t_a'\sum_j\left(
e^{-i\alpha_a}\eta_j^\dagger U_j\eta_{j+1}+{\rm H.c.}\right).
\label{eq:app_1d_mu_local_phase_hopping}
\end{equation}
This phase is removed by the Gauss-law-preserving holonomy-translation unitary
\begin{equation}
W_a=\exp\!\left(i\alpha_a\sum_jE_j\right).
\end{equation}
Indeed, $W_aU_jW_a^\dagger=e^{i\alpha_a}U_j$, while $W_a$ leaves the electric fields unchanged and commutes with
all Gauss-law generators $\mathcal G_j$. On the physical Hilbert space this yields the exact equivalence
\begin{equation}
\begin{aligned}
& W_aH_a^{\rm loc}W_a^\dagger=H_a^{\rm standard}(t_a'),\\
&2at_a' =\sqrt{v_F^2+\mu^2a^2}\to v_F.
\end{aligned}
\label{eq:app_1d_mu_standard_equivalence}
\end{equation}
Here $H_a^{\rm standard}(t_a')$ denotes the conventional nearest-neighbor compact axial gauge Hamiltonian with hopping $t_a'$ and the same electric energy and Gauss law. The unitary $W_a$ is not a gauge transformation: it translates the gauge-invariant global holonomy. Although $\alpha_a\to0$, the total shift $N\alpha_a\to\mu\ell/v_F$ remains finite.

\paragraph{Comparison theorem.}
Let $c_a=\inf\operatorname{Spec}(H_a^{\rm loc})$ be the ground-state energy of the local regulator. For fixed
$\ell,v_F,g_A,\mu$ and any $E_*>0$, there exist constants $C,a_0>0$ such that, for $a<a_0$,
\begin{equation}
\begin{aligned}
&\bigl\|(H_a^{\rm nl}-c_a+E_*)^{-1} -(H_a^{\rm loc}-c_a+E_*)^{-1}\bigr\|\\
&\quad\le Ca\left[1+\log\frac{\ell}{a}\right].
\end{aligned}
\label{eq:app_1d_mu_main_theorem}
\end{equation}
This is an operator-norm estimate on the full Gauss-law physical Hilbert space, including dynamical electric fields and compact holonomy. It neither fixes a gauge background nor assumes weak coupling. The theorem compares two regulators at the same cutoff; it does not independently establish an interacting continuum limit.

\paragraph{Gauss-law reduction and the LSM twist.}
On a ring, Gauss's law leaves a single independent global gauge degree of freedom. Define the accumulated axial charge
\begin{equation}
Q_j=\sum_{l=0}^j q_{A,l},\quad Q_{N-1}=0.
\end{equation}
The electric field is the sum of an integer global flux and the accumulated charge:
\begin{equation}
E_j=m+Q_j,\quad m\in\mathbb Z.
\label{eq:app_1d_mu_E_solution}
\end{equation}
The conjugate variable to $m$ is the compact gauge holonomy $\Phi_A\in[-\pi,\pi]$. Fourier transforming the
integer flux produces periodic, matter-valued holonomy wavefunctions $f(\Phi_A)$, for which
\begin{equation}
m\mapsto p_{\Phi_A}=-i\partial_{\Phi_A},\quad
E_j\mapsto p_{\Phi_A}+Q_j.
\label{eq:app_1d_mu_flux_fourier}
\end{equation}
In this representation, the holonomy phase $e^{i\Phi_A}$ is concentrated on one chosen boundary cut, alongside
the APBC fermion sign.

An LSM-type flux-spreading transformation distributes the boundary phase uniformly over the ring. Introduce the axial charge-polarization operator
\begin{equation}
\overline Q=\frac1N\sum_jQ_j
=-\frac1N\sum_jj q_{A,j},
\label{eq:app_1d_mu_Qbar}
\end{equation}
and change the holonomy representation by the unitary multiplication
\begin{equation}
\Psi(\Phi_A)=e^{i\Phi_A\overline Q}f(\Phi_A).
\label{eq:app_1d_mu_uniform_gauge_unitary}
\end{equation}
The rotation acts on the fermions as $e^{i\Phi_A\overline Q}\eta_je^{-i\Phi_A\overline Q}=e^{ij\Phi_A/N}\eta_j$.
Hence a forward hopping over $r$ links receives a phase $e^{ir\Phi_A/N}$. For a path crossing the chosen cut,
the polarization changes by $r/N-1$ rather than $r/N$, but the original factor $e^{i\Phi_A}$ cancels the extra
winding phase. The holonomy is therefore distributed uniformly across the links. At fixed $\Phi_A$, both matter Hamiltonians are translation invariant and diagonal in the APBC momentum basis. With
$\kappa_k(\Phi_A)=k+\Phi_A/N$, we obtain
\begin{equation}
\begin{aligned}
H_{{\rm m},a}^{\rm nl}(\Phi_A)
&=\sum_k[-2t_a\cos\kappa_k-\mu s_N(\kappa_k)]n_{\eta,k},\\
H_{{\rm m},a}^{\rm loc}(\Phi_A)
&=\sum_k[-2t_a\cos\kappa_k-\mu\sin\kappa_k]n_{\eta,k}.
\end{aligned}
\label{eq:app_1d_mu_reduced_matter}
\end{equation}
The appearance of $s_N(\kappa_k)$, rather than $\operatorname{sgn}(\sin\kappa_k)$, preserves the shortest-path gauging prescribed before introducing the holonomy.

The same LSM transformation separates the spatially uniform electric mode from the charge-induced electric
field:
\begin{equation}
E_j=p_{\Phi_A}+D_j,\quad D_j=Q_j-\overline Q,
\quad \sum_jD_j=0.
\label{eq:app_1d_mu_Dj}
\end{equation}
Because $D_j-D_{j-1}=q_{A,j}$, the nonuniform electric field is fixed by the charge distribution. Using its
discrete Fourier components $q_{A,q}=\sum_je^{-iqj}q_{A,j}$, with $q=2\pi r/N$, gives
\begin{equation}
\begin{aligned}
H_E&=\frac{g_A^2a}{2}\sum_jE_j^2
=\frac{g_A^2\ell}{2}p_{\Phi_A}^2
+\frac{g_A^2a}{2N}\sum_{q\ne0}
\frac{q_{A,q}^\dagger q_{A,q}}{4\sin^2(q/2)}\\
&=:T_{\rm hol}+V_a.
\end{aligned}
\label{eq:app_1d_mu_HE_reduced}
\end{equation}
The first term, $T_{\rm hol}$, is the kinetic energy of the global holonomy; $V_a\ge0$ is the longitudinal
Coulomb interaction. This decomposition is exact.

The flux-spreading rotation converts periodic holonomy wavefunctions into twisted ones. Since $\Psi(\Phi_A)=e^{i\Phi_A\overline Q}f(\Phi_A)$ and $f$ is periodic, the transformed wavefunction satisfies
\begin{equation}
\begin{aligned}
\Psi(\pi)&=\mathscr U_A^{-1}\Psi(-\pi),\\
\mathscr U_A^{-1}&=e^{2\pi i\overline Q}
=\exp\!\left(-\frac{2\pi i}{N}\sum_jj\,q_{A,j}\right).
\end{aligned}
\label{eq:app_1d_mu_twisted_bc}
\end{equation}
Thus the inverse of the large-gauge operator defined in Eq.~\eqref{eq:app_1d_tduality_flux_endpoint} is the LSM twist and serves as the transition function of the dynamical holonomy. Consistently, either matter Hamiltonian obeys
$H_{{\rm m},a}(\Phi_A+2\pi)=\mathscr U_A^{-1}H_{{\rm m},a}(\Phi_A)\mathscr U_A$.
Its spectrum is unchanged as a set under a full flux insertion, although individual states may exhibit spectral flow. The Coulomb interaction is invariant. The holonomy remains explicit through $k+\Phi_A/N$, while the twist identifies the endpoints of its interval.

Self-adjointness of the holonomy kinetic energy requires the same twist for the wavefunction and its derivative. The common domain of $p_{\Phi_A}^2$ is
\begin{equation}
\begin{aligned}
\mathcal D_N=\bigl\{\Psi\in H^2:\;&
\Psi(\pi)=\mathscr U_A^{-1}\Psi(-\pi),\\
&\Psi'(\pi)=\mathscr U_A^{-1}\Psi'(-\pi)\bigr\}.
\end{aligned}
\label{eq:app_1d_mu_domain}
\end{equation}
On this domain $T_{\rm hol}=(g_A^2\ell/2)p_{\Phi_A}^2$ is nonnegative and self-adjoint; its quadratic-form
domain requires only the corresponding twisted $H^1$ boundary condition. At each finite $N$, the matter
operators and $V_a$ are bounded and self-adjoint in the neutral matter sector. Thus both full gauge Hamiltonians
are self-adjoint on the same domain $\mathcal D_N$.

\paragraph{Proof: the spectral difference costs excitation energy.}
The sharp sign and its smooth replacement agree at the Dirac points but differ at UV momenta. We quantify this distinction uniformly over the fluctuating holonomy. Define
\begin{equation}
d_{N,\Phi_A}(k)=s_N(k+\Phi_A/N)-\sin(k+\Phi_A/N).
\label{eq:app_1d_mu_d}
\end{equation}
We use two uniform bounds on the finite-volume interpolant:
\begin{equation}
\begin{aligned}
\|s_N\|_\infty&\le C,\\
|s_N(x)-\operatorname{sgn}(\sin x)|
&\le\frac{C}{N|\sin x|},\quad \sin x\ne0.
\end{aligned}
\label{eq:app_1d_mu_sN_bounds}
\end{equation}
For the second estimate, compare its odd-harmonic coefficients with those of the infinite square-wave series,
$4/(\pi r)$. The coefficient correction is $(4/N)F(r/N)$, where
\begin{equation}
F(y)=\frac1{\sin(\pi y)}-\frac1{\pi y},
\end{equation}
The singularity at $y=0$ is removable, and $F$ is smooth on $[0,1/2]$. The identity
$\sum_{m=0}^{L-1}\sin((2m+1)x)=\sin^2(Lx)/\sin x$ bounds partial sums of odd sines by $1/|\sin x|$. Abel
summation, together with the bounded variation of $F$, then bounds the coefficient correction by $C/(N|\sin
x|)$. The omitted tail of the infinite square-wave series satisfies the same estimate. Finally, the uniform
bound on square-wave partial sums and the $O(1)$ sum of absolute coefficient corrections establish the first
inequality.

Near the Fermi points, the difference between $s_0(k)$ and $\sin k$ is quadratic in the distance to the node;
away from them $|\cos k|$ has a positive lower bound. Using $|\cos k_j|\ge\sin(\pi/N)\ge2/N$ on the APBC grid,
we obtain, uniformly for $|\Phi_A|\le\pi$,
\begin{equation}
|d_{N,\Phi_A}(k_j)|\le C|\cos k_j|.
\label{eq:app_1d_mu_mode_bound}
\end{equation}
Near either Fermi point, the difference is bounded by $C/N+C(|\cos k_j|+1/N)^2$. The grid separation absorbs the
$1/N$ terms into $C|\cos k_j|$, proving the estimate. Thus the modewise discrepancy is bounded by the corresponding free excitation energy.

The filled sea requires a separate estimate: modewise errors of order $N^{-1}$ could accumulate into an $O(1)$ holonomy-dependent vacuum shift. Instead, a cancellation yields
\begin{equation}
\left|\sum_{\cos k_j>0}d_{N,\Phi_A}(k_j)\right|
\le\frac CN,\quad |\Phi_A|\le\pi.
\label{eq:app_1d_mu_sea_bound}
\end{equation}
The cancellation follows from the exact geometric sum over the occupied half-filled momenta,
\begin{equation}
\sum_{\cos k_j>0}e^{irk_j}
=\frac{\sin(\pi r/2)}{\sin(\pi r/N)},
\end{equation}
which converts the interpolant sum into
\begin{equation}
\begin{split}
\sum_{\cos k_j>0}&s_N(k_j+\Phi_A/N) \\
& =\frac4N\sum_{\substack{1\le r<M\\r\,{\rm odd}}}
\frac{\sin(\pi r/2)\sin(r\Phi_A/N)}{\sin^2(\pi r/N)}.
\end{split}
\label{eq:app_1d_mu_sea_sN}
\end{equation}
To isolate its leading contribution, split the summand as
\begin{equation}
\frac{\sin(\Phi_A y)}{\sin^2(\pi y)}
=\frac{\Phi_A}{\pi^2y}+F_{\Phi_A}(y),\quad
F_{\Phi_A}(0)=0.
\end{equation}
The function $F_{\Phi_A}$ and its first derivative are uniformly bounded for $0\le y\le1/2$ and
$|\Phi_A|\le\pi$. Since $\sin(\pi r/2)=(-1)^{(r-1)/2}$ for odd $r$, its alternating sum contributes only
$O(N^{-1})$ after the overall prefactor. The singular part contributes
\begin{equation}
\frac{4\Phi_A}{\pi^2}\sum_{m=0}^{N/4-1}
\frac{(-1)^m}{2m+1}=\frac{\Phi_A}{\pi}+O(N^{-1}),
\end{equation}
whereas the smooth nearest-neighbor term gives
\begin{equation}
\sum_{\cos k_j>0}\sin(k_j+\Phi_A/N)
=\frac{\sin(\Phi_A/N)}{\sin(\pi/N)}
=\frac{\Phi_A}{\pi}+O(N^{-2}).
\end{equation}
The leading flux-dependent contributions therefore cancel, establishing Eq.~\eqref{eq:app_1d_mu_sea_bound}. The cancellation excludes a vacuum-energy discrepancy that the modewise bound alone would not control.

Let $\Omega_N$ denote the half-filled Fermi sea at zero holonomy. We measure particle--hole excitations relative
to this reference state using
\begin{equation}
K_a=\sum_k2t_a|\cos k|N_k^{\rm exc},
\label{eq:app_1d_mu_Ka}
\end{equation}
where
\begin{equation}
N_k^{\rm exc} =\begin{cases}
1-n_{\eta,k},&\cos k>0,\\
n_{\eta,k},&\cos k<0.
\end{cases}
\end{equation}
Writing $B_a:=H_a^{\rm nl}-H_a^{\rm loc}$ and normal ordering relative to $\Omega_N$ separates the
excitation-dependent part from the filled-sea contribution:
\begin{equation}
B_a(\Phi_A)=-\mu\sum_kd_{N,\Phi_A}(k):n_{\eta,k}:
-\mu\sum_{\cos k>0}d_{N,\Phi_A}(k).
\label{eq:app_1d_mu_Ba}
\end{equation}
The modewise estimate controls particle--hole excitations through $K_a$, while the sea cancellation bounds the scalar contribution. For every physical wavefunction $\Psi$ in the holonomy form domain,
this gives
\begin{equation}
\begin{aligned}
|b_a[\Psi]|&\le\frac{C|\mu|}{t_a}
\langle\Psi,K_a\Psi\rangle +\frac{C|\mu|}{N}\|\Psi\|^2,\\
b_a[\Psi]&=\langle\Psi,B_a\Psi\rangle.
\end{aligned}
\label{eq:app_1d_mu_first_form_bound}
\end{equation}
At fixed physical $\mu$, the coefficient $|\mu|/t_a=2|\mu|a/v_F$ tends to zero with the lattice spacing. It remains to control $K_a$ by the energy of the interacting local gauge theory.

\paragraph{Proof: controlling excitations in the interacting gauge theory.}
The electric and Coulomb terms enter $H_a^{\rm loc}$ but not $K_a$. To compare these energies, start from the reference sea energy
\begin{equation}
E_{{\rm sea},a}=-2t_a\sum_{\cos k>0}\cos k
=-\frac{2t_a}{\sin(\pi/N)}.
\end{equation}
At fixed holonomy write $h_{\Phi_A}(k)=-2t_a\cos(k+\Phi_A/N)-\mu\sin(k+\Phi_A/N)$ and $\delta
h_{\Phi_A}(k)=h_{\Phi_A}(k)+2t_a\cos k$. A compact holonomy shifts a lattice momentum by at most $\pi/N$, so the
change in the single-particle energy is bounded independently of the cutoff:
\begin{equation}
|\delta h_{\Phi_A}(k)|\le\frac{\pi v_F}{\ell}+|\mu|
=:\Lambda_\mu.
\label{eq:app_1d_mu_delta_bound}
\end{equation}
Normal ordering about the reference sea gives the exact decomposition
\begin{equation}
\begin{aligned}
&H_{{\rm m},a}^{\rm loc}(\Phi_A)-E_{{\rm sea},a}\\
& =K_a+\sum_k\delta h_{\Phi_A}(k):n_{\eta,k}: +\sum_{\cos k>0}\delta h_{\Phi_A}(k).
\end{aligned}
\label{eq:app_1d_mu_matter_decomposition}
\end{equation}
The excitation correction can lower the energy by at most $\Lambda_\mu N_{\rm exc}$. Since a particle or hole at
momentum $k$ costs $\varepsilon_k=2t_a|\cos k|$, we can estimate each mode by
\begin{equation}
\Lambda_\mu N_k^{\rm exc}
\le\frac{\varepsilon_k}{2}N_k^{\rm exc}
+\left(\Lambda_\mu-\frac{\varepsilon_k}{2}\right)_+.
\end{equation}
Only $O(1+\ell\Lambda_\mu/v_F)$ modes have $\varepsilon_k<2\Lambda_\mu$, a number independent of the lattice
cutoff. Summing the modewise inequality therefore gives
\begin{equation}
\Lambda_\mu N_{\rm exc}\le\frac12K_a+C.
\label{eq:app_1d_mu_Nexc_bound}
\end{equation}
The remaining filled-sea correction is explicitly
\begin{equation}
\sum_{\cos k>0}\delta h_{\Phi_A}(k)
=\frac{2t_a[1-\cos(\Phi_A/N)]-\mu\sin(\Phi_A/N)}
{\sin(\pi/N)},
\end{equation}
and stays uniformly bounded for $|\Phi_A|\le\pi$. Because the electric energy is nonnegative, we conclude
\begin{equation}
H_a^{\rm loc}-E_{{\rm sea},a}\ge\frac12K_a-C_0.
\label{eq:app_1d_mu_lower_bound}
\end{equation}
We next relate the free reference energy $E_{{\rm sea},a}$ to the interacting ground-state energy $c_a$. Control of this vacuum-energy difference is needed for the low-energy estimate.
Choose a normalized smooth holonomy wavepacket $f$ supported strictly inside $(-\pi,\pi)$ and form the physical
trial state
\begin{equation}
\Psi_N(\Phi_A)=f(\Phi_A)\Omega_N.
\label{eq:app_1d_mu_trial}
\end{equation}
Its holonomy kinetic energy is $(g_A^2\ell/2)\|f'\|^2$, and its matter energy relative to $E_{{\rm sea},a}$ is
bounded independently of $N$. The remaining contribution is the Coulomb energy of the sea's charge fluctuations.
For $q=2\pi r/N\ne0$, Wick's theorem gives
\begin{equation}
\langle\Omega_N,q_{A,q}^\dagger q_{A,q}\Omega_N\rangle
=\min(r,N-r)+\frac{N^2}{4}\delta_{r,M}.
\label{eq:app_1d_mu_structure_factor}
\end{equation}
The $r=M$ term comes from the staggered background charge and contributes $g_A^2\ell/32$ to the Coulomb energy.
With $\sin(\pi r/N)\ge2r/N$ for $1\le r\le M$, the remaining modes yield the bound
\begin{equation}
\langle V_a\rangle_{\Omega_N}
\le Cg_A^2\ell(1+\log N).
\label{eq:app_1d_mu_Coulomb_trial}
\end{equation}
Combining this trial-state upper bound with the preceding operator lower bound gives, by the variational
principle,
\begin{equation}
-C_0\le c_a-E_{{\rm sea},a}
\le C_1+C_2g_A^2\ell(1+\log N).
\label{eq:app_1d_mu_ca_bound}
\end{equation}
Hence the free excitation energy is bounded by the interacting excitation energy $L_a:=H_a^{\rm loc}-c_a\ge0$ up
to a logarithmic cutoff correction:
\begin{equation}
K_a\le2L_a+C(1+\log N).
\label{eq:app_1d_mu_K_control}
\end{equation}
Combining this with Eq.~\eqref{eq:app_1d_mu_first_form_bound} yields the central relative-form estimate, valid
for every physical wavefunction of finite local energy:
\begin{equation}
\begin{aligned}
|b_a[\Psi]|&\le\epsilon_a\langle\Psi,L_a\Psi\rangle
+\beta_a\|\Psi\|^2,\\
\epsilon_a&=O(a),\quad
\beta_a=O\!\left(a\left[1+\log\frac\ell a\right]\right).
\end{aligned}
\label{eq:app_1d_mu_relative_form}
\end{equation}
Thus the nonlocal perturbation is relatively form-small with respect to the full local gauge Hamiltonian, including dynamical electric fields, without a weak-coupling expansion.

\paragraph{Proof: norm-resolvent comparison.}
Set $M_a=H_a^{\rm nl}-c_a=L_a+B_a$. The relative-form estimate implies the two-sided inequalities
\begin{equation}
(1-\epsilon_a)L_a-\beta_a\le M_a
\le(1+\epsilon_a)L_a+\beta_a.
\label{eq:app_1d_mu_two_sided_form}
\end{equation}
For sufficiently small $a$, $\epsilon_a<1$ and $M_a\ge-\beta_a$. Given fixed $E_*>0$, we can choose $a$ so that
$\beta_a<E_*/2$; both shifted inverses then exist, with
\begin{equation}
\|(L_a+E_*)^{-1}\|\le E_*^{-1},\ \
\|(M_a+E_*)^{-1}\|\le2E_*^{-1}.
\label{eq:app_1d_mu_resolvent_bounds}
\end{equation}
Set $A_a=L_a+E_*$ and introduce the bounded, energy-weighted perturbation
\begin{equation}
R_a=A_a^{-1/2}B_aA_a^{-1/2},\ \
\|R_a\|\le r_a:=\epsilon_a+\frac{\beta_a}{E_*}\to0.
\end{equation}
For $r_a<1$, the nonlocal resolvent factorizes as
\begin{equation}
(M_a+E_*)^{-1}=A_a^{-1/2}(1+R_a)^{-1}A_a^{-1/2}.
\end{equation}
Taking the difference from $A_a^{-1}$ and using the bound on $R_a$ gives
\begin{equation}
\begin{aligned}
\|(M_a+E_*)^{-1}& -(L_a+E_*)^{-1}\|
\le\frac{r_a}{E_*(1-r_a)}\\
&=O\!\left(a\left[1+\log\frac\ell a\right]\right),
\end{aligned}
\end{equation}
This proves Eq.~\eqref{eq:app_1d_mu_main_theorem}. The same form inequalities also control the shift of the
nonlocal ground-state energy:
\begin{equation}
|\inf\operatorname{Spec}(H_a^{\rm nl})-c_a|\le\beta_a.
\label{eq:app_1d_mu_ground_shift}
\end{equation}
Consequently, the same rate holds if each Hamiltonian is centered by its own ground-state energy.

\paragraph{UV nonlocality and low-energy equivalence.}
Norm-resolvent convergence need not imply convergence of the Hamiltonians in operator norm: the inverse-energy weighting suppresses high-energy differences even when they remain large in the UV. Indeed, for fixed $\mu\neq0$, the Hamiltonian difference
$B_a=H_a^{\rm nl}-H_a^{\rm loc}$ diverges in operator norm as the cutoff is removed.

At zero holonomy, the difference is
\begin{equation}
B_a(0)=-\mu\sum_kd_{N,0}(k):n_{\eta,k}:
-\mu\sum_{\cos k>0}d_{N,0}(k).
\label{eq:app_1d_mu_Ba_zero_holonomy}
\end{equation}
Choose a fixed interval $I\Subset(0,\pi/2)$, for example $I=[\pi/6,\pi/3]$. Since $s_N$ converges uniformly to
$s_0$ away from its discontinuities,
\begin{equation}
d_{N,0}(k)\ge c_0>0,\quad k\in I,
\label{eq:app_1d_mu_d_lower}
\end{equation}
for sufficiently large $N$. Starting from the half-filled sea, transfer each occupied fermion at $k\in I$ to the
empty mode at $k-\pi$. Because $d_{N,0}(k-\pi)=-d_{N,0}(k)$, each particle--hole pair changes the expectation of
$B_a(0)$ by an amount of magnitude at least $2|\mu|c_0$. There are $O(N)$ such pairs, and the total axial charge
remains zero. The filled-sea contribution is only $O(|\mu|/N)$. A smooth holonomy wavepacket localized near
$\Phi_A=0$ extends this construction to the full physical gauge Hilbert space, giving
\begin{equation}
\|B_a\|\ge c|\mu|N=c|\mu|\frac\ell a,
\quad \mu\ne0,
\label{eq:app_1d_mu_no_operator_norm}
\end{equation}
for sufficiently small $a$. Thus the UV nonlocality remains detectable on highly excited states.

By contrast, the difference becomes small on fixed-energy subspaces. Let $P_{a,E}=\mathbf1_{[0,E]}(L_a)$ project onto states of
the local regulator with excitation energies at most $E<\infty$, independent of $a$. The relative-form estimate
gives
\begin{equation}
\|P_{a,E}B_aP_{a,E}\|
\le\epsilon_a E+\beta_a\to0.
\label{eq:app_1d_mu_low_energy_operator_norm}
\end{equation}
Thus the compressed difference vanishes in norm at every fixed energy. This controls low-energy matrix elements but does not imply $\|B_aP_{a,E}\|\to0$ or convergence of the unrestricted Hamiltonians.

Norm-resolvent closeness also controls energy-filtered dynamics. For fixed time $\tau$ and any continuous compactly supported filter $\chi$, functional calculus gives
\begin{equation}
\|e^{-i\tau M_a}\chi(M_a)-e^{-i\tau L_a}\chi(L_a)\|
\to0.
\label{eq:app_1d_mu_filtered_dynamics}
\end{equation}
Thus the two regulators have the same energy-filtered dynamics in the limit, despite their growing operator-norm difference.

The theorem compares two lattice gauge regulators; it does not independently construct their interacting
continuum limit. Any compatible strong- or norm-resolvent continuum convergence established for the local
regulator transfers to the nonlocal one. If the continuum construction uses the standard compact axial gauging
of $H_0$, the identifications must incorporate the exact holonomy translation $W_a^\dagger$. The result applies
at fixed physical chemical potential, for which $|\mu|/t_a\to0$, and does not extend automatically to fixed
nonzero $\mu/t_a$, alternative Wilson-line prescriptions, or independently gauged branch-space seams.

\paragraph{Extension to the inversion-odd spectral-sign family.}
The estimate extends uniformly to the inversion-odd signs of Eq.~\eqref{eq:1d_s_mu_universal_form}, provided the compact gauging is transported through the exact relocalization. In the adapted $\xi^{(s)}$ variables, set $d\Gamma_s(B)=\sum_{jl}\xi_j^{(s)\dagger}B_{jl}\xi_l^{(s)}$. The covariant sign operator is the same $s_N(T_U)$ defined in Eq.~\eqref{eq:app_1d_mu_sNT}:
\begin{equation}
S_N(T_U)\equiv s_N(T_U).
\label{eq:1d_s_covariant_sign}
\end{equation}
This uses the finite-volume Fourier coefficients of $s_0$ rather than a newly imposed spectral sign in the gauge
background. The transported nonlocal and local regulators are
\begin{equation}
\begin{aligned}
H_{s,a}^{\rm nl}[U]
&=-t_a d\Gamma_s(T_U+T_U^\dagger)
-\mu d\Gamma_s(S_N(T_U))+H_E,\\
H_{s,a}^{\rm loc}[U]
&=-t_a d\Gamma_s(T_U+T_U^\dagger)
-\mu d\Gamma_s\!\left(\frac{T_U-T_U^\dagger}{2i}\right)+H_E,\\
H_E&=\frac{g_A^2a}{2}\sum_jE_j^2.
\end{aligned}
\label{eq:1d_s_nl_loc_comparison}
\end{equation}
Both Hamiltonians commute with the Gauss operators $\mathcal G_{j,s}$. At zero gauge field, the nonlocal member reproduces $H_{a,\mu}[c]$ in the $\xi^{(s)}$ representation, up to the common normal-ordering scalar. By Eq.~\eqref{eq:1d_s_xi_def}, the background-independent unitary $\mathcal U_s$ transports the baseline gauge Hamiltonians and Gauss laws according to
\begin{equation}
\begin{aligned}
H_{s,a}^{\rm nl}&=\mathcal U_sH_a^{\rm nl}\mathcal U_s^{-1},\\
H_{s,a}^{\rm loc}&=\mathcal U_sH_a^{\rm loc}\mathcal U_s^{-1},\\
\mathcal G_{j,s}&=\mathcal U_s\mathcal G_{j,0}\mathcal U_s^{-1}.
\end{aligned}
\label{eq:1d_s_exact_gauged_intertwining}
\end{equation}
Consequently, $\mathcal U_s$ maps the full physical gauge Hilbert spaces into one another, including the holonomy
twists and operator domains. Since $c_a=\inf\operatorname{Spec}(H_a^{\rm loc})$, unitarity gives
\begin{equation}
\begin{aligned}
&\|(H_{s,a}^{\rm nl}-c_a+E_*)^{-1}
-(H_{s,a}^{\rm loc}-c_a+E_*)^{-1}\|\\
&=\|(H_a^{\rm nl}-c_a+E_*)^{-1}
-(H_a^{\rm loc}-c_a+E_*)^{-1}\|\\
&\le C_{E_*}a\left[1+\log\frac\ell a\right].
\end{aligned}
\label{eq:1d_s_uniform_resolvent}
\end{equation}
The constant is independent of the sign choice $s$, even when $s$ depends on the cutoff. Node matching identifies the same IR axial current, whereas uniformity of the gauge-theory estimate follows from exact canonical intertwining. The result applies to the transported shortest-path prescription, not independently gauged branch seams, and transfers any compatible continuum limit established for the local regulator without constructing one independently.

\section{2D: Compact symmetries, charge-bundle topology, and gauge locality}
\label{app:2d_gauge_locality}

This appendix provides the constructions and proofs supporting the 2D results of Sec.~\ref{sec:2d_u1}. We first establish symmetry and Hamiltonian-locality obstructions for the Kitaev--Majorana honeycomb model, then construct sharp and smooth compact charges for the auxiliary honeycomb representative and analyze gapless transitions between distinct class-D regulator sectors. We develop a general spectral-pairing construction of local compact charges and derive the square-model charge Chern number, its relation to Pfaffian parity, and the associated charge-spread and bulk--edge constraints. We then establish controlled gauge locality for Schwartz regulators and prove a general crystalline-parity criterion for odd Pfaffian parity. Finally, we analyze finite-range Onsager lifts and their IR quotients, followed by non-Abelian charge topology, crystalline constraints, and anomaly diagnostics for the two-flavor model. The separate 2D--3D bulk--surface constructions are collected in Appendix~\ref{app:2d_surface_inflow}.

\subsection{Kitaev--Majorana symmetry and Hamiltonian locality obstructions}
\label{app:2d_HK_finite_no_go}

We establish two complementary locality obstructions for the isotropic Kitaev--Majorana honeycomb Hamiltonian
$H_K$. The first excludes finite-range or exponentially local compact quadratic generators of unit Dirac number
with any finite translation period in the original honeycomb geometry. The second is stronger: even allowing
spatially nonlocal Majorana-canonical Gaussian transformations, no finite-folding relocalization can make the
compact Dirac-number symmetry onsite while retaining a periodic, continuously differentiable ($C^1$) Bloch
Hamiltonian. Both results concern finite-period or finite-folding constructions; neither excludes more general
canonical reorganizations or all possible forms of emergent gauge locality.

\subsubsection{Finite-period obstruction to local Dirac-number generators}
\label{app:2d_HK_charge_no_go}

The Bloch Hamiltonian in the $(A,B)$ sublattice basis is
\begin{equation}
\begin{aligned}
h_K(u,v)&=t\left[
\operatorname{Im}f_K\,\sigma_x
+\operatorname{Re}f_K\,\sigma_y\right],\\
f_K(u,v)&=1+e^{iu}+e^{iv},
\end{aligned}
\end{equation}
with squared dispersion
\begin{equation}
\begin{aligned}
h_K(u,v)^2&=F(u,v)\mathbf1,\\
F(u,v)&=t^2\bigl[3+2\cos u+2\cos v\\
&\hspace{18mm}+2\cos(u-v)\bigr].
\end{aligned}
\label{eq:app_HK_squared_dispersion}
\end{equation}
Its two Majorana-reality-related nodes represent a single massless complex Dirac fermion.

We first consider a compact quadratic symmetry generator invariant under primitive-cell translations. Let
$J(\mathbf k)$ denote its anti-Hermitian Majorana generator and $K(\mathbf k)=iJ(\mathbf k)$ the corresponding
Hermitian matrix. Compactness with a fixed $2\pi$ period requires $\operatorname{Spec}K(\mathbf k)\subset\mathbb
Z$. If the generator has finite-range or exponentially decaying couplings, $K(\mathbf k)$ is continuous, and its
integer-valued trace must be constant throughout the Brillouin torus. However, Majorana reality requires
\begin{equation}
\begin{aligned}
K(-\mathbf k)&=-K(\mathbf k)^*,\\
\operatorname{Tr}K(-\mathbf k)&=-\operatorname{Tr}K(\mathbf k),
\end{aligned}
\end{equation}
so this constant must vanish. Unit Dirac charge at the positive Majorana node instead requires $K(\mathbf
K_+)=\mathbf1$ and hence $\operatorname{Tr}K(\mathbf K_+)=2$, giving a contradiction.

The obstruction persists under any finite enlargement of the unit cell. Consider a supercell containing $r$
primitive cells, with folding momenta $\mathbf g_\alpha$, $\alpha=1,\ldots,r$. On the universal cover of the
reduced Brillouin torus, the folded Hamiltonian and squared dispersions are
\begin{equation}
\begin{aligned}
h_{\rm fold}(\mathbf p)&=\bigoplus_{\alpha=1}^{r}
h_K(\mathbf p+\mathbf g_\alpha),\\
F_\alpha(\mathbf p)&=F(\mathbf p+\mathbf g_\alpha).
\end{aligned}
\label{eq:app_HK_folded_dispersion}
\end{equation}
Conservation of the quadratic charge implies $[h_{\rm fold}(\mathbf p),J(\mathbf p)]=0$. In the folded basis,
its $(\alpha,\beta)$ block satisfies $h_\alpha J_{\alpha\beta}=J_{\alpha\beta}h_\beta$. Squaring the Hamiltonian
blocks therefore gives
\begin{equation}
(F_\alpha-F_\beta)J_{\alpha\beta}=0.
\label{eq:app_finite_block_intertwiner}
\end{equation}
The honeycomb dispersion has no nontrivial fixed-shift symmetry: comparing Fourier coefficients shows that
$F(\mathbf k+\boldsymbol\delta)\equiv F(\mathbf k)$ requires $e^{i\delta_u}=e^{i\delta_v}=1$. Thus, for
$\alpha\neq\beta$, the analytic function $F_\alpha-F_\beta$ is not identically zero.
Equation~\eqref{eq:app_finite_block_intertwiner} forces $J_{\alpha\beta}$ to vanish wherever these dispersions
differ, and continuity extends this conclusion to their crossing sets. Every continuous conserved generator with
a finite supercell is consequently block diagonal in the original primitive momentum sectors. Unfolding then
reduces the problem to the trace contradiction above. Hence $H_K$ admits no finite-range or exponentially local
compact quadratic generator of unit Dirac number with any finite translation period.

\subsubsection{Finite-folding obstruction to Hamiltonian-local relocalization}
\label{app:2d_HK_Hamiltonian_no_go}

We now allow the canonical transformation itself to be spatially nonlocal. Consider a Majorana-canonical
Gaussian transformation that preserves translations by a finite supercell and acts within its folded momentum
fibers, without requiring the transformation to be continuous or local. Suppose, for contradiction, that it
makes the exact Dirac-number symmetry onsite, with a momentum-independent generator $K_0$, while the transformed
Hamiltonian $h'(\mathbf p)$ remains periodic and $C^1$. Conservation of the original charge implies
$[h',K_0]=0$. Moreover, the original spectral generator acts as $+\mathbf1$ near $\mathbf K_+$ and $-\mathbf1$
near $\mathbf K_-$, so these charge assignments must be preserved. The proof proceeds by tracking the charge
multiplicities of the folded spectral branches $F_\alpha(\mathbf p)$, the squared dispersions of the original
momentum sectors brought together by supercell folding. We show that $C^1$ regularity prevents these
multiplicities from changing, while reduced-zone periodicity forces them to be equal, contradicting the opposite
valley assignments.

Let $P_+$ be the constant projector onto the charge-$+1$ eigenspace of $K_0$, and define the squared Hamiltonian
restricted to this sector,
\begin{equation}
M_+(\mathbf p)=
\left[h'(\mathbf p)\big|_{\operatorname{Ran}P_+}\right]^2.
\label{eq:app_HK_positive_sector}
\end{equation}
The canonical transformation acts unitarily within each folded momentum fiber and therefore preserves its
spectrum. Consequently, the eigenvalues of $M_+$ are selected from the folded squared dispersions
$F_\alpha(\mathbf p)$ of Eq.~\eqref{eq:app_HK_folded_dispersion}. At generic momenta these functions are
pairwise distinct, since the honeycomb dispersion has no nontrivial fixed-shift symmetry. We denote the
multiplicity of each branch in the positive-charge sector by
$
n_\alpha(\mathbf p)\in\{0,1,2\}.$ Thus $n_\alpha$ counts how many of the two states associated with the folded
momentum sector $\alpha$ carry charge $+1$. These integer multiplicities are locally constant wherever the
$F_\alpha$ are distinct. They can change only where folded dispersions cross.

Consider such a crossing at $\mathbf p_0$, and let $s$ be a momentum coordinate transverse to the crossing
curve, with $s=0$ at $\mathbf p_0$. The participating branches have expansions
$F_\alpha(s)=F_*+sv_\alpha+O(s^2)$, where $v_\alpha$ are their transverse slopes. If $M_+$ is $C^1$, the
first-order splittings of its degenerate eigenvalue $F_*$ are determined by the Hermitian matrix
$P_*(\partial_sM_+)P_*$, with $P_*$ the projector onto the corresponding eigenspace at the crossing. Hence the
slopes, counted with multiplicity, must agree on both sides:
\begin{equation}
\bigsqcup_\alpha\{v_\alpha\}^{\,n_\alpha^-}
=
\bigsqcup_\alpha\{v_\alpha\}^{\,n_\alpha^+}.
\label{eq:app_HK_slope_multiset}
\end{equation}
If the participating slopes are pairwise distinct, equality of these collections requires
\begin{equation}
n_\alpha^-=n_\alpha^+.
\label{eq:app_HK_multiplicity_continuation}
\end{equation}
Thus, provided the branches cross with different transverse slopes, $C^1$ regularity prevents their charge
multiplicities from switching across the crossing. We now verify that this condition holds generically for the
honeycomb dispersion.

For a nontrivial momentum shift $\boldsymbol\delta=(\delta_u,\delta_v)$, introduce $x=u+\delta_u/2$ and
$y=v+\delta_v/2$. Direct calculation gives
\begin{equation}
\begin{aligned}
&F(\mathbf k+\boldsymbol\delta)-F(\mathbf k)\\
&\quad=-4t^2\left[
A\sin x+B\sin y+C\sin(x-y)\right],
\end{aligned}
\label{eq:app_HK_crossing_difference}
\end{equation}
where
\begin{equation}
A=\sin\frac{\delta_u}{2},\ \
B=\sin\frac{\delta_v}{2},\ \
C=\sin\frac{\delta_u-\delta_v}{2}.
\end{equation}
Setting $z=e^{ix}$ and $w=e^{iy}$, and multiplying the bracket by the nonvanishing factor $2izw$, yields the
polynomial
\begin{equation}
\begin{aligned}
P(z,w)={}&A(z^2w-w)+B(zw^2-z)\\
&+C(z^2-w^2).
\end{aligned}
\label{eq:app_HK_crossing_polynomial}
\end{equation}
For every nontrivial momentum shift, $P(z,w)$ has no repeated polynomial factors. When $A,B\neq0$, this follows
because its highest-degree part $zw(Az+Bw)$ contains three distinct linear factors. If $A\neq0$ and $B=0$, a
repeated factor would have to be of the form $z-c$; requiring $P(c,w)\equiv0$ forces $C=0$ and $c^2=1$, leaving
$Aw(z^2-1)$, which has no repeated factors. The case $A=0$, $B\neq0$ is analogous, while $A=B=0$ corresponds to
a trivial reciprocal-lattice shift.

The absence of repeated factors implies that $F_\alpha-F_\beta$ has a nonzero gradient at generic points of each
crossing curve. Indeed, if its derivative perpendicular to an entire crossing component vanished, the difference
would vanish to second order there, requiring a repeated factor. Distinct folded dispersions therefore have
different transverse slopes at generic crossings. Equation~\eqref{eq:app_HK_multiplicity_continuation} then
prevents changes in their charge multiplicities. Since the exceptional crossing points are isolated and can be
avoided by continuous paths, each $n_\alpha$ must be constant on the universal momentum cover.

The contradiction now follows from reduced-zone periodicity. Let $\Lambda^*$ and $\Lambda_{\mathrm{sc}}^*$
denote the reciprocal lattices of the original and supercell translations, respectively. The folding momenta
represent the $r$ elements of the finite quotient $\Lambda_{\mathrm{sc}}^*/\Lambda^*$. A supercell reciprocal
translation $\mathbf G$ therefore permutes the folded branches according to
\begin{equation}
F_\alpha(\mathbf p+\mathbf G)
=F_{\pi_{\mathbf G}(\alpha)}(\mathbf p).
\end{equation}
These permutations act transitively: any folding momentum can be mapped to any other by an appropriate supercell
reciprocal translation. Since $M_+$ is periodic and the $n_\alpha$ are constant, comparison of its eigenvalue
multiplicities at generic momenta gives $n_\alpha=n_{\pi_{\mathbf G}(\alpha)}$. Transitivity then requires
\begin{equation}
n_1=n_2=\cdots=n_r.
\label{eq:app_HK_equal_multiplicities}
\end{equation}
This contradicts the prescribed Dirac-number symmetry. Both Bloch states near $\mathbf K_+$ carry charge
$+1$, whereas both near the Majorana-conjugate node $\mathbf K_-$ carry charge $-1$. Their positive-charge
multiplicities are therefore $2$ and $0$, respectively. No finite-folding relocalization satisfying the stated
assumptions can produce a periodic $C^1$ transformed Hamiltonian.

In particular, the obstruction excludes transformed hoppings satisfying
\begin{equation}
\sum_{\mathbf R}(1+|\mathbf R|)
\|T_{\mathbf R}\|<\infty,
\label{eq:app_HK_C1_locality_condition}
\end{equation}
because this condition implies a $C^1$ Bloch Hamiltonian. Thus finite-range, exponentially decaying, and
Schwartz-decaying hopping are all excluded within this finite-folding framework, independently of the locality
of the canonical transformation. This contrasts with $H_{\rm sq}$ studied in Sec.~\ref{sec:2d_sq}, whose exact
pairing of folded squared dispersions allows charge sectors to be reorganized without switching between
spectrally distinguishable branches, enabling smooth onsite relocalization and controlled asymptotic gauge
locality. The honeycomb result is not a universal obstruction to compact gauging: more general canonical
reorganizations and weaker forms of emergent gauge locality remain outside its scope.

\subsection{Honeycomb representative: sharp and smooth compact charges}
\label{app:2d_hc_charge_bundle}

The auxiliary honeycomb Hamiltonian $H_{\rm hc}$ is a second odd-Pfaffian lattice representative of one massless
complex Dirac fermion, complementing $H_{\rm sq}$. We construct a reflection-compatible sharp compact charge and
a Schwartz-local charge-onsite Hamiltonian, then obtain an exponentially local compact charge for the unchanged
$H_{\rm hc}$. The explicit reflection sewing and charge Chern number distinguish the two realizations.

Let $\mathbf a_1,\mathbf a_2$ be primitive vectors of the triangular Bravais lattice, with $\mathbf R=m\mathbf
a_1+n\mathbf a_2$, $u=\mathbf k\cdot\mathbf a_1$, and $v=\mathbf k\cdot\mathbf a_2$. The finite-range Majorana
Hamiltonian is
\begin{equation}
\begin{aligned}
H_{\rm hc}={}&-\frac{it}{2}\sum_{\mathbf R}\gamma_{A,\mathbf R}
\Big[\gamma_{B,\mathbf R}+\gamma_{B,\mathbf R+\mathbf a_2}\\
&+\frac1{\sqrt3}\gamma_{B,\mathbf R+\mathbf a_1}
-\frac1{\sqrt3}\gamma_{B,\mathbf R-\mathbf a_1+\mathbf a_2}\Big].
\end{aligned}
\label{eq:2d_hc_micro}
\end{equation}
Its structure factor is
\begin{equation}
\begin{aligned}
f_{\rm hc}(u,v)
&=1+e^{iv}+\frac1{\sqrt3}\big(e^{iu}-e^{i(v-u)}\big)\\
&=2e^{iv/2}\left[\cos\frac v2
+\frac i{\sqrt3}\sin\left(u-\frac v2\right)\right],
\end{aligned}
\label{eq:2d_hc_f}
\end{equation}
where $h_{\rm hc}=t(\operatorname{Im}f_{\rm hc}\,\sigma_x+\operatorname{Re}f_{\rm hc}\,\sigma_y)$. The only
zeros are the Majorana-reality-related nodes $(\pm\pi/2,\pi)$, describing one complex Dirac fermion. The
dispersion is isotropic in the physical honeycomb metric:
\begin{equation}
E_{\rm hc}^2(\mathbf K+\mathbf q)
=\frac{4t^2}{3}(q_u^2-q_uq_v+q_v^2)+O(|\mathbf q|^3).
\label{eq:2d_hc_cartesian_isotropy}
\end{equation}
For primitive vectors of length $a$ separated by $\pi/3$, the Dirac velocity is $v_F=|t|a$.

The microscopic reflection $\mathsf r$ along $\mathbf a_2$ acts as $(u,v)\mapsto(v-u,v)$, for which
\begin{equation}
f_{\rm hc}(v-u,v)=e^{iv}f_{\rm hc}(u,v)^*.
\label{eq:2d_hc_parity_identity}
\end{equation}
A reflection-invariant, inversion-odd spectral sign, distinct from the square-model $s_P$, is
\begin{equation}
s_{b,P}(u,v)=\operatorname{sgn}[\sin u+\sin(v-u)],
\ \  s_{b,P}\circ\mathsf r=s_{b,P},
\label{eq:2d_hc_sP}
\end{equation}
It defines an exact sharp compact charge $Q_{s_{b,P}}$, preserved by $H_{\rm hc}$ and by the reflection, and
assigns opposite unit charges to the two Dirac nodes. One seam-free finite-volume choice is $L_1=L_2=4m$, with
PBC along $u$ and APBC along $v$. On this grid, $\sin(v/2)$ and $\cos(u-v/2)$ do not vanish, so the sharp sign
is unambiguous. The chosen spin structure need not itself preserve the reflection.

The positive-charge half-zone of $s_{b,P}$ may be represented by $0<v<2\pi$ and $-\pi<2u-v<\pi$. Introduce
square momentum coordinates
\begin{equation}
p_1=2u-v,\quad p_2=v-\pi,\quad -\pi<p_i<\pi
\label{eq:2d_hc_p_square}
\end{equation}
The independent complex Majorana modes in this half-zone can then be assembled into charge-onsite fermions
$f_{\alpha,\mathbf X}$, as in 1D. After a sublattice rephasing removes $e^{iv/2}$, their matter symbol and
onsite charge are
\begin{equation}
\begin{aligned}
h_c(\mathbf p)&=2t\left[\frac{\sin(p_1/2)}{\sqrt3}\,\sigma_x
-\sin\frac{p_2}{2}\,\sigma_y\right],\\
Q_{s_{b,P}}&=\sum_{\mathbf X,\alpha}
\left(f_{\alpha,\mathbf X}^\dagger f_{\alpha,\mathbf X}-\frac12\right).
\end{aligned}
\label{eq:2d_hc_charge_square}
\end{equation}
The charge is now onsite, but the Bloch symbol $h_c$ has artificial seams at the identified square-zone
boundaries. Its squared dispersion is nevertheless smooth and periodic:
\begin{equation}
\begin{aligned}
h_c^2&=4t^2\rho^2(\mathbf p)\mathbf1,\\
\rho^2(\mathbf p)&=\frac{1-\cos p_1}{6}+\frac{1-\cos p_2}{2}.
\end{aligned}
\label{eq:2d_hc_rho}
\end{equation}
The only zero is at $\mathbf p=0$. Set $x=\sin(p_1/2)/\sqrt3$, $y=-\sin(p_2/2)$, and $\rho=(x^2+y^2)^{1/2}$.
Choose a smooth collar function $\theta(\mathbf p)$ that vanishes near the Dirac point, equals $\pi/2$ near the
square boundary, and is even under $p_1\mapsto-p_1$. The charge-preserving orbital rotation
\begin{equation}
R_\theta(\mathbf p)=\exp\!\left[-\frac{i\theta}{2\rho}
(y\sigma_x-x\sigma_y)\right]
\label{eq:2d_hc_rotation}
\end{equation}
produces the exactly isospectral Hamiltonian
\begin{equation}
h_\sharp(\mathbf p)=2t\left[\cos\theta(x\sigma_x+y\sigma_y)
+\rho\sin\theta\,\sigma_z\right].
\label{eq:2d_hsharp_explicit}
\end{equation}
Near the seams, $h_\sharp=2t\rho\sigma_z$. Since $\rho^2$ is smooth, periodic, and strictly positive there,
$\rho$ and all derivatives match across the identified boundaries. Consequently,
\begin{equation}
h_\sharp\in C^\infty(T_X^2),\quad h_\sharp^2=h_c^2,
\label{eq:2d_hsharp_smooth}
\end{equation}
while $h_\sharp=h_c$ near the Dirac point. The Fourier hopping matrices therefore satisfy
\begin{equation}
\|K_{\mathbf R}\|\le C_M(1+|\mathbf R|)^{-M}
\quad\text{for every }M>0.
\label{eq:2d_hc_schwartz}
\end{equation}
Thus Hamiltonian smoothing preserves the onsite charge and the exact spectrum while producing Schwartz-decaying
hopping. The Wilson-line prescription and operator-norm gauge-locality estimates apply as in the square model
(Appendix~\ref{app:2d_hc_asymptotic_proof}).

This smoothing does not make the microscopic reflection local. Before the rotation, the reflection acts as
$r_X:(p_1,p_2)\mapsto(-p_1,p_2)$ with orbital sewing $P_c=e^{ip_2/2}\sigma_y$. After the rotation,
\begin{equation}
\begin{aligned}
P_\sharp(\mathbf p)
&=R_\theta(r_X\mathbf p)P_c(\mathbf p)R_\theta(\mathbf p)^\dagger\\
&=e^{ip_2/2}\left[\cos\theta\,\sigma_y
+\sin\theta\left(\frac y\rho\sigma_z-i\frac x\rho\mathbf1\right)\right].
\end{aligned}
\label{eq:app_hcP_parity_seam}
\end{equation}
At the identified $p_1$ boundaries, its jump is
\begin{equation}
P_\sharp(\pi^-,p_2)-P_\sharp(-\pi^+,p_2)
=-\frac{2ie^{ip_2/2}}{\sqrt3\,\rho_\partial(p_2)}\mathbf1,
\end{equation}
where $\rho_\partial^2=\frac13+\sin^2\frac{p_2}{2}$. The nonvanishing jump prevents the original reflection from
acting as an ordinary local reflection in this charge-onsite regulator. This is consistent with the parity
anomaly of a single charged Dirac fermion.

The preceding construction makes the sharp charge onsite by smoothing the Hamiltonian. A complementary
construction leaves $H_{\rm hc}$ unchanged and instead smooths the compact charge. Its starting point is the
exact twisted pairing
\begin{equation}
f_{\rm hc}(u+\pi,v)=e^{iv}f_{\rm hc}(u,v)^*
\label{eq:2d_hc_twisted_relation}
\end{equation}
and hence the equality of paired energies,
\begin{equation}
E_{\rm hc}(u+\pi,v)=E_{\rm hc}(u,v).
\label{eq:2d_hc_spectral_matching}
\end{equation}
Fold under $u\mapsto u+\pi$ and write $h_{\rm fold}=\operatorname{diag}[h_{\rm hc}(u,v),h_{\rm hc}(u+\pi,v)]$,
with $\tau_i$ acting on the folded sectors. Define the sharp-charge numerator and the Hermitian unitary orbital
intertwiner,
\begin{equation}
\begin{aligned}
b_P(u,v)&=\frac12[\sin u+\sin(v-u)],\\
\Xi_{\rm tw}(v)&=\sin\frac v2\,\sigma_x+\cos\frac v2\,\sigma_y.
\end{aligned}
\label{eq:2d_hc_smooth_seed}
\end{equation}
The twist gives $h_{\rm hc}(u+\pi,v)=\Xi_{\rm tw}(v)h_{\rm hc}(u,v)\Xi_{\rm tw}(v)$, with $\Xi_{\rm
tw}^2=\mathbf1$. Consequently, the matrices $\tau_z\otimes\mathbf1$, $\tau_x\otimes\Xi_{\rm tw}$, and
$\tau_y\otimes\Xi_{\rm tw}$ commute with $h_{\rm fold}$, anticommute pairwise, and square to the identity. The
sharp folded involution is $q_s=\operatorname{sgn}(b_P)\tau_z\otimes\mathbf1$. A nonsingular conserved
extension, retaining its Dirac-number assignment, is
\begin{align}
q_{\rm loc}(u,v)=& \frac1{r_{\rm hc}(u,v)}\Big[b_P\,\tau_z\otimes\mathbf1
-\cos\frac v2\,\tau_x\otimes\Xi_{\rm tw}(v)\nonumber \\
&-\sin\left(u+\frac v2\right)\tau_y\otimes\Xi_{\rm tw}(v)\Big],
\label{eq:app_2d_qloc}
\end{align}
where
\begin{equation}
r_{\rm hc}^2=b_P^2+\cos^2\frac v2
+\sin^2\left(u+\frac v2\right).
\label{eq:2d_hc_smooth_norm}
\end{equation}
The bracketed numerator defines $B_{\rm hc}$, with $B_{\rm hc}^2=r_{\rm hc}^2\mathbf1$. The normalization cannot
vanish: simultaneous zeros of the transverse coefficients require $v=\pi$ and $u=\pi/2\pmod\pi$, where
$b_P=\pm1$. Moreover, the full matrix obeys the required sewing. Under $u\mapsto u+\pi$, the coefficient signs
combine with exchange of the folded sectors by $\tau_x$; under $v\mapsto v+2\pi$, the sign changes of $\Xi_{\rm
tw}$ cancel those of its half-angle coefficients. The Majorana-reality condition is also satisfied. Thus
\begin{equation}
q_{\rm loc}^2=\mathbf1,\quad [q_{\rm loc},h_{\rm fold}]=0.
\end{equation}
The globally sewn analytic matrix $q_{\rm loc}=iJ_{\rm loc}$ therefore defines an exactly compact, exponentially
local conserved charge $Q_{\rm loc}$ for the unchanged Hamiltonian.

As in the square model, the smooth and sharp involutions are related away from the sharp sign seams by the
Hamiltonian-preserving rotation
\begin{equation}
W_P=\frac{\mathbf1+q_{\rm loc}q_s}{\sqrt{2+\{q_{\rm loc},q_s\}}},
\quad \{q_{\rm loc},q_s\}=\frac{2|b_P|}{r_{\rm hc}}\mathbf1.
\label{eq:app_2d_W1}
\end{equation}
It obeys $W_Pq_sW_P^\dagger=q_{\rm loc}$ and $[W_P,h_{\rm fold}]=0$. Composing the sharp-charge relocalization
with $W_P^\dagger$ produces the full Majorana-canonical relocalization $\mathcal W$ of $Q_{\rm loc}$. Its
charge-onsite representation has the same Schwartz-local zero-field Hamiltonian and admits the compact
gauge-locality estimates described above.

The smooth charge carries nontrivial bundle topology. Keeping the transverse coefficients in
Eq.~\eqref{eq:app_2d_qloc} fixed, interpolate
\begin{equation}
b_\lambda(u,v)=(1-\lambda)\sin u+\lambda b_P(u,v),
\ \  0\le\lambda\le1.
\label{eq:2d_hc_charge_homotopy}
\end{equation}
The normalization stays nonzero: the transverse terms can vanish simultaneously only at $v=\pi$,
$u=\pi/2\pmod\pi$, where $b_\lambda=\pm1$. Folded sewing is preserved throughout, so the charge Chern number is
constant. At $\lambda=0$, let $r_0^2=\sin^2u+\cos^2(v/2)+\sin^2(u+v/2)$ and $\zeta=-\cos(v/2)-i\sin(u+v/2)$. A
positive-charge frame on $0<u<\pi$ is
\begin{equation}
V(u,v)=\frac1{\sqrt{2r_0(r_0+\sin u)}}
\begin{pmatrix}
(r_0+\sin u)\mathbf1\\[1mm]
\zeta(u,v)\Xi_{\rm tw}(v)
\end{pmatrix}.
\label{eq:app_2d_positive_frame}
\end{equation}
Across the folded boundary, the frame satisfies
\begin{equation}
\begin{aligned}
 V(\pi,v)=(\tau_x\otimes\mathbf1)& V(0,v)g(v),\\
g(v) =-e^{-iv/2}\Xi_{\rm tw}(v), &\quad \det g(v)=-e^{-iv}.
\end{aligned}
\label{eq:app_2d_seam_transition}
\end{equation}
The transition determinant winds once negatively. With the square-model orientation $dV\wedge dP$ and curvature
convention, $C_Q=-\operatorname{wind}(\det g)=+1$, consistent with the odd class-D charge-parity relation in
Eq.~\eqref{eq:2d_nu_CQ_relation}. By contrast, the onsite charge of $H_{\rm hc}^{W}$ has $C_Q=0$. An
orientation-reversing reflection preserving the smooth charge would reverse $C_Q$, so no such reflection can be
a symmetry of this nontrivial charge bundle. The reflection-invariant sharp charge and the exponentially local
smooth charge consequently have different microscopic crystalline-symmetry properties, even though both admit
the charge-onsite Schwartz-local completion. This conclusion concerns the specified compact charges, not a
universal obstruction to canonical relocalization or gauge locality.

\subsection{Gapless nodal transitions between class-D regulator sectors}
\label{app:2d_interpolation_paths}

We construct a finite-range interpolation between two gapless class-D nodal regimes with different Pfaffian
parities and numbers of complex Dirac fermions. The path preserves the microscopic honeycomb reflection and
remains gapless, passing through a quartic nodal touching. The Pfaffian distinction is stable provided no  PHIM
becomes gapless, whereas the quartic dispersion is fine-tuned.

Consider a bipartite two-sublattice Majorana Hamiltonian
\begin{equation}
h_b(u,v)=t\big[\operatorname{Im}f_b(u,v)\sigma_x
+\operatorname{Re}f_b(u,v)\sigma_y\big],
\end{equation}
where $f_b(-u,-v)=f_b(u,v)^*$. The honeycomb reflection $\mathsf r:(u,v)\mapsto(v-u,v)$ is imposed through
\begin{equation}
f_b(v-u,v)=e^{iv}f_b(u,v)^*.
\label{eq:app_path_parity_allb}
\end{equation}
The PHIM $\Gamma=(0,0)$ and $X=(\pi,0)$ are fixed by reflection, whereas $Y=(0,\pi)$ and $M=(\pi,\pi)$ are
exchanged. Since $f_b$ is real at the PHIM, the sewing relation gives
\begin{equation}
f_b(M)=-f_b(Y).
\label{eq:app_parity_TRIM_relation}
\end{equation}
We arrange the Pfaffian transition at the reflection-fixed point $X$.

Define the nonnegative trigonometric polynomial
\begin{equation}
\begin{aligned}
G_X(u,v)={}&(1+\cos u)+(1+\cos(v-u))\\
&+(1-\cos v).
\end{aligned}
\label{eq:app_path_gX}
\end{equation}
It is inversion- and reflection-invariant and vanishes only at $X$. Writing $u=\pi+x$, $v=y$, and $\mathbf
q=(x,y)$, we have
\begin{equation}
G_X(\pi+x,y)=x^2-xy+y^2+O(|\mathbf q|^4).
\label{eq:app_path_gX_quadratic}
\end{equation}
Introduce
\begin{equation}
F_\lambda^P(u,v)=(1-\lambda^2)e^{iu}G_X(u,v) +\lambda^2f_{\rm hc}(u,v),
\label{eq:app_path_FlambdaP}
\end{equation}
for $0\le\lambda\le1$, with $f_{\rm hc}$ from Eq.~\eqref{eq:2d_hc_f}. Both terms obey Majorana reality and
reflection sewing. At $\lambda=0$, $F_0^P=e^{iu}G_X$ has a quadratic zero at $X$; at $\lambda=1$, $F_1^P=f_{\rm
hc}$ has the Majorana-related nodes $(\pm\pi/2,\pi)$.

For every $0<\lambda\le1$, $F_\lambda^P$ has exactly one pair of simple Dirac nodes. To count them, set
$\alpha=1/\sqrt3$ and note that
\begin{equation}
\begin{aligned}
\operatorname{Im}(e^{-iu}f_{\rm hc})
={}&2\sin\left(u-\frac v2\right)\\
&\times\left[
\alpha\cos\left(u-\frac v2\right)
-\cos\frac v2\right].
\end{aligned}
\label{eq:app_path_im_check}
\end{equation}
On the branch $\cos(v/2)=\alpha\cos(u-v/2)$, the real part of $e^{-iu}f_{\rm hc}$ is $2\alpha>0$, excluding
zeros of $F_\lambda^P$. On the other branch, $\sin(u-v/2)=0$, only $u-v/2=\pi$ modulo $2\pi$ can support a zero.
With $r=\cos(v/2)\in[0,1]$, the remaining equation is
\begin{equation}
(1-\lambda^2)(1-r)(2+r)=\lambda^2r.
\end{equation}
It has a unique solution for each $0<\lambda\le1$, giving one Majorana-related pair of simple nodes. The pair
moves continuously from $X$ to $(\pm\pi/2,\pi)$ as $\lambda$ increases.

Fix $c\in(0,1)$ and define the interpolation for $0\le b\le1$ and $-1\le b\le0$, respectively, by
\begin{equation}
\begin{aligned}
f_b(u,v)
&=F_b^P(u,v)\big[(1-b^2)G_X(u,v)+b^2\big],\\
f_b(u,v)
&=F_{|b|}^P(u,v)e^{-iu}F_{c|b|}^P(u,v).
\end{aligned}
\label{eq:app_path_parity_z4}
\end{equation}
Both branches preserve the reflection sewing and meet continuously at
\begin{equation}
f_0(u,v)=e^{iu}G_X(u,v)^2.
\label{eq:app_path_z4critical}
\end{equation}
For $b>0$, the second factor is strictly positive, so the only nodes come from $F_b^P$: there is one complex
Dirac fermion, $n_{\rm D}=1$, and $f_{+1}=f_{\rm hc}$. For $b<0$, the factors $F_{|b|}^P$ and $F_{c|b|}^P$ have
distinct pairs of simple zeros since $c\ne1$, giving $n_{\rm D}=2$. Thus the path remains gapless on both sides
of the transition.

At $b=0$, the nodes merge at $X$. Equation~\eqref{eq:app_path_gX_quadratic} gives
\begin{equation}
E_0(\pi+x,y)
=|t|(x^2-xy+y^2)^2+O(|\mathbf q|^6).
\end{equation}
The leading dispersion is quartic, with dynamical exponent $z=4$.

The merger changes the Pfaffian parity. At $X$,
\begin{equation}
f_{b>0}(X)=2b^4>0,\quad
f_{b<0}(X)=-4c^2b^4<0,
\end{equation}
while the other PHIM remain nonsingular with unchanged signs. Consequently,
\begin{equation}
\mathcal P_D=
\begin{cases}
-1,&b>0,\quad n_{\rm D}=1,\\
+1,&b<0,\quad n_{\rm D}=2.
\end{cases}
\label{eq:app_path_gapless_pfaffian}
\end{equation}
The nodal regimes therefore differ in both Pfaffian parity and Dirac multiplicity. Within the bipartite
two-sublattice class, simple Dirac nodes are locally stable because they carry nonzero winding. More generally,
the class-D Pfaffian parity cannot change under a deformation that avoids zero-energy crossings at the PHIM,
regardless of the behavior of nodes elsewhere.

The quartic touching is not a symmetry-protected $z=4$ critical point. The finite-range perturbation $\delta
f=\varepsilon e^{iu}G_X$ preserves Majorana reality and reflection, but at $b=0$ gives
\begin{equation}
f_{0,\varepsilon}(u,v)
=e^{iu}G_X(u,v)\big[G_X(u,v)+\varepsilon\big].
\end{equation}
For $\varepsilon>0$, the leading dispersion near $X$ is quadratic. Thus the quartic exponent requires tuning
beyond reflection and bipartite symmetry. Other symmetry-preserving perturbations may separate the simultaneous
merger into simpler nodal transitions, while a change of Pfaffian parity still requires a PHIM closing.

For each $b\ne0$, the inversion-odd sign $s_{b,P}=\operatorname{sgn}[\sin u+\sin(v-u)]$ defines an exact compact
lattice charge with the Dirac-number assignment near the separated nodes. In the IR, it acts as the diagonal
particle-number $U(1)$ of the two Dirac fields for $b<0$, and as particle number for the single Dirac field for
$b>0$. At $b=0$, the nodes reach the sign discontinuity at $X$, so this identification is not uniform through
the transition. Reflection sewing alone does not imply the half-shift spectral pairing needed for the
smooth-charge construction, and no such locality claim is made for the interpolating family.

A gapless lattice regulator can change its number of complex Dirac fermions and its class-D Pfaffian parity
through a symmetry-preserving nodal merger at a particle--hole-invariant momentum, without passing through a
fully gapped phase. The present finite-range, reflection-preserving path realizes this possibility through a
fine-tuned quartic touching, whose dispersion is not itself symmetry protected.

\subsection{General construction of local compact symmetries from spectral pairing}
\label{app:2d_gauge_local_families}
The square $\pi$-flux model $H_{\rm sq}$ and the honeycomb representatives $H_{\rm hc}$ and $H_{\rm hc}^{W}$
exhibit a common mechanism: exact pairing of folded momentum sectors enlarges the quadratic Hamiltonian
centralizer and permits conserved compact charges that need not be manifest in the original lattice variables.
We first give sufficient conditions for such a charge to be exponentially local, then compare identical- and
twisted-sector realizations. Finally, we identify twisted families with precisely one complex Dirac fermion and
characterize their finite-range deformations. Spectral pairing, charge-bundle topology, Hamiltonian and gauge
locality, and Dirac-node multiplicity must be distinguished throughout.

Let $\mathbf g=\mathbf G/2$, where $\mathbf G$ is a reciprocal-lattice vector, and choose a reduced Brillouin
zone containing one representative of each pair $\{\mathbf k,\mathbf k+\mathbf g\}$. Write
\begin{equation}
h_0(\mathbf k)=h(\mathbf k),\quad h_1(\mathbf k)=h(\mathbf k+\mathbf g).
\end{equation}
Suppose the sectors are unitarily equivalent,
\begin{equation}
h_1(\mathbf k)=\Xi(\mathbf k)h_0(\mathbf k)\Xi(\mathbf k)^\dagger.
\label{eq:app_2d_general_intertwiner}
\end{equation}
Here $\Xi(\mathbf k)$ generalizes the intertwiner $\Xi(V)$ of the square model, but need not be Hermitian. For
the algebraic construction, it need only be sufficiently regular; exponential locality requires analytic
momentum-space data. We assume a folded gauge in which it obeys
\begin{equation}
\Xi(\mathbf k+\mathbf g)=\Xi(\mathbf k)^\dagger,
\label{eq:app_2d_intertwiner_sewing}
\end{equation}
possibly after a redefinition by a unitary commuting with $h_0$. This identifies the two folded charts and is
distinct from the Majorana-reality condition on the charge.

In the folded basis, set $h_{\rm fold}=\operatorname{diag}(h_0,h_1)$ and define
\begin{equation}
T_1=\begin{pmatrix}\mathbf1&0\\0&-\mathbf1\end{pmatrix},\ \
T_2=\begin{pmatrix}0&\Xi^\dagger\\\Xi&0\end{pmatrix},\ \
T_3=\begin{pmatrix}0&-i\Xi^\dagger\\i\Xi&0\end{pmatrix}.
\label{eq:app_2d_intertwiner_triplet}
\end{equation}
The intertwining relation gives $[T_a,h_{\rm fold}]=0$, while unitarity of $\Xi$ implies $
T_aT_b=\delta_{ab}\mathbf1+i\epsilon_{abc}T_c$ and $ \{T_a,T_b\}=2\delta_{ab}\mathbf1$. Thus the paired sectors
carry a Pauli algebra of conserved Hermitian involutions, independently of the detailed dispersion. For real
coefficient functions $b_a(\mathbf k)$, introduce
\begin{equation}
B(\mathbf k)=\sum_{a=1}^3 b_a(\mathbf k)T_a(\mathbf k),\ \ r(\mathbf k)=\left[\sum_{a=1}^3 b_a(\mathbf k)^2\right]^{1/2}.
\label{eq:app_2d_intertwiner_B}
\end{equation}
The Clifford identities give $[B,h_{\rm fold}]=0$ and $B^2=r^2\mathbf1$. If $r(\mathbf k)\ge\delta>0$ throughout
the real Brillouin zone, the normalized matrix
\begin{equation}
q_{\rm loc}(\mathbf k)=\frac{B(\mathbf k)}{r(\mathbf k)}
\label{eq:app_2d_intertwiner_q}
\end{equation}
is a conserved Hermitian involution, $q_{\rm loc}^2=\mathbf1$ and $[q_{\rm loc},h_{\rm fold}]=0$. This
establishes compactness at the level of each folded fiber; a physical Majorana charge also requires consistent
global sewing and reality.

Let $\tau_x$ act on the folded-sector index (with the orbital identity understood).
Equation~\eqref{eq:app_2d_intertwiner_sewing} gives
\begin{equation}
\begin{aligned}
T_1(\mathbf k+\mathbf g)&=-\tau_x T_1(\mathbf k)\tau_x,\\
T_2(\mathbf k+\mathbf g)&=\phantom{-}\tau_x T_2(\mathbf k)\tau_x,\\
T_3(\mathbf k+\mathbf g)&=-\tau_x T_3(\mathbf k)\tau_x.
\end{aligned}
\end{equation}
A sufficient coefficient sewing is
\begin{equation}
\begin{aligned}
b_1(\mathbf k+\mathbf g)&=-b_1(\mathbf k),\\
b_2(\mathbf k+\mathbf g)&=\phantom{-}b_2(\mathbf k),\\
b_3(\mathbf k+\mathbf g)&=-b_3(\mathbf k),
\end{aligned}
\label{eq:app_2d_intertwiner_b_sewing}
\end{equation}
so that $q_{\rm loc}(\mathbf k+\mathbf g)=\tau_x q_{\rm loc}(\mathbf k)\tau_x$. In addition, the charge must
obey the Majorana-reality condition
\begin{equation}
q_{\rm loc}(-\mathbf k)=-q_{\rm loc}(\mathbf k)^*.
\label{eq:app_2d_intertwiner_majorana}
\end{equation}
If the intertwiner and coefficient functions admit analytic continuation to a complex neighborhood of the real
momentum torus, the uniform nonvanishing bound ensures that $q_{\rm loc}$ is analytic in a possibly narrower
neighborhood. Its real-space kernel consequently decays exponentially. When the unnormalized matrix $B$ is a
Laurent polynomial it is finite range, but normalization generally yields exponential rather than finite-range
decay. The associated Majorana complex structure $J_{\rm loc}=-iq_{\rm loc}$ defines a compact $U(1)$ charge
with the appropriate centered quadratic normalization. These conditions provide a sufficient construction of an
exponentially local conserved charge; they do not establish onsite relocalizability or locality of its gauge
coupling.

The distinction is visible in two complementary realizations. First, consider identical folded sectors,
\begin{equation}
h(u,v)=\widetilde h(2u,v),
\label{eq:app_2d_hcstar_family_general}
\end{equation}
where $\widetilde h$ is a finite-range two-band Majorana Bloch Hamiltonian with the required reality structure.
Since $h(u+\pi,v)=h(u,v)$, the intertwiner may be chosen as $\Xi=\mathbf1$. In the corresponding Wilson--Dirac
coordinates, the sector-exchange charge is onsite, the matter Hamiltonian remains finite range, and compact
gauging is exactly local. The positive-charge bundle for this onsite construction is trivial, $C_Q=0$. A
continuous four-parameter family is
\begin{equation}
\begin{aligned}
h_{\alpha,\beta,\rho_1,\rho_2}&(u,v)=t\big[-\alpha\sin2u\,\sigma_x
+\beta\sin v\,\sigma_z\\
&+\big(\rho_1(1+\cos2u) +\rho_2(1-\cos v)\big)\sigma_y\big],
\end{aligned}
\label{eq:app_2d_hcstar_four_parameter}
\end{equation}
For $\alpha,\beta\ne0$ and $\rho_1,\rho_2>0$, the only zeros of this Hamiltonian are the Majorana-related points
$(\pm\pi/2,0)$. The nonzero linear coefficients give simple Dirac cones. Setting $\alpha=\beta=\rho_1=\rho_2=1$
recovers $H_{\rm hc}^{W}$, showing that the exactly gauge-local Wilson--Dirac representative belongs to a
finite-range family rather than being isolated.

Twisted sectors provide a contrasting realization. For a two-sublattice Majorana Hamiltonian specified by a
scalar structure factor $f$, let
\begin{equation}
h^{(f)}(u,v)=t\left[\operatorname{Im}f(u,v)\,\sigma_x
+\operatorname{Re}f(u,v)\,\sigma_y\right]
\end{equation}
and impose
\begin{equation}
f(u+\pi,v)=e^{iv}f(u,v)^*.
\label{eq:app_2d_twisted_family_condition}
\end{equation}
The twisted-sector intertwiner is the Hermitian unitary
\begin{equation}
\Xi_{\rm tw}(v)=\sin\frac v2\,\sigma_x+\cos\frac v2\,\sigma_y
\end{equation}
It satisfies $\Xi_{\rm tw}(v)^2=\mathbf1$ and implements the sector equivalence,
\begin{equation}
h^{(f)}(u+\pi,v)=\Xi_{\rm tw}(v)h^{(f)}(u,v)\Xi_{\rm tw}(v).
\label{eq:app_2d_twisted_intertwiner}
\end{equation}
Hence the same folded centralizer algebra exists independently of the hopping coefficients. For the globally
sewn analytic charge texture constructed for $H_{\rm hc}$ in Appendix~\ref{app:2d_hc_charge_bundle}, this
mechanism yields an exponentially local compact charge with $C_Q=1$ in the chosen orientation. This invariant
characterizes that charge texture, not spectral pairing in general. In particular, neither the twist nor nonzero
charge Chern number alone establishes onsite relocalizability or gauge locality.

The twisted relation also does not fix the number of Dirac nodes. To obtain precisely one massless complex Dirac
fermion, consider the subclass
\begin{equation}
\begin{aligned}
f_{F,G}(u,v)=2e^{iv/2}\bigg[&F(u,v)\cos\frac v2\\
&+iG(u,v)\sin\left(u-\frac v2\right)\bigg].
\end{aligned}
\label{eq:app_2d_one_dirac_family}
\end{equation}
Here $F$ and $G$ are real-analytic, periodic functions satisfying
\begin{equation}
\begin{aligned}
F(u+\pi,v)&=F(u,v),& G(u+\pi,v)&=G(u,v),\\
F(-u,-v)&=F(u,v),& G(-u,-v)&=G(u,v),
\end{aligned}
\label{eq:app_2d_one_dirac_AB_sym}
\end{equation}
and the strict positivity conditions
\begin{equation}
F(u,v)\ge F_0>0,\quad G(u,v)\ge G_0>0.
\label{eq:app_2d_one_dirac_AB_positive}
\end{equation}
The half-shift symmetries imply Eq.~\eqref{eq:app_2d_twisted_family_condition}, while inversion gives the
Majorana condition $f(-u,-v)=f(u,v)^*$. Trigonometric-polynomial choices of $F,G$ yield finite-range
Hamiltonians; analytic choices yield exponentially decaying hopping.

Because $F$ and $G$ are strictly positive, zeros of $f_{F,G}$ require both $\cos(v/2)=0$ and $\sin(u-v/2)=0$.
The only nodes are therefore $\mathbf K_{\pm}=(\pm\pi/2,\pi)$, a single Majorana-related pair. Their simplicity
follows directly by setting $X=F\cos(v/2)$ and $Y=G\sin(u-v/2)$. At $\mathbf K_+$, the Jacobian is
\begin{equation}
\frac{\partial(X,Y)}{\partial(u,v)}
 =\begin{pmatrix}0&-F_+/2\\G_+&-G_+/2\end{pmatrix},
\label{eq:app_2d_one_dirac_jacobian_plus}
\end{equation}
where $F_+=F(\mathbf K_+)$ and $G_+=G(\mathbf K_+)$. The determinant is nonzero, while the partner node has the
opposite Jacobian sign. More explicitly,
\begin{equation}
|f_{F,G}|^2=4\left[F^2\cos^2\frac v2
+G^2\sin^2\left(u-\frac v2\right)\right],
\end{equation}
and near $\mathbf K_+$,
\begin{equation}
|f_{F,G}|^2= F_+^2\delta v^2
+4G_+^2\left(\delta u-\frac{\delta v}{2}\right)^2+O(|\delta\mathbf k|^3).
\label{eq:app_2d_one_dirac_metric}
\end{equation}
The positive-definite quadratic form confirms a simple Dirac cone at each node. Together, the two
Majorana-related nodes describe one massless complex Dirac fermion.

For constant positive $F=a$ and $G=b$, this construction reduces to
\begin{equation}
f_{a,b}(u,v)=a(1+e^{iv})
+b\big(e^{iu}-e^{i(v-u)}\big).
\label{eq:app_2d_hc_ab_family}
\end{equation}
The auxiliary honeycomb model $H_{\rm hc}$ is obtained at $a=1$, $b=1/\sqrt3$, for which the cone is isotropic
in the physical honeycomb metric. After shifting the momentum origin appropriately, the square $\pi$-flux model
$H_{\rm sq}$ has the same two-band folding form at $a=b=1$. This correspondence is algebraic and does not equate
the two microscopic lattice embeddings or spatial metrics.

To characterize the broader finite-range twisted family, expand
\begin{equation}
f(u,v)=\sum_{m,n\in\mathbb Z}t_{mn}e^{i(mu+nv)},
\label{eq:app_2d_twisted_fourier}
\end{equation}
with finitely many real Fourier coefficients $t_{mn}$. The twisted sewing relation is equivalent to
$t_{-m,1-n}=(-1)^m t_{mn}$. The map $I:(m,n)\mapsto(-m,1-n)$ has two-element orbits and no fixed points on
$\mathbb Z^2$. Choosing one representative from each orbit in a set $\mathcal R$, every finite-range solution
takes the form
\begin{equation}
f(u,v)=\sum_{(m,n)\in\mathcal R}\lambda_{mn}\big[ e^{i(mu+nv)} +(-1)^m e^{i[-mu+(1-n)v]}\big],
\label{eq:app_2d_twisted_general_laurent}
\end{equation}
with finitely many nonzero real $\lambda_{mn}$. Allowing exponentially or rapidly decaying coefficients gives
the corresponding exponentially local or Schwartz-decaying hopping kernels. The twist constrains their Fourier
coefficients but does not, by itself, control the zeros of $f$.

Finally, the single-Dirac property is stable under sufficiently small perturbations within the twisted class.
For a reference $f_0$ with exactly two simple zeros, choose disjoint disks $D_\pm$ around them such that the
Jacobian is nonzero and has a fixed sign throughout each disk, while $f_0$ is nonzero on their boundaries.
Compactness gives $m_0=\min_{\mathbf k\notin D_+\cup D_-}|f_0(\mathbf k)|>0$. A perturbation with $\|\delta
f\|_\infty<m_0/2$ creates no zeros outside the disks. If it is also sufficiently small in $C^1$, each disk
retains its Jacobian sign and boundary winding, so degree conservation gives exactly one simple zero in each
disk. Thus strict positivity of $F$ and $G$ is a convenient sufficient condition for one complex Dirac fermion,
but not a necessary one.

\subsection{Charge topology, Pfaffian parity, and bulk--edge correspondence}
\label{app:2d_sq_charge_bundle}

We compute the charge Chern number of $Q_{\rm loc}$, establish its relation to gapped class-D topology and
massless Pfaffian parity, and derive the resulting charge-spread bound and bulk--edge correspondence. The
construction and exponential locality of $Q_{\rm loc}$ are given in Sec.~\ref{sec:2d_sq}.

\subsubsection{Integer Chern number of the smooth charge}
\label{app:2d_sq_CQ_integer}

Set $\theta=V/2$ and write the unflattened charge matrix of Eq.~\eqref{eq:2d_sq_B} as
\begin{equation}
\begin{aligned}
B_{\rm sq}=& \sum_{a=1}^3 d_a(U,V)T_a(V),\\
\mathbf d=\bigl(\cos\theta\sin(U-\theta),& \ \sin\theta,\ -\cos(U+\theta)\bigr).
\end{aligned}
\label{eq:app_sq_CQ_dvector}
\end{equation}
The Clifford relations imply $B_{\rm sq}^2=|\mathbf d|^2\mathbf1$, so the flattened charge is $q_{\rm
loc}=\sum_a n_aT_a$, where $\mathbf n=\mathbf d/|\mathbf d|$. The folded domain is $0\le U\le\pi$, $0\le
V\le2\pi$, with sewing $S=\tau_x\otimes\mathbf1$ across $U=0,\pi$. In the periodic doubled-cell basis of
Eq.~\eqref{eq:2d_sq_periodic_cell}, $\tilde q=F^\dagger q_{\rm loc}F$ and $P_+^Q=(\mathbf1+\tilde q)/2$. With
the orientation $dV\wedge dP$, $P=2U$ on the torus $T_Q^2$, and the normalization of
Eq.~\eqref{eq:2d_sq_CQ_def},
\begin{equation}
C_Q=\frac{1}{2\pi i}\int_{T_Q^2}\operatorname{Tr}\!\left(P_+^Q\,dP_+^Q\wedge dP_+^Q\right).
\label{eq:app_sq_CQ_integral_appendix}
\end{equation}
The same Chern number can be computed in the sewn folded frame, using $p_+=(\mathbf1+q_{\rm loc})/2$. In this
frame the Clifford generators are momentum dependent: $T_1=\tau_z\otimes\mathbf1$, $T_2=\tau_x\otimes\Xi(V)$,
and $T_3=\tau_y\otimes\Xi(V)$ [Eq.~\eqref{eq:2d_sq_commuting_triplet}]. To account for their derivatives, write
$A=n_2\tau_x+n_3\tau_y$, so that $q_{\rm loc}=n_1\tau_z\otimes\mathbf1+A\otimes\Xi$. The extra term in
$\partial_Vq_{\rm loc}$ is $A\otimes\partial_V\Xi$. Since $\Xi^2=\mathbf1$ and $\operatorname{tr}_\sigma\Xi=0$,
\begin{equation}
\operatorname{tr}_\sigma(\partial_V\Xi)=0,
\quad
\operatorname{tr}_\sigma(\Xi\,\partial_V\Xi)
=\tfrac12\partial_V\operatorname{tr}_\sigma(\Xi^2)=0.
\end{equation} 
The tensor-product structure then gives $\operatorname{Tr}\bigl(q_{\rm loc}[\partial_Uq_{\rm
loc},A\otimes\partial_V\Xi]\bigr)=0$: all orbital traces in this term reduce to the two identities above. The
remaining contribution is computed from $\operatorname{Tr}(T_aT_bT_c)=4i\epsilon_{abc}$, yielding
\begin{equation}
\begin{aligned}
\operatorname{Tr}\bigl(p_+[\partial_Up_+,\partial_Vp_+]\bigr)
&=\tfrac18\operatorname{Tr}\bigl(q_{\rm loc}[\partial_Uq_{\rm loc},\partial_Vq_{\rm loc}]\bigr)\\
&=i\,\mathbf n\cdot(\partial_U\mathbf n\times\partial_V\mathbf n).
\end{aligned}
\end{equation} 
Because $dV\wedge dP$ is opposite to $dU\wedge dV$, and $\mathbf n\cdot(\partial_U\mathbf
n\times\partial_V\mathbf n) =\mathbf d\cdot(\partial_U\mathbf d\times\partial_V\mathbf d)/|\mathbf d|^3$, we
obtain
\begin{equation}
C_Q=-\frac{1}{2\pi}\int_0^\pi dU\int_0^{2\pi}dV\,
\frac{\mathbf d\cdot(\partial_U\mathbf d\times\partial_V\mathbf d)}{|\mathbf d|^3}.
\label{eq:app_sq_CQ_sphere_integral}
\end{equation}
We evaluate this integral by a nonsingular homotopy. For $0\le t\le1$, define
\begin{equation}
\begin{aligned}
d_{1,t}&=(1-t)\cos\theta\sin(U-\theta)+t\sin U,\\
d_{2,t}&=\sin\theta,\\
d_{3,t}&=-(1-t)\cos(U+\theta)-t\cos U\cos\theta.
\end{aligned}
\label{eq:app_sq_CQ_homotopy}
\end{equation}
The deformation preserves the folded and Majorana sewing: under $U\mapsto U+\pi$, the first and third
coefficients change sign as required by conjugation with $S$; under $V\mapsto V+2\pi$, the second and third
coefficients change sign together with $\Xi(V)$. Moreover, $d_{2,t}>0$ for $0<\theta<\pi$, while $|\mathbf
d_t|=1$ at $\theta=0,\pi$. Thus $\mathbf d_t$ never vanishes and $C_Q$ is unchanged. At $t=1$,
\begin{equation}
\begin{aligned}
& \mathbf d_{\rm s}=(\sin U,\sin\theta,-\cos U\cos\theta),\\
& |\mathbf d_{\rm s}|^2=1+\sin^2U\sin^2\theta.
\end{aligned}
\end{equation}
Direct differentiation gives $\mathbf d_{\rm s}\cdot(\partial_U\mathbf d_{\rm s}\times\partial_\theta\mathbf
d_{\rm s}) =-(1-\sin^2U\sin^2\theta)$. Using $V=2\theta$ in Eq.~\eqref{eq:app_sq_CQ_sphere_integral},
\begin{equation}
C_Q=\frac{1}{2\pi}\int_0^\pi dU\int_0^\pi d\theta\,
\frac{1-\sin^2U\sin^2\theta}{(1+\sin^2U\sin^2\theta)^{3/2}}.
\label{eq:app_sq_CQ_elementary_integral}
\end{equation}
The normalized map $\mathbf d_{\rm s}/|\mathbf d_{\rm s}|$ covers the quarter-sphere $n_1\ge0$, $n_2\ge0$ twice
with negative orientation; the two generic preimages are related by $(U,\theta)\mapsto(\pi-U,\pi-\theta)$. Since
a quarter-sphere has area $\pi$, the total unsigned area is $2\pi$. Including the orientation in
Eq.~\eqref{eq:app_sq_CQ_sphere_integral} gives
\begin{equation}
C_Q=+1.
\label{eq:app_sq_CQ_integral_result}
\end{equation}
This determines the integer left undetermined by Pfaffian parity. Reversing $d_3$ changes the orientation of the
target sphere and gives $C_Q=-1$, without changing $H_{\rm sq}$ or the charge assignment at the positive Dirac
node.

\subsubsection{Charge topology, class-D invariant, and skyrmion shifts}
\label{app:2d_sq_gapped_characteristic}

Let $E\to T^2$ be a finite-rank Hermitian Bloch bundle equipped with Majorana particle--hole conjugation
$\mathcal C_{\mathbf k}:E_{\mathbf k}\to E_{-\mathbf k}$. Consider a fully gapped Hamiltonian $H$ commuting with
a smooth Hermitian charge involution $q$, satisfying
\begin{equation}
\begin{aligned}
&H^2\ge\Delta^2\mathbf1,\quad \Delta>0,\quad
q^2=\mathbf1,\quad [H,q]=0,\\
&\mathcal C_{\mathbf k}H(\mathbf k)\mathcal C_{\mathbf k}^{-1}
=-H(-\mathbf k),\\
&\mathcal C_{\mathbf k}q(\mathbf k)\mathcal C_{\mathbf k}^{-1}
=-q(-\mathbf k).
\end{aligned}
\label{eq:2d_charge_bundle_hypotheses}
\end{equation}
Define the occupied-energy projector $P_{\rm occ}=\mathbf1_{(-\infty,0)}(H)$ and the charge projectors
$P_\pm^Q=(\mathbf1\pm q)/2$. Let $\mathcal O=\operatorname{Im}P_{\rm occ}$ be the occupied bundle, $\mathcal
E_+=\operatorname{Im}P_+^Q$ the positive-charge bundle, and $\mathcal O_+=\operatorname{Im}(P_{\rm occ}P_+^Q)$
its occupied positive-charge subbundle. Since $[P_{\rm occ},P_+^Q]=0$, we have
\begin{equation}
\mathcal E_+=\mathcal O_+\oplus\mathcal U_+,\quad
\mathcal U_+=\operatorname{Im}[(\mathbf1-P_{\rm occ})P_+^Q],
\label{eq:2d_charge_positive_split}
\end{equation}
where $\mathcal U_+$ consists of unoccupied positive-charge states.

Similarly, let $\mathcal O_-=\operatorname{Im}(P_{\rm occ}P_-^Q)$ denote the occupied negative-charge bundle.
Particle--hole conjugation reverses both energy and charge, identifying
\begin{equation}
\mathcal O_-\simeq\iota^*\overline{\mathcal U_+},
\label{eq:2d_sq_reality_bundle}
\end{equation}
where $\iota(\mathbf k)=-\mathbf k$ is momentum inversion, $\iota^*$ denotes pullback, and the overline denotes
the complex-conjugate bundle. Thus $(\iota^*\overline{\mathcal U_+})_{\mathbf k}=\overline{(\mathcal
U_+)_{-\mathbf k}}$, and the occupied bundle decomposes as
\begin{equation}
\mathcal O\simeq\mathcal O_+\oplus\iota^*\overline{\mathcal U_+}.
\label{eq:2d_occupied_bundle_decomposition}
\end{equation}
Using $c_1(\overline F)=-c_1(F)$ together with the positive-charge decomposition gives
\begin{equation}
c_1(\mathcal O)
=(1+\iota^*)c_1(\mathcal O_+)-\iota^*c_1(\mathcal E_+).
\label{eq:2d_general_c1_identity}
\end{equation}
Since momentum inversion reverses both coordinates of the 2D torus, $\iota^*$ acts trivially on $H^2(T^2;\mathbb
Z)$. Consequently,
\begin{equation}
c_1(\mathcal O)=2c_1(\mathcal O_+)-c_1(\mathcal E_+).
\label{eq:2d_c1_bundle_theorem}
\end{equation}
Pairing with the oriented fundamental class yields $\nu_{\rm Maj}=2C_{+,\mathrm{occ}}-C_Q$, as stated in
Eq.~\eqref{eq:2d_nu_CQ_relation}. In particular, $C_Q\equiv\nu_{\rm Maj}\pmod2$. This identity requires a gap at
zero energy but allows arbitrary degeneracies within the occupied and unoccupied subspaces.

The factor of two has a geometric interpretation: at fixed occupied bundle, changes in charge topology can be
represented by particle--hole-related skyrmion textures. To make this explicit, spectrally flatten the
Hamiltonian,
\begin{equation}
H_{\rm flat}=\mathbf1-2P_{\rm occ}.
\label{eq:2d_Hflat_charge_skyrmion}
\end{equation}
which preserves the occupied bundle and $\nu_{\rm Maj}$. We keep $P_{\rm occ}$ fixed and compare distinct
smooth, Majorana-compatible charge involutions commuting with $H_{\rm flat}$.

Choose a disk $D$ around a generic momentum $\mathbf k_0$, with $D\cap(-D)=\varnothing$ and neither disk
containing a PHIM. Suppose the occupied subspace contains a two-dimensional charge-invariant sector with one
state of each charge. In a local orthonormal frame, its charge restriction is $q_{\rm occ}=\mathbf
n\cdot\boldsymbol\sigma$, with $|\mathbf n|=1$. Two textures agreeing on $\partial D$ can differ by a relative
skyrmion number $m\in\mathbb Z$, defined by the degree of the map $S^2\to S^2$ obtained by gluing their disks
along the common boundary. A representative texture is
\begin{equation}
\mathbf n(r,\phi)=
\bigl(\sin f(r)\cos m\phi,\,
\sin f(r)\sin m\phi,\,
\cos f(r)\bigr),
\end{equation}
with a smooth radial profile and boundary conditions chosen to give the specified signed degree. The occupied
positive-charge line then acquires the relative Chern-number shift
\begin{equation}
\Delta c_1^{\rm rel}(\mathcal O_+;D)=m.
\label{eq:2d_skyrmion_Oplus}
\end{equation}
Here the relative invariant compares textures with identical boundary data; the restriction of a bundle to a
disk has no nontrivial absolute first Chern class.

Because $P_{\rm occ}$ is unchanged, the complementary occupied negative-charge sector carries the opposite
shift:
\begin{equation}
\Delta c_1^{\rm rel}(\mathcal O_-;D)=-m.
\label{eq:2d_skyrmion_Ominus}
\end{equation}
The particle--hole identification in Eq.~\eqref{eq:2d_sq_reality_bundle} relates this change to the unoccupied
positive-charge sector around $-D$. Complex conjugation reverses the first Chern class, whereas momentum
inversion preserves orientation in 2D, so
\begin{equation}
\Delta c_1^{\rm rel}(\mathcal U_+;-D)=m.
\label{eq:2d_skyrmion_Wplus}
\end{equation}
The two positive-charge contributions therefore add. Using $\mathcal E_+=\mathcal O_+\oplus\mathcal U_+$, we
obtain
\begin{equation} 
\Delta C_{+,\mathrm{occ}} =m,\ \ 
\Delta C_Q=2m,\ \ 
\Delta\nu_{\rm Maj}=0. 
\label{eq:2d_skyrmion_Chern_shifts}
\end{equation}
Thus Majorana particle--hole pairing explains geometrically why the charge Chern number can differ by even
integers at fixed class-D Chern number. These are distinct charge topologies, not configurations connected by a
smooth homotopy of involutions. Moreover, compatibility with the flattened Hamiltonian does not guarantee that
the corresponding charges commute with a specified unflattened Hamiltonian or satisfy its locality constraints.

\subsubsection{Pfaffian parity and charge Chern topology}
\label{app:2d_sq_gapless_parity}

Consider a Majorana Hamiltonian $H$ commuting with a smooth compact charge $q$, both represented in a common
periodic Majorana Bloch frame. In this general statement $q$ denotes an arbitrary charge involution; for the
explicit square model below it is $\tilde q$. At a PHIM $\Lambda$, define
\begin{equation}
A=-iH(\Lambda),\quad J=-iq(\Lambda).
\label{eq:app_gapless_AJS}
\end{equation}
Since $A$ and $J$ are real antisymmetric, $[A,J]=0$, and $J^2=-\mathbf1$, they can be simultaneously reduced to
two-dimensional blocks $A_a=\lambda_aJ_a$. Here $\lambda_a$ are the energies of
$H_+=H|_{\operatorname{Im}P_+^Q}$. Taking Pfaffians in the same Majorana basis gives
\begin{equation}
\operatorname{Pf}A=\operatorname{Pf}J\,\det_{\mathbb C}H_+.
\label{eq:app_gapless_pf_identity}
\end{equation}
If $H(\Lambda)$ is nonsingular and $r_+(\Lambda)$ counts the negative-energy eigenvalues of $H_+$, then
\begin{equation}
\operatorname{sgn}\!\left(\frac{\operatorname{Pf}A}{\operatorname{Pf}J}\right)
=(-1)^{r_+(\Lambda)}.
\label{eq:app_gapless_pf_ratio}
\end{equation}

The charge Pfaffians also determine the parity of $C_Q$, independently of the physical spectrum. Majorana
reality requires
\begin{equation}
P_+^Q(-\mathbf k)=\mathbf1-\overline{P_+^Q(\mathbf k)},
\label{eq:app_gapless_projector_reality}
\end{equation}
which makes the charge Berry curvature invariant under momentum inversion. On a common periodic momentum torus
with coordinates $(k_1,k_2)$, its integral over $B=[0,\pi]_{k_1}\times S_{k_2}^1$ is therefore $C_Q/2$. To apply
Stokes' theorem, choose a smooth periodic orthonormal frame $u_a(k_2)$ of the positive-charge subspace on each
boundary circle $k_1=a$, $a=0,\pi$, and define the determinant Berry connection and phase,
\begin{equation}
\begin{aligned}
\mathcal A_a(k_2)&=i\operatorname{Tr}\!\left[u_a^\dagger(k_2)\partial_{k_2}u_a(k_2)\right],\\
\gamma_a&=\int_0^{2\pi}dk_2\,\mathcal A_a(k_2)\pmod{2\pi}.
\end{aligned}
\end{equation}
Majorana sewing quantizes $\gamma_a$ to $0$ or $\pi$. Comparing the positive-charge frame with its
Majorana-conjugate partner at opposite momenta relates this phase to the endpoint Pfaffians:
\begin{equation}
e^{i\gamma_a}=\operatorname{sgn}\operatorname{Pf}J(a,0)\,
\operatorname{sgn}\operatorname{Pf}J(a,\pi).
\label{eq:app_gapless_1d_pf_berry}
\end{equation}
With the orientation used for $C_Q$, Stokes' theorem gives $e^{i\pi C_Q}=e^{i(\gamma_\pi-\gamma_0)}$, and hence
\begin{equation}
(-1)^{C_Q}=\prod_\Lambda\operatorname{sgn}\operatorname{Pf}J(\Lambda).
\label{eq:app_gapless_charge_chern_pf}
\end{equation}
This charge identity needs no energy gap. Combining it with Eq.~\eqref{eq:app_gapless_pf_ratio} at four
nonsingular PHIM yields
\begin{equation}
\mathcal P_D(H)=(-1)^{C_Q+\sum_\Lambda r_+(\Lambda)}.
\label{eq:app_gapless_general_theorem}
\end{equation}
If the PHIM belong to one connected nonsingular region, $r_+$ is constant and its total contribution is even,
giving
\begin{equation}
\mathcal P_D(H)=(-1)^{C_Q}.
\label{eq:app_gapless_same_torus}
\end{equation}
In particular, isolated nodes away from the PHIM do not affect this relation.

For $H_{\rm sq}$, the physical parity is defined in the primitive-cell frame, while the smooth charge $\tilde q$
of Eq.~\eqref{eq:2d_sq_periodic_cell} is periodic on the doubled-cell torus $T_Q^2$. Folding brings the Dirac
nodes to $(P,V)=(\pi,0)$, as illustrated in Fig.~\ref{fig:patched}; the folded physical Pfaffian vanishes there.
In the canonical periodic basis of Eq.~\eqref{eq:2d_sq_F}, let
\begin{equation}
\mathcal H_{\rm sq}(P,V)=F(U)^\dagger h_{\rm fold}(U,V)F(U),
\label{eq:app_gapless_fold_H}
\end{equation}
where $P=2U$. At $P=0$, $F(0)$ is real orthogonal with determinant $+1$, so
\begin{equation}
\operatorname{Pf}[-i\mathcal H_{\rm sq}(0,V)]
=\operatorname{Pf}[-ih_{\rm sq}(0,V)]\,
\operatorname{Pf}[-ih_{\rm sq}(\pi,V)],
\label{eq:app_gapless_fold_pf_P0}
\end{equation}
with $ V=0,\pi$. Thus the original-cell Pfaffian parity is the product of the two folded physical Pfaffian signs
on the $P=0$ slice. Define $\mathcal J=-i\tilde q$ and
\begin{equation}
\begin{aligned}
\delta_0&=\prod_{V=0,\pi}\operatorname{sgn}\operatorname{Pf}\mathcal J(0,V),\\
\delta_\pi&=\prod_{V=0,\pi}\operatorname{sgn}\operatorname{Pf}\mathcal J(\pi,V).
\end{aligned}
\label{eq:2d_sq_delta_slices}
\end{equation}
The positive-charge negative-energy index $r_+$ is the same at the two nonsingular $P=0$ PHIM. Their factors
$(-1)^{r_+}$ cancel in Eq.~\eqref{eq:app_gapless_pf_ratio}, giving
$
\mathcal P_D(H_{\rm sq})=\delta_0.
$
Applying Eq.~\eqref{eq:app_gapless_charge_chern_pf} to $\tilde q$ on $T_Q^2$ gives
$
(-1)^{C_Q}=\delta_0\delta_\pi,
$
and hence the folded relation
\begin{equation}
\mathcal P_D(H_{\rm sq})=(-1)^{C_Q}\delta_\pi.
\label{eq:app_gapless_fold_general}
\end{equation}

The remaining factor is already fixed by the square-model charge Pfaffians in Eq.~\eqref{eq:2d_sq_q_pfaffians}.
Both $P=\pi$ PHIM have charge-Pfaffian sign $-1$, including the nodal point $(\pi,0)$, where $\tilde q$ remains
a nonsingular involution. Therefore
\begin{equation}
\delta_\pi=(-1)(-1)=+1,
\label{eq:app_gapless_delta_pi}
\end{equation}
and
\begin{equation}
\mathcal P_D(H_{\rm sq})=(-1)^{C_Q}.
\label{eq:app_gapless_folded_dirac_number_theorem}
\end{equation}
This relates the primitive-cell physical Pfaffian parity to the doubled-cell charge Chern parity at the massless
point without assigning a sign to the vanishing folded physical Pfaffian. The extra factor $\delta_\pi$ is not
generally fixed by Majorana reality; its value here follows from the square-model Dirac-number charge and the
canonical folding convention.

\subsubsection{Topological bound on charge spread}
\label{app:2d_charge_spready}

We derive the lower bound on the charge spread in Eq.~\eqref{eq:2d_charge_spread_bound}. Denote the two periodic
momentum coordinates by $\mathbf k=(k_1,k_2)$ and consider the periodic doubled-cell smooth Hermitian involution
$\tilde q(\mathbf k)$, with Fourier coefficients $q_{\mathbf R}$ as in Eq.~\eqref{eq:2d_charge_spread_def}. The
bound depends only on the charge projector and does not require an energy gap in the physical Hamiltonian.

Since $\tilde q^2=\mathbf1$, Parseval's identity fixes the denominator of Eq.~\eqref{eq:2d_charge_spread_def} to
the Bloch-fiber dimension $d_{\rm fib}$. It also gives
\begin{equation}
\begin{aligned}
&\sum_{\mathbf R}|\mathbf R|^2
\operatorname{tr}(q_{\mathbf R}^\dagger q_{\mathbf R})\\
&=\frac{1}{(2\pi)^2}\int_{T_Q^2}d^2k\,
\operatorname{tr}\!\left[
(\partial_1\tilde q)^2+
(\partial_2\tilde q)^2
\right].
\end{aligned}
\label{eq:app_charge_spread_parseval}
\end{equation}
Using $\tilde q=2P_+^Q-\mathbf1$, the integrand is four times $\operatorname{tr}[(\partial_1P_+^Q)^2+
(\partial_2P_+^Q)^2]$.

To relate this expression to the charge Berry curvature, define $A_i=(\mathbf1-P_+^Q)(\partial_iP_+^Q)P_+^Q$.
Differentiating $(P_+^Q)^2=P_+^Q$ gives $\partial_iP_+^Q=A_i+A_i^\dagger$, with $A_i^2=(A_i^\dagger)^2=0$, and
hence $\operatorname{tr}[(\partial_iP_+^Q)^2] =2\operatorname{tr}(A_i^\dagger A_i)$. The Berry curvature
$F_Q=-i\operatorname{tr}P_+^Q [\partial_1P_+^Q,\partial_2P_+^Q]$ satisfies
\begin{equation}
|F_Q|
\le 2|\operatorname{tr}(A_1^\dagger A_2)|
\le \operatorname{tr}(A_1^\dagger A_1+
A_2^\dagger A_2)
\end{equation}
by the Cauchy--Schwarz inequality. Therefore,
\begin{equation}
\operatorname{tr}\!\left[
(\partial_1P_+^Q)^2+
(\partial_2P_+^Q)^2
\right]\ge 2|F_Q|.
\label{eq:app_metric_curvature_bound}
\end{equation}
Integrating over the momentum torus and using $|C_Q|\le(2\pi)^{-1}\int_{T_Q^2}|F_Q|\,d^2k$ yields
\begin{equation}
\ell_Q^2\ge\frac{4|C_Q|}{\pi d_{\rm fib}}.
\label{eq:app_charge_spread_bound}
\end{equation}
For $H_{\rm sq}$, the momentum coordinates are $(k_1,k_2)=(P,V)$, with $d_{\rm fib}=4$ and $C_Q=1$. Thus
$\ell_Q^2\ge1/\pi$ in doubled-cell lattice units.

\subsubsection{Charge bulk--edge correspondence}
\label{app:2d_charge_bulk_edge}

We explain how the charge Chern number $C_Q$ enforces boundary spectral flow and obstructs an exactly compact
boundary charge. As in the main text, $q_{\rm loc}$ without momentum arguments denotes the real-space
one-particle charge operator. Let $R$ project onto a half-plane and define $q_{\rm cut}=Rq_{\rm loc}R$ on
$\operatorname{Ran}R$. Although $q_{\rm loc}^2=\mathbf1$ in the bulk, compression generally destroys this
involution property:
\begin{equation}
q_{\rm cut}^2
=R-Rq_{\rm loc}(\mathbf1-R)q_{\rm loc}R.
\label{eq:app_qcut_square}
\end{equation}
The positive correction $B_R=Rq_{\rm loc}(\mathbf1-R)q_{\rm loc}R$ measures charge couplings crossing the
spatial cut. For a normalized eigenstate $q_{\rm cut}|\phi\rangle=\lambda|\phi\rangle$,
Eq.~\eqref{eq:app_qcut_square} gives
\begin{equation}
1-\lambda^2
=\|(\mathbf1-R)q_{\rm loc}|\phi\rangle\|^2.
\end{equation}
Thus deviations from the bulk charge eigenvalues $\pm1$ measure the leakage of the charge eigenstate into the
discarded half-plane.

Because $q_{\rm loc}$ is exponentially local, this leakage is confined near the boundary. Indeed, if $\|(q_{\rm
loc})_{\mathbf r\mathbf r'}\|
\le Ce^{-\mu|\mathbf r-\mathbf r'|}$,
then
\begin{equation}
\|(B_R)_{\mathbf r\mathbf r'}\|
\le C'e^{-\mu'[d(\mathbf r)+d(\mathbf r')]},
\label{eq:app_boundary_local_B}
\end{equation}
where $d(\mathbf r)$ is the distance to the cut and $C',\mu'>0$. The estimate follows because each contribution
to $B_R$ connects two retained sites through an intermediate site in the discarded half-plane. The correction
also decays exponentially along the boundary. Consequently, for a translation-invariant edge, each
$B_R(k_\parallel)$ can be approximated arbitrarily accurately in operator norm by retaining only finitely many
transverse layers. Such an operator is called compact in the mathematical sense, distinct from compactness of
the $U(1)$ symmetry. It follows that the essential spectrum of $q_{\rm cut}(k_\parallel)$ remains at the bulk
values $\pm1$, while discrete boundary-localized charge levels may occur inside $(-1,1)$.

The topology determines how these boundary levels evolve. Let $P_+^Q$ denote the positive-charge projector
introduced in Eq.~\eqref{eq:2d_sq_CQ_def}; without momentum arguments, it denotes the corresponding real-space
operator. In the hybrid-Wannier picture, the positive-charge states are localized perpendicular to the boundary
and labeled by the conserved edge momentum $k_\parallel$. A nonzero $C_Q$ forces their centers to wind across
transverse unit cells as $k_\parallel$ traverses its Brillouin zone once. This winding is the charge analogue of
the hybrid-Wannier flow in a Chern band
\cite{fidkowskiJacksonKlich2011}.

A spatial cut converts this winding into boundary charge-spectrum flow. As the hybrid-Wannier centers cross the
cut, eigenvalues of the compressed projector $RP_+^Q R$ flow between zero and one. Since
\begin{equation}
q_{\rm cut}(k_\parallel)
=2RP_+^Q(k_\parallel)R-R,
\label{eq:app_qcut_RPR}
\end{equation}
these crossings correspond to charge eigenvalues flowing between $-1$ and $+1$. The bulk--edge correspondence
identifies the net number of upward minus downward zero crossings with $C_Q$, as stated in
Eq.~\eqref{eq:2d_charge_edge_spectral_flow}, with consistent bulk and boundary orientations
\cite{kellendonkEdge2002,fidkowskiEntanglement2010,
fidkowskiJacksonKlich2011}. Importantly, the relevant spectral gap is that of the charge projector, not the
physical Hamiltonian. This correspondence therefore remains valid for the gapless physical Dirac spectrum.

The enforced spectral flow obstructs an exactly compact boundary termination. Suppose an exponentially local,
boundary-translation-invariant half-space charge $q_{\rm half}$ agrees with $q_{\rm loc}$ away from the boundary
and differs from $q_{\rm cut}$ by a continuous, periodic family of boundary-localized corrections that can be
approximated uniformly by operators acting on finitely many transverse layers. Such corrections preserve the net
signed spectral flow through zero, since the essential charge gap remains open throughout the deformation. If
$q_{\rm half}^2=\mathbf1$, however, its entire spectrum is restricted to $\pm1$, leaving no zero crossings. An
exactly compact local termination under these assumptions therefore requires
\begin{equation}
C_Q=0.
\label{eq:app_no_compact_boundary_termination}
\end{equation}
The obstruction persists upon adding finitely many topologically trivial 1D boundary bands, which cannot change
the net signed crossing number. It can instead be canceled by an opposite 2D charge-topological sector or
avoided by relaxing the stated locality or compactness assumptions.

For completeness, the same bulk topology admits a mathematical characterization through a real-space index. Let
$U_{\rm w}(\mathbf r)=e^{i\theta(\mathbf r)}$ multiply a lattice wavefunction by a phase winding once around a
point away from all lattice sites. Although $U_{\rm w}$ is unitary, its projection $T_{\rm w}=P_+^Q U_{\rm
w}P_+^Q$ onto the positive-charge sector need not be. Nevertheless, exponential locality of $P_+^Q$ ensures that
$[U_{\rm w},P_+^Q]$ is compact: over any fixed lattice displacement, the winding phase varies arbitrarily slowly
far from its center, while long-distance matrix elements of $P_+^Q$ are exponentially suppressed. Thus $T_{\rm
w}$ is unitary up to compact corrections and has a well-defined integer index,
\begin{equation}
\operatorname{Ind}(T_{\rm w})
=\dim\ker T_{\rm w}
-\dim\ker T_{\rm w}^\dagger.
\label{eq:app_bulk_fredholm_CQ}
\end{equation}
The two kernel dimensions are finite, and their difference is stable under continuous deformations preserving
the index. The real-space index theorem gives
\begin{equation}
\operatorname{Ind}(T_{\rm w})=C_Q,
\label{eq:app_charge_edge_flow}
\end{equation}
with a consistent winding orientation
\cite{avronCharge1994}. Together with the
bulk--edge correspondence
\cite{kellendonkEdge2002}, this identifies the
real-space index with the boundary charge-spectrum flow. The projected winding index and the hybrid-Wannier
picture therefore characterize the same charge-bundle topology.

For $H_{\rm sq}$, $C_Q=1$ enforces a net signed charge crossing number of one at an oriented half-plane
boundary. A finite strip has two edges with compensating flows, as illustrated in
Fig.~\ref{fig:majoranaedge}(d). The distinction between charge-flow eigenstates and physical edge-energy
eigenstates, including their different responses to the Chern mass, is discussed in
Sec.~\ref{sec:2d_sq_mass_edge}.

\subsection{Gauge locality of Schwartz regulators}
\label{app:2d_hc_asymptotic_proof}

We establish the gauge-locality estimates used for $H_{\rm sq}$ in Sec.~\ref{sec:2d_sq_gauge_locality}. Once the
compact charge is onsite, each matter hopping can be dressed by a Wilson line, giving exact gauge invariance at
finite lattice spacing. The remaining question is whether the long-range hoppings can be discarded while the
physical hopping range shrinks to zero. The argument applies to smooth periodic charge-onsite Hamiltonians with
Wilson paths of at most linear length, and to compact matrix-valued links when the hopping matrices commute with
the onsite gauge action.

Take a smooth periodic dimensionless Bloch Hamiltonian $h_\sharp(\mathbf p)=\sum_{\mathbf R}K_{\mathbf R}
e^{i\mathbf p\cdot\mathbf R}$ on the relocalized 2D torus. Keep the smoothing profile fixed as $a\to0$. Repeated
integration by parts in momentum space then shows that, for every $q>0$,
\begin{equation}
\|K_{\mathbf R}\|
\le C_q(1+|\mathbf R|)^{-q}.
\label{eq:app_2d_schwartz_bound}
\end{equation}
The constants in this appendix may depend on the fixed smoothing profile and path geometry, but not on $a$ or
the hopping cutoff $L$. With the Dirac energy scale restored, the matter Hamiltonian has hopping amplitudes
$(v/a)K_{\mathbf R}$, with $K_{-\mathbf R}=K_{\mathbf R}^\dagger$.

For a displacement $\mathbf R$, choose a path $\Gamma_{\mathbf R}$ from $\mathbf0$ to $\mathbf R$ and translate
it to $\Gamma_{\mathbf X,\mathbf R} =\mathbf X+\Gamma_{\mathbf R}$. We require its number of links to satisfy
\begin{equation}
|\Gamma_{\mathbf R}|
\le C_\Gamma(1+|\mathbf R|).
\label{eq:app_2d_path_growth}
\end{equation}
Shortest lattice paths satisfy this condition. For a reversed hopping, take the adjoint of the transporter on
the reversed path. One can also average a bounded number of symmetry-related paths, $\overline
U_\Gamma=\sum_jw_jU_{\Gamma_j}$, where $w_j\ge0$ and $\sum_jw_j=1$. All paths have the same endpoints and hence
the same gauge-transformation law; moreover $\|\overline U_\Gamma\|\le\sum_jw_j=1$. The average need not itself
be a group element. Distinct path prescriptions may differ by Wilson loops, and we do not assume that they give
identical responses to arbitrary gauge backgrounds.

Each gauged hopping involves two matter sites and the links along its path. Denote this set by $S_{\mathbf
X,\mathbf R}$. Equation~\eqref{eq:app_2d_path_growth} implies that both its number of sites and links,
$|S_{\mathbf X,\mathbf R}|$, and its diameter grow at most as $1+|\mathbf R|$. For a given site or link $z$ and
displacement $\mathbf R$, there are at most $C(1+|\mathbf R|)$ translated paths passing through $z$. These two
elementary counting facts explain why gauging adds only polynomial factors to the original rapid decay of
$K_{\mathbf R}$.

To measure the locality of all hopping terms at once, write $H_a[U]=\sum_S\Phi_a(S)$, where $\Phi_a(S)$ is the
operator supported on the matter sites and gauge links in $S$. In the terminology of quantum lattice systems,
such a collection of spatially supported Hamiltonian terms is called an interaction, even when the matter
Hamiltonian is quadratic. For $s\ge0$, define
\begin{equation}
\|\Phi\|_s
=\sup_z\sum_{S\ni z}
|S|(1+\operatorname{diam}S)^s\|\Phi(S)\|.
\label{eq:app_2d_weighted_norm}
\end{equation}
This norm weights hopping terms by their support size and by a power of their spatial extent
\cite{nachtergaele2019quasilocality}.
Let $\Phi_{a,L}$ retain only the paths with $|\mathbf R|\le L$. Since the Wilson transporters have norm at most
one, the preceding support counting gives
\begin{equation}
\|\Phi_a-\Phi_{a,L}\|_s
\le C_s\frac va
\sum_{|\mathbf R|>L}
(1+|\mathbf R|)^{s+2}\|K_{\mathbf R}\|.
\label{eq:app_2d_weighted_tail_raw}
\end{equation}
The two extra powers bound the size of each path and the number of its translates passing through a given site
or link. Applying the Schwartz bound,
\begin{equation}
\|\Phi_a-\Phi_{a,L}\|_s
\le C_{s,q}\frac va
\sum_{|\mathbf R|>L}(1+|\mathbf R|)^{s+2-q}.
\label{eq:app_2d_weighted_tail_schwartz}
\end{equation}
In 2D, the number of displacements at distance of order $j$ grows at most linearly with $j$:
\begin{equation}
\sum_{|\mathbf R|>L}(1+|\mathbf R|)^{s+2-q}
\le C\sum_{j\ge L}(1+j)^{s+3-q}.
\label{eq:app_2d_shell_bound}
\end{equation}
Because $q$ can be arbitrarily large, choosing $q>s+M+4$ yields, for every $M>0$,
\begin{equation}
\|\Phi_a-\Phi_{a,L}\|_s
\le C_{s,M}\frac va L^{-M}.
\label{eq:app_2d_weighted_tail}
\end{equation}
Now take $L(a)=\lceil a^{-\beta}\rceil$, where $0<\beta<1$. The retained paths extend over a vanishing physical
distance, $aL(a)\to0$, while choosing $M>(p+1)/\beta$ gives, for every $p>0$,
\begin{equation}
\|\Phi_a-\Phi_{a,L(a)}\|_s=O(a^p).
\label{eq:app_2d_weighted_superalg}
\end{equation}
Each retained Wilson-line hopping is still exactly gauge invariant. Thus the matter--gauge Hamiltonian can be
approximated by finite-range gauge-invariant hoppings whose physical range and locality error both vanish.

On a physical torus, a hopping path may wind around a periodic direction. We define it by first drawing the path
on the covering infinite lattice and then identifying its endpoints periodically; paths of different winding
number remain distinct contributions. The sum is absolutely convergent by Eq.~\eqref{eq:app_2d_schwartz_bound}.
For $L(a)=o(a^{-1})$ and a fixed physical torus, the retained sufficiently short paths do not wind around the
torus.

The preceding norm controls the omitted hopping terms near any given site or link. On a torus of fixed physical
area $\mathcal A$, we can also control their total contribution to the complete many-body Hamiltonian. Define
\begin{equation}
B_{a,L}:=H_a[U]-H_{a,L}[U].
\label{eq:app_2d_BaL}
\end{equation}
There are $O(\mathcal A a^{-2})$ matter cells, each with hopping energy scale $v/a$. Summing the discarded
hopping norms over all cells therefore gives
\begin{equation}
\|B_{a,L}\|
\le C_{\mathcal A,M}\,v\,a^{-3}L^{-M}.
\label{eq:app_2d_finite_volume_norm}
\end{equation}
For $L=L(a)$ this is $O(a^p)$ for every prescribed $p>0$, by taking $M>(p+3)/\beta$. No assumption is made that
the fermions occupy only low-energy states or that the compact gauge links describe smooth fields.

To include gauge-field dynamics, add the same self-adjoint gauge Hamiltonian $H_{g,a}$ to both matter
regulators: $\mathbb H_a=H_a[U]+H_{g,a}$ and $\mathbb H_{a,L}=H_{a,L}[U]+H_{g,a}$. Their difference is still
$B_{a,L}$. At finite volume the matter terms are bounded, so the two total Hamiltonians have a common
self-adjoint domain. The difference between their time-evolution operators can be written as an integral with
one insertion of $B_{a,L}$:
\begin{equation}
\begin{aligned}
&e^{-it\mathbb H_a}-e^{-it\mathbb H_{a,L}}\\
&\quad=-i\int_0^t ds\,
e^{-i(t-s)\mathbb H_a}
B_{a,L}e^{-is\mathbb H_{a,L}}.
\end{aligned}
\end{equation}
Both time-evolution factors are unitary and therefore have norm one. Taking the norm immediately gives
\begin{equation}
\|e^{-it\mathbb H_a}-e^{-it\mathbb H_{a,L}}\|
\le |t|\,\|B_{a,L}\|.
\label{eq:app_2d_dynamics_bound}
\end{equation}
Thus, on every fixed physical torus, discarding long-range hoppings leaves the complete finite-time dynamics
unchanged in the continuum limit, including fluctuating gauge links.

The Hamiltonian difference also controls the energy-domain resolvents. The identity
\begin{equation}
\begin{aligned}
&(\mathbb H_a-z)^{-1}
-(\mathbb H_{a,L}-z)^{-1}\\
&\quad=-(\mathbb H_a-z)^{-1}
B_{a,L}(\mathbb H_{a,L}-z)^{-1}
\end{aligned}
\end{equation}
and the self-adjoint bound $\|(\mathbb H-z)^{-1}\|\le|\operatorname{Im}z|^{-1}$ give
\begin{equation}
\|(\mathbb H_a-z)^{-1}
-(\mathbb H_{a,L}-z)^{-1}\|
\le
\frac{\|B_{a,L}\|}{|\operatorname{Im}z|^2},
\label{eq:app_2d_resolvent_bound}
\end{equation}
for nonreal $z$. Every gauge-invariant hopping preserves the Gauss-law subspace. The Hamiltonian-difference,
time-evolution, and resolvent bounds therefore remain valid on the common physical subspace. The weighted
interaction estimates also survive restriction when each term retains its original support label. These results
do not assert an infinite-volume propagation theorem or the existence of an interacting continuum fixed point.

For completeness, finite-range truncation need not leave the massless Dirac point exactly at its original
momentum: the omitted hoppings can produce a tiny shift or gap. For a two-band Hamiltonian with a single simple
node at $\mathbf K$, where $h_\sharp(\mathbf K)=0$, set $h_L(\mathbf p)=\sum_{|\mathbf R|\le L} K_{\mathbf
R}e^{i\mathbf p\cdot\mathbf R}$ and correct the onsite matrix by
\begin{equation}
h_L^{\rm corr}(\mathbf p)
=h_L(\mathbf p)-h_L(\mathbf K).
\label{eq:app_2d_node_correction}
\end{equation}
Then $h_L^{\rm corr}(\mathbf K)=0$ exactly. The correction decreases faster than any power of $L$, and the
corrected Hamiltonian approaches $h_\sharp$ together with its first derivatives. The linear Dirac crossing
therefore remains nondegenerate; the gap away from the node excludes additional crossings for sufficiently large
$L$. The onsite charge symmetry is unchanged. In multiband models with gapped states at $\mathbf K$, only the
nodal sector may be corrected in this manner.

Schwartz hopping also controls the Dirac dispersion uniformly as the cutoff increases. For every fixed momentum
derivative $\partial^\alpha$,
\begin{equation}
\sup_L\|\partial^\alpha h_L\|_\infty
\le\sum_{\mathbf R}
|\mathbf R|^{|\alpha|}\|K_{\mathbf R}\|
<\infty.
\label{eq:2d_Wilsonian_moment_bounds}
\end{equation}
With $\widetilde h_L(\mathbf p) =h_L(\mathbf p)-h_L(\mathbf K)$, the small physical momentum $\mathbf p=\mathbf
K+a\mathbf q$ gives
\begin{equation}
\frac va\widetilde h_L(\mathbf K+a\mathbf q)
=v\sum_i\partial_i h_L(\mathbf K)q_i
+O(va|\mathbf q|^2),
\label{eq:2d_Wilsonian_uniform_Dirac}
\end{equation}
uniformly in $L$. Thus the Dirac velocity approaches that of $h_\sharp$. The same rapid decay controls
fixed-order expansions of Wilson-line gauge vertices at small momentum. These estimates establish regularity of
the free Dirac dispersion and gauge couplings, but do not by themselves determine the interacting continuum
theory.

The same gauge-locality estimates apply to the relocalized honeycomb Hamiltonian $H_{\rm hc}$.
Appendix~\ref{app:2d_hc_charge_bundle} constructs its smooth periodic charge-onsite representative and
establishes Schwartz decay of its hopping. Its compact Wilson-line gauging therefore also admits exactly
gauge-invariant finite-range approximations with vanishing physical range and operator-norm errors on fixed
physical volumes.

\subsection{Lattice reflection symmetry and odd Pfaffian parity}
\label{app:2d_dirac_pfaffian}

An odd class-D Pfaffian parity can follow directly from the ungauged Majorana spectrum and crystalline symmetry,
without assuming an exact compact or smooth lattice charge. Consider a quadratic Majorana Hamiltonian $H$ with a
periodic, at least $C^1$ Bloch matrix $h(\mathbf k)=iA(\mathbf k)$ in a globally oriented canonical Majorana
frame, satisfying $h(-\mathbf k)=-h(\mathbf k)^*$. Assume that its only zero-energy crossings are a single
Majorana-related pair $\mathbf K_\pm=\pm\mathbf K$, neither at a PHIM. At each node, exactly two bands form an
isolated, nondegenerate linear Dirac crossing,\footnote{
We require the two linearized bands to have opposite energy signs in every nonzero momentum direction.
Writing the linearized two-band Hamiltonian as
$h(\mathbf q)=(\mathbf w\cdot\mathbf q)\sigma_0+
\mathbf d(\mathbf q)\cdot\boldsymbol\sigma$,
where $\mathbf d(\mathbf q)$ is linear in $\mathbf q$,
this condition reads
$|\mathbf w\cdot\mathbf q|<|\mathbf d(\mathbf q)|$
for all $\mathbf q\ne0$. It permits a finite tilt.}
while all other bands remain separated from zero.

Suppose further that $H$ admits a unitary, orientation-reversing lattice parity with globally regular,
Majorana-compatible Bloch sewing. Its adjoint action is involutive, although its many-body implementation may
square to fermion parity. Let $\mathsf p(\mathbf k)$ denote its action on crystal momentum, possibly affine
modulo reciprocal lattice vectors, and let $\mathsf r$ be its linear part, with $\det\mathsf r=-1$. Assume that
parity fixes each node and preserves its low-energy Dirac number, rather than acting by charge conjugation or
exchanging the two Majorana-related nodes. Then
\begin{equation}
\mathcal P_D(H)=-1.
\label{eq:app_2d_dirac_pfaffian}
\end{equation}
Neither a physical mass term nor finite-range hopping is required; the stated regularity suffices.

To prove the result, choose regular local frames at the two isolated crossings. Writing $\mathbf q=\mathbf
k-\mathbf K_\pm$ for the momentum measured from either node, their linearized Hamiltonians, up to scalar tilt
terms, are
\begin{equation}
h_\pm^{\rm eff}(\mathbf q)
=q_xD_{\pm,x}+q_yD_{\pm,y}+o(|\mathbf q|),
\label{eq:app_2d_parity_linearized}
\end{equation}
where $D_{\pm,i}$ are Hermitian $2\times2$ velocity matrices whose traceless parts are linearly independent. For
the square model, Eq.~\eqref{eq:2d_sq_IR_folded} gives $D_{\pm,x}=-v\sigma_x$ and $D_{\pm,y}=\mp v\sigma_y$,
with $v=2t$. The general argument does not assume isotropic Dirac velocities.

The velocity-matrix commutator, $M_\pm=-i[D_{\pm,x},D_{\pm,y}]$,  defines an oriented Dirac mass,
\begin{equation}
\widehat M_\pm=\frac{M_\pm}
{\sqrt{\tfrac12\operatorname{tr}(M_\pm^2)}},
\quad \widehat M_\pm^2=\mathbf1.
\label{eq:app_2d_parity_oriented_mass}
\end{equation}
The matrices $\widehat M_\pm$ anticommute with the traceless kinetic matrices. Let $S_\pm$ denote the internal
action of parity in the local nodal frames. Because parity fixes each node, invariance of the massless
Hamiltonian implies
\begin{equation}
S_\pm D_{\pm,i}S_\pm^\dagger
=\sum_j\mathsf r_{ji}D_{\pm,j},
\ \
S_\pm\widehat M_\pm S_\pm^\dagger
=-\widehat M_\pm.
\label{eq:app_2d_parity_mass_odd}
\end{equation}
The second relation follows because an orientation-reversing transformation changes the sign of the commutator
of the two kinetic directions. Thus the oriented Dirac mass is necessarily parity odd.

Majorana reality relates the two nodal Hamiltonians. In compatible local frames, $D_{-,i}=D_{+,i}^*$ and
$\widehat M_-=-\widehat M_+^*$. Since the nodes are distinct and the parity sewing is regular, these local
masses can be extended using smooth, symmetry-compatible bump functions to a global Hermitian perturbation
$\delta h$ satisfying
\begin{equation}
\begin{aligned}
\delta h(-\mathbf k)&=-\delta h(\mathbf k)^*,\\
\delta h(\mathbf k)
&=-V_{\cal P}(\mathbf k)
\delta h(\mathsf p(\mathbf k))
V_{\cal P}(\mathbf k)^\dagger.
\end{aligned}
\label{eq:app_2d_parity_global_mass}
\end{equation}
Here $V_{\cal P}(\mathbf k)$ denotes the global Bloch sewing matrix of parity, in the convention of the
two-orbital reflection sewing matrices $V_i(\mathbf k)$ in Eq.~\eqref{eq:2d_sq_projective_parities}. These are
distinguished from the induced doubled-cell sewing $\mathscr S_i$ used in
Appendix~\ref{app:2d_su2_dirac_inversion}. The perturbation may be supported in arbitrarily small nodal
neighborhoods. The condition stated above ensures that a sufficiently small Dirac mass opens a full gap
near each crossing. Consequently, $h^{(\lambda)}=h+\lambda\delta h$ is fully gapped for sufficiently small
$\lambda\ne0$, with parity exchanging $h^{(\lambda)}$ and $h^{(-\lambda)}$. We denote the corresponding
quadratic Hamiltonians by $H^{(\lambda)}$. The auxiliary parameter $\lambda$ need not coincide with the physical
Chern mass $m$ of $H_m$.

The gapped descendants have integer class-D Majorana Chern numbers $\nu_{\rm Maj}(\lambda)$. As $\lambda$
changes sign, each nondegenerate Dirac crossing contributes a unit Chern jump $\chi_\pm=\pm1$. Majorana reality
reverses both momentum coordinates while relating the oriented masses as above, so the two jumps have the same
sign, $\chi_-=\chi_+$. Therefore,
\begin{equation}
\nu_{\rm Maj}(\lambda)-\nu_{\rm Maj}(-\lambda)
=\chi_++\chi_-=2\chi_+,
\label{eq:app_2d_parity_chern_jump}
\end{equation}
for $\lambda>0$. On the other hand, parity reverses the orientation of the Brillouin torus and exchanges the two
massive descendants. Hence
\begin{equation}
\nu_{\rm Maj}(\lambda)=-\nu_{\rm Maj}(-\lambda),
\ \
\nu_{\rm Maj}(\lambda)=\chi_+=\pm1.
\label{eq:app_2d_parity_chern_odd}
\end{equation}
Because the nodes lie away from the PHIM, the Pfaffian signs at the PHIM remain unchanged during this
deformation. The class-D Pfaffian--Chern relation of Eq.~\eqref{eq:2d_pfaffian_chern_parity} then gives
\begin{equation}
\mathcal P_D(H)=\mathcal P_D(H^{(\lambda)})
=(-1)^{\nu_{\rm Maj}(\lambda)}=-1,
\end{equation}
proving Eq.~\eqref{eq:app_2d_dirac_pfaffian}.

The hypotheses are essential. A node at a PHIM makes the massless Pfaffian product ill-defined, while additional
zero-energy crossings can change the Chern jump. A parity that exchanges the nodes, conjugates Dirac number, or
has singular or incompatible Bloch sewing does not satisfy the stated assumptions. The bond-centered reflections
of $H_{\rm sq}$ meet these conditions, so its odd Pfaffian parity follows independently of $Q_{\rm loc}$ and
$C_Q=1$. This does not require the chosen smooth charge $Q_{\rm loc}$ itself to be reflection invariant: the
charge-preserving reflections transport it to distinct smooth conserved charges with opposite charge Chern
number. By contrast, the charge-conjugating site-centered reflections do not satisfy the theorem's hypotheses.

\subsection{Finite-range compact lifts and Onsager quotients}
\label{app:2d_compact_lifts}

The main text constructs an exponentially local compact $U(2)$ from the orbital-dependent one-flavor complex
structure $J_{\rm loc}$. Here we examine a different, finite-range algebra generated by relative translations of
Majorana sectors. For the Wilson--Dirac-derived honeycomb model $H_{\rm hc}^{W}$, its IR quotient is $\mathfrak
u(1)_{\rm Dirac}$; for two physical square-Majorana flavors, it is $\mathfrak{su}(2)_{\rm valley}$. These
quotients describe the nodal action of the generators, not a simultaneous exact microscopic realization of the
corresponding compact symmetries.

\subsubsection{Wilson--Dirac-derived honeycomb compact lifts}
\label{app:2d_hcW_onsager}

The identity $h_{\rm hc}^{W}(u+\pi,v)=h_{\rm hc}^{W}(u,v)$ produces two identical Majorana sectors in the
doubled cell. These blocked sectors should not be confused with the two independent physical Majorana flavors of
$H_2^{\rm sq}$. With $a_{\alpha,\mathbf R}$ and $b_{\alpha,\mathbf R}$ defined in Eq.~\eqref{eq:2d_ab_blocking},
\begin{equation}
H_{\rm hc}^{W}=\frac{i}{4}(a^TBa+b^TBb),
\end{equation}
where $B$ is real antisymmetric, finite range, and translation invariant. Write $\boldsymbol\gamma_{ab}=(a,b)^T$
for the blocked Majorana column, distinct from the strip chiral operator $\Gamma$ in the main text. Define the
quadratic operator
\begin{equation}
\mathcal Q(J)=\frac{i}{4}\boldsymbol\gamma_{ab}^TJ\boldsymbol\gamma_{ab},
\quad J^T=-J,
\label{eq:app_majorana_QJ}
\end{equation}
which satisfies
\begin{equation}
[\mathcal Q(J),\mathcal Q(K)]=i\mathcal Q([J,K]).
\label{eq:app_majorana_bilinear_bracket}
\end{equation}
For a doubled-cell translation $\mathbf r=2m\mathbf a_1+n\mathbf a_2$, with $m,n\in\mathbb Z$, set
\begin{equation}
Q_{\mathbf r}=-\frac{i}{2}\sum_{\mathbf R,\alpha}
a_{\alpha,\mathbf R}b_{\alpha,\mathbf R+\mathbf r}.
\label{eq:app_Qr}
\end{equation}
Its one-particle matrix is
\begin{equation}
J_{\mathbf r}^{(Q)}=
\begin{pmatrix}
0&-T_{\mathbf r}\\
T_{-\mathbf r}&0
\end{pmatrix},
\label{eq:app_Jr}
\end{equation}
where $T_{\mathbf r}$ translates the doubled-cell coordinate. Since $T_{\mathbf r}^T=T_{-\mathbf r}$ and
$[B,T_{\mathbf r}]=0$,
\begin{equation}
\bigl(J_{\mathbf r}^{(Q)}\bigr)^2=-1,
\quad
[J_{\mathbf r}^{(Q)},\operatorname{diag}(B,B)]=0.
\end{equation}
Thus $Q_{\mathbf r}$ is conserved and generates a unit-periodic one-particle rotation. Its integer-normalized
many-body charge follows by defining
\begin{equation}
f_{\alpha,\mathbf R}^{(\mathbf r)}
=\frac{a_{\alpha,\mathbf R}-ib_{\alpha,\mathbf R+\mathbf r}}{2},
\label{eq:app_fr}
\end{equation}
for which
\begin{equation}
Q_{\mathbf r}=\sum_{\mathbf R,\alpha}
\left(f_{\alpha,\mathbf R}^{(\mathbf r)\dagger}
f_{\alpha,\mathbf R}^{(\mathbf r)}-\frac12\right).
\end{equation}
The doubled cell contains two orbitals, so the centering is integral and $e^{2\pi iQ_{\mathbf r}}=1$.
Translating the $b$ sector by $\mathbf r$ makes the charge onsite without changing the finite-range Hamiltonian.

In blocked momentum space,
\begin{equation}
J_{\mathbf r}^{(Q)}(\mathbf p)=
\begin{pmatrix}
0&-e^{i\mathbf p\cdot\mathbf r}\\
e^{-i\mathbf p\cdot\mathbf r}&0
\end{pmatrix}\otimes\mathbf1.
\label{eq:app_Jr_momentum}
\end{equation}
The Dirac node is at $\mathbf p=0$, where $J_{\mathbf r}^{(Q)}(\mathbf p)=J_{\mathbf0}^{(Q)}+O(|\mathbf p|)$ for
each fixed $\mathbf r$. Consequently,
\begin{equation}
\rho_{\rm IR}(Q_{\mathbf r})=\rho_{\rm IR}(Q_D).
\label{eq:app_Qr_same_IR}
\end{equation}
Here $\rho_{\rm IR}$ extracts the leading internal action at the node. Distinct finite-range compact charges
thus realize the same IR Dirac number.

These charges need not commute. Define
\begin{equation}
G_{\mathbf r}=\frac{i}{2}\sum_{\mathbf R,\alpha}
\left(a_{\alpha,\mathbf R}a_{\alpha,\mathbf R+\mathbf r}
-b_{\alpha,\mathbf R}b_{\alpha,\mathbf R+\mathbf r}\right),
\label{eq:app_Gr}
\end{equation}
with one-particle matrix
\begin{equation}
J_{\mathbf r}^{(G)}=
\begin{pmatrix}
T_{\mathbf r}-T_{-\mathbf r}&0\\
0&-T_{\mathbf r}+T_{-\mathbf r}
\end{pmatrix}.
\label{eq:app_JGr}
\end{equation}
The translation law $T_{\mathbf r}T_{\mathbf s}=T_{\mathbf r+\mathbf s}$ gives
\begin{equation}
\begin{aligned}
[J_{\mathbf r}^{(Q)},J_{\mathbf s}^{(Q)}]
&=J_{\mathbf s-\mathbf r}^{(G)},\\
[J_{\mathbf r}^{(G)},J_{\mathbf s}^{(G)}]&=0,\\
[J_{\mathbf r}^{(Q)},J_{\mathbf s}^{(G)}]
&=2\bigl(J_{\mathbf r-\mathbf s}^{(Q)}
-J_{\mathbf r+\mathbf s}^{(Q)}\bigr).
\end{aligned}
\label{eq:app_onsager_one_particle}
\end{equation}
By Eq.~\eqref{eq:app_majorana_bilinear_bracket}, the many-body operators satisfy
\begin{equation}
\begin{aligned}
[Q_{\mathbf r},Q_{\mathbf s}]&=iG_{\mathbf s-\mathbf r},\\
[G_{\mathbf r},G_{\mathbf s}]&=0,\\
[Q_{\mathbf r},G_{\mathbf s}]&=
2i(Q_{\mathbf r-\mathbf s}-Q_{\mathbf r+\mathbf s}).
\end{aligned}
\label{eq:app_onsager_many_body}
\end{equation}
Here $G_{-\mathbf r}=-G_{\mathbf r}$ and $G_{\mathbf0}=0$. These are the generalized Onsager relations on
$\mathbb Z^2$~\cite{pace2025parity}; a single displacement direction gives the ordinary Onsager algebra. The
charges $Q_{\mathbf r}$ agree in the IR but do not form an Abelian microscopic family.

The Onsager generators exhaust a restricted sector. Let $\mathfrak g_{\rm fr}^{\rm rel}$ consist of
finite-range, translation-invariant, Majorana-compatible generators that are traceless on the folded $(a,b)$
sector and act as the identity on its remaining orbitals. Their Bloch symbols are
\begin{equation}
\begin{aligned}
& \mathfrak g_{\rm fr}^{\rm rel}=\bigl\{\,J(\mathbf p)
=M(\mathbf p)\otimes\mathbf1:\\
&M^\dagger=-M,\quad M(-\mathbf p)=M(\mathbf p)^*,\\
&\operatorname{tr}M=0,\quad
M\ \text{a trigonometric polynomial}\bigr\}.
\end{aligned}
\label{eq:app_gfr_rel_definition}
\end{equation}
Here $M$ acts on the folded $(a,b)$ sector, with Pauli matrices $\tau_i$, and is orbital independent. Every such
generator is
\begin{equation}
J(\mathbf p)=i[x(\mathbf p)\tau_x+y(\mathbf p)\tau_y
+z(\mathbf p)\tau_z]\otimes\mathbf1,
\label{eq:app_general_rel_J}
\end{equation}
where Majorana reality requires
\begin{equation}
x(-\mathbf p)=-x(\mathbf p),\ \
y(-\mathbf p)=y(\mathbf p),\ \
z(-\mathbf p)=-z(\mathbf p).
\label{eq:app_rel_parities}
\end{equation}
The Onsager matrices provide all the allowed Fourier harmonics:
\begin{equation}
J_{\mathbf r}^{(Q)}(\mathbf p)
=-i\tau_y\cos(\mathbf p\cdot\mathbf r)
-i\tau_x\sin(\mathbf p\cdot\mathbf r),
\label{eq:app_Q_fourier}
\end{equation}
and
\begin{equation}
J_{\mathbf r}^{(G)}(\mathbf p)
=2i\tau_z\sin(\mathbf p\cdot\mathbf r),
\label{eq:app_G_fourier}
\end{equation}
with the orbital identity suppressed. Explicitly,
\begin{equation}
\begin{aligned}
i\tau_y\cos(\mathbf p\cdot\mathbf r)
&=-\tfrac12(J_{\mathbf r}^{(Q)}+J_{-\mathbf r}^{(Q)}),\\
i\tau_x\sin(\mathbf p\cdot\mathbf r)
&=-\tfrac12(J_{\mathbf r}^{(Q)}-J_{-\mathbf r}^{(Q)}),\\
i\tau_z\sin(\mathbf p\cdot\mathbf r)
&=\tfrac12J_{\mathbf r}^{(G)}.
\end{aligned}
\label{eq:app_onsager_fourier_span}
\end{equation}
Consequently,
\begin{equation}
\mathfrak g_{\rm fr}^{\rm rel}
=\operatorname{span}_{\mathbb R}^{\rm finite}
\{J_{\mathbf r}^{(Q)},J_{\mathbf r}^{(G)}\}
\simeq\mathsf{Ons}_2.
\label{eq:app_onsager_completeness}
\end{equation}
The equality holds only in this orbital-independent sector, not in the full quadratic centralizer.

Removing the traceless condition permits an additional odd scalar symbol,
\begin{equation}
J(\mathbf p)=i[w(\mathbf p)\tau_0+x(\mathbf p)\tau_x
+y(\mathbf p)\tau_y+z(\mathbf p)\tau_z]
\otimes\mathbf1,
\label{eq:app_general_sector_J}
\end{equation}
where $w(-\mathbf p)=-w(\mathbf p)$ is a real trigonometric polynomial. Since this term commutes with the
traceless generators,
\begin{equation}
\mathfrak g_{\rm fr}^{\rm sector}
=\mathfrak z_{\rm odd}\oplus\mathsf{Ons}_2,
\label{eq:app_sector_decomposition}
\end{equation}
where $\mathfrak z_{\rm odd}$ consists of $iw(\mathbf p)\tau_0\otimes\mathbf1$. A central Lie-algebra element
need not generate a compact $U(1)$. The condition $e^{2\pi J(\mathbf p)}=1$ requires the eigenvalues
$w\pm\sqrt{x^2+y^2+z^2}$ of $J/i$ to be continuous integer-valued functions on the connected Brillouin torus,
hence constants. Their sum makes $w$ constant, while oddness gives $w(\mathbf p)=0$. Their difference fixes
$x^2+y^2+z^2=n^2$ for an integer $n\ge0$, so
$
J^2=-n^2.
$
For unit Dirac charge, $n=1$, and the chosen IR orientation is $J_D=-i\tau_y$ at $\mathbf p=\mathbf0$. Thus its
finite-range compact lifts satisfy
\begin{equation}
\begin{gathered}
x(\mathbf p)^2+y(\mathbf p)^2+z(\mathbf p)^2=1,\\
x(\mathbf0)=z(\mathbf0)=0,\quad y(\mathbf0)=-1.
\end{gathered}
\label{eq:app_unit_compact_onsager}
\end{equation}
In particular, a compact unit-charge lift in this sector cannot acquire an additional odd scalar component.

At the Dirac node, Eqs.~\eqref{eq:app_Q_fourier} and \eqref{eq:app_G_fourier} give $\rho_{\rm IR}(Q_{\mathbf
r})=\rho_{\rm IR}(Q_D)$ and $\rho_{\rm IR}(G_{\mathbf r})=0$. The momentum-independent, complex-linear commutant
of $h_{\rm IR}(\mathbf q)=v_1q_1\sigma_x+v_2q_2\sigma_z$ consists only of scalar matrices. This is a
two-component Wilson--Dirac nodal frame, not the node-folded square-model frame of
Eq.~\eqref{eq:2d_sq_IR_folded}; the Pauli orientation is immaterial to the commutant. Hence
\begin{equation}
\operatorname{im}\rho_{\rm IR}\simeq\mathfrak u(1)_{\rm Dirac},
\label{eq:app_2d_IR_quotient}
\end{equation}
while the kernel restricted to the Onsager algebra is
\begin{equation}
\ker\bigl(\rho_{\rm IR}|_{\mathsf{Ons}_2}\bigr)
=\operatorname{span}_{\mathbb R}^{\rm finite}
\{Q_{\mathbf r}-Q_{\mathbf0},G_{\mathbf r}\}.
\label{eq:app_onsager_kernel}
\end{equation}
Therefore
\begin{equation}
\mathsf{Ons}_2/\ker\rho_{\rm IR}
\simeq\mathfrak u(1)_{\rm Dirac},
\label{eq:app_onsager_IR_quotient}
\end{equation}
where the kernel in the quotient is understood to be the restriction just specified. Thus the generalized
Onsager algebra contains infinitely many microscopic representatives that reduce to one IR Dirac-number
generator. Within the smaller sector commuting with $Q_D$, the compact unit-charge condition and fixed IR
orientation leave only $Q_D$. Neither statement excludes orbital-dependent quasi-local generators or nonlocal
spectral charges.

\subsubsection{Two square-Majorana flavors and the valley quotient}
\label{app:two_copy_onsager_u2}

For $H_2^{\rm sq}$ in Eq.~\eqref{eq:2d_H2}, let $a_x=\gamma_x^{(1)}$ and $b_x=\gamma_x^{(2)}$ denote its two
physical Majorana flavors, with $x$ including the one-flavor orbital index. These are the flavor degrees of
freedom acted on by $\mu_i$ in the main text, not the folded sector acted on by $\tau_i$. Define the
finite-range flavor bilinears
\begin{equation}
Q_{\mathbf r}=\frac{i}{2}\sum_xa_xb_{x+\mathbf r},\ \
G_{\mathbf r}=\frac{i}{2}\sum_x
\left(a_xa_{x+\mathbf r}-b_xb_{x+\mathbf r}\right).
\end{equation}
The sign differs from the honeycomb convention so $Q_{\mathbf0}$ counts $c_x=(a_x+ib_x)/2$. The identical
one-flavor hopping matrices commute with the Bravais translations; hence the same Onsager relations and
restricted completeness result hold, $\mathfrak g_{\rm fr,2}^{\rm rel}\simeq\mathsf{Ons}_2$. In this restricted
sector, generators act only on the two physical Majorana-flavor indices and multiply the identity on the
one-copy orbitals. Dropping flavor tracelessness yields
\begin{equation}
\mathfrak g_{\rm fr,2}^{\rm sector}
=\mathfrak z_{\rm odd}\oplus\mathsf{Ons}_2,
\label{eq:app_two_copy_sector_decomp}
\end{equation}
where $\mathfrak z_{\rm odd}$ consists of $iw(\mathbf k)\mu_0\otimes\mathbf1$, with $w$ real, odd, and a finite
trigonometric polynomial. As above, a compact continuous generator in this flavor-only sector has no such scalar
component.

The IR quotient differs from the honeycomb case because the square model has two valleys. For $\mathbf
K_+=(0,\pi/2)$ and $\mathbf r=n_1\mathbf a_1+n_2\mathbf a_2$, the displacement phase at the node is
\begin{equation}
\varphi_{\mathbf r}=\mathbf K_+\cdot\mathbf r
=\frac\pi2(n_1-n_2).
\end{equation}
The corresponding IR action is
\begin{equation}
\rho_{\rm IR}(Q_{\mathbf r})
=\sin\varphi_{\mathbf r}\,\lambda_1
+\cos\varphi_{\mathbf r}\,\lambda_2,
\end{equation}
where $\lambda_1,\lambda_2$ denote a basis of the IR valley Lie algebra, with normalization and orientation left
implicit. This abstract basis need not coincide with the fundamental-normalized Hermitian generators $S_a$ of
Eq.~\eqref{eq:2d_su2_Hermitian}. The charges $Q_{\mathbf0}$ and $Q_{\mathbf a_1}$ therefore represent two
independent valley generators, and their commutator represents the third. It follows that
\begin{equation}
\mathsf{Ons}_2/\ker\rho_{\rm IR}
\simeq\mathfrak{su}(2)_{\rm valley}.
\label{eq:app_two_copy_onsager_quotient}
\end{equation}
An odd scalar with $w(\mathbf K_+)\ne0$ supplies the central IR direction. On the enlarged flavor-only Lie
algebra this gives
\begin{equation}
\frac{\mathfrak z_{\rm odd}\oplus\mathsf{Ons}_2}
{\ker\rho_{\rm IR}}\simeq\mathfrak u(2),
\label{eq:app_two_copy_u2_quotient}
\end{equation}
where the kernel is now taken in the enlarged algebra. Neither quotient establishes a simultaneous exact compact
$U(2)$ within the restricted finite-range sector. For comparison, a corresponding two-flavor honeycomb
construction has nodal phase $\varphi_{\mathbf r}=\frac{2\pi}{3}(n_1-n_2)$~\cite{pace2025parity}: the abstract
Onsager relations are unchanged, while the IR representation depends on the nodal phases.

Every $Q_{\mathbf r}$ can nevertheless be made onsite individually. For an allowed Bravais translation $\mathcal
T_{\mathbf r}^{(b)}$ acting on the $b$ flavor alone,
\begin{equation}
\mathcal T_{\mathbf r}^{(b)}H_2^{\rm sq}
\bigl(\mathcal T_{\mathbf r}^{(b)}\bigr)^{-1}=H_2^{\rm sq},\ \
Q_{\mathbf r}=\mathcal T_{\mathbf r}^{(b)}Q_{\mathbf0}
\bigl(\mathcal T_{\mathbf r}^{(b)}\bigr)^{-1}.
\label{app:two_copy_local_splitting}
\end{equation}
The individual charges $Q_{\mathbf r}$ are compact and become onsite after a relative flavor translation, but
this property does not turn the generators $Q_{\mathbf r},G_{\mathbf r}$ into an exact microscopic
$\mathfrak{su}(2)$ with the IR commutation relations. The quotient in
Eq.~\eqref{eq:app_two_copy_onsager_quotient} describes their nodal action. It should not be identified with the
exponentially local compact $U(2)$ of Sec.~\ref{sec:2d_su2}, whose generators contain the orbital-dependent
complex structure $J_{\rm loc}(\mathbf k)$ and lie outside this flavor-only Onsager sector.

\subsection{Non-Abelian symmetry topology and anomaly checks}
\label{app:2d_su2_supplement}

We substantiate three results of Sec.~\ref{sec:2d_su2}: the quaternionic $SU(2)$ invariant survives breaking the
displayed central $U(1)$; orientation-reversing symmetries impose conditional gap and mass-inversion
constraints; and flavor-opposed mass interfaces exhibit protected charge-resolved spectral flow and a mixed
anomaly.

\subsubsection{Breaking the displayed central charge}
\label{app:2d_su2_central_breaking}

Although $\nu_{SU(2)}=C_Q\pmod2$ is expressed through a central $U(1)$ charge, its definition in terms of the
$SU(2)$ multiplicity bundle does not require that particular charge to remain conserved. We demonstrate this by
constructing an $SU(2)$-preserving perturbation that breaks the displayed central $U(1)$ while leaving the Dirac
cone unchanged to linear order.

Work in the periodic doubled-cell frame $\mathbf p=(P,V)$, with $\tilde q(P,V)$ and the generators $S_a$ defined
in Eqs.~\eqref{eq:2d_sq_periodic_cell} and
\eqref{eq:2d_su2_Hermitian}. Choose
\begin{equation}
\widetilde R=\tau_z\otimes\mathbf1_2,
\quad
f(P,V)=(1+\cos P)+(1-\cos V),
\label{eq:2d_su2_explicit_R}
\end{equation}
where $\widetilde R$ acts on the doubled-cell sector, not on flavor. It is real and onsite in this frame, while
$f$ is even and analytic. At the folded Dirac node $(P,V)=(\pi,0)$,
\begin{equation}
\tilde q(\pi,0)=\tau_y\otimes\mathbf1_2,
\quad
\{\widetilde R,\tilde q(\pi,0)\}=0.
\label{eq:2d_su2_R_nontrivial}
\end{equation}
Thus $\widetilde R$ mixes the two charge sectors. Extract its charge-off-diagonal part,
\begin{equation}
\begin{aligned}
X&=P_+^Q\widetilde R P_-^Q+P_-^Q\widetilde R P_+^Q\\
 &=\frac12(\widetilde R-\tilde q\widetilde R\tilde q),
\end{aligned}
\label{eq:2d_X_offdiag}
\end{equation}
where $P_\pm^Q=(\mathbf1\pm\tilde q)/2$. The Hermitian matrix $X$ is exponentially local and satisfies
$\{X,\tilde q\}=0$ and $X(-\mathbf p)=X(\mathbf p)^*$.

Now introduce
\begin{equation}
\delta h(P,V)=\lambda f(P,V)\,\mu_y\otimes X(P,V).
\label{eq:2d_break_U1_pert}
\end{equation}
This perturbation is periodic, exponentially local, and Majorana compatible. Because $\mu_y$ anticommutes with
$\mu_x$ and $\mu_z$, while $X$ anticommutes with $\tilde q$, it commutes with all three $SU(2)$ generators:
\begin{equation}
[\delta h,S_a]=0,\quad a=1,2,3.
\label{eq:2d_break_U1_preserve_SU2}
\end{equation}
In contrast, the displayed central charge is no longer conserved:
\begin{equation}
[\delta h,\mu_0\otimes\tilde q]
=2\lambda f\,\mu_y\otimes X\tilde q\not\equiv0.
\label{eq:2d_break_U1_commutator}
\end{equation}
Indeed, $X(\pi,0)=\widetilde R\ne0$, and $f$ is strictly positive away from the node.

Near $(\pi,0)$,
\begin{equation}
f(\pi+\delta P,\delta V)
=\frac12(\delta P^2+\delta V^2)+O(|\delta\mathbf p|^4),
\end{equation}
so $\delta h=O(|\delta\mathbf p|^2)$ and the linear Dirac Hamiltonian is unchanged. For sufficiently small
$|\lambda|$, no additional zero-energy nodes appear. The original node remains at the folded PHIM $(\pi,0)$,
where $\delta h=0$; consequently, a doubled-cell physical Pfaffian-sign product is not defined. The
primitive-cell identity $\mathcal P_D(H_2^{\rm sq})=+1$ of Eq.~\eqref{eq:2d_doubled_PD_trivial} pertains to the
undeformed Hamiltonian in its original translation convention.

Since the perturbation leaves the exact $SU(2)$ representation unchanged, it also leaves its multiplicity bundle
and quaternionic invariant unchanged:
\begin{equation}
\nu_{SU(2)}=1.
\label{eq:2d_intrinsic_SU2_gapless}
\end{equation}
Thus the nontrivial $SU(2)$ topology survives the loss of the displayed central charge; the construction does
not exclude other conserved central generators.

\subsubsection{Crystalline reflections, mass inversion, and the $SU(2)$ parity anomaly}
\label{app:2d_su2_dirac_inversion}

We next test whether the bond-centered reflections $\widetilde{\mathcal P}_{x,y}$ of
Eq.~\eqref{eq:2d_sq_projective_parities} preserve the smooth $SU(2)$ of Eq.~\eqref{eq:2d_su2_Hermitian}. In the
periodic doubled-cell frame $\mathbf p=(P,V)$, their momentum maps are
\begin{equation}
\mathsf p_x(P,V)=(P-2V,-V),\quad
\mathsf p_y(P,V)=(2V-P,V),
\end{equation}
modulo $2\pi$. Let $\mathscr S_i(\mathbf p)$ be the induced four-component sewing from the fiber at $\mathsf
p_i\mathbf p$ to that at $\mathbf p$. It follows from the two-orbital sewing in
Eq.~\eqref{eq:2d_sq_projective_parities} applied to both folded sectors, including their permutation and
reciprocal-lattice phases, followed by the periodic change of basis $F(U)$ of Eq.~\eqref{eq:2d_sq_F}. The
reflected smooth charge is
\begin{equation}
\tilde q^{(i)}(\mathbf p)
=\mathscr S_i(\mathbf p)\tilde q(\mathsf p_i\mathbf p)
\mathscr S_i(\mathbf p)^\dagger.
\label{eq:app_2d_su2_reflected_q}
\end{equation}
Since the reflections act identically on the two Majorana flavors, they normalize the specified $SU(2)$ only if
$\tilde q^{(i)}=\pm\tilde q$. The plus sign is excluded: reflection reverses the orientation of the charge
Brillouin torus, giving $C_Q[\tilde q^{(i)}]=-C_Q[\tilde q]=-1$. The minus sign is also excluded. Both
reflections preserve the sharp charge $Q_{s_P}$ and its Dirac-number action at the nodes
[Eq.~\eqref{eq:2d_sq_parity_Q_preserve}], where $Q_{\rm loc}$ agrees with $Q_{s_P}$. Hence the reflected and
original smooth charges have the same nonzero nodal action, not opposite actions. The bond-centered reflections
therefore transport the smooth $SU(2)$ to a different realization rather than normalizing the specified algebra.

A separate, conditional obstruction applies to an orientation-reversing symmetry that \emph{does} normalize this
$SU(2)$. Suppose a fully gapped quadratic Hamiltonian admits such a symmetry with smooth, periodic,
Majorana-compatible sewing that identifies the occupied multiplicity bundle as $F\simeq\mathsf p^*F$.
Orientation reversal then requires $C_{\mathcal M}=-C_{\mathcal M}$, so $C_{\mathcal M}=0$. But the
nontrivial quaternionic invariant requires $\nu_{SU(2)}=C_{\mathcal M}\pmod2=1$, a contradiction. Thus a
symmetry-preserving quadratic gap is impossible under these sewing assumptions. This result does not apply to
the original bond-centered reflections, which fail the normalization hypothesis, and is independent of the
onsite-$U(1)$ magnetic LSM constraint.

Mass inversion leads to a related, but distinct, sewing relation. Consider an $SU(2)$-preserving family
$h_m=h_0+mM_{\rm D}$, gapped for $m\ne0$, and suppose an orientation-reversing symmetry exchanges $m$ and $-m$
with smooth $SU(2)$-equivariant sewing. Its occupied multiplicity bundles then satisfy $F_{-m}\simeq\mathsf
p^*F_{+m}$, implying $C_{\mathcal M}(-m)=-C_{\mathcal M}(+m)$. Therefore
\begin{equation}
\Delta C_{\mathcal M}
:=C_{\mathcal M}(+m)-C_{\mathcal M}(-m)
=2C_{\mathcal M}(+m).
\label{eq:2d_deltaC}
\end{equation}
For the explicit doubled square-model mass in Eq.~\eqref{eq:2d_sq_U2_mass}, the one-flavor calculation gives
$C_{\mathcal M}(\pm m)=\pm1$ and $\Delta C_{\mathcal M}=2$. This explicit Chern jump does not depend on
identifying the original bond reflections with the assumed sewing.

More generally, assume that no additional UV gaps close along the mass-inversion path and that independent
fundamental $SU(2)$ Dirac doublets, counted after Majorana reality and Brillouin-zone folding, exhaust the Chern
jump. Then
\begin{equation}
\Delta C_{\mathcal M}=2\sum_a\chi_a,
\label{eq:2d_deltaC_Dirac}
\end{equation}
where $\chi_a=\pm1$ gives the orientation of each doublet. Combined with the assumed mass-inversion sewing, this
yields
\begin{equation}
\nu_{SU(2)}=n_D^{SU(2)}\pmod2,
\label{eq:2d_nuSU2_ND}
\end{equation}
where $n_D^{SU(2)}$ is the number of independent fundamental doublets. For the square model, $n_D^{SU(2)}=1$.

The same mod-two parity appears in the continuum non-Abelian parity anomaly. For $n_D^{SU(2)}$ fundamental
two-component Dirac fermions, a parity-symmetric regularization gives
\begin{equation}
\frac{Z[A^g]}{Z[A]}=(-1)^{n_D^{SU(2)}w(g)}
\end{equation}
under a large $SU(2)$ gauge transformation of winding $w(g)\in\mathbb Z$, whereas a gauge-invariant regulator
requires a Chern--Simons level with fractional part
\begin{equation}
k_{\rm frac}=\frac{n_D^{SU(2)}}2\pmod{\mathbb Z}
\end{equation}
in fundamental-trace normalization. Thus, for the explicit square-model mass family, the quaternionic invariant
and continuum parity anomaly carry the same mod-two index. Beyond this construction, the correspondence requires
the stated sewing and UV-gap assumptions; the bundle invariant alone does not imply a microscopic gauge-Pfaffian
holonomy identity.

\subsubsection{Charge-resolved interfaces}
\label{app:2d_su2_interfaces}

Consider an interface with flavor-opposed mass configurations $(+m,-m)$ on the left and $(-m,+m)$ on the right.
These masses break the mixing $SU(2)$ but preserve the commuting compact charges
\begin{equation}
q_V=\mu_0\otimes q_{\rm loc},
\quad
q_A=\mu_z\otimes q_{\rm loc}.
\end{equation}
Both are exponentially local but non-onsite in the original square geometry. A finite-range interpolation
$h_{\rm raw}$ need not conserve them exactly. Because the two flavors remain decoupled, charge conservation can
be restored in each flavor by symmetrizing with $q_{\rm loc}$:
\begin{equation}
h_{\rm DW}
=\frac12\bigl(h_{\rm raw}+q_{\rm loc}h_{\rm raw}q_{\rm loc}\bigr).
\end{equation}
The asymptotic bulk Hamiltonians already commute with $q_{\rm loc}$, so this correction is exponentially
localized near the interface and does not change either bulk phase.

The one-flavor relation $\nu_{\rm Maj}=2C_{+,\mathrm{occ}}-C_Q$, with $C_Q=1$, gives the right-minus-left
relative Chern numbers
\begin{equation}
\Delta C_{+,\mathrm{occ}}^{(1)}=-1,
\quad
\Delta C_{+,\mathrm{occ}}^{(2)}=+1.
\label{eq:2d_sq_interface_relative_chern}
\end{equation}
Thus the two positive-charge sectors have opposite nonzero interface indices, although the total class-D
Chern-number difference vanishes.

To establish their spectral flow, define the exponentially local positive-charge projector
$P_+^{Q_{\rm loc}}=(\mathbf1+q_{\rm loc})/2$ in the folded-sector basis. It represents the same charge
subspace as $P_+^Q=(\mathbf1+\tilde q)/2$ in the periodic doubled-cell basis, with
$P_+^{Q_{\rm loc}}=FP_+^QF^\dagger$. The projector commutes with the corrected interface Hamiltonian
in each flavor. For $\Lambda>0$, define the auxiliary class-A Hamiltonian
\begin{equation}
\widetilde h_{\rm DW}
=P_+^{Q_{\rm loc}}h_{\rm DW}P_+^{Q_{\rm loc}}+\Lambda(\mathbf1-P_+^{Q_{\rm loc}}).
\label{eq:2d_sq_interface_aux_classA}
\end{equation}
Its asymptotic occupied projectors are $P_+^{Q_{\rm loc}}P_{\rm occ}^{\mathrm{L/R}}$. For an infinite,
translation-invariant interface, the class-A bulk--interface correspondence equates the signed spectral flow in
sector $\alpha$ with $\Delta C_{+,\mathrm{occ}}^{(\alpha)}$, in a consistent orientation
convention~\cite{fukuiBulkEdge2012}. Hence the two sectors exhibit protected counterpropagating spectral flow,
without a global onsite trivialization of the charge bundle.

The corresponding charge vectors under $U(1)_V\times U(1)_A$ are $\mathbf t_1=(1,+1)^T$ and $\mathbf
t_2=(1,-1)^T$. Their continuum anomaly matrix in the ordered $(V,A)$ basis is
\begin{equation}
\mathcal A
=\sum_{\alpha=1}^2
\Delta C_{+,\mathrm{occ}}^{(\alpha)}
\mathbf t_\alpha\mathbf t_\alpha^T =-2\bigl(\mathbf e_V\mathbf e_A^T
+\mathbf e_A\mathbf e_V^T\bigr),
\end{equation}
where $\mathbf e_V=(1,0)^T$ and $\mathbf e_A=(0,1)^T$. Thus $\mathcal A_{VV}=\mathcal A_{AA}=0$, while $\mathcal
A_{VA}=\mathcal A_{AV}=-2$; reversing the interface orientation reverses these signs. The net chirality cancels,
but the mixed $U(1)_V$--$U(1)_A$ anomaly remains. Its coefficient and the protected spectral flow follow from
the relative Chern indices, independently of the detailed interface dispersion.

\section{2D--3D: Exact Surface Correspondences and Anomaly Inflow}
\label{app:2d_surface_inflow}

We establish the finite-grid surface correspondences and compact-gauge results used in
Sec.~\ref{sec:2d_sq_3d_parent}. The zero-field Majorana canonical identifications are exact; the magnetic
comparisons require the stated smooth-field and spectral-gap hypotheses.

\subsection{Two-root AIII--AII parent: bulk topology and exact surface correspondence}
\label{app:2d_AII_parent}

The common parent $\mathcal H_*$ of Eq.~\eqref{eq:2d_AII_bulk_H} has both AIII chiral and Kramers symmetries. We
verify its bulk topological indices and exact half-space solutions, then establish the finite-grid BdG--Majorana
identification of the selected surface Hamiltonian and compact charge with $(H_{\rm sq},Q_{s_P})$.

\paragraph{Bulk gap and topological indices.}
The anticommuting Clifford matrices give
\begin{equation}
E_3^2(\mathbf p,w)
=4t^2(\sin^2p_x+\sin^2p_y)
+\Lambda^2|\mathfrak q_*(\mathbf p,e^{iw})|^2.
\label{eq:app_2d_AII_gap}
\end{equation}
A gap closing requires $p_x,p_y\in\{0,\pi\}$ and $|a|=1$ or $|b|=1$. Evaluating the roots at these four momenta
excludes either possibility, so the bulk is fully gapped for $t,\Lambda>0$.

Inversion acts as $\mathcal I=\rho_z$. At the eight inversion-invariant momenta, the in-plane kinetic terms
vanish and $\mathcal H_*$ reduces to $\Lambda\mathfrak q_*\rho_z$. The inversion masses at $w=0,\pi$ are
$\mathfrak q_*(\mathbf p,1)=(1-a)(1-b)$ and $\mathfrak q_*(\mathbf p,-1)=-(1+a)(1+b)$. Thus the occupied Kramers
pair has inversion parity $\xi=-\operatorname{sgn}\mathfrak q_*$. Only $(0,0,0)$ has positive mass, giving one
negative occupied parity among the eight momenta. The Fu--Kane inversion criterion
\cite{fuKane2007inversion}, which determines the
strong index from the product of all eight parities and the weak indices from the products on the three
$k_i=\pi$ planes, therefore yields
\begin{equation}
(\nu_0;\nu_x,\nu_y,\nu_z)=(1;0,0,0).
\label{eq:app_2d_AII_strong_index}
\end{equation}

The parent also has AIII chiral grading $S_*=\rho_x\sigma_y$. Its four-component coefficient vector is
\begin{equation}
\mathbf d=(-2t\sin p_x,-2t\sin p_y,
\Lambda\operatorname{Im}\mathfrak q_*,
\Lambda\operatorname{Re}\mathfrak q_*).
\end{equation} 
As in the direct AIII parent, the winding number is the degree of $\widehat{\mathbf d}:T^3\to S^3$, computed by
counting the signed preimages of a regular point \cite{chiu2016classification}. The north pole has a unique
preimage at $(0,0,0)$, where
\begin{equation}
\det\frac{\partial(d_1,d_2,d_3)}
{\partial(p_x,p_y,w)}
=4\Lambda t^2>0.
\end{equation}
Hence $|\nu_3|=1$.

\paragraph{Exact half-space modes and momentum unfolding.}
Following the half-space construction for the direct AIII parent in Sec.~\ref{sec:2d_sq_3d_parent}, we solve the
two-root problem for $z\ge1$ in the $\rho_x=-1$ sector. Choose $\rho_xu_-=-u_-$ and $h_{\rm D}(\mathbf
p)\xi_\varepsilon =\varepsilon\xi_\varepsilon$. The transverse equation reduces to
\begin{equation}
\alpha_{z+1}-(a+b)\alpha_z+ab\alpha_{z-1}=0,
\label{eq:app_2d_AII_recurrence}
\end{equation}
with the open-boundary condition $\alpha_0=0$. Its characteristic roots are $a$ and $b$. The boundary condition
fixes their relative amplitudes, leaving the exact surface eigenstate
\begin{equation}
\begin{aligned}
\psi_{\mathbf p,\varepsilon}(z)
=\mathcal N(\mathbf p) & \,\alpha_z(\mathbf p)
u_-\otimes\xi_\varepsilon(\mathbf p),\\
\alpha_z=\frac{a^z-b^z}{a-b},
\ \  \alpha_0& =0,\ \
\alpha_z\big|_{a=b} =za^{z-1}.
\end{aligned}
\label{eq:app_2d_AII_surface_profile}
\end{equation}
Normalizability requires $|a|<1$ and $|b|<1$, equivalent to $\cos p_x>0$ and $\cos p_y>0$. Thus the selected
surface occupies $D=(-\pi/2,\pi/2)^2$, with the exact projected Hamiltonian $h_\partial=h_{\rm D}$. The
localization length diverges at the boundary of $D$, where the surface states merge into the bulk continuum.

The two decaying roots yield only one independent surface profile because of the open-boundary condition.
Equivalently, the transverse SSH-like winding of $\mathfrak q_*$ is $2-1=1$: two zeros inside the unit circle
minus the pole from $\zeta^{-1}$. This identifies one selected spinful Dirac sector without requiring a complete
classification of other half-space bound states.

As for $H_{\rm sq}$, we unfold the selected momentum domain by rescaling $\mathbf k_{\rm B}=2\mathbf p$. On a
$(2L)\times(2L)$ parent torus with APBC and even $L$, this maps the $L^2$ momenta in $D$ onto an $L\times L$
APBC grid. The unfolded surface symbol is
\begin{equation}
h_{\rm unf}(\mathbf k_{\rm B})
=2t\left[\sin\frac{k_{{\rm B},x}}2\,\sigma_x
+\sin\frac{k_{{\rm B},y}}2\,\sigma_z\right].
\label{eq:app_2d_AII_unfolded}
\end{equation}
The half-angle functions are discontinuous under periodic continuation across the Brillouin-zone seams,
producing algebraically decaying hopping. The canonical Bogoliubov transformation below instead yields the
finite-range Hamiltonian of Eq.~\eqref{eq:2d_AII_edge_BdG}.

\paragraph{Canonical BdG--Majorana correspondence.}
We first construct a canonical Bogoliubov transformation that converts the unfolded surface Hamiltonian into a
finite-range BdG model. In the Nambu basis
\begin{equation}
\Psi_{\mathbf k_{\rm B}}=
(d_{\mathbf k_{\rm B}\uparrow},
d_{\mathbf k_{\rm B}\downarrow},
d_{-\mathbf k_{\rm B}\uparrow}^\dagger,
d_{-\mathbf k_{\rm B}\downarrow}^\dagger)^T,
\end{equation}
the identity $h_{\rm unf}(-\mathbf k_{\rm B})^T =-h_{\rm unf}(\mathbf k_{\rm B})$ gives
\begin{equation}
h_{\rm unf}^{\rm Nambu}
=\kappa_0\otimes h_{\rm unf}.
\end{equation}
Define
\begin{equation}
\begin{aligned}
P_\pm =\tfrac12(1\pm& \kappa_x),\ \
U_0=P_++P_-i\sigma_y,\\
U_*(\mathbf k_{\rm B})
=& e^{ik_{{\rm B},x}  \kappa_y\sigma_x/4}
e^{ik_{{\rm B},y}\sigma_x/4}U_0.
\end{aligned}
\label{eq:app_2d_AII_Ustar}
\end{equation}
Direct multiplication gives
\begin{equation}
\begin{aligned}
U_*h_{\rm unf}^{\rm Nambu}(\mathbf k_{\rm B})U_*^\dagger
&=h_\partial^{\rm BdG}(\mathbf k_{\rm B}),\\
\kappa_xU_*(\mathbf k_{\rm B})^*\kappa_x
&=U_*(-\mathbf k_{\rm B}).
\end{aligned}
\label{eq:app_2d_AII_intertwiner}
\end{equation}
The second identity ensures Nambu particle--hole compatibility, so $U_*$ defines an exact canonical
transformation on compatible finite grids. Although the unfolded Hamiltonian has half-angle discontinuities, the
transformed symbol in Eq.~\eqref{eq:2d_AII_edge_BdG} is periodic and has only onsite and nearest-neighbor
couplings.

We next identify the local BdG Hamiltonian with $H_{\rm sq}$. Write $d_{\mathbf R,\sigma} =(a_{\mathbf
R,\sigma}+ib_{\mathbf R,\sigma})/2$ and $\boldsymbol\gamma_{{\rm B};m,n}
=(a_\uparrow,a_\downarrow,b_\uparrow,b_\downarrow)^T$ in each BdG cell $\mathbf R=(m,n)$. The exact Majorana
reblocking is
\begin{equation}
\begin{aligned}
& \begin{pmatrix}
\gamma_{2m,2n}\\
\gamma_{2m,2n+1}\\
\gamma_{2m+1,2n}\\
\gamma_{2m+1,2n+1}
\end{pmatrix}
=\mathcal O_n\boldsymbol\gamma_{{\rm B};m,n},\\
& \mathcal O_n
=\frac{(-1)^n}{2}
\begin{pmatrix}
1&-1&-1&1\\
1&1&1&1\\
1&-1&1&-1\\
-1&-1&1&1
\end{pmatrix}.
\end{aligned}
\label{eq:app_2d_AII_sq_O}
\end{equation}
The matrix $\mathcal O_n\in SO(4)$ combines the onsite Majorana rotation with the local $\mathbb Z_2$ gauge
transformation, while the four components are assigned to a $2\times2$ square plaquette. The gauge pattern has
period four along $y$. Substituting this map into the nearest-neighbor BdG Hamiltonian of
Eq.~\eqref{eq:2d_AII_edge_BdG} reproduces every bond of $H_{\rm sq}$ in Eq.~\eqref{eq:2d_sq_micro}, including
its $\pi$-flux signs and hopping amplitude $t$. This establishes the exact Hamiltonian correspondence of
Eq.~\eqref{eq:2d_AII_exact_sq_correspondence} on compatible finite tori.

\paragraph{Transported compact charge and Kramers sewing.}
We next show that the same canonical transformation also identifies the compact charge and Kramers symmetry. Let
$c_i=\cos(k_{{\rm B},i}/2)$ and $s_i=\sin(k_{{\rm B},i}/2)$, for $i=x,y$. The onsite particle number inherited
from the bulk becomes the BdG charge matrix
\begin{equation}
q_*(\mathbf k_{\rm B})
 =(-\kappa_yc_xc_y+s_xs_y)\sigma_y +(\kappa_yc_xs_y+s_xc_y)\sigma_z.
\label{eq:app_2d_AII_charge}
\end{equation}
It satisfies $q_*^2=1$, $[q_*,h_\partial^{\rm BdG}]=0$, and
\begin{equation}
\kappa_xq_*(\mathbf k_{\rm B})^*\kappa_x
=-q_*(-\mathbf k_{\rm B}).
\end{equation}
Thus the centered charge generates an exact compact $U(1)$ symmetry. Its discontinuity across the Brillouin-zone
seam is visible from $q_*(\pi^-,0)=\sigma_z$ and $q_*(-\pi^+,0)=-\sigma_z$.

Applying the Majorana reblocking of Eq.~\eqref{eq:app_2d_AII_sq_O} gives the charge matrix in the $2\times2$
square cell:
\begin{equation}
q_{s_P}^{(2\times2)}(\mathbf K_{\rm B})=
\begin{pmatrix}
0&0&0&\Omega_+\\
0&0&\Omega_-&0\\
0&\Omega_-^*&0&0\\
\Omega_+^*&0&0&0
\end{pmatrix},
\label{eq:app_2d_AII_sq_qsharp_fold}
\end{equation}
where $\mathbf K_{\rm B}=(k_{{\rm B},x},k_{{\rm B},y}+\pi)$ and $\Omega_\pm=e^{-i(K_x\pm K_y)/2}$. To identify
this matrix with the sharp square charge, fold the primitive $A/B$ modes into the $2\times2$ cell, placing $B$
immediately to the left of $A$. The two primitive momenta $\mathbf q=(K_x/2,K_y/2)$ and $\mathbf q'=\mathbf
q+(\pi,0)$ have opposite values of $s_P=\operatorname{sgn}(2\cos q_x\sin q_y)$. Their charge difference gives
precisely Eq.~\eqref{eq:app_2d_AII_sq_qsharp_fold}. Hence the canonical reblocking identifies the inherited
surface charge with $Q_{s_P}$ on the complete finite momentum grid, including the UV seams.

The parent Kramers symmetry is transported by the same Bogoliubov transformation:
\begin{equation}
\begin{aligned}
\widetilde U_T(\mathbf k_{\rm B})
&=U_*(\mathbf k_{\rm B})(i\sigma_y)
U_*(-\mathbf k_{\rm B})^T\\
&=e^{ik_{{\rm B},y}\sigma_x/2}i\sigma_y.
\end{aligned}
\label{eq:app_2d_AII_TR}
\end{equation}
On chosen momentum representatives it obeys
$\widetilde U_T(\mathbf k_{\rm B})\widetilde U_T(-\mathbf k_{\rm B})^*=-1$ and acts covariantly on both the
transformed Hamiltonian and charge. Importantly, $\widetilde U_T$ is not periodic across the unfolded
Brillouin-zone seams. After the local Majorana reblocking, exact canonical transport yields the operation
$\Theta_{\partial}$ of Eq.~\eqref{eq:2d_sq_induced_TR}, which preserves $(H_{\rm sq},Q_{s_P})$ and retains
the parent one-particle Kramers algebra on compatible finite grids. This does not show that
$\Theta_{\partial}$ is local in the primitive square lattice, nor that its sewing can be replaced by a
globally periodic primitive-cell operation with square $-1$. By contrast, the transported chiral operation is
the local $S_{\rm sq}=\sigma_z$ of Eq.~\eqref{eq:2d_sq_induced_chiral}.

\paragraph{Charge-preserving mass and boundary stability.}
The intermediate BdG Hamiltonian admits a finite-range mass that preserves the compact charge $q_*$. Explicitly,
\begin{align}
M_{s_P}(\mathbf k_{\rm B})=\frac14\bigl[
&(1+\cos k_{{\rm B},x})(1+\cos k_{{\rm B},y})\,\sigma_y\nonumber\\
&-\sin k_{{\rm B},x}\sin k_{{\rm B},y}\,\kappa_y\sigma_y\nonumber\\
&-(1+\cos k_{{\rm B},x})\sin k_{{\rm B},y}\,\sigma_z\nonumber\\
&-\sin k_{{\rm B},x}(1+\cos k_{{\rm B},y})\,\kappa_y\sigma_z\bigr].
\label{eq:2d_AII_charge_preserving_mass}
\end{align}
For $h_0=h_\partial^{\rm BdG}$, direct calculation gives
\begin{equation}
[M_{s_P},q_*]=0,\ \
\{M_{s_P},h_0\}=0,\ \
M_{s_P}^2=c_x^2c_y^2\mathbf1.
\label{eq:app_2d_AII_mass_algebra}
\end{equation}
Thus $h_0+mM_{s_P}$ is fully gapped for $m\ne0$, with
\begin{equation}
E_m^2=4t^2(s_x^2+s_y^2)+m^2c_x^2c_y^2.
\label{eq:2d_AII_charge_mass_spectrum}
\end{equation}
This mass is the reblocked form of $c_xc_y\sigma_y$ in the unfolded Dirac frame and defines a finite-range,
charge-preserving gapped descendant of $H_{\rm sq}$.

At the massless point, folding the nodes to a PHIM makes the doubled-cell BdG Pfaffian-sign product undefined,
although the primitive-cell result $\mathcal P_D(H_{\rm sq})=-1$ remains well defined. Enlarging the unit cell
cannot change the Chern parity of a fixed gapped Hamiltonian. Different gapping perturbations can, however,
yield different gapped descendants: the perturbation $\delta\kappa_z$ gives Pfaffian product $+1$, which
cannot be assigned to the original massless model.

Finally, the AIII and AII descendants of Eqs.~\eqref{eq:2d_AIII_chiral_Hchi}
and~\eqref{eq:2d_AII_genuine_Htheta} differ from $\mathcal H_*$ only through terms proportional to $P_{\rho,+}$.
Since the selected surface states have $\rho_x=-1$ at every layer, these perturbations leave their
wavefunctions, energies, and compact charge unchanged along the path of
Eq.~\eqref{eq:2d_AIII_AII_fixed_boundary_path}. Their strength is bounded by
$\max(\delta,4|\epsilon_\Theta|)<\Delta_*$, where $\Delta_*$ is the bulk excitation gap of $\mathcal H_*$. Hence
the bulk remains gapped throughout the path. The exact surface correspondence holds at zero gauge background and
does not imply equivalence of independently gauged Hamiltonians.

\subsection{Compact lattice gauging and parity-anomaly inflow}
\label{app:2d_surface_gauging}

At finite cutoff, the autonomous charge-onsite square regulator and the 3D bulk parents have different compact
gauge couplings: Wilson lines on Schwartz hoppings versus finite-range compact links. Although their zero-field
surface theories are exactly equivalent, their magnetic responses need only agree in the IR. We therefore
compare the models in the same smooth magnetic background, focusing on spectral isolation, UV half-charge, and
bulk--boundary indices.

\subsubsection{Compact magnetic couplings and spectral isolation}
\label{app:2d_sq_3d_flux_compactness}

The exact zero-field correspondence between the direct AIII parent and the autonomous square regulator does not
identify their independently gauged Hamiltonians. We compare them in the same smooth magnetic background to
establish a common low-energy Dirac mode. The important point is to exclude additional lattice or bulk states
near zero energy; only then can the remaining autonomous spectrum be identified as the UV contribution to the
compact gauge response.

Consider an $L\times L$ transverse torus with lattice spacing $a$ and $L\in4\mathbb Z$. We take $a\to0$ with
$La$ fixed, keeping $t,\Lambda>0$ and $\lambda=1/2$ in Eq.~\eqref{eq:2d_sq_3d_parent_H}. The smooth background
carries one flux quantum, $\int_{T^2}F=2\pi$, and is discretized by compact Wilson links, with $O(a^2)$ flux
through each plaquette. We choose compatible flat holonomies and antiperiodic Dirac-node boundary conditions.
For the autonomous regulator, the same physical field is expressed in square coordinates $k_x=V/2$ and
$k_y=(P-V)/2$. This matches the continuum magnetic problem, not the finite-cutoff gauge couplings.

\paragraph{Direct AIII magnetic coupling and Dirac limit.}
Let $T_x,T_y$ be the gauge-covariant forward shifts, generalizing the 1D shift $T_U$. For one uniform flux quantum, set $B=2\pi/L^2$ and choose
\begin{equation}
\begin{aligned}
(T_x)_{(x,y),(x+1\ {\rm mod}\ L,y)}
&=\begin{cases}
e^{-iBy},&x<L-1,\\
-e^{-iBy},&x=L-1,
\end{cases}\\
(T_y)_{(x,y),(x,y+1\ {\rm mod}\ L)}
&=\begin{cases}
1,&y<L-1,\\
-e^{+iLBx},&y=L-1.
\end{cases}
\end{aligned}
\label{eq:app_2d_3d_flux_links}
\end{equation}
The boundary minus signs implement the antiperiodic spin structure. These links obey $T_xT_y=e^{+iB}T_yT_x$ and have counterclockwise plaquette product $e^{+iB}$. For a general smooth field, $T_x,T_y$ denote the corresponding covariant shifts with the same boundary conditions.

The magnetic square-lattice hopping is
\begin{equation}
h_\Phi=i(T_x-T_x^\dagger)
-i\epsilon(T_y+T_y^\dagger),
\label{eq:app_2d_3d_flux_transverse}
\end{equation}
where $\epsilon_{x,y}=(-1)^{x+y}$ encodes the staggered sublattice structure. The zero-field mass $1-\cos
k_x\sin k_y$ becomes $M_\Phi=\mathbf1-\tfrac12\{C_x,S_y\}$, with $C_x=(T_x+T_x^\dagger)/2$ and
$S_y=(T_y-T_y^\dagger)/(2i)$. The anticommutator symmetrizes the two shortest Wilson paths; this is a convenient
gauge-covariant prescription. With $t=\Lambda=1$ for the displayed coefficients, the magnetic parent is
\begin{equation}
\mathcal H_\Phi(w)=\rho_y\sin w
+\rho_z(M_\Phi+\cos w)-\rho_xh_\Phi.
\label{eq:app_2d_3d_flux_H}
\end{equation}
For general $t,\Lambda>0$, $h_\Phi$ carries the factor $t$ and the longitudinal terms carry $\Lambda$. The
coupling is finite range and gauge covariant. It also preserves the chiral grading and the longitudinal
reflection:
\begin{equation}
\begin{aligned}
\mathcal S_\Phi&=\rho_x\epsilon,\quad
\mathcal R_z=\rho_z\epsilon,\\
\{\mathcal S_\Phi,\mathcal H_\Phi\}&=0,\quad
\mathcal R_z\mathcal H_\Phi(w)\mathcal R_z^{-1}
=\mathcal H_\Phi(-w).
\end{aligned}
\label{eq:app_2d_3d_flux_inversion}
\end{equation}

Why does this magnetic coupling retain only the selected surface Dirac fermion at low energies? At zero field,
the mass $M_0=1-\cos k_x\sin k_y$ vanishes quadratically at $\mathbf K_+=(0,\pi/2)$ and is positive away from
this sector, modulo sublattice folding. Its magnetic extension $M_a=M_\Phi$ is one eighth of a sum of four
positive squares $D_j^\dagger D_j$. Each $D_j$ compares the wavefunction with its gauge-covariant translation
along a diagonal of the square lattice, including the Bloch phase at $\mathbf K_+$. The four terms cover both
diagonal directions and both orders of the elementary hops. Since each translation is unitary, this gives
\begin{equation}
0\le M_a\le2.
\label{eq:app_2d_mag_M_squares}
\end{equation}
Let $H_a$ be the corresponding half-space Hamiltonian. The transverse kinetic energy penalizes states away from
the Dirac sector, while the longitudinal chain has no surface zero mode when its mass exceeds one. Set
$S_x=(T_x-T_x^\dagger)/(2i)$ and $C_y=(T_y+T_y^\dagger)/2$. Expanding the transverse hopping gives
\begin{equation}
h_\Phi^2=4M_a(2-M_a)+4S_xC_y^2S_x+R_a,
\qquad \|R_a\|\le Ca^2.
\end{equation}
The displayed square is nonnegative. The remainder and the commutator $[h_\Phi,M_a]$ are $O(a^2)$: each
noncommuting reordering of the finite-range covariant shifts contributes a plaquette Wilson phase differing
from unity by $O(a^2)$ for the fixed smooth field. The longitudinal half-space operator $L(M_a)$ obeys
$L(M_a)^2\ge(M_a-1)_+^2$, and its mixed terms with the transverse hopping are also $O(a^2)$. Therefore,
using $0\le M_a\le2$,
\begin{equation}
\begin{aligned}
H_a^2&\ge 4M_a(2-M_a)+(M_a-1)_+^2-Ca^2\\
&\ge\tfrac12 M_a-Ca^2.
\end{aligned}
\label{eq:app_2d_mag_coercivity}
\end{equation}
All constants are independent of $a$. Every normalized eigenstate with $|E|=O(a)$ therefore satisfies
$\langle M_a\rangle=O(a^2)$.

The same four squares also control spatial variation. Since they are all nonnegative, $\langle
M_a\rangle=O(a^2)$ forces each difference $D_j\Psi_a$ to have norm $O(a)$. In particular, the two diagonal
translations obey
\begin{equation}
\|(\mathbf1+iT_xT_y)\Psi_a\|
+\|(\mathbf1-iT_xT_y^\dagger)\Psi_a\|\le C'a.
\label{eq:app_2d_mag_H1_bound}
\end{equation}
The phases $\pm i$ are precisely those of the selected Dirac point under the two diagonal translations. After
factoring out that rapid Bloch phase, the bound says that the remaining envelope changes by only $O(a)$ over one
lattice step. These two diagonal directions generate each staggered sublattice separately. Retaining the
two Dirac-spinor components and the transverse kinetic coupling between them excludes independent
lattice-scale oscillations on the complementary sublattice. On the fixed physical torus, these bounds give
convergent continuum subsequences of normalized low-energy states.

The same small mass also confines low-energy states to the boundary. For a mass eigenvalue $m<1$, the
longitudinal surface profile is proportional to $(-m)^{z-1}$, with probability $m^2$ beyond the first layer.
Because $M_a^2\le2M_a$ and $\langle M_a\rangle=O(a^2)$, this probability is $O(a^2)$ in the low-energy sector.
The longitudinal gap suppresses admixtures of other bulk states. Thus, in wavefunction norm,
\begin{equation}
\Psi_a(z,x,y)=\delta_{z,1}u_-\,\varphi_a(x,y)+O(a),
\label{eq:app_2d_mag_first_layer_localization}
\end{equation}
where $\rho_xu_-=-u_-$ and $\varphi_a$ is a two-component surface wavefunction.

It remains to identify the dynamics of this surface state. Near $\mathbf K_+$, the hopping has the linear magnetic Dirac form, while the mass contributes only at quadratic order. Write $D_A=-2t(\sigma_x\Pi_x+\sigma_y\Pi_y)$, where $\Pi_\mu=-i\partial_\mu+A_\mu$. For a smooth continuum spinor $\chi$, let $\Psi_a^{\rm lat}[\chi]$ denote its lattice embedding, obtained by evaluating $\chi$ at the surface sites, restoring the Dirac-node phase and magnetic boundary conditions, and placing it in the $z=1$, $\rho_x=-1$ sector. Then
\begin{equation}
\|H_a\Psi_a^{\rm lat}[\chi]
-a\Psi_a^{\rm lat}[D_A\chi]\|
\le C_\chi a^2.
\label{eq:app_2d_mag_bulk_consistency}
\end{equation}
Thus a continuum Dirac zero mode yields a lattice state with an $O(a^2)$ energy residual. Conversely, the mass, surface-localization, and diagonal-difference bounds ensure that every lattice eigenstate of energy $O(a)$ converges, along a subsequence, to a continuum Dirac eigenstate. This rules out low-energy modes that an expansion around $\mathbf K_+$ alone might miss.

For one flux quantum, $D_A$ has exactly one zero mode $\varphi_0$, separated from its other eigenvalues by a positive gap $\delta$. The lattice embedding guarantees at least one lattice eigenvalue within $O(a^2)$ of zero. To see that there is only one, choose $0<c<\delta/2$. If two orthogonal lattice eigenstates lay in $(-ca,ca)$ for arbitrarily small $a$, the preceding bounds would give two orthogonal continuum zero modes, contradicting $\dim\ker D_A=1$. Therefore exactly one AIII eigenvalue lies in this window for sufficiently small $a$. The chiral symmetry in Eq.~\eqref{eq:app_2d_3d_flux_inversion} pairs every nonzero energy $E$ with $-E$; a single isolated eigenvalue must consequently be pinned exactly at zero.

\paragraph{Autonomous magnetic mode and IR correspondence.}
The autonomous regulator requires a separate check because its compact gauge coupling acts on Schwartz-local rather than finite-range hoppings. Its zero-field squared energy is
\begin{equation}
(h_\sharp^{\rm sq})^2=e^2\mathbf1,
\label{eq:app_2d_mag_aut_energy}
\end{equation}
where $e^2(P,V)=t^2\bigl( |1+e^{i(P-V)}|^2+|1-e^{iV}|^2\bigr)$, with a single zero at $(P,V)=(\pi,0)$. Replacing the two Bloch factors by their gauge-covariant translations $T_{P-V}$ and $T_V$ gives the positive operator
\begin{equation}
\mathcal L_a=t^2\bigl[
(\mathbf1+T_{P-V})^\dagger(\mathbf1+T_{P-V})
+(\mathbf1-T_V)^\dagger(\mathbf1-T_V)\bigr].
\label{eq:app_2d_mag_aut_laplacian}
\end{equation}
The operators $T_{P-V}$ and $T_V$ carry the Wilson phases of the same physical magnetic background. At zero field, $\mathcal L_a$ equals the square of the autonomous Hamiltonian. The estimate follows by expanding the square in the covariantly coupled Fourier hoppings. For two displacements $\mathbf R,\mathbf S$, distinct Wilson-path orderings differ by a phase bounded by $Ca^2(1+|\mathbf R|+|\mathbf S|)^2$ in the fixed smooth field. The corresponding weighted double sum converges uniformly by Schwartz decay of the hopping coefficients, giving
\begin{equation}
\|h_{\sharp,a}[A]^2-\mathcal L_a\|\le Ca^2.
\label{eq:app_2d_mag_aut_product_bound}
\end{equation}
Consequently, an autonomous state of energy $O(a)$ has small covariant hopping differences around the unique auxiliary Dirac point. Its envelope has the same continuum control as in the AIII argument, and the expansion on smooth states gives the same physical operator $aD_A$ with $O(a^2)$ corrections. The preceding existence-and-uniqueness argument therefore applies to this independently gauged model as well: for sufficiently small $a$, it has precisely one eigenvalue in $(-ca,ca)$, with $|E_{{\rm aut},a}|=O(a^2)$. Unlike the AIII parent, it need not have a symmetry pinning this energy exactly to zero.

The $O(a)$ comparison follows from the isolated spectral gap. For either model, let $H_a$ denote its gauged operator, let $P_a$ project onto the unique eigenstate in $(-ca,ca)$, and let $\Psi_a^{\rm lat}[\varphi_0]$ be the normalized lattice embedding of the same continuum zero mode. The $O(a^2)$ embedding residual and the separation of the remaining spectrum by at least $ca$ imply
\begin{equation}
\|(1-P_a)\Psi_a^{\rm lat}[\varphi_0]\|
\le\frac{\|H_a\Psi_a^{\rm lat}[\varphi_0]\|}{ca}=O(a).
\end{equation}
Thus the normalized low-energy eigenstate of each model differs from its embedded state by $O(a)$, after choosing consistent overall phases. Expressing their surface envelopes in the same physical coordinates gives
\begin{equation}
\|\varphi_{a,\rm AIII}^{\rm phys}
-\varphi_{a,\rm aut}^{\rm phys}\|=O(a).
\label{eq:app_2d_mag_wavefunction_correspondence}
\end{equation}
This is an asymptotic correspondence of the isolated magnetic Dirac states, not an exact intertwiner between the finite-cutoff gauge couplings. It identifies the single IR level used in the next subsection, where the remaining autonomous spectrum is shown to supply the compensating UV half-charge.

\subsubsection{Autonomous UV spectral asymmetry and temporal holonomy}
\label{app:2d_sq_3d_static_eta}

We establish the finite-lattice spectral count and temporal-holonomy limit used in
Eqs.~\eqref{eq:2d_AIII_static_eta_UV} and~\eqref{eq:2d_aut_temporal_holonomy_main}. The key ingredient is a
smooth UV orbital mass whose magnetic coupling remains nonnegative up to $O(a^2)$ corrections.

Consider the orbital rotation defining $h_\sharp^{\rm sq}$ in Eq.~\eqref{eq:2d_sq_hsharp}. Choose the rotation
angle $\theta=(\pi/2)f^2$, where $0\le f\le1$ is smooth and periodic, vanishes near the Dirac point, and equals
one in the momentum-space seam collars. The resulting nonnegative diagonal mass admits the factorization
\begin{equation}
m_W=e\sin\theta=r^2,\quad
r=f\sqrt{e\,\frac{\sin[(\pi/2)f^2]}{f^2}}.
\label{eq:app_2d_static_square_root}
\end{equation}
The quotient extends smoothly through $f=0$, and $e>0$ wherever $f\ne0$. Hence $r$ is real, smooth, and
periodic.

Let $M_a$ and $R_a$ denote the Hermitian operators obtained by coupling the Fourier hoppings of $m_W$ and $r$,
respectively, through the same Wilson-line prescription. Although $m_W=r^2$ at zero field, the composed Wilson
paths in $R_a^2$ can differ from those in $M_a$ by enclosed magnetic fluxes. For a fixed smooth background, the
corresponding phase differences are $O(a^2)$. The Schwartz decay of the hoppings makes their summed contribution
uniformly bounded, giving
\begin{equation}
\|M_a-R_a^2\|\le C_Ma^2,\quad
M_a\ge-C_Ma^2.
\label{eq:app_2d_static_mass_bound}
\end{equation}

In the $\sigma_z$ orbital basis, the gauged autonomous Hamiltonian takes the form
\begin{equation}
h_{\sharp,a}[A]
=\begin{pmatrix}
M_a&K_a\\
K_a^\dagger&-M_a
\end{pmatrix}.
\label{eq:app_2d_static_block}
\end{equation}
By the spectral-isolation result established above, it has a unique eigenvalue $|E_a(0)|\le C_Ea^2$, while all
remaining eigenvalues satisfy $|E_{n,a}|\ge ca$.

Introduce $h_{\sharp,a}(s)=h_{\sharp,a}[A]+s\sigma_z$ and choose $s_*=Ta^2$, where $T>\max(C_M,2C_E)$ is fixed.
Equation~\eqref{eq:app_2d_static_mass_bound} ensures that $B_a=M_a+s_*>0$. The Schur-complement inertia identity
then gives
\begin{equation}
\begin{aligned}
\operatorname{sig}h_{\sharp,a}(s_*)
&=\operatorname{sig}B_a
+\operatorname{sig}(-B_a-K_a^\dagger B_a^{-1}K_a)\\
&=d_a-d_a=0,
\end{aligned}
\label{eq:app_2d_static_inertia}
\end{equation}
where $d_a$ is the dimension of either orbital block and $\operatorname{sig}H=n_+(H)-n_-(H)$. Thus the shifted
Hamiltonian has equally many positive- and negative-energy states.

For the chosen positive-flux orientation, the continuum Dirac zero mode satisfies $\sigma_z\varphi_0=-\varphi_0$.
The isolated lattice eigenstate converges to this mode with $O(a)$ error. Since the gap to the remaining
spectrum is of order $a$, the perturbation $s=O(a^2)$ changes its eigenvector by at most $O(a)$. The
Hellmann--Feynman relation therefore gives
\begin{equation}
\frac{dE_a(s)}{ds}
=\langle\psi_a(s),\sigma_z\psi_a(s)\rangle
=-1+O(a)
\end{equation}
uniformly for $0\le s\le s_*$. Consequently, $E_a(s_*)=E_a(0)-Ta^2[1+O(a)]<0$ for sufficiently small $a$.

The isolated level contributes $-1$ to the vanishing total signature in Eq.~\eqref{eq:app_2d_static_inertia}, so
the UV levels have signature $+1$. Moreover, removing the shift changes each eigenvalue by at most $Ta^2$, which
is smaller than the UV gap $ca$. Their signs therefore remain unchanged, and the original autonomous regulator
satisfies
\begin{equation}
\eta_{{\rm UV},a}(A)
=\sum_{|E_{n,a}|\ge ca}\operatorname{sgn}E_{n,a}
=+1.
\label{eq:app_2d_static_eta_result}
\end{equation}
Filling the negative-energy states gives the centered UV vacuum charge
\begin{equation}
Q_{\rm UV}
=\sum_{n\in{\rm UV}}
\left(\langle d_n^\dagger d_n\rangle-\frac12\right)
=-\frac{\eta_{{\rm UV},a}}2
=-\frac12.
\end{equation}
This establishes Eq.~\eqref{eq:2d_AIII_static_eta_UV}, independently of the sign of the unshifted near-zero
eigenvalue. The result applies to the specified flux orientation, UV mass, and Wilson-line prescription.

We next establish the temporal-holonomy limit. For the centered fermion character defined in the main text,
\begin{equation}
Z_a(\alpha)=\prod_n 2\cosh\!\left(
\frac{\beta E_{n,a}-i\alpha}{2}\right),
\label{eq:app_2d_static_character}
\end{equation}
choose $\beta$ such that $\beta a^2\to0$ and $n_{{\rm modes},a}e^{-\beta ca}\to0$, where $n_{{\rm
modes},a}=O(a^{-2})$. For example, $\beta=a^{-3/2}$ satisfies both conditions. The isolated level then
contributes $\cos(\alpha/2)+O(\beta a^2)$ to the normalized character.

For a UV level, the normalized character approaches $e^{-i\alpha\operatorname{sgn}(E_{n,a})/2}$, with
corrections of order $e^{-\beta|E_{n,a}|}$. Multiplying over the UV spectrum and using
Eq.~\eqref{eq:app_2d_static_eta_result} gives
\begin{equation}
\begin{aligned}
\frac{Z_{{\rm UV},a}(\alpha)}{Z_{{\rm UV},a}(0)}
&=e^{-i\alpha\eta_{{\rm UV},a}/2}
+O(n_{{\rm modes},a}e^{-\beta ca})\\
&=e^{-i\alpha/2}
+O(n_{{\rm modes},a}e^{-\beta ca}).
\end{aligned}
\label{eq:app_2d_static_uv_character}
\end{equation}
Combining the IR and UV contributions yields
\begin{equation}
\begin{aligned}
\frac{Z_a(\alpha)}{Z_a(0)}
&=e^{-i\alpha/2}\cos\frac\alpha2
+O\!\left(\beta a^2+
n_{{\rm modes},a}e^{-\beta ca}\right)\\
&=\frac{1+e^{-i\alpha}}2+o(1).
\end{aligned}
\label{eq:app_2d_static_character_limit}
\end{equation}
This proves Eq.~\eqref{eq:2d_aut_temporal_holonomy_main}. The complete even-orbital lattice character is exactly
$2\pi$-periodic at finite $a$. The limiting statement concerns temporal holonomy at fixed magnetic background,
not a determinant-phase identity for arbitrary spacetime gauge fields.

\subsubsection{Direct AIII magnetic index and boundary parity}
\label{app:2d_sq_3d_flux}

We establish the magnetic index of the directly gauged AIII parent and its boundary consequence, as used in
Sec.~\ref{sec:2d_surface_magnetic_inflow}. Set $t=\Lambda=1$ and $\lambda=1/2$, and consider one uniform flux
quantum $\Phi=2\pi$ through an $L\times L$ transverse torus, with plaquette flux $\Phi/L^2$. The finite-range
compact-link Hamiltonian $\mathcal H_\Phi(w)$ is defined in Eq.~\eqref{eq:app_2d_3d_flux_H}, where $w$ is the
longitudinal momentum. Its exact chiral grading is $\mathcal S_\Phi=\rho_x\epsilon$, with
$\epsilon_{x,y}=(-1)^{x+y}$ and $\{\mathcal S_\Phi,\mathcal H_\Phi(w)\}=0$. In a chiral basis,
\begin{equation}
\mathcal H_\Phi(w)\simeq
\begin{pmatrix}
0&D_\Phi(w)\\
D_\Phi^\dagger(w)&0
\end{pmatrix}.
\end{equation}
When the longitudinal family is gapped, its finite-volume winding is
\begin{equation}
w_1^{(L)}(\Phi)=\frac{1}{2\pi i}
\int_{-\pi}^{\pi}dw\,
\operatorname{tr}\!\left[D_\Phi^{-1}(w)\partial_wD_\Phi(w)\right].
\end{equation}

The special structure of the direct parent makes this winding exactly reducible to endpoint spectral counts. Its
off-diagonal block can be chosen so that
\begin{equation}
\begin{aligned}
D_\Phi(w)\epsilon
&=F_\Phi(\cos w)+i\sin w\,\mathbf1,\\
F_\Phi(s)&=h_\Phi+\epsilon(M_\Phi+s).
\end{aligned}
\label{eq:app_2d_3d_flux_offdiag}
\end{equation}
Because $h_\Phi$ and $M_\Phi$ are Hermitian and $[\epsilon,M_\Phi]=0$, the matrices $F_\Phi(s)$ are Hermitian.
If $\xi_j(s)$ are their real eigenvalues, the eigenvalues of $D_\Phi(w)\epsilon$ are $\xi_j(\cos w)+i\sin w$.
They can vanish only at $w=0,\pi$. Thus invertibility of $F_{\Phi,\pm}=F_\Phi(\pm1)$ guarantees a gap for the
entire longitudinal family. Each eigenvalue curve crosses the real axis only at those endpoints, and summing its
winding gives
\begin{equation}
w_1^{(L)}(\Phi)=n_-(F_{\Phi,-})-n_-(F_{\Phi,+}).
\label{eq:app_2d_3d_flux_count_identity}
\end{equation}
Here $n_-(X)$ counts negative eigenvalues of a Hermitian matrix $X$. This identity is exact at finite volume; no
homotopy or continuum approximation is required.

At zero field, the two endpoint Bloch Hamiltonians are
\begin{equation}
\begin{aligned}
F_+(u,v)&=h_{\rm sq}(u,v)+[2-\tfrac12g(u,v)]\sigma_z,\\
F_-(u,v)&=h_{\rm sq}(u,v)-\tfrac12g(u,v)\sigma_z,
\end{aligned}
\label{eq:app_2d_3d_flux_endpoint_bloch}
\end{equation}
where $g(u,v)=\sin u-\sin v$. In the normalization used here, the $F_+$ mass is at least one, while the squared
energy of $F_-$ is at least one. Hence both zero-field endpoints have a spectral gap bounded below by one. Their
occupied-band Chern numbers in the physical square-lattice orientation are
\begin{equation}
C_-^{xy}=+1,\ \ C_+^{xy}=0.
\label{eq:app_2d_3d_flux_endpoint_chern}
\end{equation}

We next establish endpoint-gap stability at one flux quantum. Each $F_{\Phi,\pm}$ is a finite-range polynomial
in the magnetic covariant shifts. Cover the transverse torus with overlapping boxes of side $R\ll L$ and choose
a quadratic partition of unity $\sum_j\chi_j^2=\mathbf1$, with bounded overlap and discrete gradients
$O(R^{-1})$. On each box, a local gauge makes the magnetic links differ from their zero-field values by
$O(|\Phi|R/L^2)$. Finite hopping range and the zero-field bound $\|F_{0,\pm}\varphi\|\ge\|\varphi\|$ then give
$\|F_{\Phi,\pm}\chi_j\psi\|
\ge(1-\varepsilon_R)\|\chi_j\psi\|$,
where $\varepsilon_R=C|\Phi|R/L^2$. Moreover, the commutators $[F_{\Phi,\pm},\chi_j]$ are controlled by the
cutoff gradients and satisfy $\sum_j\|[F_{\Phi,\pm},\chi_j]\psi\|^2
\le C R^{-2}\|\psi\|^2$.
Writing $F=F_{\Phi,\pm}$, we obtain the following bound, where
$s_{\min}(X):=\inf_{\|\psi\|=1}\|X\psi\|$ denotes the smallest singular value:
\begin{equation}
\begin{aligned}
(1-\varepsilon_R)\|\psi\|
&\le\Bigl(\sum_j\|F\chi_j\psi\|^2\Bigr)^{1/2}\\
&\le\|F\psi\|+CR^{-1}\|\psi\|,
\end{aligned}
\end{equation}
where the second inequality uses $F\chi_j=\chi_jF+[F,\chi_j]$ and $\sum_j\chi_j^2=\mathbf1$. Consequently,
\begin{equation}
s_{\min}(F_{\Phi,\pm})
\ge1-C\bigl(R^{-1}+|\Phi|R/L^2\bigr).
\label{eq:app_2d_3d_flux_weak_field_gap}
\end{equation}
The estimate is uniform in compatible magnetic boundary twists. Fix $R=R_0$ sufficiently large that the first
error is smaller than $1/4$. For $\Phi=2\pi$, the second error is also smaller than $1/4$ for sufficiently large
$L$. Thus both endpoint gaps remain open, and the exact endpoint reduction guarantees a gap throughout the
longitudinal family.

At zero flux, each endpoint has $L^2/2$ negative-energy states, because the primitive cell contains two sites.
For $\Phi=2\pi$ of positive continuum orientation, the lattice St\v{r}eda (magnetic gap-labeling)
relation~\cite{stredaTheory1982,Dana1985}, with $U=e^{+i\int A}$, together with the open endpoint gaps
and $C_-^{xy}=1$, $C_+^{xy}=0$, gives
\begin{equation}
n_-(F_{2\pi,-})=\frac{L^2}{2}-1,\ \
n_-(F_{2\pi,+}) =\frac{L^2}{2}.
\label{eq:app_2d_3d_flux_large_L}
\end{equation}
Consequently $w_1^{(L)}(0)=0$ and $w_1^{(L)}(2\pi)=-1$.

The winding also fixes the occupied-state longitudinal Berry phase. Spectrally flatten the chiral Hamiltonian
and put $U_\Phi=D_\Phi(D_\Phi^\dagger D_\Phi)^{-1/2}$. An occupied frame is $(-U_\Phi,\mathbf1)^T/\sqrt2$, whose
Berry phase satisfies $\gamma_z=-\pi w_1^{(L)}\pmod{2\pi}$. Thus
\begin{equation}
e^{i\gamma_z(\Phi)}=(-1)^{w_1^{(L)}(\Phi)}.
\label{eq:app_2d_3d_flux_winding}
\end{equation}
For a chiral-symmetric half-space termination, the Toeplitz index fixes the difference of boundary zero modes of
opposite chiralities up to boundary orientation. The unit-flux winding therefore enforces an odd number $n_0$ of
complex zero modes. Their centered charge and character obey
\begin{equation}
Q_0=\sum_{j=1}^{n_0}(f_j^\dagger f_j-\tfrac12),\ \
e^{2\pi iQ_0}=(-1)^{n_0}=-1,
\label{eq:app_2d_3d_flux_Q0}
\end{equation}
and
\begin{equation}
Z_0(\alpha)=[2\cos(\alpha/2)]^{n_0},\ \
Z_0(\alpha+2\pi)=-Z_0(\alpha).
\label{eq:app_2d_3d_flux_Z0}
\end{equation}
For the smooth unit-flux backgrounds of Eq.~\eqref{eq:app_2d_3d_flux_links}, spectral isolation further gives
$n_0=1$ in the sufficiently fine lattice regime. This recovers the isolated Dirac character in
Eq.~\eqref{eq:2d_aut_temporal_holonomy_main}; the complete compactly gauged 3D theory remains gauge invariant.

The same proof extends to any fixed integer $n_\Phi$: replace $\Phi=2\pi$ by $\Phi=2\pi n_\Phi$ in the gap
estimate and magnetic gap labeling. For all sufficiently large compatible $L$, $w_1^{(L)}(2\pi n_\Phi)=-n_\Phi$.
The result also holds along smooth fixed-flux deformations that preserve the endpoint gaps, but is not asserted
for arbitrary finite flux density or singular concentrated fields.

\subsubsection{Two-root magnetic index and boundary parity}
\label{app:2d_sq_3d_two_root}

We now establish the analogous unit-flux index for the two-root parent $\mathcal H_*$. Unlike the direct AIII
Hamiltonian, its chiral off-diagonal block does not initially admit endpoint spectral counting. We show that a
gauge-covariant, gap-preserving homotopy reduces it to that form, after which the same magnetic gap-labeling
argument applies.

Take one uniform flux quantum $\Phi=2\pi$ through an $L\times L$ transverse torus. The gauge-covariant shifts
satisfy $T_xT_y=e^{+i\Phi/L^2}T_yT_x$ in the common Peierls convention. Define $C_\mu=(T_\mu+T_\mu^\dagger)/2$ and
$D_\mu=(T_\mu-T_\mu^\dagger)/(2i)$ for $\mu=x,y$, together with the Hermitian transverse operators
\begin{equation}
\begin{aligned}
a_\Phi&=\mathbf1-C_x,\\
b_\Phi&=\mathbf1-\tfrac12C_y-\tfrac14\{C_x,C_y\}-\tfrac12 C_y^2,\\
P_\Phi&=\tfrac12\{a_\Phi,b_\Phi\}.
\end{aligned}
\end{equation}
The anticommutators symmetrize Wilson paths with the same displacement. The corresponding finite-range,
gauge-covariant parent is
\begin{equation}
\begin{aligned}
\mathcal H_{*,\Phi}(w)={}&
-2t\,\rho_x(D_x\sigma_x+D_y\sigma_z)\\
&+\Lambda\rho_z\bigl[(1+P_\Phi)\cos w
-(a_\Phi+b_\Phi)\bigr]\\
&+\Lambda\rho_y(1-P_\Phi)\sin w.
\end{aligned}
\label{eq:app_2d_AII_magnetic_H}
\end{equation}
It reduces to Eq.~\eqref{eq:2d_AII_bulk_H} at zero field and preserves the exact chiral grading
\begin{equation}
S_*=\rho_x\sigma_y,\ \ S_*^2=1,\ \
\{S_*,\mathcal H_{*,\Phi}(w)\}=0.
\label{eq:app_2d_AII_chiral_S}
\end{equation}
Thus, when the longitudinal family is gapped, its chiral block $D_{*,\Phi}(w)$ defines the winding
$\nu_z^{(L)}(\Phi)$ in Eq.~\eqref{eq_2roots_nu} in Sec.~\ref{sec:2d_surface_magnetic_inflow}.

For convenience, set $2t=\Lambda=1$. In an $S_*$-diagonal basis, the off-diagonal block takes the form
\begin{equation}
\begin{aligned}
D_{*,\Phi}(w)
&=F_{*,\Phi}(\cos w)
+i\sin w\,\mathbf1_2\otimes(1-P_\Phi),\\
F_{*,\Phi}(s)
&=\begin{pmatrix}
u_\Phi(s)&D_y-iD_x\\
D_y+iD_x&-u_\Phi(s)
\end{pmatrix},\\
u_\Phi(s)&=(1+P_\Phi)s-(a_\Phi+b_\Phi).
\end{aligned}
\label{eq:app_2d_AII_flux_general_block}
\end{equation}
Here $F_{*,\Phi}(s)$ is Hermitian, but $1-P_\Phi$ need not be positive. The imaginary part therefore does not
exclude a singularity away from $w=0,\pi$, and the direct AIII endpoint reduction cannot yet be used. Introduce
the gauge-covariant homotopy
\begin{equation}
\begin{aligned}
D_{*,\Phi,\tau}(w)
&=F_{*,\Phi}(\cos w)\\
&+i\sin w\,\mathbf1_2\otimes[1-(1-\tau)P_\Phi].
\end{aligned}
\label{eq:app_2d_AII_flux_general_homotopy}
\end{equation}
where $0\le\tau\le1$. It joins the original block at $\tau=0$ to $F_{*,\Phi}(\cos w)+i\sin w\,\mathbf1$ at
$\tau=1$. We must establish a gap throughout this deformation before using endpoint counting.

At zero field, let $u(s,\mathbf p)=(1+ab)s-(a+b)$ for $a=1-\cos p_x$ and $b=1-\tfrac12\cos p_y(1+\cos p_x+\cos
p_y)$. Both squared singular values along the homotopy equal
\begin{equation}
\begin{aligned}
E_\tau^2={}&\sin^2p_x+\sin^2p_y
+u(\cos w,\mathbf p)^2\\
&+\sin^2w[1-(1-\tau)ab]^2.
\end{aligned}
\label{eq:app_2d_AII_flux_general_zero_gap}
\end{equation}
A zero would require $\sin p_x=\sin p_y=0$ and $u=0$. At $(0,0)$, $(0,\pi)$, $(\pi,0)$, and $(\pi,\pi)$, the
corresponding values of $u(s,\mathbf p)$ are $s+\tfrac12$, $s-\tfrac32$, $2s-\tfrac52$, and $2s-\tfrac52$. Only
$(0,0)$ allows $u=0$ for $-1\le s\le1$, at $s=-\tfrac12$. There $ab=0$ and $\sin^2w=\tfrac34$, so the final term
remains positive for all $\tau$. Compactness yields a uniform zero-field gap,
\begin{equation}
\delta=\min_{\mathbf p,w,\tau}
s_{\min}\bigl(D_{*,0,\tau}(\mathbf p,w)\bigr)>0.
\label{eq:app_2d_AII_flux_general_delta}
\end{equation}

The gap persists for one flux quantum on sufficiently large compatible tori. The same localization argument as
in the direct AIII case applies: local gauges on boxes of side $R\ll L$ give magnetic-link errors
$O(|\Phi|R/L^2)$, while a quadratic partition of unity gives a commutator error $O(R^{-1})$. Using the uniform
zero-field gap $\delta>0$, we obtain
\begin{equation}
s_{\min}(D_{*,\Phi,\tau}(w))
\ge\delta-C(R^{-1}+|\Phi|R/L^2).
\label{eq:app_2d_AII_flux_general_weak_field}
\end{equation}
This bound is uniform in $w$, $\tau$, and compatible magnetic boundary twists. Choosing a sufficiently large
fixed $R=R_0$ and then sufficiently large $L$ keeps the right-hand side positive for $\Phi=2\pi$. Thus the
entire homotopy, including its endpoints, remains gapped.

At $\tau=1$, the imaginary part is scalar, and the winding reduces exactly to the endpoint spectral imbalance,
\begin{equation}
\nu_z^{(L)}(\Phi)
=n_-(F_{*,\Phi}(-1))-n_-(F_{*,\Phi}(+1)).
\end{equation}
To evaluate it, the zero-field endpoint masses are
\begin{equation}
\begin{aligned}
u_+(\mathbf p)&=(1-a)(1-b),\\
u_-(\mathbf p)&=-(1+a)(1+b).
\end{aligned}
\label{eq:app_2d_AII_flux_general_endpoint_masses}
\end{equation}
Since $a\ge0$ and $b\ge-\tfrac12$, the negative endpoint has a strictly negative mass and is topologically
trivial. The positive-endpoint mass signs at $(0,0)$, $(\pi,0)$, $(0,\pi)$, and $(\pi,\pi)$ are $(+,-,-,-)$.
Together with the kinetic components $(\sin p_y,\sin p_x)$, they give the occupied-band Chern numbers
$C_{*,-}^{xy}=0$ and $C_{*,+}^{xy}=1$, with the transverse orientation used for the continuum flux.

For one flux quantum, the magnetic gap-labeling relation~\cite{Dana1985} applies in the endpoint gaps that
remain open by the estimate above. Each zero-field endpoint has $L^2$ occupied states; the Chern-nontrivial
positive endpoint loses one occupied state, while the trivial negative endpoint does not. Hence
\begin{equation}
n_-(F_{*,2\pi}(-1))=L^2,\ \
n_-(F_{*,2\pi}(+1))=L^2-1,
\label{eq:app_2d_AII_flux_general_result}
\end{equation}
and $\nu_z^{(L)}(2\pi)=+1$. For a chiral-symmetric half-space termination, the bulk--boundary index refines the
parity statement to
\begin{equation}
n_--n_+=\pm\nu_z^{(L)}(\Phi),\ \
n_0=n_-+n_+\equiv\nu_z^{(L)}(\Phi)\pmod2,
\label{eq:app_2d_AII_flux_edge_index}
\end{equation}
where $n_\pm$ count zero modes of opposite chiralities and the overall sign depends on the boundary orientation.
Unit flux therefore enforces odd boundary zero-mode parity and the antiperiodic centered character discussed in
the main text. The index does not exclude additional opposite-chirality pairs.

The same proof extends to every fixed integer $n_\Phi$. Replacing $\Phi=2\pi$ by $\Phi=2\pi n_\Phi$ leaves the
weak-field gap estimate valid for sufficiently large compatible $L$, and magnetic gap labeling yields
$\nu_z^{(L)}(2\pi n_\Phi)=+n_\Phi$. As for the direct parent, this does not assert the result at arbitrary finite
flux density or for singular concentrated backgrounds.

\section{3D: Charge topology, symmetry relocalization, and locality obstructions}
\label{app:3d_anomaly}

We present the finite-volume vector-flux diagnostic of the mixed $AVV$ anomaly, the boundary spectra and
nonlocal relocalization of the strong compact vector charge, and the general no-go theorem for onsite $SU(2)$
Weyl doublets. The latter establishes a class-C locality obstruction distinct from both the strong quaternionic
charge-bundle topology and the continuum Witten anomaly.

\subsection{Spectral axial charge and the mixed $AVV$ flux response}
\label{app:3D_axial}

We examine the mixed $AVV$ response of the Weyl semimetal in Eq.~\eqref{eq:3d_dirac_wsm} by threading vector magnetic flux $\Phi_{xy}=2\pi n_\Phi$ through the transverse torus and varying the continuum vector holonomy $\Phi_V=\oint_z A_V$ from $0$ to $2\pi$ (negative physical electric field). Impose APBC along $z$ with $L_z\in4\mathbb Z$ and set $\delta k=2\pi/L_z$. The allowed momenta $k_j=(j+\tfrac12)\delta k$ avoid both the discontinuities of $s(k)=\operatorname{sgn}(\sin k)$ and the Weyl nodes at $k_z=\pm\pi/2$. The spectral axial charge is
\begin{equation}
\begin{aligned}
Q_A&=\sum_k s(k)Q_{V,k},\\
Q_{V,k}&=\sum_{\mathbf x_\perp,\alpha}
\left(
c_{\mathbf x_\perp,k,\alpha}^\dagger
c_{\mathbf x_\perp,k,\alpha}-\frac12
\right),
\end{aligned}
\label{eq:app_3d_QA_finite}
\end{equation}
where $\alpha=1,2$ labels the two orbitals and $Q_{V,k}$ is the centered vector charge of the transverse momentum slice $k_z=k$.

The continuum origin of the response is the chiral zeroth Landau level of each Weyl node. Near $\mathbf K_\eta=(0,0,\eta\pi/2)$,
\begin{equation}
h_\eta(\mathbf p)
=t\left(p_x\sigma_x+p_y\sigma_y
+\eta p_z\sigma_z\right),
\quad \eta=\pm1.
\label{eq:app_3d_Weyl_LL}
\end{equation}
For a uniform magnetic field $B_V>0$ along $z$, the gauge-covariant momenta $\pi_j=-i\partial_j+A_{V,j}$ ($j=x,y$) satisfy $[\pi_x,\pi_y]=-iB_V$. Define
\begin{equation}
b=\frac{\pi_x-i\pi_y}{\sqrt{2B_V}},
\quad
b^\dagger=\frac{\pi_x+i\pi_y}{\sqrt{2B_V}},
\quad [b,b^\dagger]=1.
\end{equation}
Then
\begin{equation}
h_\eta=t
\begin{pmatrix}
\eta p_z&\sqrt{2B_V}\,b\\
\sqrt{2B_V}\,b^\dagger&-\eta p_z
\end{pmatrix}.
\label{eq:app_3d_Weyl_LL_matrix}
\end{equation}
The lowest Landau orbital $|0\rangle$, satisfying $b|0\rangle=0$, gives the Weyl-spinor eigenstate $\Psi_{0,\eta}=(0,|0\rangle)^T$ with dispersion
\begin{equation}
E_{0,\eta}(p_z)=-\eta tp_z.
\label{eq:app_3d_Weyl_zero_LL}
\end{equation}
Higher Landau levels occur in pairs, $E_{n,\pm}=\pm t\sqrt{p_z^2+2nB_V}$ ($n\ge1$), whereas each zeroth level forms a chiral 1D channel with degeneracy $B_VL_xL_y/(2\pi)=n_\Phi$. A longitudinal holonomy shifts $p_z\mapsto p_z+A_{V,z}$. For positive magnetic flux, the zeroth-level velocities $-\eta t$ are opposite to those of the corresponding 1D chiral branches, reversing the axial spectral-flow orientation for the same positive holonomy insertion. As $\Phi_V$ increases from $0$ to $2\pi$, the two nodes exhibit opposite spectral flows, but their axial-charge contributions add to $-2n_\Phi$. For negative flux, the zeroth-level polarization reverses and the signed index changes sign, reproducing the integrated Adler--Bell--Jackiw response of Eq.~\eqref{eq:3d_axial_EB}.

For the lattice calculation, use the 1D large vector gauge transformation:
\begin{equation}
\begin{aligned}
\mathscr U_{V}c_{\mathbf x_\perp,z,\alpha}\mathscr U_{V}^{-1}
&=e^{-i\delta k z}c_{\mathbf x_\perp,z,\alpha},\\
\mathscr U_{V}c_k\mathscr U_{V}^{-1}&=c_{k+\delta k}.
\end{aligned}
\end{equation}
It identifies the $\Phi_V=2\pi$ Hamiltonian with the zero-holonomy Hamiltonian. Using Eq.~\eqref{eq:app_3d_QA_finite}, we obtain
\begin{equation}
\begin{aligned}
&\mathscr U_{V}^{-1}Q_A
\mathscr U_{V}-Q_A\\
&\quad=\sum_k
[s(k+\delta k)-s(k)]Q_{V,k}\\
&\quad=2\left(
Q_{V,-\delta k/2}
-Q_{V,\pi-\delta k/2}
\right).
\end{aligned}
\label{eq:app_3d_AVV_exact}
\end{equation}
This operator identity holds exactly at finite volume. For a $z$-independent magnetic field and uniform $A_{V,z}(t)$, the Hamiltonian remains diagonal in $k_z$, so $[Q_A,H[A_V(t)]]=0$ throughout the insertion in a fixed gauge. The nontrivial transformation of $Q_A$ arises when the gauge-equivalent endpoints are identified.

To evaluate the identity, consider the 2D Wilson--Dirac Hamiltonian at fixed $k_z$,
\begin{equation}
\begin{aligned}
h_{2D}(k_x,k_y;k_z)=&t\bigl[
\sin k_x\,\sigma_x+\sin k_y\,\sigma_y\\
&+(m(k_z)-\cos k_x-\cos k_y)\sigma_z
\bigr],\\
m(k_z)&=2-\cos k_z,
\end{aligned}
\end{equation}
whose occupied-band Chern number is
\begin{equation}
C(m)=\frac12\bigl[
-\operatorname{sgn}(m-2)
+2\operatorname{sgn}m-\operatorname{sgn}(m+2)
\bigr],
\end{equation}
and
\begin{equation}
C(k_z)=
\begin{cases}
+1,&|k_z|<\pi/2,\\
0,&|k_z|>\pi/2.
\end{cases}
\label{eq:app_3d_Ckz}
\end{equation}
With the orientation $dk_x\wedge dk_y$ and the common normalization $1/(2\pi i)$, $C(-\delta k/2)=+1$ and $C(\pi-\delta k/2)=0$.

Let $|\mathrm{FS}(B_V)\rangle$ denote the state with all negative-energy magnetic bands filled. Choose a transverse magnetic torus with $L_xL_y$ cells, total flux $2\pi n_\Phi$ ($n_\Phi\in\mathbb Z$), and compatible magnetic boundary conditions and Peierls link phases. Although the full 3D Weyl semimetal is gapless, the two slices in Eq.~\eqref{eq:app_3d_AVV_exact} have finite zero-field gaps. Assume their zero-energy gaps remain open throughout the magnetic interpolation and that the occupied bands follow the same gap-labeling branches. For fixed $L_z$ and $n_\Phi$, sufficiently large transverse sizes give a weak-flux regime; when $L_z$ also varies, the field must remain small enough relative to both slice gaps.

As in 2D, the Středa formula~\cite{stredaTheory1982,Dana1985} applied to each gapped 2D slice gives $n_{\rm occ}(k)=L_xL_y-C(k)n_\Phi$, with the positive magnetic flux and Peierls-link convention used above. Since $Q_{V,k}=n_{\rm occ}(k)-L_xL_y$, we obtain
\begin{equation}
\begin{aligned}
&\langle Q_{V,k}\rangle_{\mathrm{FS}(B_V)}
=-C(k)n_\Phi,\\
&\left\langle
\mathscr U_{V}^{-1}Q_A
\mathscr U_{V}-Q_A
\right\rangle_{\mathrm{FS}(B_V)}
=-2n_\Phi.
\end{aligned}
\label{eq:app_3d_AVV_flux}
\end{equation}
Thus the exact lattice transformation of the spectral axial charge reproduces the IR mixed-anomaly coefficient $d_{AVV}=2$, with the negative spectral-flow sign appropriate to positive longitudinal holonomy and positive magnetic flux, without introducing an axial gauge field.

\subsection{Strong compact vector charge: boundary spectra and relocalization}
\label{app:3d_strong_charge}

We derive the exact half-space spectra and the nonlocal symmetry relocalization used in
Sec.~\ref{sec:3d_strong_charge}. The boundary charge-spectrum Dirac cone is distinct from the physical
surface-energy modes, which must be analyzed separately.

\subsubsection{Boundary charge and energy spectra}
\label{app:3d_strong_boundary}

We first derive the boundary charge-spectrum Dirac cone for an open half-space normal to $z$. At $k_x=k_y=0$,
the nontrivial charge block reduces to
\begin{equation}
q_{\rm AII}(0,0,k_z)=Z\cos k_z
+(\rho_x\otimes\sigma_z)\sin k_z.
\label{eq:app_3d_strong_boundary_exact_q}
\end{equation}
Writing $q_{\rm AII}(k)=Te^{ik}+T^\dagger e^{-ik}$, we have
\begin{equation}
\begin{aligned}
&T=\tfrac12(Z-i\rho_x\otimes\sigma_z),\\
&T^\dagger\varphi_s=0,\\
&\rho_y\varphi_s =\sigma_z\varphi_s=s\varphi_s,
\quad s=\pm1.
\end{aligned}
\label{eq:app_3d_strong_boundary_hopping}
\end{equation}
The Toeplitz compression to $z\ge1$ therefore has exact zero-eigenvalue states $\psi_s(z)=\delta_{z1}\varphi_s$,
forming a Kramers pair. Projecting the linear in-plane perturbation onto this pair gives
$\lambda_\pm=\pm\sqrt{k_x^2+k_y^2} +O(|\mathbf k_\perp|^2)$, establishing the boundary charge-spectrum Dirac
cone.

These charge eigenstates need not be eigenstates of the physical Hamiltonian. To analyze the surface-energy
spectrum, consider $k_x=k_y=0$ in the $\sigma_z=s$ sector. With $u_s(k)=e^{-isk}$ and $m_\delta(k)=\delta+1-\cos
k$, a chiral basis gives the off-diagonal block
\begin{equation}
\mathscr D_s(k)=
\begin{pmatrix}
-m_\delta(k)u_s(k)&
\tfrac{s\sin k}{2}[1+u_s(k)]\\
\tfrac{s\sin k}{2}[1+u_s(k)]&
m_\delta(k)
\end{pmatrix},
\label{eq:app_3d_strong_boundary_D}
\end{equation}
whose determinant factors as
\begin{equation}
\det\mathscr D_s(k)=-e^{-isk}
\left[
m_\delta(k)^2+
\sin^2k\cos^2\frac{k}{2}
\right].
\label{eq:app_3d_strong_boundary_det}
\end{equation}
For $\delta\ne0,-2$, the bracket is strictly positive on the real momentum circle, giving the chiral winding
$w_s=-s$. This establishes the physical zero-energy surface Kramers pair for the standard half-space
termination.

The penetration lengths follow from the complex zeros of the same determinant. Near $k=0$,
\begin{equation}
\det\mathscr D_s(k)\propto
-\left[
(\delta+k^2/2)^2+k^2+O(k^4)
\right],
\end{equation}
so the penetration length diverges as $\xi_z\sim|\delta|^{-1}$ when $\delta\to0$. Near $k=\pi+q$, with
$\epsilon=\delta+2$,
\begin{equation}
\det\mathscr D_s(\pi+q)\propto
-\left[
(\epsilon-q^2/2)^2+q^4/4
+O(|\epsilon|q^4+q^6)
\right],
\end{equation}
giving $\xi_z\sim|\delta+2|^{-1/2}$ when $\delta\to-2$. Thus the physical surface-energy states delocalize at
these bulk critical points, whereas the boundary charge-spectrum Dirac cone is independent of $\delta$.

\subsubsection{Nonlocal relocalization with onsite vector and Kramers symmetries}
\label{app:3d_strong_reloc}

We construct a nonlocal Gaussian transformation for the uniform Hamiltonian $h_0$ that makes the compact
vector charge onsite while keeping Kramers time reversal onsite and the transformed Hamiltonian exponentially local.
Define
\begin{equation}
\begin{aligned}
\widetilde m&=m-\frac{r^2}{2}
=\frac12\sum_j(1-\cos k_j)^2,\\
v^2&=\alpha+\frac{m+\widetilde m}{2},
\end{aligned}
\label{eq:app_3d_reloc_data}
\end{equation}
where $r^2=\sum_j\sin^2k_j$, $m=3-\sum_j\cos k_j$, and $\alpha=(1+M_s/E_s)/2$. Momentum arguments are suppressed
below. The identity $m-\widetilde m=r^2/2$ gives
\begin{equation}
v^2r^2+\widetilde m^2
=\alpha r^2+m^2=\varepsilon_0^2.
\label{eq:app_3d_reloc_spectrum}
\end{equation}
Moreover, $m,\widetilde m,\alpha\ge0$, and $m$ vanishes only at $\mathbf k=\mathbf0$, where $\alpha=1$. Hence
$v^2>0$ throughout the Brillouin torus. The positive root $v$ is therefore periodic and real analytic, ensuring
exponential locality of the target Hamiltonian in Eq.~\eqref{eq:3d_strong_reloc_h}.

The first rotation makes the charge momentum independent. For $\alpha>0$, define
\begin{equation}
U_s(\mathbf k)
=\frac{\mathbf1_4+Zq_{\rm AII}(\mathbf k)}
{2\sqrt{\alpha(\mathbf k)}}
=\sqrt{\alpha}\,\mathbf1_4
+\frac{i\rho_y\otimes S(\mathbf k)}
{2E_s\sqrt{\alpha}}.
\label{eq:app_3d_reloc_Us}
\end{equation}
Equivalently, for $r>0$, writing $\widehat S=S/r$ and $\cos\theta=M_s/E_s$, $\sin\theta=r/E_s$, we have
\begin{equation}
U_s=\exp\!\left(
\frac{i\theta}{2}\rho_y\otimes\widehat S
\right).
\label{eq:app_3d_reloc_Us_angle}
\end{equation}
This rotation satisfies $U_sq_{\rm AII}U_s^\dagger=Z$. With the flavor projectors
$P_{\mu,\pm}=(\mu_0\pm\mu_z)/2$, set
\begin{equation}
\begin{aligned}
U&=P_{\mu,+}\otimes U_s
+P_{\mu,-}\otimes\mathbf1_4,\\
q_0&=\mu_0\otimes Z.
\end{aligned}
\label{eq:app_3d_reloc_U}
\end{equation}
Then
\begin{equation}
\begin{aligned}
&Uq_VU^\dagger=q_0,\\
&Uh_0U^\dagger=q_0B_1,\\
&B_1=\sqrt{\alpha}\,
\mu_x\otimes\rho_0\otimes S
-m\mu_z\otimes\mathbf1_4.
\end{aligned}
\label{eq:app_3d_reloc_first_step}
\end{equation}
Although the charge is now onsite, $\sqrt{\alpha}$ is nonanalytic at the seven momenta where $M_s<0$ and $S=0$.
The first rotation alone therefore does not preserve exponential locality.

The second rotation removes this nonanalyticity from the Hamiltonian without changing the onsite charge. Define
\begin{equation}
\begin{aligned}
B_2&=v\,\mu_x\otimes\rho_0\otimes S
-\widetilde m\,\mu_z\otimes\mathbf1_4,\\
B_1^2&=B_2^2=\varepsilon_0^2\mathbf1_8.
\end{aligned}
\label{eq:app_3d_reloc_B2}
\end{equation}
Both $B_1$ and $B_2$ commute with $q_0$. Moreover,
\begin{equation}
\gamma=m\widetilde m+\sqrt{\alpha}\,vr^2,
\quad
\{B_1,B_2\}=2\gamma\mathbf1_8.
\label{eq:app_3d_reloc_gamma}
\end{equation}
Away from $\varepsilon_0=0$, choose
\begin{equation}
V=\frac{\varepsilon_0^2\mathbf1_8+B_2B_1}
{\sqrt{2\varepsilon_0^2
(\varepsilon_0^2+\gamma)}}.
\label{eq:app_3d_reloc_V}
\end{equation}
Since $\gamma\ge0$ and $\varepsilon_0>0$ away from $\mathbf k=\mathbf0$, the denominator is nonzero. Direct
multiplication gives
\begin{equation}
V^\dagger V=\mathbf1_8,\quad
VB_1V^\dagger=B_2,\quad
[V,q_0]=0.
\end{equation}
At the Dirac point $\mathbf k=\mathbf0$, the continuous limit is $V=\mathbf1_8$. The complete
one-particle Bloch transformation is
\begin{equation}
\mathcal W(\mathbf k)=V(\mathbf k)U(\mathbf k).
\label{eq:app_3d_reloc_W}
\end{equation}
Here $\mathcal W(\mathbf k)$ denotes the Bloch representative of the Gaussian canonical transformation
$\mathcal W$ acting on Fock space. It satisfies Eq.~\eqref{eq:3d_strong_reloc_exact}. Although
$\mathcal W$ is nonlocal, the transformed Hamiltonian is exponentially local.

The transformation also preserves the Kramers and Majorana structures. For the antiunitary operations $\Theta$
and $\mathcal C$ of Eq.~\eqref{eq:3d_strong_antiunitary},
\begin{equation}
\begin{aligned}
\Theta\mathcal W(\mathbf k)\Theta^{-1}
&=\mathcal W(-\mathbf k),\\
\mathcal C\mathcal W(\mathbf k)\mathcal C^{-1}
&=\mathcal W(-\mathbf k).
\end{aligned}
\label{eq:app_3d_reloc_covariance}
\end{equation}
Thus both operations retain their momentum-independent representations, while $q_0$ generates an onsite compact
$U(1)_V$.

At the seven momenta where $\alpha=0$, one may assign $U_s=i\rho_y\otimes\sigma_0$, preserving the algebraic and
symmetry identities. Nevertheless, $U_s$ approaches these points with a direction-dependent limit proportional
to $i\rho_y\otimes S/r$. Hence $\mathcal W$ has no globally smooth periodic Bloch representative, consistent
with the nontrivial strong quaternionic index.

The construction extends to each uniform $\delta>-1/2$, with the first rotation $U$ unchanged and
\begin{equation}
\begin{aligned}
m&\longmapsto m+\delta,\\
\widetilde m&\longmapsto\widetilde m+\delta,\\
v^2&\longmapsto v^2+\delta.
\end{aligned}
\label{eq:app_3d_reloc_uniform_mass}
\end{equation}
Indeed, $v^2\ge1/2$: when $m<1$, $\alpha\ge1/2$, while for $m\ge1$ the inequality follows directly from its
definition. Thus $v_\delta^2=v^2+\delta>0$. The second rotation also remains well defined. For $r>0$, the
coefficient vectors $
(\sqrt{\alpha}\,r,-m-\delta)$ and $
(v_\delta r,-\widetilde m-\delta)$
have kinetic components of the same positive sign and cannot be antipodal. At $r=0$, their mass components
coincide and are nonzero for $\delta>-1/2$, except at the massless Dirac point $(\mathbf
k,\delta)=(\mathbf0,0)$, where the rotation has the identity limit. Consequently,
$\varepsilon_\delta^2+\gamma_\delta>0$ whenever $\varepsilon_\delta>0$, and the denominator in
Eq.~\eqref{eq:app_3d_reloc_V} does not vanish. Each uniform $\delta$ therefore admits a nonlocal
Kramers-compatible relocalization $\mathcal W_\delta$ with an exponentially local target Hamiltonian.

The second rotation $V_\delta$, and hence $\mathcal W_\delta$, generally depends on $\delta$. Thus the
construction applies to each uniform mass separately; the limitations for spatially varying mass interfaces
are discussed in Sec.~\ref{sec:3d_strong_relocalization}.

\subsection{No-go theorem for onsite $SU(2)$ Weyl doublets}
\label{app:3d_SU2_no_go}

We present a simple proof for the locality obstruction stated in Sec.~\ref{sec:3d_weyl_doublet} in a general
setting. Consider a translation-invariant quadratic Majorana Hamiltonian with an exact compact $SU(2)$ symmetry.
Suppose a translation-preserving Majorana-canonical Gaussian transformation makes the full symmetry onsite while
the transformed Bloch Hamiltonian remains periodic and $C^1$. The transformation itself need not be local or
continuous in momentum. We show that the net Weyl charge in a half Brillouin zone must be even within each
subspace carrying copies of the fundamental $SU(2)$ representation, provided its particle--hole invariant
boundary planes are gapped. Thus the obstruction applies to a candidate relocalization whose transformed
multiplicity Hamiltonian retains odd oriented Weyl charge in that half zone, with no compensating crossings. For
a nonlocal transformation, preservation of this charge is a separate hypothesis, not a consequence of canonicity
alone.

Suppose such a relocalization exists. On the subspace carrying the fundamental $SU(2)$ representation, the
onsite generators can be written as
\begin{equation}
T_a'=\frac12\mu_a\otimes\mathbf1_{2m},
\quad a=x,y,z,
\label{eq:app_3d_nogo_onsite_generators}
\end{equation}
where $2m$ is the dimension of the multiplicity space. Since the transformed Hamiltonian commutes with all three
generators, Schur's lemma gives
\begin{equation}
\mathcal H_{\rm act}'(\mathbf k)
=\mu_0\otimes h(\mathbf k),
\label{eq:app_3d_nogo_schur}
\end{equation}
where $h(\mathbf k)$ is a Hermitian $2m\times2m$ matrix. Thus the Hamiltonian is flavor independent, with its
nontrivial momentum dependence confined to the multiplicity space.

In a canonical Majorana Bloch basis, the particle--hole operation satisfies $\mathcal C^2=+1$ and $\mathcal C
T_a'\mathcal C^{-1}=-T_a'$. The fundamental representation of $SU(2)$ is pseudoreal, with intertwiner
$J_\mu=i\mu_y$. By Schur's lemma, the restriction of $\mathcal C$ to this subspace must therefore take the form
\begin{equation}
\mathcal C_{\rm act}
=(J_\mu\otimes J_m)K,
\quad J_\mu=i\mu_y,
\label{eq:app_3d_nogo_C_factorization}
\end{equation}
where $J_m$ is a unitary matrix acting on the multiplicity space. Since $J_\mu J_\mu^*=-\mathbf1_2$, the
condition $\mathcal C_{\rm act}^2=+1$ requires
\begin{equation}
J_mJ_m^*=-\mathbf1_{2m}.
\label{eq:app_3d_nogo_quaternionic}
\end{equation}
Hence the multiplicity space carries a quaternionic antiunitary structure $\mathcal C_m=J_mK$, with $\mathcal
C_m^2=-1$. Applying Majorana particle--hole symmetry to Eq.~\eqref{eq:app_3d_nogo_schur} then yields
\begin{equation}
J_mh(\mathbf k)^*J_m^{-1}
=-h(-\mathbf k).
\label{eq:app_3d_nogo_classC}
\end{equation}
The multiplicity Hamiltonian therefore belongs to class C, irrespective of its dimension or microscopic origin.

Consider the particle--hole invariant planes $k_z=\kappa$, with $\kappa=0,\pi$, and assume that $h$ is gapped at
zero energy on both planes. Since $h$ is periodic and $C^1$, its negative-energy projectors
$P_-(k_x,k_y,\kappa)$ are also $C^1$. Their first Chern numbers are
\begin{equation}
C(\kappa)=\frac{1}{2\pi i}
\int_{T^2}dk_x\,dk_y\,
\operatorname{tr}\left(
P_-[\partial_{k_x}P_-,
\partial_{k_y}P_-]
\right).
\label{eq:app_3d_nogo_chern_degree}
\end{equation}
Each plane defines a gapped 2D class-C Hamiltonian, for which the first Chern number is necessarily even:
\begin{equation}
C(0),\,C(\pi)\in2\mathbb Z.
\label{eq:app_3d_nogo_even_chern}
\end{equation}
This is the standard 2D class-C $2\mathbb Z$ constraint in the tenfold classification
\cite{kitaevPeriodic2009}. It holds for arbitrary finite multiplicity dimension (also after addition of
trivial class-C bands) and requires no spatial symmetry beyond translation invariance. Here it applies because
the onsite fundamental $SU(2)$ generators leave a globally defined, momentum-independent quaternionic structure
on the multiplicity space.

For a two-dimensional multiplicity space ($m=1$), the evenness can be verified directly. A constant unitary
change of basis brings $J_m$ to $i\sigma_y$. Writing
\begin{equation}
h(\mathbf k)
=d_0(\mathbf k)\mathbf1_2
+\mathbf d(\mathbf k)\cdot\boldsymbol\sigma,
\label{eq:app_3d_nogo_dvector}
\end{equation}
Eq.~\eqref{eq:app_3d_nogo_classC} implies
\begin{equation}
d_0(-\mathbf k)=-d_0(\mathbf k),
\quad
\mathbf d(-\mathbf k)=\mathbf d(\mathbf k).
\label{eq:app_3d_nogo_d_even}
\end{equation}
On either invariant plane, the gap and particle--hole symmetry fix the occupied rank to one, so $\mathbf d$ is
nowhere zero. The occupied-band Chern number is therefore, up to orientation, the degree of $\widehat{\mathbf
d}:T^2\to S^2$, where $\widehat{\mathbf d} =\mathbf d/|\mathbf d|$. Since
\begin{equation}
\widehat{\mathbf d}(-k_x,-k_y,\kappa)
=\widehat{\mathbf d}(k_x,k_y,\kappa),
\end{equation}
the preimages of any regular value occur in inversion-related pairs. Inversion preserves the orientation of
$T^2$, so paired preimages contribute equal local degrees. At inversion-fixed momenta, the derivative of the
even map vanishes, excluding such points from the preimages of a regular value. The degree is consequently even,
as required by Eq.~\eqref{eq:app_3d_nogo_even_chern}. This elementary argument also shows that analyticity is
unnecessary; $C^1$ regularity suffices.

Now consider the half Brillouin zone $T^2\times[0,\pi]$. Assume that its interior contains only isolated
zero-energy crossings of $h$. Their Weyl charges $\chi_\alpha$ are defined by the first Chern numbers on small
enclosing spheres. Away from these crossings, the negative-energy states form a vector bundle, and the
Berry-flux relation gives
\begin{equation}
C(\pi)-C(0)
=
\sum_{\mathbf K_\alpha:\,
0<K_{\alpha,z}<\pi}\chi_\alpha,
\label{eq:app_3d_nogo_chern_jump}
\end{equation}
with consistent orientation conventions. Combining this with Eq.~\eqref{eq:app_3d_nogo_even_chern} establishes
the general necessary condition
\begin{equation}
\sum_{\mathbf K_\alpha:\,
0<K_{\alpha,z}<\pi}\chi_\alpha
\in2\mathbb Z.
\label{eq:app_3d_nogo_general_parity}
\end{equation}
Thus an onsite fundamental $SU(2)$ representation and Majorana reality are incompatible with an odd net Weyl
charge in a half Brillouin zone when the multiplicity Hamiltonian is periodic and $C^1$.

For the construction in Sec.~\ref{sec:3d_weyl_doublet}, Majorana reality identifies the opposite-node
representatives, leaving one independent fundamental $SU(2)$ Weyl doublet. To apply the obstruction, require the
candidate relocalization to retain the original local oriented Weyl charges and their placement relative to the
gapped planes $k_z=0,\pi$, with no additional crossings. Smooth invertibility near each node suffices to
preserve its local Chern charge, even if the transformation is nonlocal or discontinuous elsewhere. In the
specified half zone $0<k_z<\pi$, the transformed multiplicity sector then contains a single elementary crossing
at $\mathbf K_+=(0,0,\pi/2)$, with $\chi_+=\pm1$. Such a relocalization would require
\begin{equation}
C(\pi)-C(0)=\pm1,
\label{eq:app_3d_nogo_odd_jump}
\end{equation}
contradicting Eq.~\eqref{eq:app_3d_nogo_even_chern}. Hence the full compact $SU(2)$ cannot be made onsite by a
translation-preserving Majorana-canonical Gaussian transformation while retaining a periodic $C^1$ Hamiltonian
with the original Weyl content and no additional gapless modes.

In particular, this obstruction excludes exponentially local and Schwartz-local transformed Hamiltonians. It
also applies to sufficiently rapidly decaying algebraic hopping, since
$
\sum_{\mathbf r}
(1+|\mathbf r|)\, \|h_{\mathbf r}\|<\infty
$
is sufficient for $C^1$ regularity of the Bloch Hamiltonian. More fundamentally, the topological argument
requires only continuity, provided the invariant-plane gaps and Chern charges of the isolated crossings remain
well defined. The $C^1$ hypothesis permits their differential formulation through Berry curvature.

The strong quaternionic index $\nu_0(\mathcal M)=1$ of the explicit construction independently obstructs a
smooth, periodic, Kramers-compatible trivialization of the symmetry-multiplicity bundle. By contrast, the
present theorem imposes no regularity requirement on the relocalizing transformation itself; it constrains the
transformed Hamiltonian. The result is therefore independent of the particular parent projector $P_s$ and
applies to general quadratic Majorana Hamiltonians satisfying the stated assumptions. Additional compensating
Weyl crossings or transformations that abandon the assumed translation structure are not excluded. Finally, the
class-C parity argument does not compute the mod-two Pfaffian holonomy of the Witten anomaly, although the two
obstructions are consistent with the same anomalous IR Weyl content.

\bibliography{refs} 

@article{pace2024lattice,
  title={Lattice {T}-duality from non-invertible symmetries in quantum spin chains},
  author={Pace, Salvatore D and Chatterjee, Arkya and Shao, Shu-Heng},
  journal={SciPost Phys.},
  volume={18},
  number={4},
  pages={121},
  year={2025},
  doi={10.21468/SciPostPhys.18.4.121} 
}

@article{chatterjee2025quantized,
  title = {Quantized Axial Charge of Staggered Fermions and the Chiral Anomaly},
  author = {Chatterjee, Arkya and Pace, Salvatore D. and Shao, Shu-Heng},
  journal = {Phys. Rev. Lett.},
  volume = {134},
  issue = {2},
  pages = {021601},
  numpages = {7},
  year = {2025},
  month = {Jan},
  publisher = {American Physical Society},
  doi = {10.1103/PhysRevLett.134.021601},
  url = {https://link.aps.org/doi/10.1103/PhysRevLett.134.021601}
}

@Article{cheng2023lieb,
	title={{Lieb-Schultz-Mattis, Luttinger, and 't Hooft - anomaly matching in lattice systems}},
	author={Meng Cheng and Nathan Seiberg},
	journal={SciPost Phys.},
	volume={15},
	pages={051},
	year={2023},
	publisher={SciPost},
	doi={10.21468/SciPostPhys.15.2.051},
	url={https://scipost.org/10.21468/SciPostPhys.15.2.051},
}

@Article{seiberg2024majorana,
	title={{Majorana chain and Ising model - (non-invertible) translations, anomalies, and emanant symmetries}},
	author={Nathan Seiberg and Shu-Heng Shao},
	journal={SciPost Phys.},
	volume={16},
	pages={064},
	year={2024},
	publisher={SciPost},
	doi={10.21468/SciPostPhys.16.3.064},
	url={https://scipost.org/10.21468/SciPostPhys.16.3.064},
}

@article{li2023non,
  title = {Noninvertible duality transformation between symmetry-protected topological and spontaneous symmetry breaking phases},
  author = {Li, Linhao and Oshikawa, Masaki and Zheng, Yunqin},
  journal = {Phys. Rev. B},
  volume = {108},
  issue = {21},
  pages = {214429},
  numpages = {19},
  year = {2023},
  month = {Dec},
  publisher = {American Physical Society},
  doi = {10.1103/PhysRevB.108.214429},
  url = {https://link.aps.org/doi/10.1103/PhysRevB.108.214429}
}

@incollection{hooft1980naturalness,
  title={Naturalness, chiral symmetry, and spontaneous chiral symmetry breaking},
  author={Hooft, G’t},
  booktitle={Recent developments in gauge theories},
  pages={135--157},
  year={1980},
  publisher={Springer},
  doi ={10.1007/978-1-4684-7571-5_9}
}

@article{pace2025parity,
  title = {Parity Anomaly from a {Lieb-Schultz-Mattis} Theorem: {E}xact Valley Symmetries on the Lattice},
  author = {Pace, Salvatore D. and Kim, Minho Luke and Chatterjee, Arkya and Shao, Shu-Heng},
  journal = {Phys. Rev. Lett.},
  volume = {135},
  issue = {23},
  pages = {236501},
  numpages = {7},
  year = {2025},
  month = {Dec},
  publisher = {American Physical Society},
  doi = {10.1103/4x4w-pfgq},
  url = {https://link.aps.org/doi/10.1103/4x4w-pfgq}
}

@article{kitaev2006anyons,
  title={Anyons in an exactly solved model and beyond},
  author={Kitaev, Alexei},
  journal={Ann. Phys.},
  volume={321},
  number={1},
  pages={2--111},
  year={2006},
  publisher={Elsevier}, 
  doi ={10.1016/j.aop.2005.10.005}
}

@article{su2025z2, 
  title = {$\mathbb{Z}_{2}$ gauging and self-dualities of the {$XX$} model and its cousins},
  author = {Su, Lei},
  journal = {Phys. Rev. B},
  volume = {112},
  issue = {3},
  pages = {035105},
  numpages = {11},
  year = {2025},
  month = {Jul},
  publisher = {American Physical Society},
  doi = {10.1103/c16r-fdfb},
  url = {https://link.aps.org/doi/10.1103/c16r-fdfb}
}

@article{kitaev2001unpaired,
  title={Unpaired {M}ajorana fermions in quantumwires},
  author={Kitaev, A Yu},
  journal={Phys.-Usp.},
  volume={44},
  number={10S},
  pages={131},
  year={2001},
  publisher={IOP Publishing},
  doi={10.1070/1063-7869/44/10S/S29}
}

@book{nakahara2018geometry,
  title={{Geometry, Topology and Physics}},
  author={Nakahara, Mikio},
  year={2018},
  publisher={CRC press},
 doi ={10.1201/9781315275826}
}

@article{redlich1984,
  title = {Gauge Noninvariance and Parity Nonconservation of Three-Dimensional Fermions},
  author = {Redlich, A. N.},
  journal = {Phys. Rev. Lett.},
  volume = {52},
  issue = {1},
  pages = {18--21},
  numpages = {0},
  year = {1984},
  month = {Jan},
  publisher = {American Physical Society},
  doi = {10.1103/PhysRevLett.52.18},
  url = {https://link.aps.org/doi/10.1103/PhysRevLett.52.18}
}

@article{chen2013,
  title = {Symmetry protected topological orders and the group cohomology of their symmetry group},
  author = {Chen, Xie and Gu, Zheng-Cheng and Liu, Zheng-Xin and Wen, Xiao-Gang},
  journal = {Phys. Rev. B},
  volume = {87},
  issue = {15},
  pages = {155114},
  numpages = {48},
  year = {2013},
  month = {Apr},
  publisher = {American Physical Society},
  doi = {10.1103/PhysRevB.87.155114},
  url = {https://link.aps.org/doi/10.1103/PhysRevB.87.155114}
}

@article{gioia2025exact,
  title={Exact Chiral Symmetries of 3+ 1{D} {H}amiltonian Lattice Fermions},
  author={Gioia, Lei and Thorngren, Ryan},
  journal={arXiv:2503.07708},
  year={2025},
  url ={https://arxiv.org/abs/2503.07708}
}

@article{Barkeshli2019,
  title = {Symmetry fractionalization, defects, and gauging of topological phases},
  author = {Barkeshli, Maissam and Bonderson, Parsa and Cheng, Meng and Wang, Zhenghan},
  journal = {Phys. Rev. B},
  volume = {100},
  issue = {11},
  pages = {115147},
  numpages = {99},
  year = {2019},
  month = {Sep},
  publisher = {American Physical Society},
  doi = {10.1103/PhysRevB.100.115147},
  url = {https://link.aps.org/doi/10.1103/PhysRevB.100.115147}
}

@article{lew-smithInfiniteOrder2026,
  title = {Infinite-Order Lattice Chiral Anomalies and {{CPT}}},
  author = {{Lew-Smith}, Elijah and Pace, Salvatore D. and Shao, Shu-Heng},
  year = 2026,
  month = jun,
  publisher = {arXiv},
  url = {http://arxiv.org/abs/2606.12510},
  journal = {arXiv:2606.12510}
}

@article{dharanikota2026,
  title = {1+1d Lattice {D}irac Fermions from Non-Onsite Vector  and  Axial Symmetries},
  author = {Dharanikota, Tabin and Fidkowski, Lukasz},
  year = 2026,
  month = aug,
  publisher = {arXiv},
  url = {https://arxiv.org/abs/2608.02722},
  journal ={arXiv:2608.02722}
}

@article{luFermionic2026,
  title = {Fermionic {{Villain}} Model with Exact Lattice Chiral Symmetries},
  author = {Lu, Zhiyao and Shao, Shu-Heng},
  year = 2026,
  month = aug,
  publisher = {arXiv},
  url = {http://arxiv.org/abs/2608.02728},
  journal = {arXiv:2608.02728}
}

@article{yamanakaNonperturbative1997,
  title = {Nonperturbative Approach to {L}uttinger's Theorem in One Dimension},
  author = {Yamanaka, Masanori and Oshikawa, Masaki and Affleck, Ian},
  year = 1997,
  month = aug,
  volume = {79},
  number = {6},
  pages = {1110--1113},
  issn = {0031-9007, 1079-7114},
  doi = {10.1103/PhysRevLett.79.1110},
  url = {https://link.aps.org/doi/10.1103/PhysRevLett.79.1110},
  copyright = {http://link.aps.org/licenses/aps-default-license},
  journal = {Phys. Rev. Lett.}
}

@article{oshikawa2000,
  title = {Topological Approach to {L}uttinger's Theorem and the {F}ermi Surface of a {K}ondo Lattice},
  author = {Oshikawa, Masaki},
  journal = {Phys. Rev. Lett.},
  volume = {84},
  issue = {15},
  pages = {3370--3373},
  numpages = {0},
  year = {2000},
  month = {Apr},
  publisher = {American Physical Society},
  doi = {10.1103/PhysRevLett.84.3370},
  url = {https://link.aps.org/doi/10.1103/PhysRevLett.84.3370}
}

@article{nielsenAbsence1981,
  title = {Absence of Neutrinos on a Lattice: ({{I}}). {{Proof}} by Homotopy Theory},
  shorttitle = {Absence of Neutrinos on a Lattice},
  author = {Nielsen, H. B. and Ninomiya, M.},
  year = 1981,
  month = jul,
  journal = {Nucl. Phys. B},
  volume = {185},
  number = {1},
  pages = {20--40},
  issn = {0550-3213},
  doi = {10.1016/0550-3213(81)90361-8},
  url = {https://www.sciencedirect.com/science/article/pii/0550321381903618}
}

@article{nielsenAbsence1981a,
  title = {Absence of Neutrinos on a Lattice: ({{II}}). {{Intuitive}} Topological Proof},
  shorttitle = {Absence of Neutrinos on a Lattice},
  author = {Nielsen, H. B. and Ninomiya, M.},
  year = 1981,
  month = dec,
  journal = {Nucl. Phys. B},
  volume = {193},
  number = {1},
  pages = {173--194},
  issn = {0550-3213},
  doi = {10.1016/0550-3213(81)90524-1},
  url = {https://www.sciencedirect.com/science/article/pii/0550321381905241}
}

@article{nielsennogo1981,
  title = {A No-Go Theorem for Regularizing Chiral Fermions},
  author = {Nielsen, H. B. and Ninomiya, M.},
  year = 1981,
  month = oct,
  journal = {Phys. Lett. B},
  volume = {105},
  number = {2},
  pages = {219--223},
  issn = {0370-2693},
  doi = {10.1016/0370-2693(81)91026-1},
  url = {https://www.sciencedirect.com/science/article/pii/0370269381910261}
}

@incollection{wilson1975quarks,
  title={Quarks and strings on a lattice},
  author={Wilson, Kenneth G},
  booktitle={New Phenomena in Subnuclear Physics: Part A},
  pages={69--142},
  year={1975},
  publisher={Springer},
  doi ={10.1007/978-1-4613-4208-3_6}
}

@article{kogutHamiltonian1975a,
  title = {Hamiltonian Formulation of {{Wilson}}'s Lattice Gauge Theories},
  author = {Kogut, John and Susskind, Leonard},
  year = 1975,
  month = jan,
  volume = {11},
  number = {2},
  pages = {395--408},
  publisher = {American Physical Society},
  doi = {10.1103/PhysRevD.11.395},
  url = {https://link.aps.org/doi/10.1103/PhysRevD.11.395},
  journal = {Phys. Rev. D}
}

@article{susskindLattice1977,
  title = {Lattice Fermions},
  author = {Susskind, Leonard},
  year = 1977,
  month = nov,
  volume = {16},
  number = {10},
  pages = {3031--3039},
  publisher = {American Physical Society},
  doi = {10.1103/PhysRevD.16.3031},
  url = {https://link.aps.org/doi/10.1103/PhysRevD.16.3031},
  journal = {Phys. Rev. D}
}

@article{kaplanChiral2012,
  title = {Chiral Symmetry and Lattice Fermions},
  author = {Kaplan, David B.},
  year = 2012,
  month = jan,
  publisher = {arXiv},
  url = {http://arxiv.org/abs/0912.2560},
  journal = {arXiv:0912.2560}
}

@article{kaplanmethod1992,
  title = {A Method for Simulating Chiral Fermions on the Lattice},
  author = {Kaplan, David B.},
  year = 1992,
  month = aug,
  journal = {Phys. Lett. B},
  volume = {288},
  number = {3},
  pages = {342--347},
  issn = {0370-2693},
  doi = {10.1016/0370-2693(92)91112-M},
  url = {https://www.sciencedirect.com/science/article/pii/037026939291112M}
}

@article{ginspargremnant1982,
  title = {A Remnant of Chiral Symmetry on the Lattice},
  author = {Ginsparg, Paul H. and Wilson, Kenneth G.},
  year = 1982,
  month = may,
  volume = {25},
  number = {10},
  pages = {2649--2657},
  publisher = {American Physical Society},
  doi = {10.1103/PhysRevD.25.2649},
  url = {https://link.aps.org/doi/10.1103/PhysRevD.25.2649},
  journal = {Phys. Rev. D}
}

@article{neubergerExactly1998,
  title = {Exactly Massless Quarks on the Lattice},
  author = {Neuberger, Herbert},
  year = 1998,
  month = jan,
  journal = {Physics Letters B},
  volume = {417},
  number = {1},
  pages = {141--144},
  issn = {0370-2693},
  doi = {10.1016/S0370-2693(97)01368-3},
  url = {https://www.sciencedirect.com/science/article/pii/S0370269397013683}
}

@article{neubergerVectorlike1998,
  title = {Vectorlike Gauge Theories with Almost Massless Fermions on the Lattice},
  author = {Neuberger, Herbert},
  year = 1998,
  month = may,
  volume = {57},
  number = {9},
  pages = {5417--5433},
  publisher = {American Physical Society},
  doi = {10.1103/PhysRevD.57.5417},
  url = {https://link.aps.org/doi/10.1103/PhysRevD.57.5417},
  journal = {Phys. Rev. D}
}

@article{luscherExact1998,
  title = {Exact Chiral Symmetry on the Lattice and the {{Ginsparg-Wilson}} Relation},
  author = {L{\"u}scher, Martin},
  year = 1998,
  month = jun,
  journal = {Phys. Lett. B},
  volume = {428},
  number = {3},
  pages = {342--345},
  issn = {0370-2693},
  doi = {10.1016/S0370-2693(98)00423-7},
  url = {https://www.sciencedirect.com/science/article/pii/S0370269398004237}
}

@article{luscherAbelian1999a,
  title = {Abelian Chiral Gauge Theories on the Lattice with Exact Gauge Invariance},
  author = {L{\"u}scher, Martin},
  year = 1999,
  month = may,
  journal = {Nucl. Phys. B},
  volume = {549},
  number = {1},
  pages = {295--334},
  issn = {0550-3213},
  doi = {10.1016/S0550-3213(99)00115-7},
  url = {https://www.sciencedirect.com/science/article/pii/S0550321399001157}
}

@article{callanAnomalies1985,
  title = {Anomalies and Fermion Zero Modes on Strings and Domain Walls},
  author = {Callan, C. G. and Harvey, J. A.},
  year = 1985,
  month = jan,
  journal = {Nucl. Phys. B},
  volume = {250},
  number = {1},
  pages = {427--436},
  issn = {0550-3213},
  doi = {10.1016/0550-3213(85)90489-4},
  url = {https://www.sciencedirect.com/science/article/pii/0550321385904894}
}

@article{wittenFermion2016,
  title = {Fermion Path Integrals and Topological Phases},
  author = {Witten, Edward},
  year = 2016,
  month = jul,
  volume = {88},
  number = {3},
  pages = {035001},
  issn = {0034-6861, 1539-0756},
  doi = {10.1103/RevModPhys.88.035001},
  url = {https://link.aps.org/doi/10.1103/RevModPhys.88.035001},
  copyright = {http://link.aps.org/licenses/aps-default-license},
  journal = {Rev. Mod. Phys.}
}

@article{fidkowskinogo2023,
  title = {A No-Go Result for Implementing Chiral Symmetries by Locality-Preserving Unitaries in a 3 Dimensional {{Hamiltonian}} Lattice Model of Fermions},
  author = {Fidkowski, Lukasz and Xu, Cenke},
  year = 2023,
  month = nov,
  volume = {131},
  number = {19},
  pages = {196601},
  issn = {0031-9007, 1079-7114},
  doi = {10.1103/PhysRevLett.131.196601},
  url = {http://arxiv.org/abs/2306.10105},
  archiveprefix = {arXiv},
  journal = {Phys. Rev. Lett.}
}

@article{sulejmanpasicAbelian2019,
  title = {Abelian Gauge Theories on the Lattice: $\theta$-Terms and Compact Gauge Theory with(out) Monopoles},
  shorttitle = {Abelian Gauge Theories on the Lattice},
  author = {Sulejmanpasic, Tin and Gattringer, Christof},
  year = 2019,
  month = jun,
  journal = {Nuclear Physics B},
  volume = {943},
  pages = {114616},
  issn = {0550-3213},
  doi = {10.1016/j.nuclphysb.2019.114616},
  url = {https://www.sciencedirect.com/science/article/pii/S0550321319300963}
}

@article{fazzaLattice2023,
  title = {Lattice Quantum {{Villain Hamiltonians}}: Compact Scalars, {{U}}(1) Gauge Theories, Fracton Models and Quantum {{Ising}} Model Dualities},
  shorttitle = {Lattice Quantum {{Villain Hamiltonians}}},
  author = {Fazza, Lucca and Sulejmanpasic, Tin},
  year = 2023,
  month = may,
  volume = {2023},
  number = {5},
  pages = {17},
  issn = {1029-8479},
  doi = {10.1007/JHEP05(2023)017},
  url = {https://doi.org/10.1007/JHEP05(2023)017},
  journal = {J. High Energ. Phys.}
}

@article{gorantlaModified2021,
  title = {A Modified {V}illain Formulation of Fractons and Other Exotic Theories},
  author = {Gorantla, Pranay and Lam, Ho Tat and Seiberg, Nathan and Shao, Shu-Heng},
  year = 2021,
  month = oct,
  volume = {62},
  number = {10},
  pages = {102301},
  issn = {0022-2488, 1089-7658},
  doi = {10.1063/5.0060808},
  url = {http://arxiv.org/abs/2103.01257},
  archiveprefix = {arXiv},
  journal = {J. Math. Phys.}
}

@article{fidkowskiNoninvertible2025a, 
  title = {Noninvertible Bosonic Chiral Symmetry on the Lattice},
  author = {Fidkowski, Lukasz and Xu, Cenke and Zhang, Carolyn},
  journal = {Phys. Rev. X},
  volume = {16},
  issue = {4},
  pages = {041036},
  numpages = {21},
  year = {2026},
  month = {Nov},
  publisher = {American Physical Society},
  doi = {10.1103/hl1d-99hv},
  url = {https://link.aps.org/doi/10.1103/hl1d-99hv}
}

@article{luLattice2026,
  title = {Lattice Chiral Symmetry from Bosons in 3+1d},
  author = {Lu, Zhiyao and Seifnashri, Sahand and Shao, Shu-Heng},
  year = 2026,
  month = apr,
  publisher = {arXiv},
  url = {http://arxiv.org/abs/2604.06307},
  journal = {arXiv:2604.06307}
}

@article{seifnashriExactly2026,
  title = {Exactly Solvable 1+1d Chiral Lattice Gauge Theories},
  author = {Seifnashri, Sahand},
  year = 2026,
  month = jan,
  publisher = {arXiv},
  url = {http://arxiv.org/abs/2601.14359},
  journal = {arXiv:2601.14359}
}

@article{thorngrenChiral2026a,
  title = {Chiral Lattice Gauge Theories from Symmetry Disentanglers},
  author = {Thorngren, Ryan and Preskill, John and Fidkowski, Lukasz},
  year = 2026,
  month = jul,
  volume = {2026},
  number = {7},
  pages = {271},
  issn = {1029-8479},
  doi = {10.1007/JHEP07(2026)271},
  url = {https://doi.org/10.1007/JHEP07(2026)271},
  journal = {J. High Energ. Phys.}
}

@article{misumiTastesplitting2026,
  title = {Taste-Splitting Mass and Edge Modes in $3{+}1$ {D} Staggered Fermions},
  author = {Misumi, Tatsuhiro and Onogi, Tetsuya and Yamaoka, Tatsuya},
  year = 2026,
  month = aug,
  publisher = {arXiv},
  url = {http://arxiv.org/abs/2604.02078},
  journal = {arXiv:2604.02078}
}

@article{drellStrongcoupling1976,
  title = {Strong-Coupling Field Theory. {{I}}. {{Variational}} Approach to $ \varphi^4$ Theory},
  author = {Drell, Sidney D. and Weinstein, Marvin and Yankielowicz, Shimon},
  year = 1976,
  month = jul,
  volume = {14},
  number = {2},
  pages = {487--516},
  publisher = {American Physical Society},
  doi = {10.1103/PhysRevD.14.487},
  url = {https://link.aps.org/doi/10.1103/PhysRevD.14.487},
  journal = {Phys. Rev. D}
}

@article{drellStrongcoupling1976a,
  title = {Strong-Coupling Field Theories. {{II}}. {{Fermions}} and Gauge Fields on a Lattice},
  author = {Drell, Sidney D. and Weinstein, Marvin and Yankielowicz, Shimon},
  year = 1976,
  month = sep,
  volume = {14},
  number = {6},
  pages = {1627--1647},
  publisher = {American Physical Society},
  doi = {10.1103/PhysRevD.14.1627},
  url = {https://link.aps.org/doi/10.1103/PhysRevD.14.1627},
  journal = {Phys. Rev. D}
}

@article{Else2014,
  title = {Classifying symmetry-protected topological phases through the anomalous action of the symmetry on the edge},
  author = {Else, Dominic V. and Nayak, Chetan},
  journal = {Phys. Rev. B},
  volume = {90},
  issue = {23},
  pages = {235137},
  numpages = {19},
  year = {2014},
  month = {Dec},
  publisher = {American Physical Society},
  doi = {10.1103/PhysRevB.90.235137},
  url = {https://link.aps.org/doi/10.1103/PhysRevB.90.235137}
}

@article{friedanproof1982a,
  title = {A Proof of the {{Nielsen-Ninomiya}} Theorem},
  author = {Friedan, D.},
  year = 1982,
  month = dec,
  volume = {85},
  number = {4},
  pages = {481--490},
  issn = {1432-0916},
  doi = {10.1007/BF01403500},
  url = {https://doi.org/10.1007/BF01403500},
  journal = {Comm. Math. Phys.}
}

@article{craiggaugeinvariant2026,
  title = {A Gauge-Invariant Measure for Lattice Chiral Fermions},
  author = {Craig, Nathaniel},
  year = 2026,
  month = oct,
  publisher = {arXiv},
  url = {http://arxiv.org/abs/2610.09093},
  journal = {arXiv:2610.09093}
}

@article{dempsey2022discrete,
  author = {Dempsey, Ross and Klebanov, Igor R.
            and Pufu, Silviu S. and Zan, Bernardo},
  title = {Discrete Chiral Symmetry and Mass Shift in the
           Lattice {H}amiltonian Approach to the {S}chwinger Model},
  journal = {Phys. Rev. Res.},
  volume = {4},
  pages = {043133},
  year = {2022},
  doi = {10.1103/PhysRevResearch.4.043133}
}

@article{qi2006,
  title = {Topological quantization of the spin {H}all effect in two-dimensional paramagnetic semiconductors},
  author = {Qi, Xiao-Liang and Wu, Yong-Shi and Zhang, Shou-Cheng},
  journal = {Phys. Rev. B},
  volume = {74},
  issue = {8},
  pages = {085308},
  numpages = {7},
  year = {2006},
  month = {Aug},
  publisher = {American Physical Society},
  doi = {10.1103/PhysRevB.74.085308},
  url = {https://link.aps.org/doi/10.1103/PhysRevB.74.085308}
}

@article{cho2017anomaly,
  title = {Anomaly manifestation of {Lieb-Schultz-Mattis} theorem and topological phases},
  author = {Cho, Gil Young and Hsieh, Chang-Tse and Ryu, Shinsei},
  journal = {Phys. Rev. B},
  volume = {96},
  issue = {19},
  pages = {195105},
  numpages = {16},
  year = {2017},
  month = {Nov},
  publisher = {American Physical Society},
  doi = {10.1103/PhysRevB.96.195105},
  url = {https://link.aps.org/doi/10.1103/PhysRevB.96.195105}
}

@article{cobanera2015equivalence,
  title = {Equivalence of topological insulators and superconductors},
  author = {Cobanera, Emilio and Ortiz, Gerardo},
  journal = {Phys. Rev. B},
  volume = {92},
  issue = {15},
  pages = {155125},
  numpages = {16},
  year = {2015},
  month = {Oct},
  publisher = {American Physical Society},
  doi = {10.1103/PhysRevB.92.155125},
  url = {https://link.aps.org/doi/10.1103/PhysRevB.92.155125}
}

@article{luFillingenforced2020,
  title = {Filling-Enforced Constraint on the Quantized {{Hall}} Conductivity on a Periodic Lattice},
  author = {Lu, Yuan-Ming and Ran, Ying and Oshikawa, Masaki},
  year = 2020,
  month = feb,
  volume = {413},
  pages = {168060},
  issn = {0003-4916},
  doi = {10.1016/j.aop.2019.168060},
  url = {https://www.sciencedirect.com/science/article/pii/S000349161930315X},
  journal = {Ann. Phys.}
}

@article{hsuStandard2026,
  title = {The {{Standard Model}} on a Finite Bosonic Lattice with Continuum {{Weyl}} Fermions},
  author = {Hsu, Stephen D. H.},
  year = 2026,
  month = oct,
  publisher = {arXiv},
  url = {http://arxiv.org/abs/2610.29019},
  journal = {arXiv:2610.29019}
}

@article{wittenNote2026,
  title = {A Note On Lattice Definition of Gauge Theories With Chiral Fermions},
  author = {Witten, Edward},
  year = 2026,
  month = nov,
  publisher = {arXiv},
  url = {http://arxiv.org/abs/2611.05051},
  journal = {arXiv:2611.05051}
}

@article{schulzShastry1998,
  author  = {Schulz, H. J. and Shastry, B. S.},
  title   = {A New Class of Exactly Solvable Interacting
             Fermion Models in One Dimension},
  journal = {Phys. Rev. Lett.},
  volume  = {80},
  pages   = {1924--1927},
  year    = {1998},
  doi     = {10.1103/PhysRevLett.80.1924}
}

@article{Dana1985,
  title = {Quantised {H}all conductance in a perfect crystal},
  author = {Dana, Itzhack and Avron, Yosi and Zak, J.},
  journal = {J. Phys. C: Solid State Phys.},
  volume = {18},
  number = {22},
  pages = {L679--L683},
  year = {1985},
  doi = {10.1088/0022-3719/18/22/004},
}

@article{avronCharge1994,
  title = {Charge deficiency, charge transport and comparison of dimensions},
  author = {Avron, Joseph E. and Seiler, Ruedi and Simon, Barry},
  journal = {Commun. Math. Phys.},
  volume = {159},
  pages = {399--422},
  year = {1994},
  doi = {10.1007/BF02102644},
}

@article{banks1976strong,
  title = {Strong-coupling calculations of lattice gauge theories: (1+1)-dimensional exercises},
  author = {Banks, T. and Susskind, Leonard and Kogut, John},
  journal = {Phys. Rev. D},
  volume = {13},
  pages = {1043--1053},
  year = {1976},
  doi = {10.1103/PhysRevD.13.1043},
}

@article{carroll1976lattice,
  title = {Lattice gauge theory calculations in 1+1 dimensions and the approach to the continuum limit},
  author = {Carroll, Allen and Kogut, John and Sinclair, D. K. and Susskind, Leonard},
  journal = {Phys. Rev. D},
  volume = {13},
  pages = {2270--2277},
  year = {1976},
  doi = {10.1103/PhysRevD.13.2270},
}

@article{chenImpossibility2014,
  title = {The impossibility of exactly flat non-trivial {C}hern bands in strictly local periodic tight binding models},
  author = {Chen, Li and Mazaheri, Tahereh and Seidel, Alexander and Tang, Xiang},
  journal = {J. Phys. A: Math. Theor.},
  volume = {47},
  pages = {152001},
  year = {2014},
  doi = {10.1088/1751-8113/47/15/152001},
}

@article{deNittisSchulzBaldesFlux2016,
  title = {Spectral flows associated to flux tubes},
  author = {de Nittis, Giuseppe and Schulz-Baldes, Hermann},
  journal = {Ann. Henri Poincar{\'e}},
  volume = {17},
  pages = {1--35},
  year = {2016},
  doi = {10.1007/s00023-014-0394-5},
}

@article{delplace2011,
  title = {Zak phase and the existence of edge states in graphene},
  author = {Delplace, P. and Ullmo, D. and Montambaux, G.},
  journal = {Phys. Rev. B},
  volume = {84},
  pages = {195452},
  year = {2011},
  doi = {10.1103/PhysRevB.84.195452},
}

@article{fidkowski2019interacting,
  title = {Interacting invariants for {F}loquet phases of fermions in two dimensions},
  author = {Fidkowski, Lukasz and Po, Hoi Chun and Potter, Andrew C. and Vishwanath, Ashvin},
  journal = {Phys. Rev. B},
  volume = {99},
  pages = {085115},
  year = {2019},
  doi = {10.1103/PhysRevB.99.085115},
}

@article{fidkowskiEntanglement2010,
  title = {Entanglement Spectrum of Topological Insulators and Superconductors},
  author = {Fidkowski, Lukasz},
  journal = {Phys. Rev. Lett.},
  volume = {104},
  pages = {130502},
  year = {2010},
  doi = {10.1103/PhysRevLett.104.130502},
}

@article{frasertaliente2025nonlocal,
  title = {A nonlocal {S}chwinger model},
  author = {Fraser-Taliente, Ludo and Herzog, Christopher P. and Shrestha, Abhay},
  journal = {J. High Energ. Phys.},
  volume = {2025},
  number = {6},
  pages = {252},
  year = {2025},
  doi = {10.1007/JHEP06(2025)252},
}

@article{fuKane2007inversion,
  title = {Topological insulators with inversion symmetry},
  author = {Fu, Liang and Kane, C. L.},
  journal = {Phys. Rev. B},
  volume = {76},
  pages = {045302},
  year = {2007},
  doi = {10.1103/PhysRevB.76.045302},
}

@article{fujita1996,
  title = {Peculiar Localized State at Zigzag Graphite Edge},
  author = {Fujita, Mitsutaka and Wakabayashi, Katsunori and Nakada, Kyoko and Kusakabe, Koichi},
  journal = {J. Phys. Soc. Jpn.},
  volume = {65},
  pages = {1920--1923},
  year = {1996},
  doi = {10.1143/JPSJ.65.1920},
}

@article{gioiaBurkovHughes2026,
  title = {No-go theorem for single time-reversal invariant symmetry-protected {D}irac fermions in {(3+1)D}},
  author = {Gioia, Lei and Burkov, Anton A. and Hughes, Taylor L.},
  journal = {Phys. Rev. B},
  volume = {114},
  pages = {245105},
  year = {2026},
  doi = {10.1103/g4h4-kb3c},
}

@article{hamer1982massive,
  title = {The massive {S}chwinger model on a lattice: Background field, chiral symmetry and the string tension},
  author = {Hamer, C. J. and Kogut, John and Crewther, D. P. and Mazzolini, M. M.},
  journal = {Nucl. Phys. B},
  volume = {208},
  pages = {413--438},
  year = {1982},
  doi = {10.1016/0550-3213(82)90229-2},
}

@article{kellendonkEdge2002,
  title = {Edge Current Channels and {C}hern Numbers in the Integer Quantum {H}all Effect},
  author = {Kellendonk, Johannes and Richter, Thomas and Schulz-Baldes, Hermann},
  journal = {Rev. Math. Phys.},
  volume = {14},
  number = {1},
  pages = {87--119},
  year = {2002},
  doi = {10.1142/S0129055X02001107},
}

@article{kitaevPeriodic2009,
  title = {Periodic table for topological insulators and superconductors},
  author = {Kitaev, Alexei},
  journal = {AIP Conf. Proc.},
  volume = {1134},
  pages = {22--30},
  year = {2009},
  doi = {10.1063/1.3149495}
}

@article{kobayashiInamuraShiozaki2026,
  title = {Remarks on invertible phases with non-onsite symmetry},
  author = {Kobayashi, Ryohei and Inamura, Kansei and Shiozaki, Ken},
  journal = {arXiv:2609.12042},
  year = {2026},
  url = {https://arxiv.org/abs/2609.12042},
}

@article{lu2017lsmSPT,
  title = {{Lieb-Schultz-Matti}s theorems for symmetry-protected topological phases},
  author = {Lu, Yuan-Ming},
  journal = {Ann. Phys.},
  volume = {470},
  pages = {169806},
  year = {2024},
  doi = {10.1016/j.aop.2024.169806} 
}

@article{nakada1996,
  title = {Edge state in graphene ribbons: Nanometer size effect and edge shape dependence},
  author = {Nakada, Kyoko and Fujita, Mitsutaka and Dresselhaus, Gene and Dresselhaus, Mildred S.},
  journal = {Phys. Rev. B},
  volume = {54},
  pages = {17954--17961},
  year = {1996},
  doi = {10.1103/PhysRevB.54.17954},
}

@article{niemiSemenoff1983,
  title = {Axial-Anomaly-Induced Fermion Fractionization and Effective Gauge-Theory Actions in Odd-Dimensional Space-Times},
  author = {Niemi, A. J. and Semenoff, G. W.},
  journal = {Phys. Rev. Lett.},
  volume = {51},
  pages = {2077--2080},
  year = {1983},
  doi = {10.1103/PhysRevLett.51.2077},
}

@article{readGreen2000,
  title = {Paired states of fermions in two dimensions with breaking of parity and time-reversal symmetries and the fractional quantum {H}all effect},
  author = {Read, N. and Green, Dmitry},
  journal = {Phys. Rev. B},
  volume = {61},
  pages = {10267--10297},
  year = {2000},
  doi = {10.1103/PhysRevB.61.10267},
}

@article{ryuHatsugai2002,
  title = {Topological Origin of Zero-Energy Edge States in Particle-Hole Symmetric Systems},
  author = {Ryu, Shinsei and Hatsugai, Yasuhiro},
  journal = {Phys. Rev. Lett.},
  volume = {89},
  pages = {077002},
  year = {2002},
  doi = {10.1103/PhysRevLett.89.077002},
}

@article{suSchriefferHeeger1979,
  title = {Solitons in Polyacetylene},
  author = {Su, W. P. and Schrieffer, J. R. and Heeger, A. J.},
  journal = {Phys. Rev. Lett.},
  volume = {42},
  pages = {1698--1701},
  year = {1979},
  doi = {10.1103/PhysRevLett.42.1698},
}

@article{thakurathiSenguptaSen2014,
  title = {Majorana edge modes in the {K}itaev model},
  author = {Thakurathi, Manisha and Sengupta, K. and Sen, Diptiman},
  journal = {Phys. Rev. B},
  volume = {89},
  pages = {235434},
  year = {2014},
  doi = {10.1103/PhysRevB.89.235434},
}

@article{witten1982su2,
  title = {An {SU(2)} anomaly},
  author = {Witten, Edward},
  journal = {Phys. Lett. B},
  volume = {117},
  number = {5},
  pages = {324--328},
  year = {1982},
  doi = {10.1016/0370-2693(82)90728-6},
}

@article{farrelly2020review,
  title   = {A review of Quantum Cellular Automata},
  author  = {Farrelly, Terry},
  journal = {Quantum},
  volume  = {4},
  pages   = {368},
  year    = {2020},
  doi     = {10.22331/q-2020-11-30-368}, 
  archivePrefix = {arXiv}
}

@article{schnyder2008classification,
  title   = {Classification of topological insulators and superconductors in three spatial dimensions},
  author  = {Schnyder, Andreas P. and Ryu, Shinsei and Furusaki, Akira and Ludwig, Andreas W. W.},
  journal = {Phys. Rev. B},
  volume  = {78},
  pages   = {195125},
  year    = {2008},
  doi     = {10.1103/PhysRevB.78.195125}
}

@article{ryu2010tenfold,
  title   = {Topological insulators and superconductors: tenfold way and dimensional hierarchy},
  author  = {Ryu, Shinsei and Schnyder, Andreas P. and Furusaki, Akira and Ludwig, Andreas W. W.},
  journal = {New J. Phys.},
  volume  = {12},
  pages   = {065010},
  year    = {2010},
  doi     = {10.1088/1367-2630/12/6/065010}
}

@article{chiu2016classification,
  title   = {Classification of topological quantum matter with symmetries},
  author  = {Chiu, Ching-Kai and Teo, Jeffrey C. Y. and Schnyder, Andreas P. and Ryu, Shinsei},
  journal = {Rev. Mod. Phys.},
  volume  = {88},
  pages   = {035005},
  year    = {2016},
  doi     = {10.1103/RevModPhys.88.035005}
}

@article{wangYou2022SMG,
  title   = {Symmetric Mass Generation},
  author  = {Wang, Juven and You, Yi-Zhuang},
  journal = {Symmetry},
  volume  = {14},
  number  = {7},
  pages   = {1475},
  year    = {2022},
  doi     = {10.3390/sym14071475} 
}

@article{seifnashriShirley2026,
  title   = {Disentangling Anomaly-Free Symmetries of Quantum Spin Chains},
  author  = {Seifnashri, Sahand and Shirley, Wilbur},
  journal = {Phys. Rev. Lett.},
  volume  = {136},
  pages   = {216603},
  year    = {2026},
  doi     = {10.1103/bscj-r5tg}
}

@article{kapustinHigher2025,
  title  = {Higher Symmetries and Anomalies in Quantum Lattice Systems},
  author = {Kapustin, Anton and Xu, Shixiong},
  year   = {2025},
  journal   = {arXiv:2505.04719}, 
  url    = {https://arxiv.org/abs/2505.04719}
}

@article{kawagoeShirley2025,
  title  = {Anomaly Diagnosis via Symmetry Restriction in Two-Dimensional Lattice Systems},
  author = {Kawagoe, Kyle and Shirley, Wilbur},
  year   = {2025},
  journal  = {arXiv:2507.07430}, 
  url    = {https://arxiv.org/abs/2507.07430}
}

@article{tuLongElse2026,
  title   = {Anomalies of Global Symmetries on the Lattice},
  author  = {Tu, Yi-Ting and Long, David M. and Else, Dominic V.},
  journal = {Phys. Rev. X},
  volume  = {16},
  pages   = {011027},
  year    = {2026},
  doi     = {10.1103/m188-w1ct}
}

@article{shirleyZhangJiLevin2026,
  title   = {Anomaly-Free Symmetries with Obstructions to Gauging and Onsiteability},
  author  = {Shirley, Wilbur and Zhang, Carolyn and Ji, Wenjie and Levin, Michael},
  journal = {Phys. Rev. Lett.},
  volume  = {136},
  pages   = {216602},
  year    = {2026},
  doi     = {10.1103/zrs1-j2xd}
}

@article{fidkowskiJacksonKlich2011,
  author  = {Fidkowski, Lukasz and Jackson, T. S. and Klich, Israel},
  title   = {Model Characterization of Gapless Edge Modes
             of Topological Insulators Using Intermediate
             {Brillouin-Zone} Functions},
  journal = {Phys. Rev. Lett.},
  volume  = {107},
  pages   = {036601},
  year    = {2011},
  doi     = {10.1103/PhysRevLett.107.036601}
}

@article{nachtergaele2019quasilocality,
  author  = {Nachtergaele, Bruno and Sims, Robert
             and Young, Amanda},
  title   = {Quasi-locality bounds for quantum lattice
             systems. {I. Lieb-Robinson} bounds,
             quasi-local maps, and spectral flow
             automorphisms},
  journal = {J. Math. Phys.},
  volume  = {60},
  number  = {6},
  pages   = {061101},
  year    = {2019},
  doi     = {10.1063/1.5095769}
}

@article{luscherLattice2000,
  title = {Lattice Regularization of Chiral Gauge Theories to All Orders of Perturbation Theory},
  author = {L{\"u}scher, Martin},
  year = 2000,
  month = jul,
  volume = {2000},
  number = {06},
  pages = {028},
  issn = {1126-6708},
  doi = {10.1088/1126-6708/2000/06/028},
  url = {https://doi.org/10.1088/1126-6708/2000/06/028},
  journal = {J. High Energ. Phys.}
}

@article{denittisClassification2015,
  title = {Classification of ``Quaternionic" {B}loch-Bundles},
  author = {De Nittis, Giuseppe and Gomi, Kiyonori},
  year = 2015,
  month = oct,
  volume = {339},
  number = {1},
  pages = {1--55},
  issn = {1432-0916},
  doi = {10.1007/s00220-015-2390-0},
  url = {https://doi.org/10.1007/s00220-015-2390-0},
  journal = {Comm. Math. Phys.}
}

@article{fukuiBulkEdge2012,
  title = {Bulk-Edge Correspondence for {C}hern Topological Phases: {A} Viewpoint from a Generalized Index Theorem}, 
  author = {Fukui, Takahiro and Shiozaki, Ken and Fujiwara, Takanori and Fujimoto, Satoshi},
  year = 2012,
  month = nov,
  journal = {J. Phys. Soc. Jpn.},
  volume = {81},
  number = {11},
  pages = {114602},
  publisher = {The Physical Society of Japan},
  issn = {0031-9015},
  doi = {10.1143/JPSJ.81.114602},
  url = {https://journals.jps.jp/doi/10.1143/JPSJ.81.114602}
}

@article{qi2008topological,
  title = {Topological field theory of time-reversal invariant insulators},
  author = {Qi, Xiao-Liang and Hughes, Taylor L. and Zhang, Shou-Cheng},
  journal = {Phys. Rev. B},
  volume = {78},
  issue = {19},
  pages = {195424},
  numpages = {43},
  year = {2008},
  month = {Nov},
  publisher = {American Physical Society},
  doi = {10.1103/PhysRevB.78.195424},
  url = {https://link.aps.org/doi/10.1103/PhysRevB.78.195424}
}

@article{stredaTheory1982,
  title = {Theory of Quantised {{Hall}} Conductivity in Two Dimensions},
  author = {Streda, P.},
  year = 1982,
  month = aug,
  journal = {J. Phys. C: Solid State Phys.},
  volume = {15},
  number = {22},
  pages = {L717},
  issn = {0022-3719},
  doi = {10.1088/0022-3719/15/22/005},
  url = {https://doi.org/10.1088/0022-3719/15/22/005}
}

@article{fu2007topological,
  title = {Topological Insulators in Three Dimensions},
  author = {Fu, Liang and Kane, C. L. and Mele, E. J.},
  journal = {Phys. Rev. Lett.},
  volume = {98},
  issue = {10},
  pages = {106803},
  numpages = {4},
  year = {2007},
  month = {Mar},
  publisher = {American Physical Society},
  doi = {10.1103/PhysRevLett.98.106803},
  url = {https://link.aps.org/doi/10.1103/PhysRevLett.98.106803}
}

@article{gross2012index,
  author  = {Gross, D. and Nesme, V. and Vogts, H. and Werner, R. F.},
  title   = {Index Theory of One Dimensional Quantum Walks and Cellular Automata},
  journal = {Commun. Math. Phys.},
  volume  = {310},
  pages   = {419--454},
  year    = {2012},
  doi     = {10.1007/s00220-012-1423-1}
}

@article{kourkoulouOnsager2026,
  title = {{O}nsager Algebras and Lattice Fermions},
  author = {Kourkoulou, Ioanna and Tong, David},
  year = 2026,
  month = nov,
  publisher = {arXiv},
  url = {http://arxiv.org/abs/2610.31653},
  journal = {arXiv:2610.31653}
}

\end{document}